# Complex Dynamics of Real Quantum, Classical and Hybrid Micro-Machines

A.P. KIRILYUK[*]

Institute of Metal Physics, Kiev, Ukraine

**ABSTRACT.** Any real interaction process produces many equally possible, but mutually incompatible system versions, or realisations, giving rise to the omnipresent, purely dynamic randomness (chaoticity) and universally defined complexity (sections 3-4). Since quantum behaviour dynamically emerges as the lowest level of unreduced world complexity (sections 4.6-7, 5.3), quantum interaction randomness can only be relatively strong (explicit), which reveals the causal origin of quantum indeterminacy (sections 4.6.1, 5.3(A)) and true quantum chaos (sections 4.6.2, 5.2.1, 6), but rigorously excludes the possibility of unitary quantum computation, even in an 'ideal', noiseless system (sections 5-7). Any real computation is an internally chaotic (multivalued) process of system complexity development occurring in different regimes at various complexity levels (sections 7.1-2). Unitary quantum machines, including their postulated 'magic', cannot be realised as such because their dynamically single-valued scheme is incompatible with the irreducibly high dynamic randomness at quantum complexity levels (sections 4.5-7, 5.1, 5.2.2, 7.1-2) and should be replaced by the explicitly chaotic, intrinsically creative machines already realised in living organisms and providing their quite different, realistic kind of magic. The related concepts of reality-based, complex-dynamical nanotechnology, biotechnology and intelligence are outlined, together with the ensuing change in research strategy and content (sections 7.3, 8). The unreduced, dynamically multivalued solution of the quantum (and classical) many-body problem reveals the true, complex-dynamical basis of solid-state dynamics, including the origin and internal dynamics of macroscopic quantum states (section 5.3(C)). The critical, 'end-of-science' state of conventional, unitary knowledge and the way to positive change are causally specified within the same, universal concept of complexity (section 9).

---

[*] Address for correspondence: Institute of Metal Physics, Solid State Theory Department, 36 Vernadsky Av., Kyiv 03142, Ukraine. E-mail address: Andrei.Kirilyuk@Gmail.com.

# Complex Dynamics of Real Quantum, Classical and Hybrid Micro-Machines

## From Causally Complete Quantum Mechanics to the Efficient Nanotechnology and Development Concept

**Back-cover book description.** The lasting stagnation of fundamental science becomes an increasingly urgent problem, especially by contrast to apparent success of its previous discoveries. In particular, the unsolved problems of quantum mechanics and particle theory strangely correlate with the stable absence of key advance in quantum computation and full-scale nanotechnology, despite the high promises and huge efforts applied. In this book the decisive extension of conventional science basis is introduced in the form of unreduced solution to the arbitrary interaction problem. It is shown how the qualitatively new, "dynamically multivalued" structure of the unreduced solution and real system behaviour leads to the intrinsically universal and problem-solving concept of dynamic complexity and chaos. It is applied here to problems of full-scale quantum computation and nanomachines, where the impossibility of unitary quantum machines is demonstrated together with the high potentialities of their extended, complex-dynamical version, already realised in natural living systems. The book is a mixture of rigorous basis, popular explanation and new paradigm discussion oriented to a wide range of educated readers.

**CONTENTS**

---

*Il me paraît plus naturel et plus conforme aux idées qui ont toujours heureusement orienté la recherche scientifique de supposer que les transitions quantiques pourront un jour être interprétées, peut-être à l'aide de moyens analytiques dont nous ne disposons pas encore, comme des processus très rapides, mais en principe descriptibles en termes d'espace et de temps, analogues à ces passages brusques d'un cycle limite à un autre que l'on rencontre très fréquemment dans l'étude des phénomènes mécaniques et électromagnétiques non linéaires.*

Louis de Broglie, *Les idées qui me guident dans mes recherches* (1965) [381]

---

# ABBREVIATIONS
used in the text

EP *for* Effective Potential (introduced in Section 3.2)

SOC *for* Self-Organised Criticality (introduced in Section 4.5.1)

# 1. Introduction

The end of a big enough stage of evolution of any complex system is characterised by critically sharp instabilities marking the 'last breath' of the disappearing regime and possible beginning of transition to a qualitatively new level of development. This universal expression of the 'generalised phase transition' [1] in the complex system dynamics is readily observed for the whole variety of real world phenomena, from a brightly blinking and then forever fading light source (a lamp or a star) to the modern 'sudden' revival of activity in the fundamental 'new physics' as if repeating the excitement of its birth period a hundred years ago (see e. g. [2-4] and references therein). Concentrating especially around 'quantum' phenomena (though they are often arbitrary extended to any weird-looking, 'mysterious' things and ideas), this latest 'advance' of 'quantum computation', 'quantum teleportation', 'entanglement' and other announced 'quantum miracles' produces much promise of 'fantastic', *extraordinary* new possibilities, but uses only old, always *unsolved* quantum 'mysteries' and 'paradoxes', *without* any *qualitatively new* concept and *deeper* (more consistent) understanding of the *unreduced* physical reality behind them, and therefore provides, alas, just another characteristic example of the clearly visible "end" [5] of *conventional* fundamental science, involving the "ironic", explicitly speculative and imitative 'play of words' of its stylish 'innovations' without novelty (conventional 'science of complexity' provides another relevant example) [1,6,7]. In this respect today's 'quantum magicians' resemble too closely any ordinary swift-handed jugglers (the latter being in average much more honest), since both pretend to obtain something 'very interesting' from virtually nothing, without any realistic, *causal creation* of the promised new properties and new understanding.

At the same time the *causally complete*, i. e. free from any physical 'mysteries' and mathematical ambiguities, totally consistent and realistic understanding of micro-world phenomena and objects becomes indeed the more and more indispensable with growing *real* advance of *practical (micro-) technology* that *cannot* rely any more on its conventional, *purely empirical* way of development. In this work, initially written in 2002, we propose the unreduced analysis of real micro-object behaviour, which involves both the *qualitatively new*, rigorously derived, reality-based concept of (universal) *dynamic complexity* of *any* unreduced interaction process and clear explanation of the fundamental, *unavoidable* failure of conventional science approaches to theoretical description of such artificial or natural systems as 'quantum computers' or any other micro-devices and 'nanomachines'. To avoid confusion, note that we shall use the prefix 'micro-' as a generalised designation of all 'sufficiently small' structures where quantum effects or other explicit manifestations of unreduced dynamic complexity become important, so that our 'micro-machines', for example, include both 'nanomachines' and their microscale assemblies, etc.

We show that any elementary interaction process leading to a non-trivial change of real system state and indispensable for its useful operation is characterised by the *purely dynamic* uncertainty emerging in the form of *dynamically multivalued, or redundant*, interaction result [1-4,8-13] even in a totally closed micro- or macro-system protected from any 'environment' and related speculative 'decoherence'. The difference between micro- and macro-system scale is that the *relative* magnitude of this irreducible dynamic uncertainty tends to be larger for smaller systems and becomes indeed irreducibly large for the ultimately small, essentially quantum systems (which provides the causal explanation for the famous 'quantum indeterminacy' [1-4,10-13] and reveals the source of *true quantum chaos* [1,8,9]).

It follows, correspondingly, that the *dynamically single-valued*, or *unitary*, always essentially *perturbative* analysis of conventional quantum (and classical) theory, including all its 'post-modern' modifications and 'interpretations', leads to basically *wrong* predictions that cannot describe real system operation in principle, irrespective of evoked non-universal, easily adjustable factors, such as environmental influences or control/correction procedures. Since the *unreduced dynamic complexity* of real system behaviour is consistently determined by the total number of rigorously derived, *incompatible* versions, or *realisations*, of emerging system configuration in *real*, physical space and equals to zero for the unrealistic case of only one system realisation exclusively considered by conventional science, the whole approach and 'paradigm' of the latter are basically limited to that artificially over-simplified, perturbative reduction of complex-dynamical reality down to its effectively *zero-dimensional (point-like) projection* characterised by the *zero* value of genuine, unreduced complexity [1] (the latter should be distinguished from its abstract, postulated imita-tions by 'quantum/computational complexity', information, entropy, etc., all of them referring to the same unitary projection of reality).

Whereas the obtained conclusions are universally applicable to any real system (interaction process), the case of microscopic, 'quantum' systems corresponds to the lowest sublevels of world complexity, where any 'control of chaos', always using lower complexity levels to produce the desired change of higher-complexity dynamics in the direction of *external* regularity (pseudo-unitarity), cannot be efficient in principle [1,3,4]. The main illusion behind the idea of unitary quantum devices comes from the apparent unitarity of standard quantum mechanics, but this attitude does not want to take into account the fact that the nonunitary, complex-dynamical (multivalued) character of actual quantum dynamics is trickily hidden in the 'inexplicable' postulates of the standard theory, which explains their unique 'weirdness' and provides totally realistic solution of the canonical 'quantum mysteries'. The unreduced analysis of underlying interaction processes in the universal science of complexity [1-4,9-13] gives rise to the causally complete, essentially nonunitary and nonlinear extension of quantum mechanics free from any para-scientific 'mysteries' and therefore suitable for efficient practical applications. On the contrary, it is difficult to expect that *any* theory so much relying on the 'mystique' as conventional quantum mechanics can ever be useful in practical applications, where all those 'inexplicable' details do matter and contribute directly to the main expected results through the system dynamics.

Interaction between microscopic entities with essentially quantum behaviour refers basically to the field of *quantum chaos*. The scholar quantum chaos theory, being an integral part of the conventional, unitary science paradigm, falls totally within its dynamically single-valued approach and therefore cannot find its own main subject, the true randomness of purely dynamic origin, being obliged to replace it with various, always inconsistent imitations, such as *postulated*, abstract 'signs' of *statistically averaged* randomness as if appearing 'indirectly', not in real system behaviour (that remains perfectly regular within its unitary imitation), but rather as 'energy level ergodicity', 'phase-space intermittency', 'random matrix' properties, etc. In this respect the conventional quantum chaos theory only emphasises the basic limitations of the equally single-valued, and thus unitary, theory of classical chaos and other related branches of the scholar

'science of complexity' (like various 'turbulence scenarios' in abstract spaces or postulated geometrical constructions of the 'catastrophe theory'). Similar to the particular case of 'chaos control', all those axiomatic imitations of the true origin of randomness may *seem* indeed to be more consistent in the world of macroscopic, classical phenomena, since at the corresponding, higher complexity levels one has much more possibilities to 'approximately' maintain the desired property, be it either regularity or randomness, with the help of lower, more fundamental levels of complex dynamics (which are absent for the case of lowest-level, or 'quantum', systems). Thus, randomness is 'obtained' in conventional description of classical chaos as a result of its postulated 'exponential' amplification from a 'small' form, 'discreetly' borrowed from the outside of the system ('random initial conditions', 'influence of the environment', etc.). The resulting inconsistency becomes explicitly evident at the *lowest*, i. e. 'quantum' levels of complexity, where it cannot be hidden any more behind 'amplified small influence' effects (although such attempts exist), since *any* real change is relatively big (and only eventually unpredictable) in essentially quantum dynamics, already according to the standard quantum postulates, causally explained now by the unreduced complex dynamics of the underlying interaction process [1-4,8-13].

The basic absence of the 'ultimate', purely dynamic origin of randomness in any dynamically single-valued approach of conventional science, whether referring or not to an imitation of 'complexity', can correlate only positively with the possibility and properties of unitary quantum computation, irrespective of any plays of words around 'quantum ergodicity' and 'energy level mixing' that have been advanced within the unitary quantum chaos description as explanation of 'possible' inefficiency and problems of quantum computers [14]. By contrast, the dynamic multivaluedness concept of the universal science of complexity proves the impossibility of large-scale unitary quantum computation exactly in the same way in which it provides the purely dynamic origin of true randomness at the level of interacting quantum objects [1,8,9] (the 'genuine' quantum chaos), or any other level of world dynamics [1-4,10-13]. Whereas unitary description of quantum devices vainly tries to simulate creation of new entities in any unreduced interaction process, the unitary quantum chaos theory fails to explain its genuine dynamical randomness, which is indeed inherent to any real entity emergence in the interaction development process. Therefore, both these 'branches' of the same, basically limited unitary projection of reality lead to similar, qualitatively wrong results, as we shall show in detail below (see also [1,3,4,13]). Dealing with the perspectives of real micro-device development, it is especially important to avoid that technically sophisticated, but basically trivial, effectively one-dimensional hierarchy of unitary imitations of the essentially nonunitary reality.

In return, the unreduced, dynamically multivalued description and related causally complete understanding of micro-system interaction processes provide unlimited possibilities for their practical creation and control, in both 'physical' and 'biological' applications, as well as the inseparable entanglement of the two, just inherent in the unreduced concept of dynamic complexity [1]. The dynamically multivalued, *irreducibly chaotic* behaviour of real micro-objects, considered as a basic 'obstacle' within the unitary approach and way of thinking in general, in reality opens the way to the new kind of machinery and technology with qualitatively extended possibilities met until now only within natural living structures (such as intrinsic, interactive adaptability, the capacity for autonomous progressive development, etc.). Practical realisation of these possibilities necessitates, however, a decisive transition from the dynamically single-valued imitations of the canonical 'quantum mysteriology' to the causally complete understanding of real system dynamics within the dynamic redundance paradigm and related universal concept of dynamic complexity. In this book we shall briefly review, further develop and specify the details of this causally complete description of real micro-device dynamics within the dynamic redundance paradigm, as well as the perspectives of its practical application (see also [1,3,4,9,13]). These results clearly demonstrate, in particular, the big conceptual and practical difference between our truly dynamic, intrinsic origin of randomness inherent in any real system and the 'stochastic' kind of irregularity which is mechanistically, artificially added to a basically regular dynamics in the unitary science and gives rise to single-valued imitations of 'chaotic' or 'probabilistic' effects in micro-system behaviour. Those purely abstract imitations of dynamic randomness within the dynamically single-valued picture of conventional science may include technically and terminologically sophisticated constructions of the unitary 'science of complexity', such as 'unstable periodic orbits' or 'multistability', but those imitative plays of words do not change the essential, fundamental difference between the unreduced, dynamically multivalued interaction process and over-simplified, effectively zero-dimensional projection of its observed results to mechanistically fixed and disrupted abstract 'spaces' of conventional unitarity.

It is also important that the same transition to the universally nonperturbative method of dynamic redundance paradigm provides the causally complete version of entire quantum mechanics, including reality-based and rigorously derived solution of all its canonical 'mysteries', 'paradoxes' and 'contradictions', as well as intrinsic unification with the causally extended 'relativity' and 'unified field theory' [1-4,11-13] qualitatively exceeding the most optimistic expectations of unitary science. Indeed, it would be difficult to expect highly efficient applications from a kind of fundamental knowledge that suffers itself from a whole series of deepest contradictions and separations actually reducing it to a sort of 'magic', or 'para-science', as it is the case for the micro-world picture within conventional 'theoretical' and 'mathematical' physics, where one cannot present any consistent and realistic, at least general, description of the simplest real-world entities, the elementary particles and their 'intrinsic' properties, actually giving rise to all higher-level entities and properties of the world. For the evident reason mentioned above, the *illusive*, purely empirical 'control' and 'understanding' of the macroscopic levels of complexity performed by conventional, dynamically single-valued science, fail *explicitly* and totally at the lowest complexity levels, which were *empirically* revealed as the 'new physics' a century ago just in relation to those difficulties [2-4], but only now become fully, practically accessible to modern technology (the same is actually true for the highest levels of explicit complexity [1]). It shows why the crucial, qualitative extension to the universal science of complexity, being evidently necessary for the truly scientific, conscious understanding of the otherwise totally 'mysterious' micro-world behaviour, actually reveals itself as being *equally* indispensable to the causally complete, adequate description of real *macro*-world phenomena, where the explicit manifestations of the unreduced dynamic complexity also quickly come to the foreground of development of the *technically* powerful, but intellectually *blind* civilisation which, being *potentially conscious*, remains *practically unaware*, dark-minded and therefore inevitably self-destructive, within the ex-hausted zero-dimensional projection of the dominating, unitary level of knowledge and corresponding 'calculative' way of thinking (cf. the "shadows of the mind" image of conventional science proposed by a prominent 'mathematical' physicist [15] or a similar "veiled reality" concept [16]). An interesting, qualitative novelty of the emerging field of micro-devices, nanomachines, etc. is that it explicitly, directly unifies all those diverse levels and properties of the real world that could yet be artificially separated into their single-valued projections within 'usual', relatively 'simple' applications of conventional science, but now the intrinsically unified manifestations of the unreduced, multivalued real-world dynamics re-enter from its lowest levels and enforce the irreversible transition to the genuine dynamic complexity in the entire body of human knowledge.

# 2. Impossibility of unitary quantum computation: Qualitative considerations

The truly dynamical, causally consistent and omnipresent source of randomness at various levels, from the fundamental quantum indeterminacy [1-4,10-13], higher-sublevel quantum chaos and quantum measurement dynamics [1,3,4,8-10] to unpredictability of detailed conscious brain activity and all its results [1], is provided simply by the truly rigorous, universally non-perturbative analysis of underlying interaction processes, revealing the phenomenon of system splitting into many incompatible versions, or realisations, which are completely ignored and projected into a single, 'averaged' realisation by the invariably perturbative interaction reduction of conventional science. However, before considering the details of the unreduced analysis (Chapters 3-7) and its applications to micro-device dynamics (Chapters 5-8), it would be expedient to consider first more general, already qualitatively evident difficulties of the idea of *unitary*, dynamically single-valued quantum computers (and other micro-machines), revealing its basic contradictions even within the narrow framework of conventional quantum mechanics (remaining multiply confirmed by experiment), which clearly shows the necessity for some *qualitatively extended*, causally complete kind of description of real quantum system dynamics that can be efficiently used for practical purposes.

The theory of unitary quantum computation [17-89] has quickly grown during the last period into a major branch of quantum mechanics, including the first versions of the idea by P. Benioff [17] and R. Feynman [18]; its further intensive development in various directions and aspects (see e. g. reviews and textbooks [19,20,23,25,28,30,31,33-41]); first apparent experimental realisations (see [37,38] and references in other latest reviews) remaining, however, basically limited (see Chapters 5-7); 'standardisation' in scholar courses [34,36,40] and encyclopaedia articles [39]; practically unlimited generalisation to 'unconventional' quantum computing devices [42], 'linear-optic devices' [43], 'quantum-like systems' [44], 'computation using teleportation' [32] and even 'ground state quantum computation' [45]; extensions to other 'quantum machines', 'automata' and 'robots' (e. g. [26,27]); unitary 'quantum memory' design [47-49]; 'exciting', 'hot-issue' popularisation in scientific media [19,20,25,30,33,35];

pronounced tendency towards as if practically oriented, but always characteristically 'exotic' aspects of 'quantum information', such as 'quantum programming' [50] or 'quantum internet' [51,52]; related applications to 'quantum communication', 'quantum cryptography' and other 'quantum gambling' [20,28,30,33-37,39-41,53-57] (we shall not analyse these in detail considering them as a part of 'quantum computation'); extensions to '(quantum) complexity' within either its simulation by 'non-computability' [58,77,81] and 'Kolmogorov/computation complexity' [59-61] issues, or a general approach relating quantum computation to the entire universe dynamics [62,63], or expected involvement/use of hypothetical 'nonlinear quantum evolution' [64] and conventional (quantum and classical) chaos [65-67], or 'quantum games' referring to evolution [55,56] and market economy [57] (that necessarily acquire 'quantum' flavour), or 'quantum brain dynamics' [68-71] (all those 'complexity' manifestations described within the *unitary* dynamics framework); and finally, the 'deepest' group of connections to 'foundational' issues of quantum mechanics [72-76] (preserving, however, all its canonical 'mysteries' in their primal state) and the very meaning of 'logic'/'truth' and the laws of this world [77,78].

In general, the explosive proliferation of unitary 'quantum informatics' can be compared only to that of a world-wide epidemic, so that the remaining limited expressions of doubt [79-88] sound rather weakly (and more like reference to only 'practical' and thus, in principle, 'resolvable' problems), even despite occasional, but explicit acknowledgement of their validity by many 'quantum magicians'.[1] Thus, quantum information processing has entered as a major, highly 'advanced' and top-fashion discipline in the most prestigious universities, physical institutes and 'solid' professional journals, despite the evident, explicitly emphasised air of 'quantum mystification' and characteristic 'post-modern', para-scientific and senseless plays of words around 'quantum teleportation', 'quantum entanglement', 'Schrödinger-cat states', etc. that serve *not* to clarify, but to *preserve and amplify* the irreducible 'mystique' of the canonical quantum the-ory. It is now implied that 'quantum computers' can provide the incredible, *fantastic* increase of computational possibilities *just due to* the fact that they are based on 'inexplicable', 'quantum' *magic*, whereas any system with causally explainable, 'banal' dynamics can realise only 'ordinary', basically limited performance, even though it may be quantitatively sufficient for solution of some 'uninteresting', low level, 'computable' tasks. Needless to say, it is precisely the 'magic' part of 'rigorous' and 'exact' official science that obtains ever growing support and attention of the most educated and elitist parts of the 'developed' society, its governing 'decision makers' and related 'high priests of science' (see also Chapter 9).

---

[1] The true weakness of existing arguments both for and against quantum computers is in their dynamically single-valued, unitary basis, so that the opponents of quantum computation can rely at maximum only on the standard quantum postulates, which do provoke some doubts (see item (ii) below), but still cannot help to find the consistent problem solution because of the intrinsic, irreducible 'mystique' of unitary theory (Sections 4.6-7 and 5.3 provide a general description of the causally complete, dynamically multivalued extension of quantum mechanics involved with the new, positive solution of the quantum computation problem).

However, the extraordinary resistance of quantum mysteries to all the enormous efforts of their solution, including extremely elaborated computational and experimental tools of modern science, may leave a trace of unpleasant feeling that the whole fuss of unitary 'quantum magic' and its promised 'miraculous' applications can be but another *artificial* mystification, gigantically amplified by the modern unlimited power of money and media-made publicity, and in reality result simply from the subjective absence of 'normal', logically consistent knowledge about the true origin of 'quantum' phenomena. This 'unorthodox' attitude towards 'officially supported' manipulations around 'quantum mysteries' can be appreciated especially by those who preserve intrinsic attachment to the 'old-fashioned' principle that actually provided all the successes of the 'developed' civilisation: 'first understand, then use'. However, the essential, new feature of the last, 'post-developed' stage is that now the necessary quality of understanding objectively exceeds all the possibilities of the canonical, basically empirical, 'postulated' knowledge of conventional science, which should therefore give place to the truly consistent, totally first-principles and causally complete picture of reality within a superior kind of knowledge. Before describing the major lines of such qualitatively extended kind of description of elementary dynamical processes hidden behind the *externally* 'mysterious' behaviour of quantum systems and deriving the conclusions for operation of real 'quantum' devices, one should properly specify the rather evident doubts emerging already within the conventional, unitary theory of quantum systems. Note that the 'doubts' described below have quite fundamental, irreducible origin, despite their 'qualitative' expression and absence of a 'good' solution *within the unitary theory itself*. Therefore, leav-ing the detailed presentation of the complete solution to the next Sections, we include here some comments and general ideas hinting on the actual source of difficulties and way of their elimination.

(i) <u>Non-creative character of unitary dynamics *vs* real computation nature and demands</u>. Every real process is realised as a sharply inhomogeneous sequence of discrete *events*. This empirically based idea of conventional science has been especially emphasised within recent pronounced advance of the 'science of complexity', even though its conventional, scholar version always fails to give the truly consistent, rigorously derived description of 'events' and their natural 'emergence' (see [1] for more details and references). In any case, the basic role of 'events', those sharply inhomogeneous changes of system state is evident for any 'computation', or 'information processing', since it deals, by definition, just with those *highly* discrete 'portions of information', or 'bits', and their equally discrete transformations, irrespective of their exact meaning and physical realisation. In other words, a physically tangible and relatively large change (event) should necessary *happen* in each elementary interaction act within a *useful* computing system (where the particular case of external 'absence of change' is possible, but cannot dominate and practically always hides within it some internal, externally 'invisible' or transient change/event). By contrast, every unitary evolution inherent in the canonical mathematical basis of quantum mechanics (and in 'mathematical physics' in general) means that *no* event, nothing *truly* 'inhomogeneous' can ever happen within it ('unitary' *means* 'qualitatively homogeneous'). Therefore, the fundamental contradiction between unitary theoretical schemes of 'quantum information processing' and nonunitary character of any real computation process is evident: unitary quantum computation tries to obtain 'something from nothing', which is directly related to its suspiciously priceless, 'miraculously' increased efficiency with respect to classical, presumably nonunitary computation.

The same contradiction can be expressed as explicit violation of the 'energy degradation principle', or (generalised) 'second law of thermodynamics', by the unitary computation: any system, or 'machine', producing a measurable, non-zero change (like actual computation) should also produce a finite, and strongly limited from below, amount of 'waste'/'chaos', or 'dissipation'/'losses', or 'heat energy', which result, whatever its particular manifestation is, cannot be compatible with the unitary quantum evolution. This 'something-from-nothing' problem is the main defect of conventional quantum computation theory underlying its other contradictions described below.

The conventional scheme of quantum computation tries to overcome this contradiction by one or another 'combination' of the unitary evolution and nonunitary 'measurement' (or 'decoherent interaction') stages presented as a sort of 'punctuated unitarity' (see e. g. [62,89]). In that way the conventional quantum computation theory tries to reproduce the corresponding structure of the standard quantum mechanics itself, where the unitarity of the 'main' dynamics is trickily entangled with explicitly nonunitary 'quantum measurement' processes, though this 'connection' and its components remain quite 'mysterious' and are imposed only formally by the canonical 'quantum postulates'. It is evident, however, that unitarity of a total computation scheme will be violated by the unavoidable 'measurement' stages, which invalidates, though in an unpredictable and 'inexplicable' way, the conclusions based on the unitary system dynamics. The difference from the simplest dynamical systems considered within the canonical, unitary quantum mechanics is that any realistic quantum computation process should include much more involved, *dynamically emerging* configurations of participating systems with interaction, which belong to a higher sublevel of (complex) quantum dynamics and cannot be considered only 'statistically': all the 'dynamical' details hidden in 'statistical' postulates that describe the standard, 'averaged' quantum dynamics do matter at this higher sublevel of quantum microsystem dynamics. It is only the unreduced, dynamically multivalued description of genuine quantum chaos [1,9] that can provide the causally complete, detailed picture of irreducibly probabilistic quantum computation dynamics (Chapters 3-6).

In terms of mathematical analysis of interaction processes within quantum computing systems, the problem is reduced to the necessity to obtain the *unreduced, provably complete* solution of Schrödinger equation for the quantum system wavefunction (as opposed to a much less consistent, incorrectly used density matrix formalism, see [1,4]). In any case, one needs to solve a dynamic equation (partial differential equation of at least second order) containing practically arbitrary, strong interaction between multiple system components. The standard theory is not only unable to provide the unreduced solution for this type of problem, but cannot even predict what it could be like: any nontrivial interaction is 'nonintegrable', or 'nonseparable', in the conventional theory, which means here that its result is absolutely unknown, even qualitatively or approximately. This certainly refers to the rigorous description of any real problem of quantum computation. What the conventional quantum theory invariably tries to do is to replace the unknown true, unreduced solution with a heavily reduced version of perturbation theory corresponding to the assumed *weak* interaction influence (which *cannot* produce any essential change/event by definition). The difficulties arising from such unjustified simplification are quite serious even in the case of relatively simple systems of standard quantum mechanics: the axiomatically fixed 'mysteries' of quantum behaviour, inevitable empiricism and mechanistic adjustment of assumed structure parameters to experimental results. It is evident that such 'methods' of the canonical 'exact' science become totally senseless for much more involved interaction processes of quantum computation: the diversity of possible results (system configurations) cannot be covered by 'postulates' and one cannot even approximately guess what *are* the parameters/configurations to be adjusted. The truly 'exact', causally complete solution becomes indispensable at this higher sublevel of dynamics, and it is provided by the universally nonperturbative analysis of the unreduced science of complexity called *quantum field mechanics* at these lowest complexity levels [1-4,9-13]. The unreduced problem solution contains a qualitative novelty with respect to the perturbation theory results, the *dynamic redundance and entanglement* phenomena that change the very character of dynamics, providing its *essential* nonlinearity, nonunitarity, *purely dynamic*, ultimate and universal origin and *meaning* of randomness (probability), *a priori* determined probability values, *dynamically* emerging *new* entities/configurations (*creativity*), and show thus what the problem 'nonintegrability'/'nonseparability' *actually* means. Since the conventional, unitary theory of quantum computation is reduced instead to regrouping of linear superpositions of inevitably perturbative solutions, all the essential, qualitatively important details of real computation dynamics are definitely lost, including, in particular, processes of emergence and transformation of the main computation elements, discretely structured, inhomogeneous physical units of information, bits, or 'quantum bits (qubits)' in this case.

This fundamental property of the unitary theory of quantum computation/interaction can also be described as its basically *non-dynamic* (non-interactional) character, which is evident within any its presentation (see references in the beginning of this Chapter) always reduced to an oversimplified, linear and abstract scheme of intuitively guessed and arbitrary assumed interaction results, whereas consistent solution of the unreduced dynamic equations is not even tried. Such unconditional reduction looks especially strange taking into account simultaneous explicit acknowledgement by many leading 'quantum computer engineers' of the irreducible involvement of 'nonlinearity' and '(dynamic) complexity' in quantum computation process (see e. g. [62,64-67,81]). It seems that when one needs to emphasize a (general) relation to the popular 'complexity issues' one readily underlines this aspect, but when one needs to provide particular results of the dynamically complex computation process, one finds it more convenient to return to the 'good old' linear scheme of the standard quantum mechanics as if 'forgetting' about all the 'conceptual' novelties brought in by the essential nonlinearity. We emphasize that the latter should evidently appear even for the case of a totally linear underlying formalism (like the standard Schrödinger equation), due to the *dynamically* emerging nonlinearity of interaction process itself, just providing the 'computation' as such. Here again one can see how the expected 'magic' advantages of quantum computation are 'obtained' in a *too* easy, 'no-cost' way: when either essentially 'quantum' features (like 'linear superposition of states') or 'dynamic complexity' properties (like physical creation of 'bits') are needed by the theory, they are *exclusively* and intuitively *assumed* in the respective moments, but then equally easily dismissed at other moments, whereas in reality all of those contradictory and 'mysterious' properties of quantum *and* complex dynamics are permanently present and intrinsically 'mixed' within any real interaction process. This is but a particular, though probably the most convincing, example of fundamental deficiency of conventional science ideas on *both* quanticity and dynamic complexity/nonlinearity, which can be intrinsically *unified*, in their *causally extended* versions, only within the unreduced concept of dynamic complexity [1-4,9-13].

In this connection, it is important to note the fundamental difference between this concept of dynamic complexity/chaos based on the *intrinsic, purely dynamic* phenomenon of dynamic redundance/entanglement and various conventional estimates of *externally* imposed influence of 'decoherence' [90-92] or other stochastic 'randomness' effects and related 'quantum chaoticity' [14,65-67] of the *unitary* (dynamically single-valued) evolution on the quantum computer operation. In particular, we show that *real*, truly chaotic (dynamically multivalued) micro-devices of *any* type can *never* correctly operate as unitary machines, even in the absence of any external influence (i. e. apart from any 'decoherence' and stochastic 'chaoticity' effects), but *can* instead produce a useful, and even 'fantastically efficient', result of *another* origin just *due* to their complex-dynamical operation that should be described, however, by the unreduced, universally nonperturbative theory of real interaction processes (Chapter 3).

The unreduced dynamic complexity involvement in any interaction process shows also that all hopes of unitary approach to compensate appearing 'deviations' from the desired unitary evolution through 'error correction' (software) [92-99] or 'dynamical (chaos) control' (hardware) [14,100-107] are vain. Indeed, any change-bringing interaction leads to intrinsically probabilistic (partially unpredictable), quantised structure creation and since the step of quantisation of a system or any its 'correcting' part is fixed and relatively big at the quantum level of complexity (see item (ii)), no any 'correcting'/'controlling' procedure can influence the quantum system dynamics in a desirable way, without creating some other, undesired and relatively big changes. One could probably relatively (but *not* arbitrarily) reduce probabilistic scatter of one variable at the expense of other one(s), but this does not solve the problem and may not be readily permissible for a generic system configuration (usually only very modest 'deformations' of the system 'phase space' can be acceptable without destruction of the main computation dynamics), while the unitary, dynamically single-valued theory cannot even approximately describe the real result of any such 'intervention', from either outside or inside of the quantum computer core. This is because the conventional theory cannot (and actually does not try to) provide the complete, nonperturbative solution to the main dynamic equation describing many-body system with interaction. Instead of this, the unitary 'control' and 'correction' schemes tend to manipulate with mechanistically fixed, postulated mathematical structures ('algebras' etc.) supposed to represent the result of system dynamics, but being in reality its over-simplified caricature (effectively zero-dimensional projection [1-4,11-13], see Chapter 3).

Note also that the existing 'experimental demonstrations' of quantum information processing (see e. g. [38,97,107,108]) do not really contradict the fundamental impossibility of unitary quantum computation following from the causally complete analysis of the quantum field mechanics. None of those experiments can be considered as approaching a full scale quantum computation, and their actual realisation and precision are such that they can well represent nonunitary processes, considered as 'approximately unitary'. They can be compared to experiments with some imperfect elements of ordinary, classical computer performing separate elementary operations, like multiplication of small numbers, with a precision within, say, 10 % and said therefore to be 'promising' for the full-scale calculation. The essential feature of the quantum computer case is that its elements are *already* as perfect as they can be, basically, and they can be further 'controlled' only by the elements of the same, 'quantum' level of precision limited by the multiply verified quantum postulates. A purely external resemblance between real processes and their unitary imitation can be better for quasi-classical situations and other 'more controllable', ordered, or 'self-organised' regimes of quantum dynamics (Section 4.5.1) [1,9], but in such cases one deviates from an 'essentially quantum' regime, thus losing the expected 'advantages' of unitary quantum computation (whereas quantum uncertainty and coherence break-up due to the intrinsic multivaluedness cannot be arbitrarily reduced even in those relatively 'regular' cases). Therefore, approximate 'self-organisation' of quantum dynamics is not impossible, but it can never even approach the 'ideal precision' of deterministic, 'coherent' dynamics, which is absolutely necessary for the correct realisation of unitary computation scheme and its characteristic 'advantages'. Quite another, explicitly nonunitary (dynamically multivalued) regime of real quantum device operation can indeed be useful and efficient in applications, but it should be described within the unreduced, universally nonperturbative interaction analysis of the quantum field mechanics [1-4,9-11], which will be specified below, starting from Chapter 3.

(ii) <u>Contradiction between unitary description of quantum devices and quantum postulates</u>. The truly consistent, or 'causally complete', description of quantum behaviour within the quantum field mechanics [1-4,9-13] shows that in reality the formally unitary scheme of standard quantum mechanics masks its intrinsic nonunitarity (the *internal*, dynamical ran-domness, *not* a noisy 'decoherence') that manifests itself through conventional 'quantum postulates', their 'mysterious' and 'unprovable' character representing just the unavoidable payment for the artificial, dynamically single-valued simplification of real, always dynamically multivalued processes within their unitary scheme of the standard theory. The severely limited and purely abstract logic of the standard scheme can be relatively 'successful' only for simple enough systems, where one can empirically fix a small number of individual, qualitatively specific events and emerging system configurations and then treat them 'collectively', i. e. statistically, using the necessary number of adjustable 'free parameters' (like potential shapes, energy level positions and widths). Any really 'computing', practically efficient, many-body quantum system with interaction certainly falls outside of this narrow framework for the evident reasons outlined above (see item (i)). Therefore, even if one does not want to advance towards the causally complete understanding of quantum behaviour before proposing a real quantum computer scheme (which seems to be a strangely superficial approach by itself), one can see, nevertheless, several evident contradictions between the fundamental, multiply verified quantum postulates and the expected/desired properties of such more complicated, but still allegedly unitary quantum system.

The intrinsic 'indeterminacy', or 'probabilistic character', of any quantum dynamics strictly implies a basically indeterministic, probabilistically distributed result of any generic interaction process, even in the absence of external 'noise'. This creates immediately a fundamental problem for conventional, digital computation (like factoring and other arithmetical operations with numbers), since not only some 'final' result of a computational chain, but also that of each intermediate, elementary operation will contain an irreducible uncertainty that cannot be arbitrarily diminished, according to the 'probability postulate'. All the 'unitary fantasies' of conventional quantum programmers are wrong in this respect, since they suppose, explicitly or implicitly, a qualitatively small, perturbative, 'integrable' influence of the interactions involved, which therefore cannot produce a qualitatively new structure and thus any elementary computation result, even though they may give an illusion of relative regularity. Once the universally nonperturbative, unreduced analysis of the quantum field mechanics is applied to obtain the complete problem solution (Chapter 3), the genuine dynamic randomness (i. e. *true* quantum chaos [9]) inevitably emerges in the form of dynamic multivaluedness (≠ unitary 'decoherence') of every elementary interaction result. It should be noted that here one deals with a manifestation of the fundamental dynamic uncertainty at a higher sublevel of (complex) interaction dynamics with respect to the most fundamental quantum indeterminacy existing even for a 'free' particle (in the form of its 'quantum beat' dynamics) [1-4,11-13]. However, the two sublevels of dynamic uncertainty are dynamically related to each other and actually unified within the unreduced science of complexity, while the total and unconditional omission of the higher sublevel uncertainty does not seem to be consistent even without knowing the causal origin of the fundamental quantum indeterminacy.

The problem with quantum probabilism exists even if it can appear only during the 'final' measurement process of the postulated unitary computation result. Conventional quantum programmers are quite aware of the problem (cf. e. g. [109]), but they prefer to concentrate on the unitary imitation of 'internal' system interactions (reduced to mechanistic manipulation with various linear superposition versions) while counting upon something like 'multiple runs of the same calculation process' as a solution of the 'probability problem'. It is evident, however, that the 'fantastic' efficiency of quantum computation can be arbitrarily reduced in that way (showing once more that nothing can be obtained for free in this world!), and what is even more important, the actual degree of the 'final measurement' uncertainty cannot be properly estimated without the complete, nonperturbative problem solution, which is inaccessible to the conventional, unitary approach.

The probability postulate of standard quantum mechanics is complemented by a group of duality-measurement-reduction postulates endowed with a large ambiguity, but still entering in contradiction with the unitary computation scheme. Thus, the famous 'wave-particle duality' implies that particles/waves participating in an interaction (e. g. computation) process show a dynamically driven alternation of both their states (i. e. localised 'particle' and extended 'wave'), somehow regulated by a 'quantum measurement' involving 'wave reduction' to the 'particle' state. Although conventional theory leaves a large space for speculations about the exact origin, reality and structure of all those states and transitions between them [15,16] (contrary to the quantum field mechanics eliminating all the ambiguities [1-4]), it becomes clear that dealing exclusively with wave-like, linear type of behaviour (expressed, in addition, exclusively by postulated 'configurations' and abstract 'state vectors' from simplified mathematical 'spaces'), while artificially pushing the equally important corpuscular state and reduction process to the 'outside' of the assumed unitary evolution as it is done in the conventional quantum computation description, cannot be consistent in principle. Not only the well-known, provocatively irresolvable 'mysteries' of quantum duality and measurement remain in their untouched, 'para-scientific' state, but the unitary quantum programmers insist on their right to arbitrary manipulate with them by using or neglecting each 'mystery' when it is respectively needed or not for a 'promising' guess promotion. The resulting 'would-be' mode of the bankrupt unitary theory is a mixture of a science fiction caricature and deliberately fraudulent, 'postmodern' play of words in the style of 'ironic' science [5] (Chapter 9).

Another entity that should be postulated during formulation of any quantum-mechanical problem is system 'configuration' forming eventually the 'configuration space', in which the system evolves according to the main dynamic equation (basically Schrödinger equation). There is no general recipe in the standard theory for possible configurations and their space for each particular system, but conventional quantum mechanics dealing with the simplest systems resolves this difficulty in a semi-empirical or intuitive way, so that in most cases the configuration space can simply be 'guessed' with a reasonable degree of realism. However, for more involved and especially 'dynamically creative' quantum systems, including efficient quantum computers, such approach may lead to serious mistakes and in any case loses any reliability. Indeed, it can be practically impossible to 'guess' real configurations for a many-body quantum system with unreduced, creative interactions, which stops the unitary description of quantum computation at the very beginning (apart from some unpredictably unrealistic approximations, actually used in the existing unitary imitations of quantum dynamics). The problem can instead be regularly and universally resolved within the unreduced science of complexity that just gives explicit expressions for emerging system configurations as an integral part of the complex-dynamic interaction process development [1-4,9-13], as we shall demonstrate below (Chapter 3).

One of the main quantum postulates related to the very emergence of quantum mechanics 100 years ago [3] and determining its name, is the fundamental discreteness of all occurring dynamical processes described by the *absolutely universal* value of Planck's constant, $h = 6.6261 \times 10^{-27}$ erg·s. Since the information treated by a computer is also discretely structured in bits, it seems to be evident that the physical realisation of information unit in an essentially quantum computer, or 'qubit', is determined by the same universal Planck's constant, which implies also much more profound relations between 'information', (quantised) mechanical action and dynamic complexity, within its unreduced concept [1]. Surprisingly, however, this inevitable and physically transparent relation between Planck's constant (quantum discreteness) and extremely widely used qubit concept somehow escapes the attention of unitary quantum information theory, including its 'reality-based critics' and studies of 'physical realisation' of 'quantum information' (see e. g. [35,38,39,80,110-113]), as if 'quantum computing' could exist apart from 'quantum computer', which is a real, essentially quantum system performing the computation, a process that should necessarily be quantised by *h*, as any other one. A closer examination may reveal the origin of such 'strange' omission of quantisation in conventional quantum computing theory: explicit acknowledgement of qubit emergence in physical quantisation determined by Planck's constant necessarily implies acknowledgement of other related consequences of *real* quantum *dynamics*, which contradict the idea of unitary computation (such as quantum indeterminacy and wave-particle duality mentioned above). In this sense, the characteristic 'ignorance', or rather intentional neglect, by the official 'quantum computer science' of the direct qubit-*h* relation is another manifestation of its abstract-linear, 'anti-dynamical' approach (see item (i)) that indeed *actually* considers 'quantum computing' *apart from* any 'quantum computers', *even abstract ones*, determined exclusively by conventional quantum postulates. The scholar theory of quantum computation becomes thus a branch of *pure mathematics* [39] (and a conceptually trivial one, since it deals with something as simple as 'Hilbert space geometry' or elementary 'combinatorial analysis'), but at the same time it continues to insist intensely, and with a spectacular financial success, on its direct relation to creation of *real* quantum devices. This 'contradictory' state, as well as its 'strangely' persisting support, is not specific to the scholar quantum computation, but is the inevitable result of the taken dead-end direction of the 'new physics' development in the twentieth century, known as 'mathematical physics' and unfortunately dominating over the entire physical theory (cf. [114]), despite the ensuing absence of solutions to any real, nontrivial problem (see also Chapter 9).

The involvement of $h$-determined physical quantisation in any quantum system dynamics is related also to the ideas of 'error correction' and dynamical (chaos) control in quantum computers [14,92-107] as the means to compensate all externally or internally induced 'deviations' of real quantum dynamics from the desired unitary scheme (see also item (i)). Indeed, any interaction/action in a quantum system is quantised into probabilistically appearing results, according to the main postulates, and therefore any imposed change, whatever its origin is, can be neither arbitrarily small nor deterministic (cf. the canonical 'uncertainty relations'). This means that, especially for a more complicated many-body structure of the full-scale quantum device, one cannot hope to be able to really eliminate unpredictable 'errors' without creating yet greater and less predictable ones. Again it seems strange that unitary quantum programming does not take into account the evident difference (or actually even *any* qualitative difference) between quantum and classical levels of dynamics when they try to directly extend the methods of classical 'control theory' to quantum systems. At higher-complexity levels of classical systems control actions can be performed with the help of relatively fine-grained (and 'self-organised') lower-level dynamics, so that their unavoidable small fluctuations can be of minor importance. But for quantum dynamics the 'fluctuations' of controlling interactions are always as big as any result of the 'controlled' interaction dynamics, just because of the quantisation postulate. At the *lowest*, quantum level of world dynamics there can be no basic difference between 'noisy fluctuations' of an 'influence' on a system and the intrinsic, 'quantum' fluctuations of the system dynamics: this is the basis of the well-known quantum system 'sensitivity to external influences' appearing e. g. through multiply verified 'uncertainty relations' and causally explained in the quantum field mechanics by the intrinsic dynamic instability of the underlying interaction process (starting from *essential*, dynamic nonlinearity and ending with irreducible dynamic chaoticity) [1-4,11-13].

(iii) <u>Contradiction between unitary computation and entropy growth law</u>. Without entering here in the detailed rigour of possible generalisations of the entropy growth law, or 'second law of thermodynamics' (see Chapter 7), we can accept its following apparently irreducible general formulation: if a system (device) is designed to produce some ordering action, then this ordering can never be the single result of system operation (a larger disorder appearance in the system and/or its environment is implied, so that the 'net' effect is definitely an increase of disorder or 'generalised entropy'). If one accepts also that any computation is a sort of ordering action (i. e. simply a meaningful action producing some real results), then it follows that unitary quantum computation certainly contradicts this very general formulation of the second law. One cannot find this contradiction for classical, macroscopic computers, since for any computation process there can be many even very rough, thermal effects (contributing eventually to the heat flow from computer to the environment and thus 'ensuring' the second law), let alone various more subtle, but also unavoidable disorders in computation dynamics, which appear explicitly e. g. when the computer suddenly stalls (gets to a 'hangup' state). As explained above (see item (ii)), all such disordering fluctuations in classical computers can be decreased to relatively small magnitudes with the help of many available fine-grained, lower levels of (complex) dynamics, which are absent, by definition, for the essentially quantum computer dynamics.

A particular case of this contradiction between unitary quantum computation and impossibility of 'purely useful' result (net increase of order) encompasses various widely discussed schemes of unitary, or 'coherent', or '(purely) quantum' control, or 'error correction' of quantum computation systems or any other quantum devices often evoked as the means to suppress possible disordering deviations of quantum system dynamics (often designated as 'decoherence') from its 'normal', unitary (totally regular) evolution [92-107]. Whatever is the origin and the way of increase of this announced 'quantum robustness' of computation dynamics, it should always involve *real* interactions leading to 'positive', order-increasing result, which actually cannot be attained within the (unitary) quantum dynamics itself, since it occurs at the lowest sublevels of (complex) world dynamics and therefore does not contain the necessary *separate* 'sink' for the excessive, compensating disorder. If we take the refrigerator analogy as a canonical example of second law realisation, we can say that dealing with purely quantum dynamics one can never close the refrigerator door and any effort to decrease, or 'control', the 'inside' temperature is useless, already because there can be no good separation between the 'inside' and the 'outside' (the needed minimal structure is always *too* complex for the lowest, quantum levels of world dynamics/complexity).

One could suppose also that the second law demand could be satisfied during the 'final measurement' stage of reversible unitary computation, but this would mean that this quantum measurement process should introduce chaotic uncertainty that exceeds and thus totally destroys any order attained during unitary computation stages, making it useless. In reality, even such distinct separation into stages of 'pure' unitary computation and 'dirty' quantum measurement cannot be possible because of the same entropy growth principle, but applied now to the sequence of stages. Moreover, the unitary stage, considered separately, should necessary produce irreducible, and relatively large, randomness, if it involves any state-changing interaction process (which is necessary for every real computation). The dynamic multivaluedness of any real interaction result [1-4,9-13] (Chapter 3) simply specifies the detailed (and universal) mechanism of this inevitable randomness creation.

Note also that the generalised second law is correctly satisfied in the case of ordinary quantum dynamics which is opposite with respect to the above classical computer case: it is precisely the irreducible 'quantum indeterminacy' which, though remaining 'mysterious' by origin, is really observed and provides the elementary 'support' for the generalised second law at the corresponding levels of dynamics. However, the unitary quantum computation as such is free from any indeterminacy: the basic quantum uncertainty is silently expelled either to lower sublevel of dynamics of individual participating quantum objects or to some 'final' measurement process, external for the unitary computation itself. Therefore, violation of the 'quantum uncertainty' postulate by the unitary quantum devices, discussed above (item (ii)), can be considered as another expression of the present contradiction to the generalised second law.

This contradiction between the necessary dynamical randomness and the persisting regularity of unitary science can be consistently resolved only if one finds the detailed, and *purely dynamic*, mechanism of uncertainty (or 'disorder') emergence at the level of interacting quantum objects (which are elementary 'components' of a quantum computer/device, *including* its interaction with the 'environment') that should be *added* to the fundamental quantum indeterminacy at the lowest sublevel of component dynamics. This is actually done within the causally complete theory [1,9] of (Hamiltonian) quantum chaos revealing the strictly dynamic, internal system randomness (Chapters 3, 6), contrary to any version of conventional, unitary 'quantum chaos' description that either denies genuine randomness or tries to trickily 'postulate' or insert it from the outside (e. g. [14]). In addition, the *same* dynamic redundance mechanism that gives genuine quantum chaoticity provides the causal, totally dynamic and realistic origin of the fundamental quantum indeterminacy at the lowest sublevel of world dynamics (e. g. for free elementary particles) [1-4,11-13], which is an important closure of the theory, while the unitary imitation of dynamic randomness by the externally driven, ambiguous 'decoherence' [90-92] cannot be consistent with respect to its unambiguous attribution to one or another sublevel of quantum dynamics (whose true origin also remains 'mysterious'). Finally, the consistent description of the 'quantum measurement' process is obtained [1,10] as a slightly dissipative version of the unreduced quantum chaos dynamics at the same sublevel of interacting quantum objects. We obtain thus a consistent, internally unified, 'dynamically multivalued' extension of the conventional unitary, 'dynamically single-valued' projection of real quantum dynamics, applicable at *all* its involved sublevels (this extension continues in the same fashion for classical and all the highest levels of complexity [1], which is also an essential conceptual argument in favour of this description, even if one concentrates on quantum levels of reality).

It is a surprising fact, to be added to other 'surprising omissions' of the conventional theory of quantum computation, that its evident contradiction to the second law remains 'invisible' not only to the 'main', explicitly linear approach, but actually also to 'physically oriented' studies involving doubts about feasibility of the purely coherent scheme, inquiries into 'physical nature of information', emphasis on 'nonlinearity' and 'complexity' of quantum computation and definition of 'limits of system control' related to entropy and second law [58-62,79-81,113,115] (which can strangely coexist with coherent control schemes [100]). This basic and 'unlimited' defi-ciency of conventional analysis reflects its inability to obtain clearly specified relation between 'dynamics' and 'thermodynamics', i. e. the inevitable absence of consistent 'foundations of thermodynamics' within the scholar, dynamically single-valued theory of both quantum and classical behaviour (contrary to the dynamically multivalued theory [1]).

(iv) <u>Contradiction between quantum computer coherence and its irreducible structure.</u> The above dynamic randomness deficiency within the unitary quantum evolution reflects its insufficiently rich temporal structure. A complementary aspect of this contradiction consists, naturally, in insufficient spatial richness (i. e. inhomogeneity) of the coherent state of a hypothetical unitary quantum device. Indeed, the unbroken 'quantum coherence' of the unitary device implies that the whole many-body, many-element structure of a full-scale quantum computer is in a 'macroscopic' (or at least 'spatially extended') quantum state, similar to that of a superconducting, or superfluid system, or Bose-condensate. But it is a well-known fact that the spatial structure of a coherent component of any such system is always rather simple and 'symmetric', it can hardly contain any arbitrary, large and asymmetric density variations of well-defined (though interacting) structural elements. In other words, it is 'either structure or coherence', but not both simultaneously,[2] which specifies another aspect of fundamental impossibility of unitary quantum computer realisation as a material, 'hardware' device (and not just a numerical simulation scheme). In this connection one should note that the existing 'experimental confirmations' of unitary quantum computation (e. g. [38,97,107,108]) are nothing but its basically limited imitations, where one need not have either true coherence or the involved structure of the full scale device, or rigorously confirmed temporal stability of a larger, realistic computational process.

The basic spatio-temporal 'simplicity' of coherent quantum states and their unitary evolution, observed e. g. in real 'macroscopic quantum states', is indeed a universal property that can be consistently explained within the unreduced concept of dynamic complexity [1-4]: what is called 'essentially quantum' (coherent) behaviour and actually simply postulated in conventional science appears to be an unreduced interaction *process* limited to several lowest sublevels of the universal hierarchy of dynamic complexity of the world, and this *low* dynamic complexity cannot produce more involved/asymmetric spatial and temporal structures by definition. The same argument shows, by the way (see also below), why any quantum computation *cannot* reproduce any higher-level, including any 'classical', micro- and macroscopic, system dynamics in principle, contrary to what is often presented as a 'proven' property of universal computing in the conventional (unitary) theory of quantum computation ('universal quantum computers/simulators' etc.).

---

[2] This basic incompatibility can be viewed as manifestation of the famous quantum 'complementarity', which remains a real property despite the air of ambiguity and 'mystery' around it in conventional quantum mechanics (disappearing within the causally complete understanding of complementarity in the quantum field mechanics [1] as a standard property of any complex, dynamically redundant interaction process). The structure-coherence complementarity is more directly related to such manifestations of quantum complementarity as wave-particle duality and coordinate-momentum uncertainty relation that force one to choose between well-structured, localised, 'particle-like' system state and its loosely structured, delocalised, undular state.

The ultimate, but actually correct, expression of coherent structure of unitary quantum computation is 'quantum computation in the ground state' [45]. Indeed, any truly coherent quantum state is the ground, lowest state of a system, whereas any excited system state is always a chaotic [1,9] and thus not totally coherent one (though it can be only slightly irregular). Since the ground system state is the least structured one (because of its lowest complexity [1,9]), we arrive at another formulation of the above contradiction between coherence (unitarity) and structure: computation in the ground state, i. e. actually *any* unitary computation, is impossible because of the ultimate poorness of its spatial and temporal structure. Whereas the unreduced complexity concept provides the causally complete substantiation for this statement, it should seem to be basically valid even within the standard quantum mechanics.

A similar objection should provoke fundamental doubts about feasibility of any 'quantum memory' [47-49] within a unitary device: any 'memorisation' act is realised as a basically irreversible transition to another, stable and inhomogeneous enough, system state, which cannot be compatible with unitary dynamics. Any state of a coherent quantum system cannot be so isolated as necessary to prevent spontaneous and unpredictable system transitions to it and back, while any excited system state is both unstable and incoherent.

Although coherent, ground state of unitary quantum system cannot contain sufficiently inhomogeneous spatial structure, it does have some simple (symmetric) structure and an 'internal life' (temporal structure) within it. This internal dynamics consists in permanent chaotic transitions between hierarchically organised individual and collective states of system elements, which appear externally as a slightly inhomogeneous, statistically averaged 'mixture', or 'ground state' as such [1]. This dynamically chaotic, and therefore only partially coherent, mixture of many-body system states within its 'embedding' meta-state (see also Section 5.3) is imitated, within the unitary theory, by the notorious 'quantum entanglement' of superimposed states of a linear system/evolution and the related 'teleportation', imitating spontaneous transitions between internal system states and supposed to realise (unitary) quantum computation (e. g. [116]). However, unitarity is incompatible with the irreducibly *chaotic* internal structure of a real 'coherent quantum computer' (persisting even in the total absence of external influences), while its basic simplicity prevents any useful output emergence.

(v) <u>Transition from quantum to classical computation and back</u>. Although the link between quantum and classical behaviour expressed by the 'quantum measurement' procedure remains obscure and subject to ambiguous speculations within all 'officially permitted', unitary versions of quantum mechanics, one cannot avoid its explicit involvement in quantum computation, already because the latter supposes eventual production of a humanly readable, 'classically' structured result of essentially quantum process, similar to experimental observation of any quantum effect. However, the difference of quantum-classical transition for more complicated, computing machinery from that for canonical quantum systems with the simplest, empirically known and globally fixed configurations is that 'essentially quantum' (including 'semiclassical') and totally 'classical' types of behaviour and configuration may be not so well separated from each other for more intricate processes as it is postulated in the textbook version of 'quantum measurement' for the simplest systems (remaining, however, quite 'mysterious' and thus only partially, indirectly 'confirmed' experimentally, even for that case). In particular, as mentioned above (item (i)), any computing system is a *creative* system with changing, *dynamically emerging* configuration, which is especially important for quantum systems whose 'soft' (dualistic) structure and intrinsic 'vulnerability' (sensitivity) are fixed as well-confirmed facts in the main quantum postulates. Therefore one cannot be sure to be able to guess 'semi-empirically' the dynamically evolving configuration of a computing quantum system (which would mean, in particular, to guess the 'quantum calculation' result before obtaining it). As a matter of fact, a *classical* configuration can dynamically and 'unpredictably' emerge, at least transiently, during 'coherent' quantum computation, thus immediately destroying its coherence and the whole 'architecture' of unitary, linear-wave imitation of real interaction dynamics. In the case of a full-scale version of computing micro-device with nontrivially structured units, this nonunitary 'dynamical reduction' of quantum computer elements to a classical configuration will inevitably intervene in the middle of expected 'coherent' computation and not only during the unavoidable 'final measurement' stage, also remaining the evident 'weak point' of unitary quantum computer (see item (ii)).

The situation can be fully clarified only within the causally complete picture of 'quantum measurement' obtained within the dynamic multivaluedness paradigm [1-4,10]. It appears that *any* nontrivial, change-bringing interaction process (necessarily occurring in the course of 'quantum computation') breaks down 'quantum coherence' in the form of either (Hamiltonian) 'quantum chaos' [1,9], for a closed system (vanishingly small dissipativity), or 'quantum measurement' [1,10], for a slightly dissipative, open system. In the latter case, the occurring excitation of dissipative ('open') degrees of freedom is accompanied by the physically real, highly nonlinear (catastrophic) system localisation (or 'collapse', or 'reduction') around the dynamically (and probabilistically) chosen 'centre' of excitation event, which is equivalent to a transient emergence of classical (localised, or 'particle-like') state. This classical state becomes more stable with growing 'bounding force' of interaction, and when it is sufficient for a stable bound state formation, the true classicality emerges in the form of elementary bound system (like atom). Therefore *some* classicality is always present as a part of almost any interaction process involving real excitation of interacting elements, even though usually it lasts for a very short time, being quickly replaced by the reverse self-amplified expansion of the system to a delocalised, undular state. However, the system coherence, or unitarity, is lost in principle, in any such 'measurement' event, as well as during the complementary, 'quantum chaos' mode of interaction development (but in this latter case the system does not need to pass by a spatial localisation phase).

The conclusion following from both this causally complete analysis of the quantum field mechanics and the above qualitative analysis within conventional quantum theory is that in reality one may have not 'purely quantum', but only 'hybrid', 'quantum-and-classical' computer, or any large enough device, irrespective of any external or internal 'noise' exploited in speculative theories of 'decoherence' or (conventional, unitary) 'quantum chaos'. An important element of 'quantum computer' (and part of the quantum computation process) that should necessarily involve irreducible dissipativity and more stable, classical state formation is (quantum) computer memory (and memorisation/erasure/reading processes) which is, therefore, rather a classical than quantum part of a 'quantum' device, contrary to the conventional theory [47-49] trying to preserve its unitary imitation of reality even for such evidently nonunitary function as memory (cf. item (iv)). Some other, usually very short, stages of quantum computation can preserve their 'non-classical', delocalised character, while still being subject to another inevitable source of decoherence, (genuine) quantum chaos [1,9]. Therefore, a real 'quantum' computation process in the whole can be described as a fine dynamic entanglement between short stages of undular coherent (rare), undular chaotic and classical (always internally chaotic) dynamics that can be consistently analysed only within the *unified* description of fundamental dynamic multivaluedness (Chapters 3-7) changing completely the very character of 'computation' and confirming the basic insufficiency of unitary, dynamically single-valued imitation of reality. In particular, the spatial and temporal sequence of those chaotic 'classical' and 'quantum' phases of computation dynamics is dynamically chaotic itself because of the same intrinsic 'undecidability' of unreduced, multivalued interaction processes.

It can be useful to consider the quantum-classical transition within a computing system in the reverse order, i. e. starting from an ordinary, classical (and quasi-regular) computer and decreasing its elements down to characteristic 'quantum' sizes. This is the situation which is much more practically important than futuristic quantum experimentation with the 'original' quantum computers, since that kind of transition to quantum limit is going to happen soon to actually produced computers with further increase of element density within their microchips. It is easy to see, already within a qualitative consideration, that what will practically emerge in a re-al device with 'subquantum' sizes of at least some of its elements is the explicitly multivalued, chaotic, quantum-and-classical (hybrid) dynamics described above, rather than any version of unitary quantum computation. Indeed, it is clear that the emergence of quantum effects with decreasing element sizes will produce some 'smearing' of the normally quite 'distinct' classical computation dynamics, this 'quantum blur' including both 'undular' and 'probabilistic' aspects of quantum behaviour, which are intrinsically connected among them, according to the standard postulates. The resulting computer 'errors' will appear first with a relatively small probability, but it will progressively grow with diminishing element size, until the computer dynamics will become totally 'indistinct', that is chaotic (and 'hybrid'), irrespective of 'wiring' or 'soft-ware' details. Now, the unitary theory of 'pure' quantum computation implies that one should have, instead of this, a certain internal 'ordering', regularisation of the emerging quantum dynamics that should be comparable, as noted above (item (iv)), to a sudden 'phase transition' analogous to a '(generalised) Bose-condensation', or a 'superconducting' phase appearance. Apart from structural problems involved with such peculiar states (item (iv)), their origin, mechanism and degree of generality remain subject to strong doubt, which shows again that the intrinsic dynamic multivaluedness of the essentially quantum dynamics, hidden behind the 'quantum mysteries' of its standard, dynamically single-valued description, will necessary appear in its explicit form for more involved, higher-level interactions within a quantum-size computing structure. In particular, the influence of *any* quantum effects upon *any* microstructure dynamics can be correctly analysed only within the unreduced, dynamically multivalued theory (Chapter 3), while any perturbative 'cutting' of essential dynamical links in the name of questionable 'simplicity' of 'exact', unitary solutions will produce a qualitatively big, 'fatal' deviation from reality.

It is also clear that the real, intrinsically chaotic and 'hybrid' dynamics of a computing quantum system is closer to 'analogue' computation, or 'simulation', than to the 'digital' operation mode of ordinary, regular computers. Contrary to various versions of ('fantastically' fast) operation with numbers considered by the unitary theory of quantum computation (e. g. [23,117]), it is difficult to imagine how the chaotic, dynamically multivalued interaction process within a real quantum device could preserve the regularity of a digital simulation of reality. And since the reality itself is universally chaotic, rather than regular [1], and is certainly 'analogue', rather than 'digital', there is no sense to insist on its unnatural, digital simulation, incorrectly realised within a regular quantum computer (actually unrealistic). In this sense, the real quantum computer will be much closer by the character of its operation to the brain dynamics [1,4], though the latter belongs to a much higher level of dynamic complexity (contrary to what is implied by its existing reduction to the unitary quantum dynamics [68-71]). Those 'natural', intrinsically chaotic 'computers' are efficient in operation with unreduced 'images' of chaotic reality, but inefficient (for the same reason) in operation with digital, artificially regularised (and thus strongly simplified) representation of reality. In the 'digital' computer case the 'result', a number, is basically separated from the real system dynamics actually producing it, whereas in the dynamically multivalued, unreduced mode of 'chaotic analogue' computers one has intrinsic unification of complex computation dynamics and the result which is simply a (multivalued) momentary image of that unreduced dynamics. This explains the qualitative difference between description/understanding of digital/unitary and 'natural'/complex-dynamical devices, the latter necessitating the use of unreduced, dynamically multivalued (nonunitary) description, even for an approximate representation. The difference is applicable also to the case of the proposed formally analogue mode of 'simulation' of quantum (or classical) dynamics by a unitary quantum computer [24,25,29,31]: the dynamics of the latter remains basically unitary, regular, dynamically single-valued (including the false 'quantum chaos' simulation [65-67]) and therefore it is 'infinitely far' from any real, dynamically multivalued system behaviour, contrary to the claimed 'universality' of such unitary 'quantum simulators' [25]. Moreover, the universal science of complexity clearly shows [1] that any purely quantum, even dynamically chaotic, device could not correctly simulate any higher-level dynamics, starting already from the simplest classical systems, because the latter emerge dynamically from quantum systems as a *higher* level of complexity [1-4,11-13] (simulation could be better performed, at least for certain cases, with the help of slightly dissipative, 'hybrid' microcomputers). This fundamentally substantiated and transparent conclusion of the unreduced science of complexity totally escapes unitary theory, for it cannot consistently describe the complex-dynamical relation between quantum and classical behaviour, as well as the unreduced, dynamically multivalued origin of either of them (cf. [118]).

Note finally that the described long series of evident, basic contradictions, items (i)-(v), in the conventional science of (unitary) quantum computation and closely related scholar quantum mechanics (including all its modern 'versions' and 'interpretations') is a characteristic sign of the deep impasse in the canonical, purely 'mathematical' (abstract) approach in fundamental physics, marking the 'sudden' fall from the previously triumphantly announced (but actually deceitful and misleading) 'unreasonable efficiency of mathematics in natural sciences' to the really unreasonable agglomeration of vain and fruitless abstractions and related arbitrary guesses about a possible 'mathematical basis of reality' (see also Chapter 9). Thus, 'feeling' the more and more clearly that something very serious is missing in its simplified imitation of reality, the unitary quantum theory, instead of looking for the truly consistent, qualitatively new solution, tries to 'repaint' its façade and compensate the absent elementary causality/realism by artificial, ever more speculative, abstract and inconsistent additions within the same, unitary (dynamically single-valued) projection of real system dynamics, such as multiple versions of 'quantum histories' (or 'path integrals'), ambiguous 'decoherence' of 'state vectors' due to varying 'external influences', or mechanistic fixation (postulation) of the observed multiple 'quantum potentialities' in various 'multiverse'/'many-worlds' interpretations of the same, linear and abstract, imitation of 'quantum reality', only amplifying its para-scientific 'mysteries'. The conventional, unitary theory of 'quantum computation' (similar to 'quantum field theory' and 'cosmology') forms a 'point of concentration' of those evident blunders of externally 'solid', 'rigorously proven' and 'well-established' scholar theory dominated by 'mathematical physics', where its futility is only emphasised by the exponentially growing pressure of meaningless, 'post-modern' plays of words taking the form of esoteric, as if 'very special', terminology that hides, in reality, the absence of elementary consistency. Without going into the detailed structure of 'quantum' linguistic exercises, it is enough to recall the infinite flux of all those 'extremely' quantum 'teleportations', 'entanglements', 'distillations', 'holographies', 'tomographies', 'Schrödinger-cat states', 'Bell/EPR states/nonlocality' and many other 'new' and 'quantum' terms, which are actually devoid of any realistic, causal meaning, but describe, in a highly speculative and deliberately perverted form, exactly the same 'inexplicable', but empirically 'well-established' knowledge as the canonical quantum postulates. In addition, similar 'interpretations' that can, by their construction, 'explain' and 'justify' everything are created in each branch of 'mathematical' physics and then arbitrarily superimposed upon each other, like e. g. conventional theories of quantum mechanics and 'complexity' in unitary quantum computation, which gives the really intractable mixture of verbal decorations covering the underlying deficiency and rough mistakes, but dominating in the most 'ambitious' and 'solid' scientific establishments. While this peculiar 'quality' of fundamental science content is a sign of its modern profound 'bifurcation' (Chapter 9) [1], the described situation in micro-device description and understanding is ripe for a definite clarification and transition to a superior level of understanding, especially in view of the quickly growing potentialities of practical application. The causally complete description of unreduced interaction dynamics is presented in the next Chapter, in the form of generalised theory of 'true' dynamical chaos (purely dynamic randomness in quantum and classical systems with interaction [1,9,10], see also Sections 4.6.2, 5.2.1, Chapter 6), with the following application to real 'quantum computers' and other micro-machine dynamics (Chapters 5-8).

# 3. Dynamic origin of randomness in a noiseless system
## 3.1. Generalised many-body problem and micro-machine dynamics

Consider a system of *N* interacting elements, or subsystems, each of them possessing a perfectly known internal dynamics in the absence of interaction with other elements. Although we intend to apply the results to description of quantum computing systems or other micro-systems, we do not impose any limitations at the beginning and shall specify 'quantum' or 'computational' properties later, at suitable moments. Therefore, we are actually dealing with the general case of 'many-body problem', i. e. we consider arbitrary (real) system of interacting 'bodies' (or 'elements') and want to obtain the general, unreduced (and in particular nonperturbative) solution to this problem. Not only such solution is unknown to canonical science, but it cannot even predict what this solution can be like, i. e. what can be a qualitative, expected result, or structure, of arbitrary interaction process (apart from the implied common-sense response, "everything can happen"). As pointed out above, in the case of quantum computation one cannot avoid the unreduced many-body problem solution, since any perturbative, 'exact-solution' approach of conventional theory cannot describe just the most important, properly 'computational' system function consisting in purely dynamic ('creative') emergence of qualitatively new system configurations ('calculation results') for actually arbitrary interactions, which correspond to various possible 'calculations' (or 'simulations') by a maximally 'universal' computing system.

We generalise various particular equations describing compound system behaviour to what we call the (system) 'existence equation', a basic dynamic relation that actually does nothing more, in its starting form, than simple statement of the problem conditions, i. e. it expresses the fact of unreduced, 'nonseparable' interaction between the system components or, in other words, it describes the system existence (in its starting composition) by stating that its given interacting components form a 'single whole'. For a quantum system, the Schrödinger equation is normally implied behind the existence equation, but one does not need to be limited to it from the beginning, within our universal analysis. The system existence equation can be presented thus in the following universal form:

$$\left\{ \sum_{k=0}^{N} \left[ h_k(q_k) + \sum_{l>k}^{N} V_{kl}(q_k, q_l) \right] \right\} \Psi(Q) = E\Psi(Q), \qquad (1)$$

where $h_k(q_k)$ is the 'generalised Hamiltonian' for the $k$-th system component in its 'free', 'integrable' state (in the absence of interaction), $q_k$ denotes the degree(s) of freedom of the (free) $k$-th component, $V_{kl}(q_k, q_l)$ is the 'interaction potential' between the $k$-th and $l$-th components, $\Psi(Q)$ is the system 'state function', i. e. the function characterising completely the compound system state/configuration and depending, in a 'nonseparable' way, on all the participating degrees of freedom (brought by the interacting components), so that $Q = \{q_0, q_1, ..., q_N\}$ by definition, $E$ is the value, or 'eigenvalue', of the quantity expressed by the 'generalised Hamiltonian' in the compound system state $\Psi(Q)$, and the summations are performed over all system components numbered from $k, l = 0$ till $k, l = N$ (the total number of interacting entities). As shown in the universal science of complexity [1], any correct equation can be considered as a particular case of the single universal equation expressing the generalised Hamilton formalism (see Section 7.1) and therefore $h_k(q_k)$ can indeed be described as particular forms of the generalised Hamiltonian (which is a realistic extension of conventional 'operator'). However, one does not need to insist on this particular interpretation from the beginning. A suitable form of the 'generalised Hamiltonian' should represent an exhaustive characteristic of the system, which actually means that it should express a form of the unreduced dynamic complexity, such as (extended) mechanical action, energy, momentum and space structure [1] (it is self-consistently confirmed by the results of the unreduced interaction analysis, Sections 4.1, 7.1).

It is clear that eq. (1) describes the general many-body problem. In particular, it is not restricted by its 'quantum' or 'mechanical' version and refers, in general, to an arbitrary interaction between arbitrary (given) entities. The pairwise interaction potential in eq. (1) is not a limitation either, since the *unreduced* development of this 'elementary' interaction process can give rise to any more complicated combinations of nonseparable interacting entities (Chapter 4). In particular, they can form a hierarchy of interaction levels, such as internal atomic electron interactions, interactions between different atoms, their various agglomerates (like molecules), etc., but all those possibilities are included in the universal formulation and following solution of eq. (1). Moreover, we shall see that this 'many-body' interaction can be reduced itself to a yet simpler case of interaction between two (structured) entities leading to the main, universally applicable formalism of the unreduced science of complexity [1-4,9-13].

Several particular cases of eq. (1) are worthy of mentioning at the very beginning, especially taking into account the case of quantum (computing) devices considered in this paper. The properly 'quantum' problem character appears formally through the form of at least some of the 'free-component' Hamiltonians $h_k(q_k)$. Thus, those of them which represent localised 'elements' of the device can be generally considered as particular cases of 'potential-well' system configuration (with the known, complete set of 'eigen-solutions'):

$$h_k(q_k) = -\frac{\hbar^2}{2m}\frac{\partial^2}{\partial q_k^2} + U_k(q_k), \qquad (2)$$

where $\hbar = h/2\pi$ is Planck's constant, introducing the 'quantum' problem scale, $m$ is the 'working' particle (usually electron) mass (or its suitable 'effective' version), $q_k$ is the vector of spatial coordinates, $U_k(q_k)$ is a binding potential well with discrete (known) energy levels and the operator of eq. (2) is supposed to act upon the system wavefunction (in general, a dynamically discrete version of partial derivative can be implied in the complex-dynamical extension of the kinetic energy 'operator', Section 7.1 [1,11-13]). The Hamiltonians of eq. (2) can represent 'atom-like' elements of a quantum device, such as real atoms (molecules) or their artificial imitations like 'quantum dots' or 'heterostructures'. For other element types, including (generalised) spins or photons, respective 'free-element' Hamiltonians have other well-known expressions and solutions bearing signatures of corresponding, quantum or classical, scales of system structure.

It can often be useful to separate explicitly one of the participating degrees of freedom, say $q_0 \equiv \xi$, from other variables $q_k$ ($k = 1,...,N$), after which eq. (1) takes the following (equivalent) form

$$\left\{ h_0(\xi) + \sum_{k=1}^{N}\left[ h_k(q_k) + V_{0k}(\xi,q_k) + \sum_{l>k}^{N} V_{kl}(q_k,q_l) \right] \right\} \Psi(\xi,Q) = E\Psi(\xi,Q),$$

(3)

where the summations start from $k,l = 1$ and $Q = \{q_1,...,q_N\}$. This form of the starting existing equation corresponds to the situation where the separated variable $\xi$ (or group of several of them, in general) describes the really existing 'common' (extended) degree(s) of freedom, like position coordinate(s) varying along inhomogeneous system structure that contains its 'elements' and 'connections'. A simpler form of eq. (3) can be pertinent in the case of vanishingly small direct interaction between the 'localised' elements ($V_{kl}(q_k,q_l)$), each of them interacting only with the 'common' degrees of freedom ($\xi$):

$$\left\{ h_0(\xi) + \sum_k \left[ h_k(q_k) + V_{0k}(\xi,q_k) \right] \right\} \Psi(\xi,Q) = E\Psi(\xi,Q) \ . \quad (4)$$

The simplest nontrivial case of eqs. (1), (3), (4) arises when we consider only two interacting entities and degrees of freedom, say $q_0$ and $q_1$, corresponding, for example, to an element (atom) interacting with the surrounding radiation field or, in general, to an elementary interaction act between any two entities (like any elementary action of a computer 'gate') within the encompassing interaction/computation process:

$$\left[ h_0(q_0) + V_{01}(q_0,q_1) + h_1(q_1) \right] \Psi(q_0,q_1) = E\Psi(q_0,q_1), \quad (5a)$$

or

$$\left[ h_e(q) + V_{eg}(q,\xi) + h_g(\xi) \right] \Psi(q,\xi) = E\Psi(q,\xi) \ , \quad (5b)$$

in different notations ($q_0 \equiv \xi$, $q_1 \equiv q$), in which form the existence equation corresponds exactly to its 'canonical' version from the universal science of complexity and its various applications [1-4,9-13]. As we shall see later, the unreduced interaction results are essentially the same for the 'many-body' and 'two-body' cases (eqs. (1) and (5), respectively), which means that it is the unreduced interaction development itself that introduces all the essential, 'nonperturbative' properties, rather than the number of interaction components and other details determining particular manifestations of those general properties.

Now let us proceed in the unreduced interaction description using its most general expression, eqs. (1), (3). According to the problem conditions, the system constituents are perfectly known entities, with the complete sets of their eigen-solutions (in the absence of interaction), $\{\varepsilon_{nk}, \varphi_{knk}(q_k)\}$:

$$h_k(q_k)\varphi_{kn_k}(q_k) = \varepsilon_{n_k}\varphi_{kn_k}(q_k) \;, \tag{6}$$

where $\{\varphi_{kn_k}(q_k)\}$ are the eigenfunctions and $\{\varepsilon_{n_k}\}$ the corresponding eigenvalues of the 'free-element' Hamiltonian $h_k(q_k)$, forming the complete set of solutions/functions. In the case of binding potential well, eq. (2), the lower-state (most important) eigenfunctions are represented by a set of well-localised functions which can often be approximated by the δ-function (or δ-like functions) centred at the element position, while the corresponding eigenvalues form a discrete set of energy levels. Note that if the eigen-solutions of a component are not 'perfectly known', this means that one deals with an 'incorrectly posed problem'. In this case one should first descend one level lower in the system description and obtain the complete picture of the internal, generally also complex, element dynamics (or choose another, indeed 'perfectly known', system composition).

Since one can be interested eventually in the state of 'structural' degrees of freedom summarising, in particular, the elementary 'calculations', it would be expedient to express the problem in terms of their variables. It can be done by expanding the total system state-function $\Psi(q_0,q_1,...,q_N)$ over the complete sets of eigenfunctions $\{\varphi_{kn_k}(q_k)\}$ for the 'functional' degrees of freedom $(q_1,...,q_N) = Q$ describing the 'internal dynamics' of 'operating' (computing) system elements (see eq. (6)), which leaves one with functions depending only on the selected 'structural' ('distributed') degrees of freedom $q_0 \equiv \xi$:

$$\Psi(q_0,q_1,...,q_N) \equiv \Psi(\xi,Q) = \sum_n \psi_n(\xi)\Phi_n(Q) \;, \tag{7}$$

where the summation is performed over all possible combinations of eigenstates $n \equiv (n_1,n_2,...,n_N) \equiv \{n_k\}$ and for brevity we have designated $\Phi_n(Q) \equiv \varphi_{1n}(q_1)\varphi_{2n}(q_2)...\varphi_{Nn}(q_N)$. It will be convenient, in most cases, to interpret variables $\xi$ as (generalised) common coordinates of the system configuration (element distribution), so that $\psi_n(\xi)$ from eq. (7) characterises the (eventually probabilistic) coordinate distribution of the *n*-th internal state of the elements, similar to corresponding electron wavefunction representations in solid state theory. The system of equations for $\{\psi_n(\xi)\}$ is obtained from eq. (3) after substitution of expansion of eq. (7), multiplication by $\Phi_n^*(Q)$ and integration over all variables $Q$ (taking into account the orthonormality of eigenfunctions $\{\varphi_{kn_k}(q_k)\}$):

$$h_0(\xi)\psi_n(\xi) + \sum_{k,n'}\left[V_{0k}^{nn'}(\xi) + \sum_{l>k}V_{kl}^{nn'}\right]\psi_{n'}(\xi) = \eta_n\psi_n(\xi), \qquad (8)$$

where

$$\eta_n \equiv E - \varepsilon_n, \quad \varepsilon_n \equiv \sum_k \varepsilon_{n_k}, \qquad (9)$$

$$V_{kl}^{nn'} = \int_{\Omega_Q} dQ\, \Phi_n^*(Q) V_{kl}(q_k, q_l) \Phi_{n'}(Q) =$$

$$= \left(\prod_{g\neq k,l}\delta_{n_g n_{g'}}\right)\int_{\Omega_q} dq_k dq_l\, \varphi_{kn_k}^*(q_k)\varphi_{ln_l}^*(q_l) V_{kl}(q_k,q_l)\varphi_{kn_{k'}}(q_k)\varphi_{ln_{l'}}(q_l), \qquad (10)$$

$$V_{0k}^{nn'}(\xi) = \int_{\Omega_Q} dQ\, \Phi_n^*(Q) V_{0k}(\xi, q_k) \Phi_{n'}(Q) =$$

$$= \left(\prod_{g\neq k,l}\delta_{n_g n_{g'}}\right)\int_{\Omega_q} dq_k\, \varphi_{kn_k}^*(q_k) V_{0k}(\xi, q_k)\varphi_{kn_{k'}}(q_k). \qquad (11)$$

It would be convenient to separate, in eqs. (8), the terms with $n = n'$, corresponding to the 'mean-field approximation':

$$H_n(\xi)\psi_n(\xi) + \sum_{n'\neq n} V_{nn'}(\xi)\psi_{n'}(\xi) = \eta_n\psi_n(\xi), \qquad (12)$$

$$H_n(\xi) = h_0(\xi) + V_{nn}(\xi), \qquad (13)$$

$$V_{nn'}(\xi) = \sum_k\left[V_{0k}^{nn'}(\xi) + \sum_{l>k}V_{kl}^{nn'}\right], \qquad (14)$$

where $H_n(\xi)$ is the mean-field Hamiltonian. Taking into account the expressions of eqs. (10), (11), we can present the mean-field interaction potential, $V_{nn}(\xi)$, in the following form:

$$V_{nn}(\xi) = \sum_k\left[V_{n_k}(\xi) + \sum_{l\neq k}V_{n_k n_l}\right], \qquad (15)$$

$$V_{n_k}(\xi) = \int_{\Omega_q} dq_k\, |\varphi_{kn_{k'}}(q_k)|^2 V_{0k}(\xi, q_k),$$

$$V_{n_k n_l} = \int\limits_{\Omega_q} dq_k dq_l \left|\varphi_{k n_k{}'}(q_k)\right|^2 \left|\varphi_{l n_l{}'}(q_l)\right|^2 V_{kl}(q_k, q_l) \; ,$$

which demonstrates explicitly the origin of averaging in the 'mean' interaction structure. Note also that due to the pairwise interaction character in the initial existence equation leading to δ-symbols appearance in eqs. (10), (11), the summation over $n'$ in eqs. (12) will contain only the corresponding sums over $n_k{}'$ and $n_l{}'$. However, we shall not show it explicitly, so that eqs. (12) apply formally to any kind of initial many-body interaction. It is important that the obtained problem formulation in terms of the system of equations for the functions, $\psi_n(\xi)$, describing all possible interactions between the 'normal modes' of system components remains valid not only for the main case of quantum-mechanical system considered here, but also for any system with interaction, where the above canonical expressions for quantum 'scalar products' and 'matrix elements' should be replaced by respective expressions for that particular system, which should always exist for the correctly formulated problem (including perfectly known dynamics of the system components and their interaction within the system).

In view of this universality, it is not surprising that the obtained many-body problem formulation, eqs. (12), is mathematically equivalent to the one for the unreduced interaction between only two (structured) entities, eqs. (5), taken as the basis for the universal science of complexity and its various applications [1-4]. Indeed, it is sufficient to consider that indexes $n$, $n'$ in eqs. (12) number the 'normal modes' of one of the two interacting components (corresponding to the generalised Hamiltonian $h_e(q)$ in eq. (5b)) in order to interpret eqs. (12) in terms of this simplest case of unreduced interaction process. On the other hand, one can easily imagine that the normal modes of each of the two interacting entities are subdivided into arbitrary groups of 'subentities', so that one returns to the general case of the many-body problem. The key point here is that 'everything interacts with everything', within *any* real, *unrestricted* interaction process, so that the formulation of a problem of mutually interacting 'entities' and their 'modes', remains always the same, while the particular 'mode' grouping in 'entities' and their numbering change for each individual case, reflecting the initial system configuration or other convenient choice. One important aspect of this equivalence is that all the essential, qualitatively important effects of the unreduced interaction that will be revealed below in the general solution for the arbitrary many-body problem should emerge already in the unreduced interaction between two entities, which implies that the irreducible dynamic randomness and nonunitarity of the system evolution (see below, Sections 3.3, 4.1) characterise already *each* interaction act, determining operation of an elementary computer 'gate', and then hierarchically re-emerge in interaction cycles involving all higher levels of dynamics and the computing system in the whole.

This universal hierarchy of universal properties of unreduced interaction processes has also a more general sense: it provides the totally adequate, 'exact' representation of the real world structure, from its most fundamental entities (elementary particles and fields) to the most elaborated dynamical systems (living and conscious beings) [1], where each lower-level interaction results provide the 'interaction components' for the next higher level. In the case of quantum devices, this dynamically emerging, and consistently derived, hierarchy of interaction processes has at least two more direct implications. The first one involves the causally understood classical behaviour that naturally emerges from complex quantum dynamics as a higher complexity level determined approximately by the elementary bound system formation (Sections 4.7, 5.3) [1-4,11-13], which permits us to naturally include those dynamic quantum-classical transitions into description within the same approach and formalism. The second direct manifestation of the universal hierarchy of complexity involves the complementary dynamic relation of quantum computation to the underlying (complex) interaction dynamics at the lowest quantum sublevels (internal elementary particles dynamics) providing the causally complete explanation of canonical 'quantum mysteries' (Sections 4.6, 5.3) [1-4,9-13] and thus preventing their arbitrary, incorrect and speculative 'use' in the unitary imitation of real, dynamically multivalued dynamics of quantum devices.

Before advancing towards the solution of eqs. (12), note that the above formalism does not contain any explicit dependence on time, which would correspond to a basically closed system that actually 'generates' its essential temporal changes by passing from one its emerging state-

'realisation' to another (Sections 4.3, 7.1) [1-4,12,13]. In the case of a computing system, this would correspond to computer dynamics considered in between the input and output operations involving time-dependent connections to the exterior world. One can also include the (classical) input/output units into the system, using analysis universality mentioned above, which seems to be reasonable taking into account the 'sensitive' dependence of quantum/complex system dynamics on 'external' influences and may not permit the definite separation between the 'computer' itself and the first, immediate input/output structures. Thus, a quantum computer, or 'intelligent sensor', used in practice for a micro-structure/environment monitoring at the level of single atomic/molecular species and respective 'quantum' events, will certainly constitute an 'indivisible whole' with the medium/process it controls, so that the usual division between 'production', its 'control' and related 'computation' disappears in principle, transforming the 'quantum computation' process into a qualitatively new kind of adventure closely related to its dynamically multivalued (chaotic) character (see also Section 7.3). This sort of dynamic disappearance and reappearance of time in system description is partially imitated by a relation between 'time-dependent' and 'time-independent' approaches in usual quantum mechanics, considered to be generally equivalent and used, in particular, in scattering theory (see e. g. [119]).

However, as we discussed above (Chapter 2), the correct description of quantum computation is possible only through extension of the conventional unitary projection to the causally complete, dynamically multivalued picture leading to the intrinsic time emergence as a result of time-independent interaction development [1]. This does not prevent one from including the explicit time dependence into the same formalism in cases where the corresponding separation from the (changing) outside world dynamics is possible. In such cases one or several variables $q_k$ ($k \geq 0$) in the initial existence equation, eqs. (1), (3), will represent explicit dependence on time, $t$, entering through the corresponding interaction potential dependence and reflecting e. g. the input/output processes. The right-hand side of the existence equation is proportional to the time derivative of the state-function (for example, it equals to $i\hbar(\partial \Psi/\partial t)$ for the time-dependent

Schrödinger equation). When, however, we perform, according to eq. (7), expansion over eigenfunctions of the corresponding operator, like simple harmonics of the Fourier integral/series for the time derivative, we recover the above systems of equations, eqs. (8), (12), where the respective eigenvalues, $\varepsilon_{n_k}$, $\varepsilon_n$, $E$ and $\eta_n$ (see eq. (9)), will be proportional to frequencies of those temporal harmonical components (cf. refs. [1,9] and Section 5.2). Some of the variables $q_k$ are therefore replaced by time and the resulting system of equations preserves the same general form representing the usual 'Fourier analysis' of a time-dependent problem (other suitable system of time-dependent eigen-solutions can also be used).

## 3.2. Universally nonperturbative problem solution by the unreduced effective potential method

We proceed with the analysis of the system of equations, eqs. (12), describing the unreduced interaction between all elementary modes of the many-body system (numbered by $n,n'$), by separating the partial state-function, $\psi_n(\xi)$, for one of the modes and trying to express the problem in terms of equation for this function alone. In many cases it will be convenient, for example, to choose the system 'ground state' (the state with the lowest energy and dynamic complexity [1-4,11-13]) as this separated system mode, and we shall designate it, without limitation of generality, as $\psi_0(\xi)$ and suppose, correspondingly, that $n,n' \neq 0$ below. Separating the equation for $\psi_0(\xi)$ in eqs. (12), we can rewrite the main system of equations in the following form:

$$H_0(\xi)\psi_0(\xi) + \sum_n V_{0n}(\xi)\psi_n(\xi) = \eta\psi_0(\xi) , \qquad (16a)$$

$$H_n(\xi)\psi_n(\xi) + \sum_{n' \neq n} V_{nn'}(\xi)\psi_{n'}(\xi) = \eta_n\psi_n(\xi) - V_{n0}(\xi)\psi_0(\xi), \quad (16b)$$

where we have designated $\eta \equiv \eta_0$.

Keeping in mind the exact analogy of the main system of equations, eqs. (12), (16), with the basic analysis and particular cases of the universal science of complexity mentioned above, we can proceed by expressing $\psi_n(\xi)$ through $\psi_0(\xi)$ from eqs. (16b) with the help of Green functions for the 'homogeneous' parts of those equations and then substituting the result into eq. (16a) [1-4,9-13]. The mentioned Green functions are obtained for the 'truncated', or 'auxiliary', system of equations, not containing the terms with $\psi_0(\xi)$ on their right-hand side, contrary to eqs. (16b):

$$H_n(\xi)\psi_n(\xi) + \sum_{n'\neq n} V_{nn'}(\xi)\psi_{n'}(\xi) = \eta_n \psi_n(\xi) \ . \qquad (17)$$

The Green function for the equation for $\psi_n(\xi)$ from this system is given by the standard expression:

$$G_n(\xi,\xi') = \sum_i \frac{\psi_{ni}^0(\xi)\psi_{ni}^{0*}(\xi')}{\eta_{ni}^0 - \eta_n} \ , \qquad (18)$$

where $\{\psi_{ni}^0(\xi)\}$ and $\{\eta_{ni}^0\}$ are the complete sets of eigenfunctions and eigenvalues, respectively, for the truncated system of equations, eqs. (17). The well-known property of the Green function thus defined is that the solution of the full equation for $\psi_n(\xi)$ from system (16) can be expressed through the 'inhomogeneous' term on the right (containing $\psi_0(\xi)$) with the help of $G_n(\xi,\xi')$:

$$\psi_n(\xi) = -\int_{\Omega_\xi} d\xi' G_n(\xi,\xi') V_{n0}(\xi')\psi_0(\xi') \ , \qquad (19)$$

where $\Omega_\xi$ is the domain of the function under the integral. The direct substitution of this expression into eqs. (16b) confirms that $\psi_n(\xi)$ defined by eq. (19) is a solution of the unreduced system of equations.

Now if we substitute this expression for $\psi_n(\xi)$ through $\psi_0(\xi)$ into eq. (16a), we obtain a formally 'closed' equation for $\psi_0(\xi)$ that inevitably involves, however, the (unknown) eigen-solutions of the truncated system of equations:

$$[h_0(\xi) + V_{\text{eff}}(\xi;\eta)]\psi_0(\xi) = \eta \psi_0(\xi) \ , \qquad (20)$$

where the operator of the *effective* (interaction) *potential* (EP), $V_{\text{eff}}(\xi;\eta)$, is given by

$$V_{\text{eff}}(\xi;\eta) = V_{00}(\xi) + \hat{V}(\xi;\eta), \ \hat{V}(\xi;\eta)\psi_0(\xi) = \int_{\Omega_\xi} d\xi' V(\xi,\xi';\eta)\psi_0(\xi'), \quad (21a)$$

$$V(\xi,\xi';\eta) \equiv \sum_{n,i} \frac{V_{0n}(\xi)\psi_{ni}^0(\xi)V_{n0}(\xi')\psi_{ni}^{0*}(\xi')}{\eta - \eta_{ni}^0 - \varepsilon_{n0}} \ , \ \varepsilon_{n0} \equiv \varepsilon_n - \varepsilon_0 \ , \quad (21b)$$

and we have used the definitions of eqs. (9) and (13). The obtained 'effective existence equation' for the quantum many-body system, eq. (20), can be considered as an extended 'mean-field formulation' of a problem, where the EP $V_{\text{eff}}(\xi;\eta)$, eqs. (21), plays the role of *exact* 'mean field' produced by other system modes, which leads to its nonlinear dependence on the eigenvalues to be found, $\eta$, and the unknown eigen-solutions of the truncated problem, $\{\psi_{ni}^0(\xi),\eta_{ni}^0\}$. The latter dependence means that the problem is reduced to solution of a similar, but simpler problem, while the effective problem formulation itself, eqs. (20)-(21), expresses the remaining, essential part of the unreduced system dynamics through the mentioned interaction dependence on the problem eigenvalues, absent in the initial problem formulation.

Since the effective existence equation, eq. (20), is an equation for a single function depending on one variable, it can be solved (provided we have a suitable approximation for the auxiliary system solutions, see below, Section 4.4). Its eigenfunctions, $\{\psi_{0i}(\xi)\}$, and eigenvalues, $\{\eta_i\}$, should then be substituted into eqs. (19) to obtain other state-function components:

$$\psi_{ni}(\xi) = \hat{g}_{ni}(\xi)\psi_{0i}(\xi) \equiv \int_{\Omega_\xi} d\xi' g_{ni}(\xi,\xi')\psi_{0i}(\xi') ,$$

$$g_{ni}(\xi,\xi') \equiv V_{n0}(\xi') \sum_{i'} \frac{\psi_{ni'}^0(\xi)\psi_{ni'}^{0*}(\xi')}{\eta_i - \eta_{ni'}^0 - \varepsilon_{n0}} ,$$

(22)

after which the total system state-function $\Psi(q_0,q_1,...,q_N) = \Psi(\xi,Q)$, eq. (7), is obtained as

$$\Psi(\xi,Q) = \sum_i c_i \left[ \Phi_0(Q) + \sum_n \Phi_n(Q)\hat{g}_{ni}(\xi) \right] \psi_{0i}(\xi) , \qquad (23)$$

where the coefficients $c_i$ should be found from the state-function matching conditions at the boundary (and/or time moment) where the effective interaction vanishes, so that the system has a well-known configuration. The state-function of eq. (23) provides thus the 'general solution' for the many-body problem in terms of eigen-solutions of the effective dynamic (existence) equation rigorously derived from the starting existence equation, eqs. (1), (2)-(5) (taking the form of Schrödinger equation for the quantum sys-tem of interacting entities). The main observed quantity represented by the (generalised) system density, $\rho(\xi,Q)$, is obtained then from the total state-function as its squared modulus, $\rho(\xi,Q) = |\Psi(\xi,Q)|^2$ (for 'quantum' and other 'wave-like' levels of complexity), or as the state-function itself, $\rho(\xi,Q) = \Psi(\xi,Q)$ (for 'classical', 'particle-like' levels of complexity) [1].

The presented problem 'solution' is nothing but its another formulation, since to obtain it explicitly one needs to know the solutions of the auxiliary system of equations, eqs. (17), which is only slightly simpler than the full system of equations, eqs. (12), (16). However, this new problem formulation in terms of effective dynamic equation, eqs. (20)-(21), has a nontrivial significance, since it does reveal a qualitative novelty introduced into system dynamics by the unreduced interaction development and hidden in the starting, 'general' problem formulation, eqs. (1), (2)-(5), (12), (16). This fundamental feature, the dynamic redundance, or multivaluedness, phenomenon [1-4,8-13], is expressed by the self-consistent, nonlinear EP dependence on the eigenvalues to be found, eqs. (21), reflecting the unreduced interaction development and considered in detail below. It is interesting to note that the EP method and its expressions similar to those obtained above are well-known in scattering theory and its various applications, including solid state theory [120,121], under the name of optical, or effective, potential method. However, the fundamental meaning of the unreduced EP dependence on the eigenvalues remains undiscovered and the related general problem solution is replaced by perturbative approximations to EP that reduce interaction influence to trivial 'small corrections' for the observed quantities and thus 'kill' any complexity manifestations, including the feedback EP dependence on eigenvalues. In particular, such is the conventional mean-field approximation widely applied in various versions of the many-body problem, where the exact 'mean field' of the unreduced EP, possessing the intrinsic structure-creating properties (Sections 4.2, 4.3, 7.1), is simplified down to interaction values averaged over fixed state-functions for absent, or simplified, interaction. This basic deficiency of the canonical, invariably perturbative, theory and approach gives the false unitarity, i. e. uniformity, of the resulting system evolution and underlies, in particular, illusive possibility of unitary quantum device operation (in the low-noise limit).

## 3.3. Dynamically multivalued interaction result as the unified origin of randomness and the a priori probability values

If now we consider the properties of the unreduced EP solutions, eqs. (20)-(21), it will not be difficult to see [1,2,8-13] that the presence of the multi-branch EP dependence on the eigenvalue to be found, $\eta$, leads to multiplication, or dynamical splitting, of eigen-solutions of eq. (20), and thus of the whole problem, with respect to their 'normal' quantity, implied by the ordinary problem formulation (e. g. eqs. (1), (2)-(5), (12), (16)). Indeed, if $N_\xi$ is the number of eigen-solutions of eq. (20) for a generic, ordinary potential that does not depend on $\eta$ (such as $V_{00}(\xi)$, the first term in the unreduced EP expression, eq. (21a)), then the total solution number for the unreduced, $\eta$-dependent EP of eqs. (21), $N_{q\xi}$, should be at least as great as $N_{q\xi} = N_\xi N_{\text{eff}}$ (without special, 'transient' realisation solutions, see section 4.2), where $N_{\text{eff}}$ is the number of terms in the sum over $n$ and $i$ in the unreduced EP expression, eq. (21b). This follows from the fact that each such term increases the maximum power of the characteristic equation for eigenvalues, $\eta$, by $N_\xi$ and adds a series of branches of potential dependence on $\eta$ in the graphical representation of this equation, providing separate solutions by intersection with the linear dependence on $\eta$ from the right-hand side of eq. (20). It is clear also that $N_{\text{eff}} = N_\xi N_q$, where $N_\xi$ and $N_q$ are the numbers of terms in summation over $i$ and $n$ in eq. (21b) determined, respectively, by the number of solutions for $\psi_n(\xi)$ of the auxiliary system of equations, eqs. (17) (generally equal to that for $\psi_0(\xi)$ found from the first equation of the total system, eqs. (16)), and the eigen-solution number for non-interacting system elements (see eqs. (6), (9)).

In the obtained total number of solutions for the unreduced EP formalism, $N_{q\xi} = (N_\xi)^2 N_q$, each corresponding subset of $N_\xi N_q$ solutions reflects a 'normal' solution multiplication in the system of equations with respect to solution number, $N_\xi$ (or $N_q$), for a single equation, due to appearance of additional degrees of freedom. It is clear from the above interpretation of $\xi$ as system element coordinates (see eq. (7)) that in general $N_\xi = N$, the number of system elements and therefore the 'normal' number of system solutions, $NN_q$, is simply equal to the number of all individual element states times the number of (interacting) elements, which seems as natural as any ordinary perturbative 'state splitting' effect. However, further multiplication of this 'normal' solution number by $N_\xi$ cannot be explained by any such 'evident' reasons or eliminated as a particular, spurious effect. Therefore, we see that the *unreduced* interaction development leads to emergence of a *redundant* number of eigen-solutions forming $N_\xi$ sets of *complete* 'normal' solutions, each of them giving the exhaustive system state description (this algebraically derived conclusion is confirmed by the graphical analysis of the same characteristic equation [1,9,10]).

Despite its nontrivial character, the *dynamic redundance (or multivaluedness)* phenomenon should look not as unnatural or too specific, but rather standard, inevitable manifestation of the unrestricted dynamics of (any) real interaction, though it is invariably omitted by the canonical theory. Indeed, already a rough estimate shows that if we have $N$ ($= N_\xi$) interacting system elements with $N_q$ states/modes per element (which corresponds to $NN_q$ of 'normally' splitted states), then the unreduced interaction between the modes ('everything interacts with everything') gives $NN_q$ states per element, or $N^2 N_q$ for the total number of states. However, the physical reality, including individual element capacities, does not change after the interaction is turned on, which means that each element and the system in the whole is forced to bear an excessive, *N*-fold redundant number of its states in the normal regime of fully developed interaction. By comparison, any conventional, perturbative description, such as various canonical 'mean-field' approximations, will always give $N_q$ states per element (or $NN_q$ 'splitted' states in the total) as a result of elements interaction with the *single* mean field artificially replacing *many* (*N*) influences of individual elements, which reveals the elementary, obvious origin of severe, fundamental deficiency of canonical, dynamically single-valued *imitations* of natural interaction dynamics within the unitary theory. Most curious is the fact that those unitary approaches readily recognise evident limitations of their perturbative schemes (absence of the general problem solution, divergence of perturbation series), but after that they implicitly yield to a naive, unjustified 'hope' that the unreduced, real problem solution is a certain 'continuous', purely quantitative refinement of the unitary scheme, preserving at least its 'main', qualitative features.[3]

In reality, the new *quality*, dynamic multivaluedness, does appear after transition from perturbative imitations to the unreduced interaction analysis. The latter shows, in particular, that each of the redundant solutions describes a complete, normal state of the compound system and therefore can be called system *realisation*, i. e. its really existing version. It is obtained by the unreduced EP method presented above, eqs. (20)-(23), in the form of the corresponding set of eigen-solutions of the effective existence equation, eqs. (20)-(21), which is substituted then into the expression for the total system state-function, eqs. (22)-(23). We shall attach index *r* to values referring to the *r*-th system realisation ($1 \leq r \leq N_\xi$), so that the complete set of eigen-solutions for the EP equation, eq. (20), can now be presented as $\{\psi_{0i}(\xi), \eta_i\} = \{\psi_{0i}^r(\xi), \eta_i^r\}$, describing, for variable *r* and *i*, the whole redundant set of eigen-solutions and, for each fixed *r* and variable *i*, the (ordinary) 'complete' set of eigen-solutions forming the *r*-th realisation. All the explicitly obtained system realisations have 'equal rights' for their appearance driven by the same interaction, but at the same time, being (locally) complete, they are mutually *incompatible* and therefore can appear only 'one by one'. This leads us to the *rigorously substantiated* conclusion that a system of interacting elements can exist only as the process of *permanent realisation change* driven by the main interaction itself and performed in the *dynamically, or causally, random* order. This dynamic, *causal randomness* of system realisation emergence originates in the dynamic redundance phenomenon and *means* that any of the equally probable, *explicitly obtained* system realisations *should* appear, with the corresponding *probability*, but cannot be stable with respect to other realisation emergence driven by the same, nonseparable interaction process.

---

[3] This kind of mechanistic 'intuition' leads eventually to the widely imposed belief in 'unreasonable', but actually only illusive, 'effectiveness of (unitary) mathematics in natural sciences' that finally turns out to be indeed unreasonable and gives rise to persisting 'mysteries' and inevitable impasse of the canonical fundamental physics (see also [1] and Chapter 9). A related another example of this illusive 'efficiency' of conventional 'mathematical physics', hiding a huge and real deficiency, is provided by the conventional 'uniqueness theorems' for the main dynamic equations that enter in direct contradiction with the above dynamic multivaluedness of the unreduced solutions of the same equations. The origin of this fatal 'mistake' of conventional theory can be easily revealed in the form of a characteristic vicious circle in the underlying logic: the 'desired' property, such as uniqueness (single-valuedness), is first *silently assumed* in the problem formulation (in this case by assuming the potential function single-valuedness) and then as if 'rigorously derived', or 'proved', by simple rearrangement of problem expression. One always obtains therefore exactly what one inserts from the beginning, within that kind of 'unreasonably efficient' but still evident trickery, which is a manifestation of the fundamental fruitlessness of the whole conventional, unitary science, as it was acutely emphasised by H. Bergson yet at the beginning of the twentieth century (see Chapter 9). One vicious circle inevitably gives rise to the whole growing family of similar logical tricks of unitary science, so that one is obliged, for example, to artificially insert the fundamentally absent randomness into conventional, dynamically single-valued chaos theory (Chapter 6) or formally postulate the unceasing flow and irreversibility of time (cf. Section 4.3), as well as all other natural properties of the unreduced dynamic complexity (Chapter 4), for the whole range of various observed systems and patterns of behaviour. The underlying deficiency of 'uniqueness theorems' clearly shows up e. g. as invariable perturbation series divergence and other problems of the same origin appearing at any attempt to explicitly *obtain* that 'unique' solution. However, the 'unique' and 'chosen' status of conventional science, definitely putting itself beyond any critics, leads to infinite persistence of all those evident blunders and the related 'unsolvable' problems and 'mysteries', in all courses of the 'omnipotent' scholar science (cf. Chapter 9) [1].

The dynamic multivaluedness of (any) unreduced interaction process provides thus the universal, purely dynamic and omnipresent origin and meaning of randomness in the world, at all its levels, from quantum systems to classical and higher-level objects [1], a situation that is very different from the formal, empirically based, 'intuitive' introduction of randomness in any field of conventional, unitary science, including all its imitations of 'dynamical chaos' and 'complexity'. In particular, the dynamic redundance and causal randomness emerge even in the total absence of 'noise' or 'dissipation' and for any real, even simplest and totally closed, system configuration. At the level of quantum systems with interaction (like 'quantum computer' or other quantum machine), causal randomness takes the form of the *true* quantum chaos (i. e. truly random quantum system evolution, as opposed to the 'intricate regularity' of conventional quantum chaos concept) [9] and causal 'quantum measurement (reduction)' process [10] avoiding the inconsistency and 'mysteries' of their conventional versions (see also Sections 4.6, 4.7, 5.3 and Chapter 6 for more details).

The derived dynamic multivaluedness and randomness of the unreduced problem solution can be expressed mathematically by presenting the measured system density, $\rho(\xi,Q)$, as the special, *causally, or dynamically, probabilistic* sum of respective individual realisation densities, $\{\rho_r(\xi,Q)\}$:

$$\rho(\xi,Q) = \sum_{r=1}^{N_\Re} {}^\oplus \rho_r(\xi,Q), \qquad (24)$$

where the sum is performed over the total number, $N_\Re$, of actually emerging (observed) system realisations numbered by $r$ (as follows from the above analysis, in general $N_\Re = N_\xi = N$, the number of interacting system elements) and the sign $\oplus$ serves to designate the special, dynamically probabilistic meaning of the sum, described above. The latter implies that (i) each *explicitly obtained* realisation, represented by its density $\rho_r(\xi,Q)$ in eq. (24), appears in a causally random order among other system realisations, (ii) realisation change process for a given system can never stop because it is *permanently* maintained by the main, driving system interaction and constitutes the very *essence* of any system *existence* (in the *causally obtained* space and time, Section 4.3), and (iii) transitions between normal, regular system realisations are performed through a special transient state (also obtained as a particular system realisation among causally complete set of solutions of the effective existence equation), which dynamically binds those regular realisations into the *single whole* of (complex) system dynamics and can be specified as the causal, unified extension of the conventional wavefunction, density matrix and distribution function concepts (Section 4.2) [1,4,12,13]. The density, state-function and EP for the *r*-th realisation are obtained by substituting the corresponding eigenvalues, $\eta_i^r$, and eigenfunctions, $\psi_{0i}^r(\xi)$, for their general versions in eqs. (21)-(23):

$$\rho_r(\xi,Q) = |\Psi_r(\xi,Q)|^2 ,$$

(25a)

$$\Psi_r(\xi,Q) = \sum_i c_i^r \left[ \Phi_0(Q) + \sum_n \Phi_n(Q) \hat{g}_{ni}^r(\xi) \right] \psi_{0i}^r(\xi) ,$$

$$\psi_{ni}^r(\xi) = \hat{g}_{ni}^r(\xi) \psi_{0i}^r(\xi) = \int_{\Omega_\xi} d\xi' g_{ni}^r(\xi,\xi') \psi_{0i}^r(\xi') ,$$

(25b)

$$g_{ni}(\xi,\xi') = \sum_{i'} \frac{\psi_{ni'}^0(\xi) V_{n0}(\xi') \psi_{ni'}^{0*}(\xi')}{\eta_i^r - \eta_{ni'}^0 - \varepsilon_{n0}} ,$$

$$V_{\text{eff}}(\xi;\eta_i^r) \psi_{0i}^r(\xi) = V_{00}(\xi) \psi_{0i}^r(\xi) +$$

(25c)

$$+ \sum_{n,i'} \frac{V_{0n}(\xi) \psi_{ni'}^0(\xi) \int_{\Omega_\xi} d\xi' \psi_{ni'}^{0*}(\xi') V_{n0}(\xi') \psi_{0i}^r(\xi')}{\eta_i^r - \eta_{ni'}^0 - \varepsilon_{n0}} .$$

Each system realisation is physically represented by, and observed as, a 'normal', inevitable interaction process product, or result, providing an observed system configuration. We just show that there are *always many* such *incompatible* realisations (interaction results or elementary system configurations) and therefore they are forced to permanently replace one another with an average frequency, which is usually comparable to (but somewhat lower than) lowest characteristic system frequencies within each realisation, since the realisation change process is governed by the same interaction that gives rise to the internal realisation dynamics. The plurality of realisations and their resulting change can be more or less visible externally, depending upon greater or smaller difference between the corresponding system configurations, which gives rise to two limiting cases of dynamic complexity, (global, or uniform) dynamical 'chaos' and 'self-organisation', or 'self-organised criticality', respectively (Section 4.5).

Consider a general case of quantum machine represented, for example, by a spatial arrangement of interacting 'elements', each of them possessing its own internal dynamics (in the absence of interaction), as described by eqs. (6). Then the state-function $\psi_n(\xi)$, defined by eq. (7), can describe the spatial distribution of the *n*-th system state (including, in principle, those of all elements), and the *r*-th realisation for all $\{\psi_n(\xi)\}$ can correspond to the obtained *r*-th version of (dynamically correlated) spatial distributions for all states taking the form, for example, of *n*-th state localisation around the *r*-th system element or another, dynamically emerging kind of centre (other, e. g. delocalised, types of realisation configuration can also emerge, of course, for various system interactions). It is not difficult to see that the *r*-th EP realisation, eq. (25c), will dynamically produce a confining potential well for the *n*-th state just around the *r*-th element (centre) corresponding to actually obtained concentration of the *r*-th realisation state-function, eqs. (25a,b) [1,10,11] (see also Section 4.3). Now, causally random transitions between realisations will correspond to dynamically chaotic, 'spontaneous' jumps of system states (represented e. g. by its considered *n*-th state) between various possible centres of localisation, correlated for different states and centres, as described by our *unreduced* general solution of eqs. (24), (25). This *causally complete, dynamically probabilistic general solution* and its interpretation provide also the detailed, causally complete understanding of the *generic* many-body system dynamics that can be useful especially in certain interesting cases of larger quantum system behaviour, such as atomic Bose-condensation and other highly 'collective', macroscopic quantum phenomena in condensed matter, including superconductivity and superfluidity, that remain 'mysterious' and basically 'nonintegrable' within the conventional theory (see Section 5.3(C)).

It is important to emphasize that the causally probabilistic sum of eq. (24) means not only that the emergence of each system realisation occurs in the dynamically probabilistic fashion, but also that realisations *unceasingly* replace one another, in the causally random order, under the influence of the same, driving interaction between system elements $V_{kl}(q_k,q_l)$, eq. (1), that forms each realisation structure, so that the universal and unique way of (any) system existence is a *dynamically random process* specified by the above results. It should be clearly distinguished from the widely spread imitations of the unitary 'science of complexity' and 'chaos theory' (such as 'unstable periodic orbits' or 'multistability' concepts) which, being unable to provide the truly dynamic origin and structure of randomness within their dynamically single-valued projection of reality, replace them with a formally inserted, postulated randomness, or stochasticity, of equally formally introduced, purely abstract 'states'. In the case of quantum system, or 'machine', the dynamically probabilistic sum of eq. (24) can be considered as the causally complete, essential extension of the formally introduced, abstract and fundamentally deficient 'density matrix' concept from conventional quantum theory. Contrary to the conventional density matrix, defined through purely mathematical, formally postulated elements from abstract (and linear) 'spaces', our dynamically probabilistic system density is a physically real quantity subject to an interaction-driven, permanent and chaotic change. Similar to those stochastic imitations of randomness from the conventional complexity theory, the canonical density matrix is nothing but a formal mathematical model describing a certain, arbitrarily guessed and postulated 'statistical mixture' of observed, but remaining unspecified, 'pure states'. Either dynamic origin of these states or chaotic, interaction-driven transitions between them fall definitely outside of the framework of unitary, artificially inserted 'stochasticity'.

Another important feature of the obtained general solution, eq. (24), is that the changing system realisations always remain *dynamically* joined together by a physically real transitional state, forming a special, interme-diate system realisation that provides the causally complete extension of the conventional wavefunction and related 'Born's probability rule' (as well as various abstract 'distribution functions' from the unitary stochasticity concept, including the density matrix, see Section 4.2). The involved internal structure of the unreduced general solution is a consistently *derived* result of the interaction process *dynamics* described by the Schrödinger equation, eqs. (1)-(5) (which is also rigorously derived from lower-level interaction dynamics [1,4,11-13], Section 7.1), and therefore the fundamental origin of system density randomness, hidden behind formally postulated and ambiguous 'decoherence' (or 'trajectory divergence') of the conventional theory, is in the purely dynamic, strictly internal and universal phenomenon of interaction-induced system splitting into many incompatible realisations. It can be compared to arbitrary postulation of mysterious 'transformation' of an ill-defined 'coherent' wavefunction into 'incoherent' mixture of conventional density matrix 'magically' induced by some ill-defined (and evidently quite variable) *external* 'noise', which creates, in particular, vicious circles characteristic of unitary science (cf. footnote 3 above in this Section): the postulated, external randomness is the source of 'obtained', but also unspecified density matrix randomness, or 'decoherence' of abstract space 'vectors', while the accompanying, equally postulated wavefunction reduction is 'confirmed' by the resulting density matrix 'localisation'. Correspondingly, the widely used conventional equations for the density matrix can at best provide extremely simplified, incorrect, regular imitations of the dynamically random realisation change process. The correct equation for the dynamically probabilistic system density distribution takes the form of extended, generally nonlinear Schrödinger equation for the generalised system wavefunction (or distribution function) of the corresponding complexity sublevel determining realisation probability distribution (Section 7.1) [1,4], and this equation should also be solved with the help of the same unreduced EP method leading to the new dynamic redundance and randomness, and so on. The fundamental deficiency of the conventional density matrix approach reflects the basic inconsistency of the unitary description of quantum computer dynamics often using this approach, since both are based on the perturbational, single-valued projection of real system dynamics.

According to the dynamic origin of randomness in the unreduced interaction development, the causally probabilistic sum over realisations of eq. (24) is accompanied by the *dynamically defined probability* concept and causally derived, *a priori* values of probabilities of the causally specified *events* of realisation emergence. Indeed, it becomes clear from the obtained unreduced solution of eq. (24) that all elementary, primary realisations have identical probabilities of emergence, equal to $1/N_\Re$, where $N_\Re$ is the total realisation number (generally equal to $N_\xi = N$, the number of interacting units or system elements). However, in the majority of practical cases these elementary realisations are distributed inhomogeneously (because of initial system inhomogeneity) and one often measures actually dense groups of elementary realisations containing their different numbers and forming the really observed, compound system 'realisations'. Therefore, in the general case the probability of *r*-th realisation emergence, $\alpha_r$, is determined by the number, $N_r$, of elementary realisations it contains:

$$\alpha_r(N_r) = \frac{N_r}{N_\Re} \left( N_r = 1,...,N_\Re; \sum_r N_r = N_\Re \right), \sum_r \alpha_r = 1 . \qquad (26)$$

These realisation probabilities are determined eventually by the dynamical boundary/initial conditions of the interaction process and therefore are related to the coefficients, $c_i^r$, in the total state-function expression, eqs. (25a): $|c_i^r|^2$ is proportional to $\alpha_r$, with the coefficient that is less important and determines only various internal eigen-state contributions within the *r*-th realisation [1,10-13]. The distribution of system realisation probabilities thus specified is related also, by the generalised 'Born's probability rule', to the system 'generalised wavefunction', or distribution function, from the higher sublevel of complexity that just describes 'averaged' system dynamics during its transition between realisations and satisfies the generalised Schrödinger equation (Section 7.1) [1,4] (it can be useful for $\alpha_r$ calculation in the case of large numbers of different realisation groups).

Note that the inhomogeneous distribution of realisation probabilities of eq. (26), as well as the detailed internal structure, or *texture*, of each particular realisation, eqs. (25), provide the way for *dynamic* and irreducibly *probabilistic* 'self-organisation' (structure formation) of real system configuration in the universal science of complexity [1] (see also Section 4.5.1). The dynamic redundance phenomenon realises thus the unique way of intrinsic and 'harmonious' mixture of regularity and randomness, order and diversity, function and probability within the *same* natural system structure, as opposed to their mechanistic insertion and 'superposition' in the unitary, dynamically single-valued imitations of complexity (including conventional, dynamically single-valued 'self-organisation').

As follows from eqs. (24), (26), the expectation, averaged value of the observed density distribution is obtained as

$$\rho_{\exp}(\xi,Q) = \sum_{r=1}^{n_{\Re}} \alpha_r \rho_r(\xi,Q) \, , \qquad (27)$$

where $r$ enumerates the actually observed (generally compound) system realisations with causal emergence probabilities $\alpha_r$ determined by eqs. (26) and $n_{\Re}$ is the observed realisation quantity.

It is important to emphasize the extended meaning of probability emergence in eqs. (24)-(27) related to its explicit dynamic origin and derivation. In particular, these relations, contrary to conventional versions of probability/randomness, have a well-defined dynamical meaning and can be rigorously specified not only for long enough observation including many individual *events* (each of them now being *rigorously* defined as individual realisation emergence), but also for any small number of events, each individual event, or even expectation of future events in their total absence (*a priori* probability determination from the first principles). This causally complete definition of *future* expectation value and probability of the *emerging* system configuration (*event*) or measured quantity, *irrespective* of the observed number of events, is a qualitative advance of the universal science of complexity with respect to the canonical unitarity (dynamically single-valued science), with its well-known and inevitable confusion about the ultimate origin of randomness and probability.[4]

---

[4] One should emphasize, in particular, the essential difference between the *dynamic* probability and randomness in the system behaviour thus obtained, eqs. (20)-(27), and mechanistic, rigidly fixed version of 'probabilistic', or 'indeterministic', properties introduced in conventional quantum computation schemes which try then to get rid of that uncertainty and move towards a quasi-deterministic calculation result. However, even the smallest dynamic probability, inevitably intervening in a developed interaction process, will necessarily destroy artificial correlations between mechanistic probabilities and revive the true, unpredictable randomness of the result (unless the quantum system is dynamically transformed into a classical, less uncertain configuration, now lacking, however, all 'advantages' of 'quantum' computation, see also Section 7.2).

# 4. Universal dynamic complexity, its properties and manifestations in micro-system dynamics

## 4.1. Universal concept of complexity and chaoticity by the dynamic redundance paradigm

The above first-principles derivation of the dynamic probability concept and values is only one manifestation of the essential extension with respect to the scholar theories of 'complexity', 'chaos' and probability, obtained in our approach due to the consistent dynamical, unreduced analysis of driving interaction processes. Since conventional theories always use perturbative, reductive, 'exact-solution' analysis of interaction, they inevitably get the same, dynamically single-valued, mechanistically fixed projection of reality devoid of any intrinsic randomness, creation and development. In order to compensate this evident deficiency with respect to their promise and observed real system behaviour, scholar studies on complexity, quite similar to other, 'non-complex' approaches of conventional, unitary science they pretend to transcend, insert the missing plurality, randomness, irreversibility and other readily observed properties artificially, by defining and postulating the 'corresponding', but actually fatally simplified, purely mathematical 'models' of the missing degrees of freedom, supposed to reproduce the lacking complexity features.

In the absence of truly dynamic, reality-based, universal and clearly specified origin of complex behaviour, the formally imposed mathematical rules and symbols can provide only an over-simplified, practically useless (or even dangerously misleading) imitation of natural system properties, confirming the fundamental limitation of canonical, unitary science already known from its 'non-complex' system description and speculatively 'rejected' for its 'reductive' character by the scholar 'science of complexity', which is actually based on the same paradigm of unitarity (dynamic single-valuedness). Such is the conventional chaos concept based on 'exponentially diverging' trajectories (incorrectly extended perturbation theory result [1]) supposed to 'amplify' the unexplained, externally introduced randomness, or 'noise'; or that of 'multiple attractors' ('multistability'), or 'unstable periodic orbits', *coexisting* in an *abstract* space artificially composed from continuous system trajectories (all the 'attractors' and 'unstable trajectories' of the canonical science of complexity are postulated, rather than consistently derived, starting from computer simulation results or simple 'mathematical intuition'); or that of 'self-organisation', or 'synergetics', operating with 'exact', perturbative and dynamically single-valued type of solutions; or that of (ordinary) 'fractals' obtained from mathematical recursive processes containing no intrinsic randomness, interaction dynamics and real system matter at all; or other irreducibly separated concepts of the unitary science of complexity showing evident inconsistency and inability to provide even a clear definition for at least the main quantities, like complexity itself, (true) randomness and chaos (further details can be found in refs. [1,5,6]). It is important to keep in mind the essential difference between those mechanistic, basically incorrect simulations of complexity by the official unitarity and the reality-based, consistently derived and intrinsically unified concept of complexity and chaos within the dynamic redundance paradigm, especially when such 'sensitive' applications as quantum interaction dynamics are involved. In that way one can avoid the 'strange' combination of a general 'complexity talk' and explicit unitarity domination that occurs unfortunately just in the conventional theory of quantum information processing, despite the well-known, fundamental deficiencies of conventional 'science of complexity' (see also Chapter 7). A clear manifestation of the imitative character of both conventional 'complexity' (or 'chaos') and its 'applications' in conventional quantum computation appears in the fact that all those theories and approaches try eventually to *get rid* of the true complexity and chaoticity (i. e. *avoid any deviation from unitarity*) and reduce system behaviour and description to the explicitly predictable, effectively zero-dimensional, 'exact-solution' type with the zero value of genuine dynamic complexity (we leave apart yet more ambiguous, 'post-modern', purely verbal 'interpretations' of pseudo-philosophical or computer-assisted empiricism densely entangled with the conventional theory of quantum computation, see Chapters 5, 7, 9).

    Our *universally* applicable *definition of the unreduced dynamic complexity* of a (real) system (of interacting entities) is based on the main, consistently derived property of dynamic multivaluedness of emerging system realisations. Dynamic complexity, $C$, can be defined [1-4,9-13] by any

growing function of the number of system realisations $N_\Re$ (Section 3.3), or related rate of their change, equal to zero for the (unrealistic) case of only one realisation, just exclusively considered in conventional science, including its complexity imitations. Mathematically, $C = C(N_\Re)$, $dC/dN_\Re > 0$, $C(1) = 0$, where realisations and their actually observed number (usually at a given complexity level) are determined by the unreduced system interaction dynamics, in agreement with the generalised, nonperturbative EP method (Chapters 3, 4, 5) [1,9,11,13]. Note in this connection, that any formal, postulated 'counting' of the observed structure elements, often used for 'intuitive' complexity definition in scholar theories, cannot replace the number of incompatible system realisations obtained by consistent analysis of its unreduced interaction, since it is the internal dynamic origin and connection between realisations that are important for any nontrivial manifestation of complexity. One can also be easily mistaken by formally counting observed structure elements, since the unreduced interaction development gives rise to the whole hierarchy of dynamically multivalued objects (the unreduced dynamical fractal, Section 4.4) and one should clearly understand which structure elements should be taken into account at the current (considered) level of complexity. At the same time, each realisation has its own internal structure that forms, however, a dynamically single whole, one irreducible complexity element (at the respective level of complexity). Therefore quantities proportional to $\ln(N_\Re)$ (generalised dynamic entropy), or to $\partial N_\Re/\partial t$ (generalised energy-mass), or to $\partial N_\Re/\partial x$ (generalised momentum) [1-4,9-13] constitute correct measures of complexity only if the (observable) realisation number, $N_\Re$, is explicitly obtained together with realisations themselves, the relevant time and space structure, in the unreduced analysis of the driving interaction.

As follows from the above dynamic randomness concept, the dynamic complexity thus defined expresses *simultaneously* the property of (dynamic) *chaoticity* involving the *intrinsic*, true randomness and present within *any* real dynamical system (*always* having a non-zero dynamic complexity), as opposed to mechanistic division of the world into 'complex' and 'non-complex', 'chaotic' and 'regular' behaviour, systems and their parts, within the conventional 'science of complexity'. For example, the genuine dynamic randomness has been revealed, within the dynamic redundance paradigm, in the behaviour of essentially quantum systems with interaction [1,8-10] having direct relation to the quantum computer case and described by the above formalism, if the Schrödinger equation is used as the starting existence equation, eq. (1) (see also Section 5.2.1). One obtains, in that way, simultaneous resolution of the intrinsic inconsistency of the canonical 'quantum chaos' theory opposed to (true) randomness emergence in quantum dynamics (see Chapter 6 for more details). The real dynamical chaos becomes therefore another, synonymous expression of the unreduced dynamic complexity of a system (interaction process), emphasising the property of (genuine) dynamical randomness inherent in any real, complex entity. This does not prevent the intrinsic randomness from being dynamically 'squeezed', confined to an externally quasi-regular structure, in the 'self-organisation' regime of unreduced complexity/chaos [1] (highly inhomogeneous realisation probability distribution in eq. (27)), as opposed to the conventional, basically regular version of 'self-organisation' containing only one, averaged 'realisation' and excluding the complementary manifestation of the same dynamic complexity, true randomness (see also Section 4.5.1).

## 4.2. Dynamic entanglement, causal wavefunction and the internal structure of real interaction processes

Several other important properties of the unreduced dynamic complexity are inseparably related to the above dynamic redundance (causal randomness) phenomenon, thus confirming the nontrivial, unifying and universal meaning of the dynamically derived interaction complexity. Most important is the phenomenon called *dynamic entanglement* of interacting entities [1-4,10-13] and consisting in the physically real, dynamically driven entanglement (mixing, intertwining) of interacting entities as they are represented by all their elementary modes. We deal here with the causally specified, extended meaning of the unreduced, 'nonseparable' character of a real *interaction* process and its result, where 'everything interacts with everything else' and therefore the emerging 'fine mixture' of interaction components within the system cannot be artificially separated or represented as a 'simple sum of system parts' (the last property is often quoted, but never specified, in the canonical 'science of complexity' and 'systems theo-ry'). Dynamic entanglement is expressed mathematically by the sums of products of eigenfunctions for individual modes of interacting components, depending on one of $\xi$ and $Q$ variables, in the expression for the total system state-function, eqs. (23), (25). However, contrary to the canonical, mechanistic version of 'quantum entanglement', these 'cross-products' of the component eigenfunctions also contain irreducible and dynamically meaningful resonance factors just reflecting the *interaction-driven*, essentially *nonlinear* (Section 4.3) origin of the *real* entanglement phenomenon unifying it with the accompanying properties of causal randomness, essential nonlinearity, dynamic instability, fractality and reduction (catastrophic dynamical collapse, or squeeze) of system configuration towards that of each current realisation (see Sections 4.2-4.4 for further details). The 'natural', dynamic origin of this entanglement, as well as its nonperturbative, nonlinear character can be seen explicitly from eqs. (23), (25) based on the eigen-solutions of the main dynamic equation, in its effective form, eq. (20). The direct relation with the dynamic redundance phenomenon becomes thus also evident.

Taking into account the simple physical origin of dynamic multivaluedness mentioned above (Section 3.3), we can describe the essence of any real, unreduced interaction process as *dynamically redundant entanglement* (of interacting components) meaning that as interacting entities and their modes are driven to intertwine with one another, one inevitably obtains the redundant number ($N_\xi = N = N_\Re$) of possible versions (detailed configurations, or realisations) of this entanglement, which leads to unceasing dynamical change of entanglement realisations within the system, constituting the essence (and the sense) of system existence, the unreduced internal 'life' of its 'organism' (including all levels of the emerging dynamical fractal, see Section 4.4). The realisation change can only happen through the reverse process of transient *disentanglement* of the last realisation structure followed by re-entanglement into the next, equally probable realisation 'selected' by the system at random among $N_\Re$ possible versions of entanglement.

During the realisation change process, constituting the true content of each real interaction, the system needs thus to pass each time though a particular state, where the interaction components become transiently dis-entangled and thus recover their quasi-free state existing, hypothetically, in the degenerated 'system' state with 'separated' components, i. e. in the absence of interaction between them. This special transient state of a system, called its *main*, or *intermediate, realisation* [1,4,11-13], is given by the same unreduced version of the effective existence equation, eq. (20), in addition to other, 'regular' realisations that we have been counting before, in Section 3.3 (their number is $N_\xi = N = N_\Re$ and each of them forms a 'genuine', characteristic system configuration with 'strong', irreducible interaction, where the system components are tightly intertwined, or 'entangled'). Indeed, it is easily seen from eqs. (20)-(21) that the highest power of the characteristic equation for $\eta$, determining the *maximum* number of solutions, $N_{\max}$, is given by $N_{\max} = N_\xi(N_{\text{eff}} + 1) = N_{q\xi} + N_\xi$, where $N_{\text{eff}} = N_\xi N_q$ and $N_{q\xi} = N_\xi N_{\text{eff}} = (N_\xi)^2 N_q = N_\Re N_\xi N_q$ is the redundant solution number considered above (Section 3.3), originating from the EP dependence on $\eta$ (that contributes the value of $N_{\text{eff}} = N_\xi N_q$ to the maximum power of $\eta$ in the characteristic equation) and giving rise to emergence of $N_\Re = N_\xi$ 'regular' realisations (each of them contains the locally complete number, $N_{\text{eff}} = N_\xi N_q$, of elementary eigen-solutions). However, the really complete number, $N_{\max}$, of elementary solutions (and maximum $\eta$ power) includes, in addition to those $N_{q\xi}$ 'regularly redundant', mutually equivalent solutions giving $N_\Re$ regular realisations, a supplementary set of $N_\xi$ solutions appearing due to the presence of $\eta$ also in the right-hand side of eq. (20), at its 'normal' place occurring in any conventional form of the existence equation. As it is especially clearly seen in the graphical representation of the characteristic equation for eq. (20) (we do not show it here for brevity, see [1,9,10]), these additional solutions are characterised by small values of the essential EP part described by the nonlocal operator $\hat{V}(\xi;\eta)$ in eqs. (21), so that for them $V_{\text{eff}}(\xi;\eta) \cong V_{00}(\xi)$ and the effective system dynamics is described rather exactly by the first-order mean-field approximation corresponding to the practically free, though maybe 'renormalised', motion of non-interacting system components. Therefore these particular solutions form a special, intermediate, system realisation that just corresponds to chaotic system jumps between regular realisations implementing the essential, 'strong' part of the interaction process and representing its full-scale, properly entangled, 'nonseparable' results. This specific, transient and quasi-free character of intermediate realisation dynamics is confirmed by the exceptionally low number of elementary solutions it contains, $N_\xi$ instead of $N_\xi N_q$ for regular realisations, which reflects the effectively absent interaction/entanglement. At the same time the system in the intermediate realisation moves in the mean field obtained by interaction averaging over all the (localised) component states, which points to the delocalised system configuration in this transitional realisation. It is not difficult to see also that it corresponds approximately to the single, averaged realisation exclusively taken into account in the conventional, dynamically single-valued, perturbative interaction analysis (that's why we also refer to the intermediate realisation as the 'main' system realisation), which explains both limited correlations of the averaged unitary projection with reality and its irreducible faults formally expressed by perturbative expansion divergence and physically reflecting the neglected, but inevitably occurring emergence of many regular, internally entangled realisations. The unreduced interaction picture of the universal science of complexity considerably extends this effectively zero-dimensional projection of the canonical theory by revealing the true role of its unique solution as that of only transitional, chaotically fluctuating system state during its jumps between the 'normal', 'strong' results of interaction development represented by regular realisations with localised configuration and 'seriously', inseparably entangled system components.

With this character of system motion in the main realisation, it is not difficult to understand that it provides the causal, totally *realistic* extension of the conventional *wavefunction* notion, which can be generalised to *any* interaction process and thus any level of (complex) world dynamics. The fundamental quantum-mechanical wavefunction itself is obtained as the intermediate realisation at the lowest level of the world interactions, that of the two coupled, initially structureless protofields (represented by two physically real, electromagnetic and gravitational, media), which gives rise to the 'embedding' space and time, elementary particles/fields and all their 'intrinsic' properties, such as 'quantum' duality, relativistic motion dynamics, inertial/gravitational mass-energy, electric charge, spin and the four unified 'fundamental interaction forces' (all these features emerge dynamically, as intrinsically unified, physically real results of unreduced interac-tion development) [1-4,11-13]. In this work we apply the same analysis to higher sublevels of complexity corresponding to interaction between elementary particles/fields thus formed and their emerging simplest agglomerates. At these sublevels of complexity we obtain the true quantum chaos (described basically by the same causally probabilistic realisation change process, eqs. (24)-(27)) [1,8,9], causally complete picture of quantum measurement/reduction for slightly dissipative (open) systems [1,10] and the dynamic emergence of classical (essentially localised) micro-systems [1-4,12,13]. The generalised wavefunction always corresponds to causally random system jumps between its configurations in regular realisations and provides, for the mentioned sublevels of 'quantum' dynamics, the causally complete extension and modification of the canonical 'density matrix' postulated, together with its formalism, as a simplified, purely mathematical entity and having no realistic interpretation (similar to the conventional wavefunction at the lowest complexity level).

We can see now that the consistently obtained wavefunction, or distribution function, at those higher sublevels of quantum dynamics is given not by the system density itself in the regular realisations, $\{\rho_r(\xi,Q)\}$, but rather by the probability distribution of those (permanently changing) realisations. The latter is directly determined by the generalised wavefunction, according to the generalised Born's probability rule, while the wavefunction satisfies the universal, generally nonlinear Schrödinger equation (Section 7.1) [1,4]. The detailed form of this equation depends on the particular system considered (although a small number of standard forms can be sufficient for any real system description) and we shall not further develop here this aspect of the universal formalism of the unreduced theory of complexity [1,4]. It becomes quite evident, however, that the standard density matrix formalism gives only a fatally simplified, single-valued and incorrect imitation of real quantum system dynamics and therefore cannot provide any reliable, or even qualitatively correct, result in description of any quantum device (or a natural 'micro-machine'), which should be compared to its extensive use in the canonical theory of quantum computation (see references from Chapter 2) and other modern applications of scholar quantum theory to various versions of quantum many-body problem. It is important to emphasize that the causally extended wavefunction of the universal science of complexity has not only its 'central', probabilistic inter-pretation (which is now *dynamically* derived by the unreduced interaction analysis), but also a tangible, *physically real* (causal) meaning of the unified, quasi-free system state during its *chaotic* jumps between realisations. All these unique properties of the generalised wavefunction (distribution function), explicitly derived and causally understood within the dynamic multivaluedness paradigm, clearly demonstrate the essential difference of the universal concept of complexity from its various imitations in the canonical science, where such quantity cannot even appear (i. e. be realistically interpreted or even related to the underlying complex interaction dynamics), as it happens in the canonical quantum mechanics, where it is simply postulated empirically – and inconsistently.

It is important to emphasize, in connection to the above concepts of dynamic entanglement and generalised causal wavefunction, that the notion of 'quantum entanglement', extensively used in the conventional, unitary theory of quantum computation as one of its basic ideas, is different from the physically real entanglement of interacting components not only by its purely abstract, mechanistically fixed and therefore 'mysterious' origin (including that of the canonical 'wavefunction' or 'density matrix'), but also by its limitation to a much more narrow class of phenomena, where some interaction may at best be only implied behind the postulated character of its results. As noted above, our approach leads to the causally complete, realistic extension of 'quantum entanglement', where one can clearly see the mechanism and result of the interaction-driven entanglement between the system components and related chaotic change of its multiple realisations (see also Section 5.3).

## 4.3. Omnipresent dynamic instability, essential nonlinearity, generalised dynamical collapse, physical space, time and quantisation

We have shown that the reality-based approach of the universal science of complexity and its applications at lower complexity levels, designated as quantum field mechanics, provide a physically transparent picture of the origin of dynamically redundant multivaluedness of any unreduced interaction process, where 'everything interacts with everything else'. The detailed mechanism of gradual emergence of unstable, permanently chang-ing system realisations can also be traced in the unreduced problem solutions, eqs. (20)-(27). Namely, the obtained 'effective', essentially nonlinear problem formulation, eqs. (20)-(21), shows that any real, even externally 'linear' and 'simple' interaction process (cf. eqs. (1)-(5)) is characterised by the irreducible, omnipresent *dynamic instability*, which naturally emerges with interaction process development in the form of dynamic *feedback loops* of interaction described by the self-consistent EP dependence on the eigen-solutions to be found ($\eta, \psi_n(\xi)$), eqs. (21). Such (positive) feedback existence leads to system instability with respect to its self-amplified collapse (reduction, or squeeze) towards a 'spontaneously' (dynamically) emerging and randomly chosen configuration, or 'realisation', when any small fluctuation of a 'free', delocalised system state in the direction of one of (future) realisations leads, in agreement with eqs. (25), to self-consistent formation of the EP 'seed' that tends to increase the fluctuation it results from, which leads to further growth of EP, and so on, until the fully developed EP well and the corresponding realisation it confines are formed in this self-amplifying, avalanche-like process, or 'collapse' of the system (it is limited eventually by 'reaction' forces always implied by a physically real interaction between system components). The process looks like effective 'self-interaction' in a quasi-homogeneous system, which leads to 'spontaneous' violation of homogeneity and 'catastrophic', essentially nonlinear structure formation.[5]

In that way one obtains the well specified, dynamically based and universal origin and definition of nonlinearity due to the interaction feedback loops formation revealed due to the 'effective' formulation of the unreduced problem, eqs. (20)-(25) and remaining hidden within its ordinary, starting formulation (see e. g. eqs. (1)-(5)). We call this universal mechanism and process of dynamic instability of *any* interaction process the *es-*

---

[5] This 'dynamical self-interaction', consistently derived from the ordinary, simple interaction between several entities, should be distinguished from its imitation in conventional theories by the direct, mathematically postulated 'self-interaction' of a single entity (e. g. a field) and related ill-defined 'nonlinearity' whose dynamic origin, mechanism and detailed structure remain unknown, and together with them the unreduced results of interaction development, such as dynamic redundancy, entanglement and fractality (see below). The same refers to 'feedback' connections between the interacting entities or processes, which are artificially, explicitly inserted into postulated 'models' of conventional 'science of complexity' in their 'ready-made', non-dynamical and therefore simplified form that does not reveal the main property of *creativity* of natural interaction processes (see also below).

*sential, or dynamic, nonlinearity*, in order to distinguish it from the ill-defined, mechanistic 'nonlinearity' of conventional, unitary science (and its versions of 'complexity'), which is usually inserted into perturbative, dynamically single-valued analysis artificially, through a particular, more 'uneven' functional dependence of an equation term, etc. Any such ordinary 'nonlinearity' looks quite *linear* in our approach, since it gives rise to the dynamically single-valued, perturbative reduction of real system behaviour, even though it can modify the 'zero-order approximation' and provide an externally 'curved' and mechanically 'intricate' (but always artificially inserted) imitation of the dynamically emerging, internally multivalued structure of real entities (a characteristic example is provided by the canonical solitons, being *exact*, single-valued solutions of formally 'nonlinear' equations).

The unambiguously defined, essential nonlinearity is a property of the *unreduced development* of *any* interaction process that leads inevitably (through the dynamic instability) to the fundamental dynamic multivaluedness of interaction results and can be considered therefore as the universal *mechanism* of dynamic redundance and causal randomness. It is easy to see from the above physical picture of each realisation formation by the self-amplifying, avalanche-like development of the omnipresent interaction instability that the 'essential' nonlinearity, due to its *purely dynamic* origin, is indeed qualitatively 'more nonlinear' and 'obtrusive' (unavoidable) than any its unitary imitation. It becomes clear also why and how the essential nonlinearity and related dynamic instability, causal randomness and autonomous structure formation are invariably and totally 'killed' by the perturbative interaction reduction in the conventional theory: the latter, including its mechanistic *imitation* of nonlinearity, just cuts the essential dynamical links within the interaction development (including reduction of EP dependence on the eigen-solutions) and tends to inconsistently 'jump' immediately to its observed results postulated with the help of mechanistic 'models', after which nothing 'interesting' can happen to the system with effectively destroyed interaction. Unfortunately, but not surprisingly, it is just such false, mechanistic 'nonlinearity' (and similar zero-dimensional 'complexity') that is used invariably in the conventional, unitary theory of quantum computation (see e. g. [43,58,60,62,64,81]), with the fatal conse-quences for its relation to real physical systems (see Chapters 5-7 for more details).

This dynamically 'orchestrated' formation of structure of each of system's realisations is directly reflected in the obtained state-function and EP expressions for individual realisations, eqs. (25): one can see that the expressions for both state-function and EP for a given realisation contain essentially the same singular terms combining resonant denominators with 'cutting' integrals in the numerators, which express mathematically the self-consistent and physically real system 'collapse', i. e. its catastrophic dynamical squeeze, towards this particular realisation [1,10-13]. This transient 'reduction' of system configuration to that of its currently taken realisation can be considered as a result of the dynamic entanglement phenomenon (Section 4.2): the more the system components entangle with each other, the more is their mutual attraction, and vice versa, which gives the dynamic instability of any real interaction process and resulting system collapse towards its current realisation. The whole process is replayed in the reverse direction during the following phase of dynamic disentanglement, when the last system realisation is catastrophically destroyed by attraction towards other possible configurations and centres of reduction, passing first by the intermediate realisation (generalised wavefunction), after which the next realisation of dynamic entanglement emerges (it is always described by the same expressions of eqs. (21), (23), (25) but with the eigenvalues and eigenfunctions corresponding to the new, currently formed realisation). The observed, real system structure always consists thus of many internal 'sub-structures' corresponding to individual realisations that permanently replace one another in a dynamically random order, even when this change cannot be explicitly discerned within the observed system configuration. Such is the essential, qualitative extension of conventional 'self-organisation' (structure formation) description in the universal science of complexity (see also Section 4.5.1).

Since we did not explicitly introduce spatial and temporal inhomogeneities created by the emerging realisations, it becomes clear also that the realisation formation and change process provides the detailed, purely dynamic mechanism of emergence, and the *causal meaning* itself, of physically real *space* and *time* of the corresponding level of complexity: space structure is dynamically 'woven' from the dynamically entangled interac-tion participants, especially within the emerging inhomogeneities of regular system realisations (connected by the 'shuttle' of intermediate realisation, or generalised wavefunction), whereas the unceasing, dynamically random and qualitatively 'strong' realisation *change* creates the permanently advancing, *intrinsically irreversible* sequence of well-defined *events* constituting the perceived 'flow of time', which measures itself by the essentially nonlinear 'pendulum' of changing system realisations. The real, complex-dynamical time is thus naturally, *dynamically irreversible* by its very origin inseparably related to the causal randomness emergence, whereas space is *dynamically discrete*, or 'quantised' (see also below in this Section). It becomes evident also that, in contrast to conventional fundamental physics (and especially its 'relativity' version), the real space thus obtained is a physically tangible, 'material' entity just determining the perceived 'texture' of reality, while time, being equally real, is not a material entity, since it characterises a real *action* (realisation change) within an object, rather than the object matter itself. It is clear then that any direct mixture between space and time within a single, mechanistically fixed (though conveniently 'deformed') 'space-time manifold', constituting one of the corner-stones of conventional relativity and cosmology, has no physical sense and can be considered only as a technical tool of questionable validity and universality [1-4,12,13].

In correspondence with the hierarchical structure of complex interaction processes, the physically real space and time are also organised in a hierarchical sequence of levels, where the most fundamental, 'embedding' space and time (causal extension of respective 'Newtonian' concepts) result from the same lowest level of interactions, the protofield interaction process, that gives rise to elementary particles, fields and all their intrinsic properties [1-4,11-13]. The natural units of dynamically emerging space, determining its quantisation and dynamic origin, are given by configurations of emerging realisations and their closely spaced, 'self-organised' groups described, respectively, by the eigenvalue spacing within individual realisations (the intrinsic system 'size') and probability and eigenvalue distribution for different realisation groups (characteristic size of the average dynamic tendency, such as the de Broglie wavelength [1,2,12,13]). The natural unit of time is determined, of course, by the rate ('frequency', characterising 'intensity') of realisation change process, while its relation to the units of space gives the essential, observed system 'dynamics', in the form of generalised 'dispersion relations'. The practically useful measures of dynamical space and time can be quantitatively specified with the help of extended mechanical action representing a natural universal (integral) measure of complexity and information [1,3,12,13], but we shall not develop these details at this point (see Section 7.1).

The essential nonlinearity and dynamic instability within any real system with interaction result in a natural system auto-squeeze (or reduction, or collapse) towards the realisation it actually takes, which cannot be separated from the dynamic entanglement of interacting system components. The gradually entangling system components, forming eventually the dynamically fractal internal structure of each emerging realisation (Section 4.4), are the more attracted to each other, the more they are intermixed, and the reverse, which evidently leads to the dynamic instability considered above and the resulting realisation structure formation. We deal here with the *holistic* process of unreduced interaction development leading to quasi-periodic system 'fall' into one of its realisation, accompanied by the generalised, essentially nonlinear configurational squeeze and component entanglement and followed by the reverse disentanglement and extension towards a quasi-free state of generalised wavefunction until it 'collapses' again into the next realisation, and so on. An important consequence of this holistic dynamics, and a universal property of unreduced complexity, is the fundamental *dynamic discreteness, or (generalised) quantisation*, of realisation formation and change process. It results, basically, from the unreduced character itself of the driving interaction, where 'everything interacts with everything else' and therefore arbitrary, 'infinitesimal' changes cannot be fixed, even transiently (like the eventually emerging realisations), since they will immediately produce other displacements of system components, and so on, until the avalanche-like development of a randomly initiated tendency arrives at the completely formed system realisation (which is also 'stable' only transiently, with respect to nearby, yet less stable configurations).

As can be seen from the unreduced problem solution, eqs. (20)-(25), the resulting finite difference between the emerging structural elements (realisations) is determined by the driving interaction itself and therefore should be clearly distinguished from the simplified, mechanistically im-posed and arbitrarily structured 'discreteness' often used in conventional, dynamically single-valued theory and always present in its dominating, computer-assisted version. Such artificially inserted discreteness can rather be considered as an additional periodic influence arbitrarily introduced into the studied system and leading to the corresponding, more or less chaotic, system modification [1,9].

Another difference of natural dynamic discreteness (quantisation) from conventional, 'mathematical' discretisation is that the former, contrary to the latter, contains within it the unbroken dynamic and structural *continuity*: the dynamic discreteness is rather a manifestation of high, 'essential' inhomogeneity of the real interaction process closely related to the essential nonlinearity and expressing its causally specified *nonunitarity*. The internal continuity of the unreduced complex dynamics is directly expressed by the generalised system wavefunction (Section 4.2), which just 'joins together' the discretised system realisations by implementing system transitions between them. It is not surprising that the dynamic discreteness of the unreduced complex dynamics provides the causally complete explanation for the micro-world quantisation [1,3], including the physical origin and universality of Planck's constant and discrete character of all fundamental structures and properties, which are only axiomatically fixed and remain 'mysterious' within the conventional quantum mechanics and related theories. Since the dynamically emerging structure of physically real, tangible space and non-material, but equally real time is determined, as mentioned above, by realisation parameters, it becomes clear that all the above conclusions about the dynamically discrete, but internally continuous structure of the underlying interaction process refer also to causal space and time of the universal science of complexity (Section 7.1) [1-4,11-13].

The same group of inter-connected, dynamically emerging properties, including dynamic entanglement, generalised quantisation and essential nonlinearity, constitutes the direct expression of the crucial property of *creativity* of unreduced (complex) interaction dynamics. The multivalued dynamic entanglement reveals the internal 'content', or 'fabric', of any real structure, while the explicitly obtained discreteness of emerging objects is a necessary attribute of a real creation process, including the mentioned internal continuity and unceasing (self-developing) character (see also the next Section).

## 4.4. Probabilistic dynamic fractality, interactive adaptability and universal complexity development

One can express the main result of our analysis by saying that the basic origin of the well-known (and now rigorously substantiated) divergence of typical perturbation theory expansion in the conventional interaction analysis is the existence of the omnipresent 'singularity' of system 'branching' into multiple realisations, persisting however at *any* 'point' (moment) of system evolution. The 'convergence' of perturbative expansions can be as if reconstituted in our theory, but only at the expense of explicitly obtained dynamic multivaluedness that can be considered, in this sense, as the universal result of consistent 'summation' of perturbative expansions introducing, however, the *qualitative novelty*, dynamic redundance, in the whole system *mode d'existence* and thus providing any real system evolution with the *nonunitary*, uneven, jump-like, essentially nonlinear and dynamically chaotic character. Once this major result of unreduced interaction development is explicitly established, one can safely use various suitable approximations for secondary, minor process details that determine fine structural features becoming eventually smaller than the practical measurement resolution (at the current level of dynamics). This conclusion concerns, first of all, the auxiliary system solutions, $\{\psi_{ni}^0(\xi), \eta_{ni}^0\}$ (eqs. (17)), entering the EP and state-function expressions, eqs. (21)-(23), (25). Taking into account the 'averaged' influence of these fine details on the principal emerging structure of redundant realisations, one can try to use, for example, a simple mean-field approximation for the truncated system of equations, eqs. (17), in which case it splits into separate equations:

$$[h_0(\xi) + V_n(\xi)]\psi_n(\xi) = \eta_n \psi_n(\xi), \qquad (28)$$

where the mean-field potential, $V_n(\xi)$, changes approximately between two its limiting values,

$$V_{n1}(\xi) = V_{nn}(\xi) \quad \text{and} \quad V_{n2}(\xi) = \sum_{n'} V_{nn'}(\xi). \qquad (29)$$

Various other approximations to the mean-field potential can be suitable for different particular systems. The obtained perturbative solutions of the auxiliary system of equations are qualitatively close to the simplest separable dynamics for the Hamiltonian $H_0(\xi) = h_0(\xi) + V_{00}(\xi)$, which can be found from the effective existence equation, eq. (20), by neglecting the most important, 'complexity-bringing' part of the unreduced EP, eqs. (21). However, now its essential influence at the current complexity level is already taken into account in the dynamically redundant structure of the general solution, eqs. (23)-(25), and one can use the approximations of eqs. (28), (29) for choosing a quantitatively correct form of the dynamically multivalued solution, without the risk to make a big, qualitative mistake (like that of the unitary reduction).

This does not mean that one cannot proceed in obtaining the detailed, exact expressions for the fine structure formation of the unreduced interaction process by applying the same, unreduced EP method to finding the 'effective' auxiliary solution through yet more truncated system of equations and so on, until one is left with only one, integrable equation. At each level of this hierarchical solution one obtains the dynamical splitting into redundant and therefore chaotically changing realisations, but the relative magnitude of splitting diminishes towards finer structural scales. The obtained hierarchical system of randomly changing realisations of ever finer scale forms what we call the *dynamical fractal* of a problem. It provides an essential extension of canonical fractals [122-126], including 'quantum fractals' [127,128], which are not obtained as a result of a real interaction development and correspondingly do not possess the key properties of dynamic multivaluedness and related causal randomness and dynamic entanglement (at any level of fractal branching). Canonical fractals represent therefore a dynamically single-valued, unitary imitation of natural, dynamically multivalued fractality that actually incorporates *any* kind of real structure and not only some particular, 'fractal-looking' and 'scale-invariant' structures.

The irreducibly probabilistic character of real fractals, exactly reproduced in our approach, means that they behave as 'living creatures' performing permanent interaction-driven chaotic (but *not* 'stochastically' random) search of the best way of further interaction development with the help of their unceasingly moving, dynamically changing and splitting 'branches'. Such kind of behaviour, directly obtained in our approach, pro-vides the unique, totally consistent explanation for various properties of natural structure formation processes, such as *dynamic adaptability*, meaning that the probability density of structure development is determined by the local intensity of interaction processes, due to the 'generalised Born's probability rule' [1,10-13].

In addition, the unreduced dynamical fractal is explicitly 'made of' a tangible spatial 'flesh' probabilistically changing in real time and emerging together with it as a result of dynamic entanglement between interaction components, occurring at each level of branching, quite similar to the first level considered above (Section 4.2) and contrary to the canonical fractals represented just by abstract mathematical 'functions' without any 'material quality', or 'texture'. Note that finer levels of fractal dynamic entanglement contribute, in a self-consistent manner, to the development of its lower, coarse-grained levels and the reverse, which is reflected in the basic *non-separability* of the problem equations acquiring now the direct *physical* meaning that involves also system's essential nonlinearity, dynamic instability and reduction (collapse) to the current realisation (Section 4.3). One obtains thus the unified, dynamically fractal and probabilistic, adaptable and self-developing hierarchy of structure formation by dynamic entanglement and realisation emergence.

We shall not reproduce here the results of EP method application (cf. Sections 3.2, 3.3) to the truncated system of equations, eqs. (17), because here we are interested mainly in the general character of the problem solution thus obtained. It can be summarised by the same causally probabilistic sum over explicitly obtained system realisations which was derived above for the first level of dynamical splitting, eq. (24), but should in general include summation over all levels of dynamic fractality:

$$\rho(\xi,Q) = \sum_{j=1}^{N_\text{f}} \sum_{r=1}^{N_{\Re j}} {}^\oplus \rho_{jr}(\xi,Q), \qquad (30)$$

where the external summation (index $j$) is performed over levels of fractal hierarchy until the finite or some desired level number $N_\text{f}$ is attained, and $\rho_{jr}(\xi,Q)$ is the measured (generalised) density of the system in its $r$-th realisation at the $j$-th level of the hierarchy containing $N_{\Re j}$ realisations (so that $\rho_r(\xi,Q) \equiv \rho_{1r}(\xi,Q)$ and $N_\Re \equiv N_{\Re 1}$ for the case of one-level splitting, eq. (24)). Fractal 'levels' are not rigidly fixed or clearly separated, either by their origin or method of derivation; they may correspond to consecutive orders of truncation of the 'auxiliary' system of equation. Realisation densities $\rho_{jr}(\xi,Q)$ are accompanied by the respective probabilities, $\alpha_{jr}$, obtained similar to those for the first level of fractality (see eqs. (26)). Correspondingly, the expression for the expectation value of $\rho(\xi,Q)$, eq. (27), is directly generalised to the whole fractal structure:

$$\rho_{\exp}(\xi,Q) = \sum_{j=1}^{N_\mathrm{f}} \sum_{r=1}^{N_{\Re j}} \alpha_{jr} \rho_{jr}(\xi,Q), \qquad (31)$$

Finally, the integral measure of total dynamic complexity is determined, according to rules established above (Section 4.1), by the total realisation number, $N_\Re^{\mathrm{tot}}$:

$$N_\Re^{\mathrm{tot}} = \sum_j N_{\Re j}, \qquad (32)$$

where $N_{\Re j}$ is the measurable realisation number at the *j*-th level of fractality. It should be taken into account, however, that usually one deals with observable realisations from a limited number of levels and summation in eq. (32) should involve only relevant levels of fractality.

It is important to remember also that the dynamic fractal under study is obtained from a particular level of interactions entering the full hierarchy of complex dynamics of the world and therefore represents, with all the intricacy of its own hierarchy of levels, only a small part of that universal hierarchy of complexity. At the same time the dynamical fractal emerging in the development of an interaction process is indispensable for further transformation of these results into structures from other complexity levels. It is not difficult to see that finer branches of dynamical fractal play the role of higher-level 'interaction' between more 'tangible', 'coarse-grained' products of interaction of the current complexity level made up by relatively 'thick', lower-level branches of the same dynamical fractal. This *self-developing* hierarchy of system dynamics forms the transparent, realistic basis of *complexity development (or transformation)* process making the true sense of any system evolution and providing the *unified*, causally complete extension of the first and second 'laws of thermodynamics' (conservation of energy and growth of entropy) now universally applicable to any system, in the form of *universal law of conservation and transfor-* *mation, or symmetry, of complexity* [1] (see Section 7.1 for details). The underlying property of autonomous system development, or *creativity*, is another expression of the interactive probabilistic adaptability described above.

Of course, the difference between the two parts of the dynamical fractal is not quite distinct and one of the consequences is the equivalence between potential energy and mass simply postulated in the canonical relativity, but now causally explained in the quantum field mechanics and applicable to dynamical processes at any level of complexity [1,11-13]. The finely splitted, chaotically moving fractal web of a given interaction process creates also its links to lower complexity levels and ensures 'interaction' between the current system realisation and its other, potential realisations, maintaining thus the permanency of realisation change process. In particular, finer parts of the unreduced fractal, due to their inbred chaoticity, play the role of *intrinsic* system 'noise', which is a necessary component of its irreducible instability/change and should only be imposed from the outside in conventional imitations of complexity, thus completely neglecting the omnipresent origin of internal, multi-scale chaoticity provided by dynamic multivaluedness of the externally totally regular (closed) interaction process.

One should not be deceived, therefore, by the apparent simplicity of the expression of eq. (30) for the general problem solution, since we have seen above how many intricate details are contained only within a single level of fractality (Sections 3.3-4.3). However, all those details can be obtained, in principle, and completely understood by the unreduced analysis of interaction process described above and the resulting dynamically probabilistic fractal just demonstrates explicitly the full complexity of the *complete many-body problem solution*. It is important to emphasize, in particular, that although eqs. (30), (31) contain summation over multiple scales and realisations, which implies a number of practical approximations, they basically describe the real system structure, contrary to perturbative expansions of the unitary theory that makes a qualitative mistake from the beginning, so that any larger, 'more exact' summation does not approach one to reality. Dynamically probabilistic summation over multiple levels of fractal hierarchy in eqs. (30), (31) corresponds to real existence of those levels (hierarchical system structure) and taking into account each subsequent level considerably increases the result accuracy.[6] We deal here with the genuine, unreduced complexity of natural interaction processes, giving the observed diversity and intricacy of the world in their truly 'exact', unreduced version, which also shows *what one can ever obtain, or expect*, 'in general' within the *unreduced* solution of *any* problem.

The properly developed general solution of the many-body problem, eqs. (20)-(27), (30)-(31), contains thus the multilevel hierarchy of dynamically emerging and permanently, chaotically moving (changing) system realisations unified in the single complex dynamics by the corresponding hierarchy of generalised wavefunctions (or distribution functions). By its dynamically redundant, entangled, multilevel (fractal) and emerging (self-developing) character, this causally complete problem solution provides, in particular, the consistent meaning of 'nonintegrability' and 'nonseparability' notions remaining uncertain within any canonical constructions because of their fundamental dynamic single-valuedness. We see that it is still possible to obtain the causally complete, adequate description of any system behaviour, but only at the expense of qualitatively new properties listed above and transforming dramatically the very notion of system existence, from its fixed zero-dimensional projection in conventional science to the self-developing, 'living' dynamical fractal in the universal science of complexity. One can then neglect those parts of the obtained unreduced complexity which are of less interest for a particular study, but it is important that such omission, with its meaning and consequences, can now be totally understood and justified, as opposed to the effectively blind, 'trial-and-error' approach of unitary science, often presented under the deceitful cover of 'theory verified by experiment' and inevitably becoming inefficient for systems with explicitly complex behaviour. The most meaningful, qualitatively important system properties, dynamic redundance and entanglement, will always be present, directly or indirectly, in the causally complete understanding of system behaviour, contrary to their unavoidable lack in the unitary science projection.

---

[6] Thus, the extended mean-field, integrable approximation of the unreduced EP formalism (Section 3.2) can correspond to local two-body interactions, which are split then into a hierarchy of three-body, four-body and further many-body interactions giving eventually the complete problem solution in the form of eqs. (30), (31).

## 4.5. Generic types of system behaviour as particular cases of dynamically multivalued interaction process

## 4.5.1. Dynamically multivalued self-organisation and control of chaos

The unlimited *universality* of the dynamic multivaluedness concept is its intrinsic advantage which manifests itself not only in the unrestricted applicability of the approach and its results to any kind of system, but also in actual reproduction of the same dynamical splitting and entanglement mechanism at all explicitly emerging levels of the dynamical fractal for a given system and universal hierarchy of complexity in general. Therefore, despite the high intricacy of the full problem solution, with all levels of dynamical fractal (Section 4.4), one can obtain the most important and universal properties already from the 'principal' part of the general solution, eqs. (20)-(27), describing the first level of dynamic fractality.

One of such universal properties of irreducibly complex, dynamically multivalued behaviour is related to existence of its two limiting, and characteristic, regimes containing between them all possible kinds of behaviour that will actually appear for a particular system in the corresponding, well-defined intervals of its parameter values. One of these universal limiting regimes of multivalued dynamics emerges when the resonance between the internal motion frequencies of system elements not related to the compound system structure (see eqs. (6), Section 3.1) and those introduced by the driving interaction between elements (system structure as such) is absent, so that, for example (see also eqs. (16), (17), (20), (21), Section 3.2), $\Delta\eta_i \ll \Delta\eta_n \sim \Delta\varepsilon$, or $\omega_\xi \ll \omega_q$, where $\Delta\eta_i$, $\Delta\eta_n$ are the separations between neighbouring eigenvalues $\eta_{ni}^0$ with changing $i$ and $n$ respectively, $\Delta\varepsilon$ is the eigenvalue separation for structure-independent element spectra, eqs. (6), and $\omega_\xi$, $\omega_q$ are the (characteristic) frequencies of structure-dependent and structure-independent (internal) motions, respectively, proportional to the corresponding eigenvalue separations $\Delta\eta_i$ and $\Delta\eta_n$. In this case we can neglect, to a reasonable approximation, the $\eta_{ni}^0$ dependence on $i$ in the denominators of EP expression, eq. (21b), after which the potential becomes local, due to the property of completeness of the eigenfunction set $\{\psi_{ni}^0(\xi)\}$:

$$V(\xi,\xi';\eta) = \delta(\xi-\xi')\sum_n \frac{|V_{0n}(\xi)|^2}{\eta-\eta_{ni}^0-\varepsilon_{n0}} ,$$

(33)

$$V_{\text{eff}}(\xi;\eta) = V_{00}(\xi) + \sum_n \frac{|V_{0n}(\xi)|^2}{\eta-\eta_{ni}^0-\varepsilon_{n0}} ,$$

where $\eta_{ni}^0$ does not actually depend on $i$, designating here the value averaged over different $i$, and we considered the driving interaction to be Hermitian, so that $|V_{0n}(\xi)|^2$ stands for $V_{0n}(\xi)V_{n0}(\xi)$ in the numerator. Correspondingly, the state-function expressions, eqs. (22), (23), take the form:

$$\psi_{ni}(\xi) = \frac{V_{n0}(\xi)\psi_{0i}(\xi)}{\eta_i - \eta_{ni'}^0 - \varepsilon_{n0}} ,$$

(34a)

$$\Psi(\xi,Q) = \sum_i c_i \left[ \Phi_0(Q) + \sum_n \frac{\Phi_n(Q)V_{n0}(\xi)}{\eta_i - \eta_{ni'}^0 - \varepsilon_{n0}} \right] \psi_{0i}(\xi) .$$

(34b)

The number of eigen-solutions of the effective existence equation, eq. (20), with the local EP from eq. (33) is reduced to its ordinary, non-redundant value, $N_\xi N_q$, because of disappearance of summation over $i$ introducing higher eigenvalue power in the characteristic equation. It seems that we return to the unitary description with only one system realisation. In reality, however, eq. (33) is only an approximate expression of the exact EP, eqs. (21), and although it is a generally correct approximation, its difference with the exact expression hides important *qualitative* property, the dynamic redundance. Indeed, if we do not simplify the original EP expression, eqs. (21), then we obtain additional, though quantitatively small, splitting of eigenvalues, endowing the system with a new quality, internal chaotic change of redundant realisations. It is clear that in the considered limiting case realisations are very similar and 'densely packed' within the single, 'enveloping' (averaged) realisation of the approximate description and therefore may remain unobservable individually (though they will contribute to quantities like 'energy level width', which are explained usually in terms of 'stochastic' approach, with its *extrinsic* randomness insertion). The small splitting into permanently changing realisations can be demonstrated in eq. (33) if we use in it real values of $\eta_{ni}^0$, slightly differing for dif-ferent *i*, and consider, correspondingly, that the 'infinitely narrow' δ-function is split into $N_\xi = N_\Re$ slightly diverging components with a finite width, reflecting slightly different EP configurations for individual realisations. We call this limiting case *dynamically multivalued (extended) self-organisation*, or *self-organised criticality (SOC)*, since it is not difficult to see [1] that it provides the *unified*, dynamically multivalued (and thus always *intrinsically chaotic*) extension of conventional notions of *both* self-organisation (or 'synergetics', or structure formation) [129,130] and SOC [131] (the multivalued extension of the latter case includes, of course, more explicit manifestations of dynamically probabilistic fractality of the unreduced complexity structure, Section 4.4). Whereas usual self-organisation looks for some dynamically single-valued approximation to the external system 'envelope', eq. (33), the conventional SOC applies a 'statistical' (often computer-assisted), but also basically single-valued description to a particular, simplified 'model', so that the internal, dynamically probabilistic 'life' of *any* 'self-organised' system (i. e. explicit realisation *emergence* and *chaotic change* around the average observed pattern) remains hidden behind those formally postulated, unjustified unitary approximations to the corresponding real interaction process (which leads, in particular, to intrinsic incompatibility of canonical self-organisation/SOC with chaos [132-134]). The *unceasing* transitions between realisations within any 'self-organised' structure, either more or less externally static, provide the universal, purely dynamic and first-principles origin of those permanent 'fluctuations' around the 'equilibrium' system shape (e. g. average sand-pile slope) that constitute the empirically observed and then simply postulated basis of the conventional SOC concept missing, however, the key, essentially nonunitary system properties and their direct, analytically substantiated emergence from the driving system interaction.

On the other hand, the obtained regime of unreduced, dynamically multivalued SOC can be considered as the causal extension of another application of conventional 'science of complexity', known as 'control of chaos' [135] and implying a possibility of transformation of a chaotic dynamical regime into a regular one (or another chaotic, but *predetermined*, desired dynamic behaviour). Since our multivalued SOC describes a *universal* way of a more distinct shape emergence in the corresponding limit of *any* interaction process (including the 'controlling' influences and schemes), it comprises this application and unifies all separated conventional cases of (speculatively) 'controlled' or 'synchronised' dynamics in *one* description showing, in particular, that the absolute system control, suggested by the dynamically single-valued projection of canonical theories and oriented towards *total* elimination of *random* deviations from the desired, regular dynamics, is impossible in principle, for any kind of system (cf. [136,137]). Any system 'control' or 'monitoring' can lead it therefore only from one intrinsically chaotic regime to another, where the transition itself is also subjected to irreducible, though dynamically changeable, randomness. This conclusion has especially important consequences for lowest, essentially quantum levels of micro-system dynamics, since here one cannot have a large difference between the participating energy quanta or frequencies (see the above analysis in this Section) and a distinct SOC structure cannot appear in principle, which implies a very low efficiency of conventional 'control'. Note that *all* cases of 'dynamic synchronisation' and 'chaos control', separated between them and from 'self-organisation' or 'SOC' within their perturbative imitations in conventional theory, are extended and *naturally unified* now within the *single* limiting regime of chaotic (dynamically multivalued) SOC, which is intrinsically unified, in its turn, with other cases of unreduced complex dynamics (see the next Section) by the universal formalism of effective dynamical functions, eqs. (20)-(31).

The unreduced, dynamically multivalued SOC regime is universally obtained thus as the limit of closely resembling and weakly separated system realisations, so that their unceasing change creates only small variations of the average, externally observed system configuration and can therefore remain unnoticed, despite the dramatic internal transformation of the system involved in its realisation change process (this picture of internal system dynamics is confirmed by the geometric analysis of the effective existence equation in the SOC limit [1,9,10] which is not reproduced here). The eigenvalues (and thus also eigenfunctions) constituting the 'self-organised' system states (or its actually observed, 'compound' realisations) are well separated among them in the SOC regime [1,9,10], forming compact groups of eigen-solutions that constitute respective realisation dynam-ics (this can be directly seen from eq. (21b) for $\Delta\eta_i \ll \Delta\eta_n$). Note that in a general case one may have almost any number, $N_{SOC}$, of 'self-organised', pseudo-regular system configurations ($1 \leq N_{SOC} < N_\xi$, but usually $N_{SOC} \ll N_\xi$ and includes just several essentially different system configurations), with more frequent chaotic transitions between elementary realisations within each configuration and relatively rare system jumps between different 'regular' configurations (actually forming 'compound realisations' of a next, thus emerging complexity sublevel).

In terms of the above frequency conditions, we can say that we obtain indeed the 'enslavement' of the high-frequency part of system configuration by its low-frequency part, where the former conforms adiabatically to the latter, but contrary to the canonical 'synergetics' [129] actually equivalent to the standard 'motion in a rapidly oscillating field' [138], this 'control-of-chaos' regime of a real system dynamics [1] always involves permanent chaotic change of its slightly different modes within the observed, externally almost regular pattern (as well as less frequent, but equally inevitable and chaotic switch between several realisations of the latter, if $N_{SOC} > 1$). Thus, in the case of interacting elements of a quantum machine, the dynamically multivalued SOC regime means that not only the energy levels of individual element dynamics are slightly splitted due to the interaction between elements (as the canonical, perturbative analysis would suggest), but also that there is in general $N_\xi = N$ versions (realisations) of this splitting, each of them corresponding to a particular version of individual and collective element dynamics. It is especially important that whereas each individual realisation, irrespective of its internal involvement, represents a coherent, unitary system evolution, different realisations are not (totally) coherent among them, and their unavoidable and dynamically random change constitutes the irreducible and purely dynamic (internal) source of true chaoticity, nonunitarity and *temporal irreversibility* (non-stop and inimitable character) of the system evolution, irrespective of proximity between realisations (quasi-regularity of external system configurations). In other words, although quantitatively the additional splitting into different realisations can be small for the SOC type of dynamics, it always introduces the qualitatively new, essential property of dynamical randomness, or chaoticity, or nonunitarity, into the system evolution providing thus a really intrinsic, irreducible source of (substantial) errors for any unitary operation scheme.

This purely dynamic source of errors has nothing to do with its simulations by either ambiguous 'quantum decoherence' [90-92] or canonical 'quantum chaos' [14] (see also Chapter 6), since both of them do not propose any internal source (and meaning) of randomness and should rely therefore on the external, arbitrarily varying source of 'stochasticity', or 'noise', introducing randomness artificially, 'by hand' (even when the source of 'noise' is physically placed within the system). This is a crucially important difference because the extrinsic noise can be, in principle, reduced by creating less noisy conditions of system operation (like decreasing temperature or eliminating imperfections in the system structure), or else by using computational schemes that can separate useful signal from noise (usually at the expense of a much longer calculation multiply repeating the same algorithm, then comparing the results, etc.) [92-107]. The truly intrinsic randomness produced by dynamic redundance of any elementary interaction process will not decrease even in ideal, noiseless conditions and can only grow in 'error-correcting' procedures involving many new interactions. This means that the possibility of a totally 'coherent', or 'nondestructive' control of quantum dynamics assumed by the conventional theory of quantum computation is but an illusion caused by the evident neglect of real interaction dynamics replaced by its perturbative, unitary 'model', without any particular consideration of details.

Being derived by a universal analysis, these conclusions about irreducible nonunitarity remain true for both quantum and classical device, but in the latter case each 'self-organised' structure element contains many smaller constituents (like atoms) subject to noisy influences directly and *independently*, which makes the probability of their simultaneous macroscopic 'fluctuation' in one direction, leading to a fault, vanishingly (exponentially) small, contrary to the essentially quantum machine which operates, according to definition, just with the smallest, indivisible units, where useful 'quantum bits', 'noisy influences', 'controlling actions' and chaotic realisation change are all of the same order of magnitude determined by Planck's constant. This means also that the pronounced SOC regime of the unreduced interaction dynamics, giving a well-defined shape of the emerging system configuration, can hardly be compatible with the essentially quantum dynamics providing all the expected advantages of quantum computation. Any nontrivial, well-defined, long-living patterns, including those necessary for the memory function, can emerge only starting from a semi-classical and usually fully classical regime, which is *causally explained* itself, in the quantum field mechanics, as a (multivalued) SOC type of state (see also Sections 4.7, 5.3) [1-4,11-13]. In addition, the true, dynamic randomness of real interaction processes is always densely mixed with regularity, by its very origin, which makes the essence of the unreduced dynamical chaos concept (see Section 4.1). This property, expressed formally by the inhomogeneous realisation probability distribution, eq. (26), and fractal internal configuration of each particular realisation, eqs. (30), (31), can make inapplicable the canonical idea of useful signal 'filtration', which is based on the assumption about 'purely random' noise mechanistically added to the signal.

## 4.5.2. Uniform (global) chaos, its universal criterion and physical origin

The second limiting case of complex (multivalued) dynamics, opposite by its character to the above SOC regime, emerges when the internal element dynamics and interaction-induced motion enter in resonance, so that their characteristic energy level separations and frequencies are close enough to each other: $\Delta \eta_i \approx \Delta \eta_n \sim \Delta \varepsilon$, or $\omega_\xi \ll \omega_q$. As can be seen from the unreduced EP expression, eq. (21b), in this case the eigenvalues forming individual realisations become intermingled and therefore the corresponding realisation configurations, determined by eigenvalue separations, can change considerably and unpredictably from one realisation to another, which leads to absence of any distinct system 'shape' in the process of permanent realisation change. No reduction of general EP and state-function expressions, analogous to eqs. (33), (34), can be found in this case, which means that individual realisation contributions to the general system behaviour are comparable among them and confirms the absence of any global 'dynamical order' in the emerging system pattern. In other words, realisation probability distribution is a rather homogeneous one (tending in the limit to equal probabilities, $\alpha_r = 1/N_\Re$), so that sufficiently differing realisations emerge in a random sequence and with comparable probabili-ties. Therefore this limiting case of unreduced complex dynamics is called *uniform chaos* and corresponds to the regime of 'global' (intense) chaoticity considered for particular systems [1,8,9].[7] As pointed out in the preceding Section, any essentially quantum system with interaction, including canonical quantum computer, should find itself just in this regime of complex dynamics demonstrating the possibility and meaning of genuine quantum chaos/randomness [1,8-13], as opposed to its absence and inconsistent imitations in the canonical quantum chaos concept (see Chapter 6). The implication for the quantum computer theory is evident: any unitary description of a full scale quantum computation is totally, basically incorrect, while any really existing micro-machine or device operating at the quantum level can only be described as dynamically multivalued, truly chaotic system (any unitary, *stochastic* imitation of 'quantum chaoticity' [14] is clearly as deficient as straightforward regularity of 'pure' unitary evolution). In addition, any intricate enough, quantum or classical, micro-machine with sufficiently fine structure and complicated function should contain many strongly interacting and close enough frequencies, so that multiple resonances among them are practically inevitable and thus the true, uniform, global chaos for at least a part of essential operation stages.

A distinctive property of our description of the two opposite regimes of unreduced interaction dynamics, the externally 'regular' (but internally multivalued) SOC and 'totally' irregular, uniform chaoticity, is its universality, so that both cases are consistently derived within the unified analysis and correspond to different parameters of basically the same picture of multiple, incompatible system realisations replacing one another in a dynamically random order. This means that *any* other, intermediate regime of unreduced interaction dynamics can also be derived and understood as more or less homogeneous distribution of realisation parameters (including especially dynamic probabilities of their emergence), which is quite different from the situation in the canonical, dynamically single-valued 'science of complexity' (let alone conventional 'quantum chaos'), where various allegedly 'unified' cases and properties, such as usual 'self-organisation' (synergetics), 'chaos', 'self-organised criticality', 'adaptability' and 'fractality', remain actually separated and often incompatible, in addition to the imitative, deficient character of each property (which is clearly expressed by the absence of unified and consistent definition of the underlying concept of dynamic complexity itself [1,5-7], cf. Section 4.1). The continuous, causally traced transition between uniform chaos (explicit randomness) and dynamically multivalued self-organisation (external 'regularity') revealed within the dynamic redundance paradigm can be explicitly observed for those systems which allow of their parameter change in a wide enough range covering both limiting regimes, without destruction of the system as a whole. In practice, the natural evolution of a complex enough dynamical system is governed not by artificial 'parameter' variation (analysed within simplified system 'models'), but rather by the universal dynamic symmetry of complexity (including its development) that guides system evolution through an 'optimal' sequence of roughly alternating regimes of quasi-regular SOC and global chaos [1] (see also Sections 4.4, 4.7, Chapter 7).

---

[7] Note the fundamental difference between this dynamically multivalued, 'truly random' chaos and its conventional imitations at either quantum or classical level, where randomness as such is either absent or introduced 'by hand', as external 'noise' (sometimes together with its *postulated* and totally *regular* 'amplification'), without any specification of its ultimate, dynamic origin. It turns out eventually that any conventional 'chaos' is reduced to a *regular*, though *apparently* 'intricate' motion with a finite, but very long, *practically* infinite period, which is not surprising taking into account the effectively one-dimensional (dynamically single-valued) character of the 'models' and analysis used (see Chapter 6 for details).

In any case, a more definite transition from a global dynamical 'order' providing a discernible system shape/configuration (SOC regime) to the global chaos of virtual, changing 'shapes' (uniform chaos regime) is determined by the universal criterion of resonance between the characteristic internal dynamics of system components and its interaction-driven, inter-element motion:

$$\kappa \equiv \frac{\Delta \eta_i}{\Delta \eta_n} = \frac{\omega_\xi}{\omega_q} \simeq 1 \ , \tag{35}$$

where the parameters of inter- and intra-element dynamics, $\Delta \eta_i$, $\omega_\xi$ and $\Delta \eta_n \sim \Delta \varepsilon$, $\omega_q$, were defined above (Section 4.5.1) and we have introduced the parameter of (system) *chaoticity*, $\kappa$, equal to their ratio and determining the (approximate) 'point' of 'order-chaos' transition.[8] Correspondingly, for $\kappa \ll 1$ one obtains, as we have seen above, the dynamically multivalued SOC regime with decreasing external signs of probabilistic re-alisation change within the system. One could also define the parameter of (system) *regularity*, $r \equiv 1/\kappa$, with the evident meaning opposite to that of $\kappa$.

---

[8] It is not difficult to express $\omega_\xi$ through the specific interaction parameters, $\omega_q$ through the system element parameters and thus $\kappa$ through the relevant system parameters for each particular system (see e. g. [1,8,9] and Section 5.2.1). Specific features of each particular system can also be taken into account, such as the existence of several characteristic values of $\omega_\xi$, $\omega_q$ and $\kappa$ for more complicated systems (interaction processes).

It is not difficult to see that if $\kappa$ grows substantially over unity, $\kappa \gg 1$, one should obtain again a generally ordered system state, where now the low-frequency intra-element motion 'enslaves' the quick inter-element (structural) dynamics and determines the observed system configuration. In practice, however, a system with 'interesting' dynamics can more rarely possess this kind of symmetry between the two components of its dynamics and the case $\kappa \gg 1$ will usually correspond to some trivial kind of order, like a uniform energy-level shift (e. g. quasi-ballistic element motion). Loosely speaking, the limit of $\kappa \ll 1$ refers to relatively 'weak' effects of interaction when the resulting system configuration is still determined by some (more rigid) elements from the initial configuration, while at $\kappa \gg 1$ one deals, in general (for a strong enough interaction), with a 'destroyed' initial system, which may be of little interest (although this case is also correctly described by our universal analysis, it is probably better to choose the 'complementary' system configuration as the starting point of unreduced interaction analysis, cf. Section 5.2.1). Therefore the most interesting events happen when $\kappa$ grows from its small values ($\kappa \ll 1$) to one and around this point a rather abrupt change from a remaining order to the global chaos occurs in the form of 'generalised phase transition' [1]. The reverse transition 'chaos-order', occurring with decreasing $\kappa$ around the same point, $\kappa \approx 1$, corresponds to explicit appearance of universally defined 'structure formation', i. e. causal, *natural* (purely dynamic, autonomous, interaction-driven) *creation* (emergence) of qualitatively new 'entities' forming a new (sub)level of complexity as a result of unreduced, dynamically multivalued interaction development (including dynamic entanglement of lower level components).

This general picture of continuous (but uneven) transition between (relative) order and chaos contains also other interesting details that can only be mentioned here, such as step-like partial transitions between order and randomness around higher-order resonances ($\omega_q \approx n\omega_\xi$, or $\kappa \approx 1/n$, $n = 2,3,\ldots$) with the complete 'inversion' of system structure in the main resonance point ($\omega_q \approx \omega_\xi$, or $\kappa \approx 1$) through the global chaos explosion [1]. Such description can be considered, in the whole, as a causally complete, universal (including quantum) extension of the canonical, essentially perturbative KAM theory dealing only with small chaoticity values and establishing conditions for the *absence* of any essential change in the given, initial system configuration (in agreement with the general character of unitary science).

On the other hand, the unreduced interaction analysis gives the universal, nonperturbative picture of the phenomenon of *resonance*, which has been extensively mentioned and analysed in both 'regular' and 'chaotic' (or 'statistical') versions of conventional mechanics, but could not be provided, within the dynamically single-valued approach, with a causally complete description showing its unreduced effect and role in chaoticity and dynamic complexity due to the universally emerging phenomenon of dynamic redundance (multivaluedness) of internally entangled, 'resonant' system realisations. Thus, the popular criteria of 'overlapping resonances' or 'positive Lyapunov exponents' for the global chaos onset in the conventional theory of classical chaos [139-145] appear to be erroneous by both their definition of 'chaos' ('rapid', but *regular* and actually *incorrect* law of 'exponential amplification' of *postulated*, ill-defined 'randomness' of *external* 'noise') and artificial complication of chaos criterion, which in reality is directly determined by the *resonance itself* representing the very fact of unreduced interaction within the system. The developed, 'global', or 'uniform' chaos is the main, final and inevitable *result* of resonance between various parts of system dynamics, which happens *always* in *certain* its parts, but involves the entire system dynamics when the resonance/chaos condition, eq. (35), is fulfilled for principal, characteristic frequencies of 'perturbation' and free-element dynamics. Thus, *every* resonance is internally chaotic and every chaotic behaviour is the direct result of real, dynamically multivalued resonance.

This finding provides a physically transparent explanation for the origin of uniform (global) chaos, the most pronounced form of dynamically random behaviour. The resonance criterion of uniform chaos onset, eq. (35), implies that in this case the interaction partners with equal 'forces' (or dynamical 'sizes') of their characteristic modes collide and the dynamic instability produced by this almost *equal partners* encounter takes the form of omnipresent and relatively big changes between the redundant system configurations (realisations). In other words, the resonant 'equality' of in-teracting modes leads inevitably to the locally 'coarse-grained', and therefore globally shapeless, 'very irregular' type of chaotic dynamics. In the opposite case, where the system is far from the main internal resonances, its 'larger', lower-frequency components naturally 'encompass' (or 'enslave') 'smaller', higher-frequency modes and we obtain a SOC type of behaviour (see the previous Section) with a rather distinct shape determined roughly by larger components. The irreducible dynamic instability of mode encounter arises in this case as well (due to inevitable higher-order resonances between relatively 'marginal' mode frequencies) and leads to dynamic redundance and internal chaoticity, but now it is a relatively 'fine-grained', 'hidden' chaos involving random rearrangements of only internal, high-frequency modes covered by a relatively inert (though still slightly chaotic) 'envelope' of low-frequency components.

It is important to emphasize the difference between the resonance-driven true chaoticity of any unreduced interaction process and unitary imitation of the dynamic redundance phenomenon in conventional 'science of complexity' by such concepts as 'multistability' and 'unstable periodic orbits' representing formal variations (though separated among them) of a more general idea of multiple 'coexisting attractors' of certain, *special* systems usually analysed in the form of unrealistically deformed, symbolical 'models' and distinguished from other, 'regular' and 'non-complex' systems. Those purely abstract, postulated constructions of unitary science are not obtained by analytical solution of dynamic equations, but instead are inserted artificially as an attempt to account for the empirically *observed* chaotic change of system states (in natural experiments or computer simulations). Since the perturbative, dynamically single-valued 'approximation' of conventional science cannot provide any source of *true* randomness in principle, the scholar theory of 'complexity' inserts additional, purely mathematical 'dimensions' and simply places there, 'according to definition', a contradictory, poorly defined imitation of the 'necessary', additional system states (which are *absent* in the effectively zero-dimensional projection of real system dynamics). However, even apart from the incorrect, arbitrary logic of such manipulations with purely abstract, postulated symbols and rules, one obtains in that way multiple system states, or 'attractors', in the form of system *trajectory* 'shapes' *coexisting* in those *abstract* spaces (e. g. 'phase spaces'), so that the (closed) system can remain in each of them for a long (or even infinite) time, as opposed to our *incompatible* system realisations which are analytically, *consistently derived* from *unreduced* dynamic equations as *equally* possible *elements* of *real*, thus *emerging* space of the corresponding level, so that the system *needs* to *permanently* change them in *fundamentally random*, or 'non-computable', sequence thus defined (together with the real, dynamically obtained space and time).

Returning to the case of *essentially* quantum system with many interacting components, it is important to emphasize once more that one always deals here with conditions close enough to the uniform, pronounced chaoticity, eq. (35). Indeed, in the opposite case the essential difference between the characteristic frequencies in the system dynamics is necessary for the SOC configuration emergence, but such difference implies realisation of at least semiclassical conditions in the system, in contradiction to the demand of its 'essentially' quantum character. In addition, it will be very difficult to impede the natural transformation of such quasi-classicality in a relatively complicated system with interaction into full classicality, which can form an inherent part of operation of the dynamically multivalued machine (see Section 7.2, Chapter 8), but cannot participate in the unitary quantum machine dynamics. Finally, even those parts of 'essentially quantum' dynamics where the emergence of SOC type (e. g. semiclassical) dynamics is possible will inevitably contain, as we have seen, irreducible dynamic randomness destroying the unitary quantum evolution, even in the ideal, zero-noise and 'fault-tolerant' system configuration.

## 4.6. Causally complete description of quantum behaviour as the lowest level of unreduced dynamic complexity

### 4.6.1. The dynamic origin of elementary particles, their properties and wave-particle duality

Every real many-body system, and especially a more complicated 'machine' producing diverse enough results (like 'universal quantum computation'), will certainly possess complex, multivalued dynamics that forms at least several (or even many) related levels and sublevels. They naturally emerge in the unreduced interaction process development described above, where more 'coarse-grained' parts of the growing dynamical fractal represent the new level of 'interacting system elements', while its fine-structured 'foliage' constitutes the physical basis for 'interaction' between the elements. Each (sub)level of complexity can contain, in principle, dynamic regimes of both types described above, SOC and uniform chaos, as well as their combination or intermediate cases. However, the naturally emerging levels of growing complexity tend to alternate between the two limiting cases, so that a predominantly 'chaotic' level of complexity gives rise to the emerging higher level of 'self-organised' structures that interact and form the next sublevel of a uniformly chaotic behaviour and so on. Although this sequence of alternating relative 'chaos' and 'order' is rather irregular itself and may contain various deviations and 'mixed' regimes, its existence as a tendency is essential for maintaining the dynamic, autonomous creativity of natural interaction processes and related diversity of created forms. It is important that the dominating type of behaviour from a lower complexity level does not simply disappear in favour of the opposite behaviour at the emerging higher level, but takes a new, 'compact' form hidden within the entities of the new level and playing the crucial role in their very existence. It is impossible to understand and control the operation of a 'quantum', explicitly complex-dynamical (multivalued) machine without a well-specified picture of such multi-level hierarchy of creative interaction dynamics. The unreduced science of complexity, including quantum field mechanics, provides the *universal* framework of this natural complexity development [1-4,9-13] that can then be easily specified for each particular system (see also Sections 4.4, 7.1).

The approximate sequential transformation between (dynamically multivalued) order and chaos in the interaction complexity development can be traced starting from the lowest, properly 'quantum', sublevels of world dynamics, where the consistent description of progressively emerging entities [1-4,9-13] provides the causally complete solution of canonical quantum 'mysteries' concentrated just around 'impossible' combination of randomness and order, as well as other related pairs of 'incompatible' properties (cf. the 'principle of complementarity' proposed by Niels Bohr), such as nonlocality and locality (including 'wave-particle duality'), or continuity and discreteness.

The very first level of world dynamics, designated as 'isolated elementary particles and fields', emerges explicitly from the attractive interac-tion between two physically real, homogeneous (practically structureless) 'protofields', one of them having the electromagnetic (e/m) nature that shows up eventually as e/m basis of the observed entities and another one being of a less transparent, explicitly hidden gravitational origin (it gives rise to the universal gravitation, now causally explained). Application of our universal analysis of interaction shows [1-4,11-13] that this simplest possible starting configuration is inevitable as such (i. e. it cannot be further simplified) and, on the other hand, it can and *does* actually produce (we show how exactly) all the observed elementary entities endowed with all their intrinsic, 'quantum' and 'relativistic' properties, which are now causally, realistically *derived* and explained, without any para-scientific mystification, abstraction and irreducible separations of conventional 'quantum mechanics', 'field theory', 'relativity', etc. (see also Section 5.3). This is possible simply due to the *unreduced*, non-perturbative analysis of this first level of the protofield interaction dynamics giving, quite similar to the above results, eqs. (20)-(27), the essentially nonlinear formation and unceasing change of multiple, incompatible, dynamically unstable system realisations, each of them taking the form, in this case, of the locally highly squeezed state of the protofields *homogeneously* attracted to each other in the initial system configuration.[9] Whereas this dynamically squeezed state of the 'entangled' protofields forms the observed 'corpuscular', localised state of the elementary field-particle thus obtained, its transitional, 'disentangled' state during chaotic jumps between the localised realisations constitutes the realistic extension of both quantum mechanical 'wavefunction' (automatically provided with the probabilistic and now dynamically based interpretation) and 'undular', wave-like, nonlocal state of the same object, thus ensuring the causally complete explanation for the 'wave-particle duality' as a standard manifestation of the unreduced (multivalued) interaction dynamics (cf. the above analysis in Chapters 3, 4).

---

[9] The revealed dynamically multivalued origin of 'quantum strangeness' shows also why one cannot obtain a realistic and consistent picture of the fundamental physical reality within the conventional, dynamically single-valued theory in principle, irrespective of the efforts applied. The single-valued, effectively *one-dimensional* (or even zero-dimensional, point-like) and static *projection* of the multivalued, intrinsically creative reality, determining the 'method' of canonical science as such, can only produce an abstract, grotesquely simplified and irreducibly ruptured image of real world dynamics, which may have relatively heavier or easier consequences for various particular systems, depending on the more or less explicit and numerous manifestations of the unreduced dynamic complexity in their observed behaviour. Whereas in the case of simplest quantum systems the standard manifestations of complexity can still be hidden behind the inexplicable 'postulates' and para-scientific 'mysteries', comfortably tolerated by the 'rigorous' scholar science for almost a century [1-4], the modern *direct* applications of the theory to much more involved, *explicitly* creative (e. g. computing) systems, evoked by the developing empirical technology, cannot be realised within the same simplification in principle, since the number of the necessary 'postulates' and their 'mysteries' quickly diverges, together with the number of possible system versions and states (i. e. actually its unreduced dynamic complexity, just determined, as we have seen in Section 4.1, by the number of such explicitly obtained 'states', or realisations).

Since we deal here with the very first level of any structure emergence in the initially homogeneous system of interacting protofields, the characteristic eigenvalue separations, $\Delta\eta_i$ and $\Delta\eta_n$ (describing here spatial dimensions of the emerging protofield inhomogeneities), are of the same magnitude, $\Delta\eta_i \approx \Delta\eta_n$, giving the regime of uniform chaos, eq. (35), with maximally irregular distribution of realisation probabilities (for the field-particle at rest), which could be expected because of the uniform initial configuration of the system. This true, dynamical randomness hidden 'within' the elementary particle plays an indispensable role as the current-level manifestation of unreduced dynamic complexity and accounts for the intrinsic, universal property of mass, rigorously defined as temporal rate of realisation change, being thus naturally 'equivalent' to extended, complex-dynamical energy, and unifying its (relativistic) inertial and gravitational aspects [1-4,11-13]. We deal therefore with a basically 'chaotic' (irregular, distributed) type of behaviour at this first level of world's complexity, where it is represented mainly by the unceasing process of *quantum beat* constituting the essence of each (massive) elementary particle and consisting of periodic cycles of protofield reduction-extension around randomly chosen, neighbouring centres. At the same time, the minimum regularity is present even within such highly irregular dynamics, in the form of individual realisation configuration, determining the characteristic particle 'size' and 'shape', the causally understood property of spin, and temporal synchronisation between realisation change processes (or 'quantum beats') for different field-particles accounting for the discreteness and universality of elementary electric change (now causally explained as intrinsic property of particle dynamics), as well as exactly two opposite values ('signs') of its 'kind' [1,11-13]. Note that different species of (massive) particles correspond to different EP amplitudes in our description, and therefore the observed diversity of elementary particles can be explained as a result of the same, universal kind of dynamical splitting into multiple realisations in the system of two interacting protofields. The number (four) and *intrinsically,* *dynamically unified* origin of the fundamental interaction forces between particles obtain a transparent, causally complete explanation within the same physical picture [2-4,11-13].

## 4.6.2. Complex-dynamical particle interaction, genuine quantum chaos and quantum measurement dynamics

The next sublevel of the world complexity appears, according to our general picture, when the entities of the first level, the isolated field-particles, start interacting among them through the mentioned inter-particle forces, which emerge dynamically as a less dense, 'fractal' part of the same protofield perturbation that constitutes particles themselves in its denser parts and appears as a result of homogeneous protofield attraction. The four fundamental interactions among particles can therefore be described as higher-order 'remnants' of the *same* basic protofield attraction whose main, zero-order part is transformed into (massive) elementary particles themselves represented by the respective quantum beat processes that account, in particular, for the major property of mass-energy [1-4,11-13]. This 'allocation' of a smaller part of the current level interaction to the emerging higher sublevel, in the form of a fine-grained part of the dynamical fractal, continues through all progressively appearing levels of (complex) world dynamics, so that all the existing structures and processes, including the most advanced, 'anthropic' ones, are explicitly obtained as progressively unfolding, integral parts of the single interaction process between two world-forming protofields, the same one that gives the elementary particles of the very first level of complexity. This intrinsically unified world dynamics is rigorously described by the universal law of transformation and conservation, or symmetry, of complexity [1] just expressing the progressive transformation of the potential (informational) form of complexity at the beginning of interaction process to the form of complexity-entropy describing the interaction results (see also Section 7.1). In particular, the Schrödinger and Dirac equations for the emerging field-particles of the first sublevel of complexity can be consistently *derived* and also generalised to all higher levels of complexity, together with the causally explained wave-function and its probabilistic interpretation ('Born's rule') [1,4,11-13], as opposed to artificial and 'puzzling' postulation of all those laws for the sin-gle, quantum level of world dynamics in conventional theory, where it remains mysteriously and irreducibly separated from higher-level, 'classical' dynamics (the latter being equally *postulated*, though in an intuitively more 'natural' way).

The entities emerging at the second (sub)level of complexity are moving particles and elementary bound states of particles, like atoms. Contrary to the ambiguous, intuitive empiricism of conventional science, the *phenomenon and state of motion itself* can now be *rigorously defined* as a state with the unreduced dynamic complexity value (measured by mass-energy) that exceeds its minimum possible value for the given system, which is always well-defined, finite (for massive systems) and determines its *state of rest*. It is not difficult to show that both any unbound motion and bound state emerge as partially ordered, SOC type of *spatial structure*, where the chaotic quantum-beat pulsation of individual particles is 'packed' within a regular 'envelope'. The simplest example of such regular structure induced by global motion is the famous 'de Broglie wave' of a particle, now causally derived and totally understood as another standard feature of the unreduced complex dynamics, in agreement with original expectation of Louis de Broglie and by contrast to conventional mystification by Niels Bohr and his followers (see [1,2,12,13] for further details and references).[10] This *partial* spatial order in the internal dynamical structure of

---

[10] It is important to emphasize, in this connection, the essential difference of both our explicitly complex-dynamical quantum field mechanics and the original de Broglie approach, designated by him as the 'double solution' and including, though implicitly, the unreduced dynamic complexity, from the so-called 'Bohmian mechanics' first introduced by de Broglie himself (under the name of 'pilot-wave interpretation') as a technically convenient, but fatally simplified version of the full double solution (1927), but then 'rediscovered' by David Bohm (1952) and now extensively imposed as the 'causal de Broglie-Bohm interpretation' of quantum mechanics (see e. g. ref. [196]). The pilot-wave interpretation, or Bohmian mechanics, is indeed only another *interpretation* of basically the same, formally postulated and contradictory scheme of standard quantum mechanics, obtained from the canonical form of Schrödinger equation by a simple, identical replacement of variables followed by a pure 'philosophy', which provides only 'plausible', or maybe even 'attractive', *assumptions*, but *not* their consistent, causally complete substantiation (which should inevitably contain a *conceptual*, qualitative novelty in its rigorous basis). As a result, the main 'mysteries' and 'unsolvable' problems of conventional quantum mechanics remain unsolved, sometimes changing only their formulation. Thus, the physical origin of the 'particle' itself allegedly 'guided' by the 'attached' wave remains totally mysterious within this 'causal (realistic)' approach, together with the nature of the 'wave', its 'link' to the 'guided' particle and the observed permanent *transformation* between wave and particle (*rather than* their simple *coexistence*). There is no any underlying, developing interaction in the Bohmian mechanics and thus no essential nonlinearity and *dynamical* randomness (i. e. intrinsic chaoticity or 'indeterminacy'): it is a unitary, dynamically single-valued theory, quite similar to other officially permitted 'interpretations'. Correspondingly, it is *actually* formulated in terms of

'quantum' objects, always containing a truly random component, accounts, in particular, for the property of 'quantum' *coherence*, much referred to in the canonical theory of quantum computers, but without any clear idea of its real physical origin. Since the relative proportion of order and randomness in the unreduced complex dynamics depends, as shown above, on the particular interaction parameters, one may have, at this higher sublevel of 'quantum' complexity, various situations, changing between global chaos and multivalued self-organisation, even though the sublevel as a whole emerges rather as a somewhat more ordered structure from the uniformly chaotic lowest sublevel of 'free' particles. The large diversity of results of complex-dynamical interaction between elementary particles and their simplest agglomerates can be classified, nevertheless, into a small number of qualitatively different cases, the most important and practically relevant among them being (genuine) *quantum chaos* [1,8,9], dynamic *quantum measurement* (including 'wave reduction') [1,10] and complex-dynamical emergence of *classical*, truly localised (trajectorial) type of behaviour [1-4,12,13]. While the first two interaction results (quantum chaos and measurement) are obtained as particular manifestations of the uniform chaos regime (Section 4.5.2), the genuine classicality emerges in the form of typical SOC structures (Section 4.5.1) actually represented by the simplest bound systems (such as atoms).

Quantum chaos emerges even in the noiseless, strictly conservative (Hamiltonian) quantum dynamics when an elementary, quantum system (like particle in a symmetric, 'integrable' potential) with a regular type of dynamics is subjected to additional perturbation (time-dependent or static) with a characteristic frequency $\omega_\xi$ close to that of internal system dynamics, $\omega_q$: $\omega_\xi \cong \omega_q$, in agreement with eq. (35). This shows, in particular, that genuine quantum chaos is at the origin of all ordinary 'resonant excitation' phenomena (e. g. in atoms) and 'excited states' dynamics, which leads to a considerably extended understanding of detailed dynamics of such 'standard' quantum systems forming the new, explicitly 'chaotic' and totally

---

the same kind of purely abstract, mathematical entities, which give only an *illusion* (though partially justified *as such*) of being closer to reality. Other consistently *derived* correlations of the complex-dynamical picture, such as causal relativity and gravity intrinsically unified with quantum behaviour, together with space and time, mass-energy and the four 'fundamental interactions' [1-4,11-13], cannot even be formulated within the narrow unitary framework of Bohmian or any other 'interpretation' of standard quantum postulates.

causal framework of the whole conventional 'quantum mechanics' of systems with interaction [1,9]. The extended, dynamically multivalued quantum theory includes also causal complex-dynamical explanation of quantum tunneling effect and particle energy level discreteness representing two canonical 'quantum mysteries' in the standard interpretation.

As mentioned above (Section 4.5.2), the case of unreduced quantum chaos is of special importance for any essentially quantum machine, since it provides truly 'quantum' and uniquely efficient way of its productive operation (involving change of state), which means that any essentially quantum machine can usefully operate only in a highly chaotic regime, *excluding any unitarity*. The same conclusion remains basically valid for partially ordered (SOC) regimes whose internal chaoticity will still create irreducible, and 'fatal', errors during 'massive' machine operation (even in the total absence of noisy, 'decohering' influences) with respect to the assumed ideal regularity of unitary evolution.

The quantum measurement case is different from the quantum chaos situation only by the presence of small enough system dissipativity related to its minor, 'non-destructive', but non-vanishing openness towards the exterior, 'macroscopic' world. Note that this small dissipativity, appearing through some 'excitable' degrees of freedom, does not play any 'decoherence' role inducing chaos itself. The true chaoticity appears, in the form of multiple incompatible realisations, in the course of the same unreduced and fundamentally conservative interaction process involving the given, 'measured' quantum system and another, 'measuring', but *also quantum*, system (like the excited detector atom), which is related, however, to a hierarchy of larger systems through amplified excitation processes. The 'dissipative' function of the latter is reduced to transient 'measured' system binding to the 'measuring' system for a long enough time (during which initial excitation act really occurs). The normally extended quantum beat process of the measured system (like a projectile interacting elastically with multiple slits) becomes transiently confined to a close vicinity of the occurring excitation event (situated in one of the slits) thus necessarily losing its distributed, undular properties for a time period exceeding a characteristic interaction time (after that the normal, extended quantum beat is reconstituted, but the measured undular effects, such as interference between wave scattering by different slits, cannot happen any more). It is clear, therefore, that the real spatial 'reduction' (squeeze) of the normal measured system behaviour does happen during quantum measurement, as well as subsequent system extension to its unperturbed dynamics, which constitutes the causally complete picture of 'wavefunction collapse' remaining unrealistic and 'mysterious' within conventional quantum theory and all its scholar modifications (see refs. [1,10] for details).

It is important that the dynamical collapse process thus understood does not need any modification of the standard Schrödinger equation like it is done in various unitary 'theories of explicit collapse' (or a yet more inconsistent use of arbitrarily postulated equations for various unitary imitations of probabilistic 'distribution function' for a quantum system, such as density matrix). Indeed, it is the 'ordinary', but now consistently derived and causally (realistically) interpreted Schrödinger equation that *results* from the dynamically multivalued quantum beat process including unceasing cycles of real system 'collapse' to one of its realisations and subsequent extension. Those 'background', fundamental reduction events of a measured quantum system, constituting the essence of its existence, simply 'conform' dynamically to the next-level interactions during the measurement process, but remain basically unaltered and usually determine the discrete set of possible quantum measurement results.

In other terms, the slight dissipativity of the quantum measurement configuration, concentrated around certain location, simply chooses, by 'creating preference' (higher probability), a small part of possible system realisations, concentrated around that location, thus temporally 'disabling' other realisations and related quantum nonlocality. At the same time, the 'measuring' excitation can choose in that way only among already existing system realisations, determined by its main, non-dissipative interaction dynamics. The criterion of resonance and thus maximum chaoticity, eq. (35), is automatically fulfilled in the essential stage of quantum measurement (dynamical choice of realisations), since the efficient measuring system should have a substructure in its spectrum that closely resembles the measured interaction spectrum (hence $\omega_\xi \cong \omega_q$), which does not prevent it from having other parts of the spectrum, with $\omega_\xi \ll \omega_q$ ($\kappa \ll 1$), which are responsible for subsequent pseudo-classical localisation (multivalued SOC regime).

Applying this general picture to quantum machine operation, one should take into account the fact that any useful operation involves omnipresent events of quantum measurement, since it occurs practically at every essential energy exchange between elements changing their states through excitation, which again demonstrates a totally illusive character of unitary description of real functional, multi-component quantum systems with interaction. Actually the normal operation of a quantum machine consists of a set of 'sequential' and 'parallel' events of quantum chaos (for nondissipative, elastic interactions) and quantum measurement (for dissipative interactions), each of them introducing essential, true, dynamical and irreducible randomness, which disables any unitary scheme of quantum dynamics, but opens quite interesting possibilities for the unreduced system dynamics actually already realised and successfully used by natural micro-systems (see Sections 7.2,7.3, Chapter 8).

## 4.7. Interaction complexity development and dynamic origin of classical behaviour in noiseless micro-systems

The *classical* type of behaviour is a special case of partially ordered, SOC type of regime realised by the universal mechanism of dynamic multivaluedness (Section 4.5.1) in elementary bound systems (like atoms) and then persisting in larger agglomerates of elementary particles. The fundamentally dynamic, universal and physically transparent origin of this truly (permanently) localised behaviour, appearing already for microscopic, but bound configurations of several (at least two) particles in the absence of any external 'decoherence', is due to the same phenomenon of dynamic redundance that accounts for nonlocal behaviour of a free particle at the lowest complexity sublevel. Indeed, each of the bound particles preserves the dynamically *random* (probabilistic) character of its quantum-beat jumps, since the largest part of the fundamental protofield interaction (accounting for the particle rest mass) is always preserved. At the same time, the probability of jumps separating the bound particles by more than the (small) average system size quickly (exponentially) drops with growing separation. The only possibility for the elementary bound system to perform a nonlocal chaotic wandering as a whole (as it happens to its free 'quantum' compo-nents) could be realised if the components could perform *many random* jumps in almost *one* direction. But since each of them is *independently* random (being driven by the *strong* protofield interaction), the probability of a large sequence of such highly correlated jumps is extremely low.

Therefore the bound system, in the irregular, 'distributed' part of its dynamics, exhibits only small random jerks while remaining almost at one place as a whole because each of the bound components tries to 'pull' its partners in a randomly chosen direction of its current jump, which gives vanishing average result (this irregular part has, of course, its regular 'envelope' of bound motion, but it plays only secondary role in the mechanism of classicality emergence). This kind of behaviour represents a sort of 'dynamical', strictly internal 'Brownian motion', as opposed to its usual, stochastic, externally driven version. It can be rigorously described by the universal theory and criterion of dynamically multivalued SOC/chaos emergence, eq. (35), since the quantum beat frequency of each component, $\omega_q$, is much greater than any bound motion frequency $\omega_\xi$, so that $\kappa = \omega_\xi/\omega_q \approx U_\xi/m_q c^2 \ll 1$ (where $U_\xi \approx \hbar\omega_\xi$ is the binding energy and $m_q c^2 = \hbar\omega_q$ is the total mass-energy of a component) and one deals with a pronounced case of dynamically multivalued SOC. Thus, in the case of hydrogen atom $\omega_\xi$ coincides with Bohr's frequency (the reciprocal to the atomic unit of time), while $m_q$ is the electron mass, and we obtain $\kappa = \alpha^2$, where $\alpha \approx 1/137$ is the fine structure constant, which further extends its complex-dynamical interpretation as the reciprocal of the total realisation number for the electron [1] and explains why and how the smallness of $\alpha$ ensures the basic stability of the atomic structure of matter. It becomes also clear that elementary classical system localisation may disappear only in a hypothetical case of 'relativistic' interaction, for which the binding energy is as high as the mass-energy, $U_\xi \sim m_q c^2$, though system components themselves can be transformed by such strong interaction and with them maybe the system (however, the elementary hadrons and nuclei provide examples of such kind of 'internally relativistic' and still externally stable system whose quantum behaviour may have thus a nontrivial explanation and structure alternating chaotically with periods of classical localisation).

It is interesting that the same explanation and estimate apply also for quite another, essentially quantum state from the same sublevel of com-plexity, that of a uniformly moving elementary particle, where $U_\xi$ in the above estimate should be understood as the 'regular' part of the total particle energy, $U_\xi \sim m_q v^2$ ($v$ is the particle velocity) [1,2,12,13], so that $\kappa = v^2/c^2$, and the dynamically ordered structure of a moving quantum particle (represented by its de Broglie wave) can be destroyed in favour of a more randomly structured entity only for relativistic velocities, $v \approx c$, where particle transformation processes become increasingly probable (this emerging irregular regime of relativistic field-particle dynamics can correspond, at least partially, to classical, rather than quantum behaviour [3]). This universal validity of the same picture of both globally chaotic and quasi-regular dynamic regimes and their quantitative description, eqs. (20)-(27), (33)-(35), for different situations and sublevels of complexity responsible for the explicit emergence of the most fundamental entities and properties of the world demonstrates convincingly the universality of the dynamic redundance paradigm that continues at all higher complexity levels [1], which is to be compared with the efficiency of conventional 'science of complexity' that should restart its 'unified' analysis for every new case and level of 'macroscopic' complexity and cannot propose any consistent involvement of complexity in the fundamental, microscopic levels of the really, physically unified world dynamics.

Returning to the practically important case of classicality emergence as the 'generalised phase transition' [1] to a higher than purely quantum level of complexity, coinciding approximately with elementary bound system formation, we should emphasize once more the universality and purely dynamic origin of this mechanism and the related meaning itself of the property of classicality, which does not depend now on any changing 'environment' or external 'noise', contrary to the officially accepted idea of 'decoherence' as the origin of classicality [92,146-149], containing much other ambiguity, evident inconsistency and unfortunately widely used in such applications as quantum computation. Thus, it applies to *abstract* space 'vectors', which are supposed to be 'perturbed' by quite *real* influences. The problem is 'resolved' by resorting to various special, arbitrarily postulated, 'suitable' equations for a kind of 'distribution function' (e. g. 'density matrix') with an equally postulated meaning that *should* replace the basic wavefunction because of ill-defined 'decoherence', which then is indeed 'obtained' within such vicious-circle kind of theory contradicting, in addition, the well-established Schrödinger formalism. The 'classical' behaviour thus 'explained' is also *understood* in a correspondingly peculiar way, not as a causally obtained, persistent and internal localisation of a real system, but rather as a formal 'selection' of a small number of particular *quantum* states (in a purely abstract 'space'), having 'classical' behaviour due to ambiguous 'predictability sieve' (classicality is interpreted thus rather as an inevitable *illusion* of the observer, in that 'post-modern' kind of science, where everything is finally reduced to an illusion). It remains finally unclear what a mathematical 'decoherence' of purely abstract entities could mean in terms of real entities evolution, why it should matter for *certain* real micro-system dynamics, but not for other, often much larger quantum systems preserving their quanticity in the same 'noisy environment', and why, in general, that 'quantum decoherence' is so different from proportionally diminished dynamics of macroscopic, classical wave, or particle, perturbed by a macroscopic noise. In return, it becomes clear that 'decoherence', similar to other 'post-modern interpretations' of quantum mechanics, such as various 'quantum histories' approaches, is reduced to nothing but a tricky, speculative reformulation of the old, well-known 'quantum mysteries' remaining *unsolved* and deliberately hidden behind 'very special', artificially inserted terminology and abstract, meaningless, postulated mathematics.

A practical consequence for quantum machine description is that 'decoherence' theory applications result inevitably in a demand for a 'pure enough', low-noise environment, where a unitary quantum computer could remain as 'coherent' as other known 'essentially quantum' systems, at least with the help of a 'fault-tolerant' quantum programming [92-107]. We have seen, however, that the truly fundamental, and thus necessarily *internal (dynamic)*, origin of classicality constitutes also the universal origin of genuine, *intrinsic* randomness within *every* quantum and classical state and in the world dynamics in general, so that any real, productive operation, including calculation, cannot be realised as a unitary process in principle, irrespective of technical 'tricks' applied. Note that noisy environment can, of course, exert *additional* influence on quantum and classical state dynamics and in particular modify, in certain cases and to a certain degree, classical behaviour emergence, but even in the absence of any such external noise one always has the major, dynamic, omnipresent and unavoidable source of *strong* randomness and the above-specified realistic classicality, which provides the 'main effect' in the form of dynamic redundance of any real interaction. This comparison illustrates the fundamental and practically important difference between the unitary, dynamically single-valued imitation and dynamically multivalued description of reality in general: the unreduced interaction analysis of the latter approach gives the *unified* dynamic origin of *both* regularity *and* randomness, which appear as *one*, indivisible, *chaotic* process of probabilistic realisation change with dynamically varying parameters, while the single-valued projection of reality can only *postulate* its unitary, *absolutely regular* imitation and is forced therefore to mechanistically superimpose upon it some equally *unexplained*, external 'noise', 'predictability sieve', etc., in order to account for the observed dynamical, changing, but inseparable mixture of order and randomness (this unitary science 'method' is used also by conventional, dynamically single-valued 'science of complexity', despite all its 'dynamical' and 'nonlinear' terminology, cf. Sections 4.1, 4.5.2 and ref. [1]).

As concerns recently reported and apparently 'successful' experimental demonstrations of extremely simplified versions of conventional 'quantum computation', it may finally appear that they actually reproduce, at a small scale, the visible unitarity of macroscopic, classical computers due to a relatively large number of participating units/states, which realise a 'semi-classical' operation and classical measurement procedures at the expense of some relatively small 'inaccuracy', as if 'acceptable' for these 'first demonstrations' (we leave apart a possibility of direct errors, or 'loopholes', known to happen in such rather sophisticated experiments and their result interpretation). Those 'small deviations' are not acceptable, however, in a full-scale unitary computation and cannot be decreased, while all the 'semiclassical' concessions should be abandoned for the full realisation of the promised quantum computer advantages, which can only increase the 'small errors'. As a result, one does not see any larger-scale demonstration of unitary quantum computation, despite the confirmed high possibilities of modern experimental techniques in quantum physics and sufficiently old age of 'miraculous' theoretical expectations for unitary computers. Note, by the way, that the origin of irreducible 'imperfections' observed in real quantum (including optical) devices and 'generally' at-tributed to that ambiguous 'decoherence' (see e. g. [79]) will often be reduced just to the dynamic redundancy of the 'main', driving interaction processes (apart from cases of direct noisy influences). The persisting uncertainties with practical realisation of unitary quantum computation, as well as hidden doubts about its possibility in principle, can explain a 'strange' and growing tendency in last-time publications on quantum information processing towards 'purely theoretic (mathematical)' development of this *computation* scheme, which has grown up exclusively due to its theoretically 'proven' and widely publicised advantages in *practical* applications (such as 'miraculous' efficiency and unlimited universality).

Another experimental effect contradicting the official decoherence theory and confirming our mechanism of dynamic classicality emergence is the existence of quantum, undular behaviour for rather large systems, such as various 'quantum condensates' (superfluidity, superconductivity, gaseous atomic Bose-condensates) and diffraction of large atomic agglomerates/molecules (see [3,12] for further details and references), which should normally have a classical type of dynamics. This 'return' of quanticity in certain, rather exotic, cases of a next level of interaction between elementary classical systems looks absurd within decoherence theory (since there is always more of decoherence in a larger system), but can be naturally explained in the quantum field mechanics. Indeed, in our theory classicality is obtained as a purely dynamic result of the *main* (binding) system *interaction* and therefore another interaction with a suitable magnitude, to which the system thus formed participates, can modify, at least partially, the result of the first interaction and provoke, in particular, the 'magic' reappearance of undular effects. Due to the property of intrinsic dynamic instability of any complex dynamics described above and appearing especially during unceasing system 'jumps' between its realisations, the former 'classical' system (e. g. atom or molecule) does not need to be destroyed in order to reproduce the essentially quantum behaviour of its components. It is sufficient for the (small) 'classical' system dynamics to enter in proper resonance condition (cf. eq. (35)) with another such system or external 'interaction potential' and the quantum jumps of its components can become correlated 'in the direction of force' (for the period of essential interaction), thus cancelling the above dynamical cause of classical localisation. Such more special interaction configuration becomes less and less probable with grow-ing number of components (system mass) and therefore 'quantum revivals' become the more and more exotic for ever larger systems, demonstrating the *dynamically fractal* structure of the border between quantum (lower) and classical (higher) levels of complexity characteristic for all structures and borders produced by the real, unreduced complexity development (the dynamically fractal 'density distribution' of a border is often imitated in the unitary science by the statistically averaged 'asymptotic limit', including in particular the conventional '(semi)classical limit') [1,12,13]. The dynamically fractal character of the quantum-classical boundary confirmed experimentally for various system kinds provides, therefore, an additional support for the direct involvement of unreduced complexity (dynamic multivaluedness) in quantum behaviour, classical dynamics and transition between them, further amplifying thus the existing noncontradictory system of other correlations within the quantum field mechanics (see e. g. [13] and section 3 in ref. [3]).

The dynamical, interaction-driven switches from quantum to classical behaviour and back, occurring at neighbouring sublevels of microsystem dynamics, should naturally happen within any nontrivial micro-machine operation, determining the links between its lowest, essentially quantum levels and the surrounding macroscopic, purely classical world. The above unified picture of complex interaction dynamics shows that one cannot, and actually should not, avoid such part quantum, part classical, or 'hybrid', character of micro-machine dynamics, contrary to the unitary theory illusions, where the calculation as such is performed exclusively by (unrealistically) 'coherent' quantum dynamics, which is mechanistically separated from its unavoidable final transformation into a classical, practically accessible output. We have seen above that in reality the essential nonunitarity is present at every stage and interaction act of micro-machine operation and simply takes different forms for the characteristic cases of Hamiltonian (but truly random) quantum chaos (representing the realistic extension of unitary 'computation' as such), slightly dissipative quantum measurement (causally complete extension of the canonical version) and the SOC regimes of emerging semiclassicality and classicality, where all types of behaviour appear dynamically and therefore inseparably mixed in a unified complex dynamics outlined above. It is the unreduced complex dynamics itself that properly attributes the 'roles' and 'functions' to differ-ent regimes that arise (probabilistically) just there where they are necessary (it is the universal property of *dynamic adaptability* of multivalued dynamics [1], see also Section 4.4). Thus, the strongly random and distributed quantum chaos regime (Sections 4.6.2, 5.2.1, Chapter 6) performs the main work of efficient change of system states in the course of its complexity development (Chapter 7), while quantum measurement 'collects' the 'ripe' results of this development and transfers them into macroscopically useful states of multivalued classicality forming the directly readable 'output'.

Note in this connection that the above complex-dynamical pictures of real quantum measurement and classical state dynamics show that the key stage of quantum measurement process can be considered as a 'transient', 'half-made' classical state, which is later transformed into fully classical states of other instrument components by further interaction development within the instrument [1-4,10-13]. This result provides the causally complete explanation for the instrument 'classicality' emphasised in the conventional theory, but remaining inconsistent (we show, in particular, what the exact physical meaning of classicality is and how it dynamically emerges in *purely quantum* instrument components, starting from its *noiseless* interaction with the measured quantum object). On the other hand, the 'coherent', 'essentially quantum' dynamics of the lowest level of complexity exists mainly within elementary components of a quantum machine (with some occasional appearances at a higher sublevel of interaction results) and also has the multivalued, chaotic internal structure which, however, can be better 'synchronised' and hidden within the observed 'coherent' envelopes at this *exceptional*, lowest complexity level.

Another indispensable role of dynamic classicality and quantum measurement in the real quantum machine operation is related to the function of *memory*. The unreduced interaction analysis of the universal science of complexity confirms the evident conclusion, almost deliberately neglected by the unitary computation theory, that *any* memory, including both memorisation and storage, can be realised as such only by the strongly irreversible, and thus certainly nonunitary, dynamics [1]. The above detailed description of characteristic results of a complex interaction process shows that any reasonably stable, useful memory (including memorisation, storage and retrieval) can be realised in the form of a well-defined SOC type of states, and thus actually as classical states, for a long-term memory and SOC-type (e. g. excited) quantum states for a short-term memory (indeed, every 'object' stored in memory is a 'bound' state, by definition, and our dynamical picture of classicality, and SOC in general, only provides this general notion with a well-specified meaning). In other words, any memory can only be based on a rather well-defined (distinct and stable enough) structure, already because by definition memory stores *determinate* structures and even imposes them during retrieval. This evident property of memory is in contradiction with the uncontrolled uncertainty of essentially quantum system.[11] These fundamental, and now dynamically substantiated, properties of memory demonstrate the unavoidable way of unreduced classicality and quantum measurement emergence in the dynamics of a quantum machine with memory and also show how far from reality and elementary consistency the conventional quantum computation theory can go in its attempt of artificial memory insertion into the unitary quantum evolution.

Actually the whole dynamically multivalued picture of real microsystem dynamics, basically *unpredictable* in detail and therefore *creative*, but also *dynamically* probabilistic and therefore rigorously described, leads to quite another understanding of the essence and purpose of natural and artificial machines performing both 'calculation' and 'production' within the same, internally 'liberated' (adaptable) dynamical process (see also Chapters 5, 7, 8). The unitary schemes of conventional quantum computation cannot reproduce the most essential features of real system dynamics and its results even approximately; by avoiding the real interaction complexity within their invariably perturbative analysis, they so to speak 'throw away the baby with the water' and neglect just the unreduced calculation result emergence, replacing it with a grotesquely simplified, dynamically single-valued, unchangeable projection. The same refers to various artificial, a posteriori modifications of the canonical unitary scheme, trying as if to take into account 'real system dynamics', 'influence of noise', or 'quantum measurement', but remaining hopelessly abstract, perturbative and dynamically single-valued. Indeed, instead of finding the unreduced solution of dynamic equations describing the underlying dynamical process, the 'quantum programming' schemes operate with assumed results of abstract, actually unrealistic, unitary evolution by constructing various 'suitable' linear compositions of symbolical 'state vectors', but when an essential stage of creation or transformation of those 'quantum bits' intervenes, they just rely on arbitrary postulates about its results equivalent to a perturbative, evidently incorrect approximation (we have seen above, in Sections 3.2-3.3, 4.2, 4.4, why perturbative expansions, like those for the unitary evolution operator, are qualitatively different from reality and what will be the result of the unreduced, nonperturbative analysis, see also Section 5.1 below).

---

[11] As shown above, the unitary control of unitary quantum dynamics [100,105,106] is but another, vicious-circle illusion of the perturbative approach closed within its own, evident limitations, since any real, and in particular 'controlling' (change-inducing), interaction gives rise to the irreducible dynamic redundance and randomness, leading to a relatively large uncertainty at the lowest, quantum levels of dynamics. However, even if one assumes that purely quantum control can be efficient, then it follows that for memorising some distinct enough, and in particular classically localised, state structure the 'essentially quantum' memory should reproduce classical configuration, which reveals a contradiction. Rigorous analysis of the unreduced science of complexity confirms the fundamental origin of this contradiction, since it shows that everything 'classical' emerges from 'quantum' as a higher level of the well-defined dynamic complexity and therefore no purely quantum system can correctly reproduce a classically complex behaviour without violating the complexity conservation law (see also below, Sections 5.2.2, 7.1, 7.2).

Note also that the 'probabilistic', or 'indeterministic' elements or stages sometimes introduced into unitary quantum computation schemes are quite different from the *dynamically* probabilistic effects of unreduced interaction: the former are characterised by rigidly fixed probabilities, which are actually eliminated from the process of computation as such, ensuring its 'miraculous' efficiency, irrespective of mechanistically added, *stochastic* randomness. Even without the detailed dynamic analysis it is not difficult to see, however, that a small deviation in the expected value of thus introduced probability, which cannot be avoided in practice, will ruin the whole scheme by creating the dynamically 'jumping' and therefore uncontrollable randomness, with fatal consequences for the massive unitary calculations. The false, mechanistic 'indeterminism' is therefore nothing but another manifestation of the general 'method' of artificial randomness insertion into the regular unitary dynamics, basically similar to 'quantum chaos' imitations [14] (see also Chapter 6) and other cases of such stochastic, non-dynamical and inconsistent, introduction of empirically observed randomness. By contrast, the real, dynamically multivalued micromachines, including their successfully operating natural versions, obtain their main *advantages* just from the omnipresent dynamic chaoticity, which uniquely ensures the true miracle of creation (of interaction results), at the expense of some unavoidable, but rather *positive* uncertainty (see also Chapters 7, 8). Thus, irregularity of emerging natural micro- and macrostructures, including those of living organisms, is a sign of their *higher*, ra-ther than lower, dynamical symmetry (the universal symmetry of complexity [1]) underlying their real possibilities (like self-reproduction and very large, dynamic and creative adaptability), which exceed qualitatively the illusive efficiency of 'regularly symmetric' products of unitary thinking, even though there is always an appreciable probability of failure of every complex-dynamical function (Chapters 7-9).

In a similar way, the property of dynamic entanglement of unreduced interaction process, accompanying its dynamic multivaluedness (Section 4.2), is fundamentally different from the unitary 'quantum entanglement', by both irreducible universality of the former and unrealistic, and even 'mystical', interpretation of the latter. While the dynamic entanglement involves real, fractally structured, inseparable and permanently changing mixing between the interaction components, constituting the essence of chaotic realisation change, conventional quantum entanglement, concerning exclusively interaction between several quantum particles, implies a mechanistically fixed, non-dynamical, purely mathematical and separable (linear) combination of 'state vectors', closely related to the 'exact-solution' logic of unitary paradigm (see Section 5.3). It is the illusive regularity of mystical 'quantum entanglement' that determines much of the expected 'magic' properties of conventional quantum computation and it is the probabilistic fractality and inseparability of real, essentially nonlinear dynamic entanglement that replaces the impossible 'free' advantages by quite real magic of distributed complex-dynamical creation.

Taking into account the above necessarily brief, but basically complete description of realistic, unreduced system evolution obtained by rigorous solution of the underlying many-body problem, eqs. (20)-(31), we can see now what a huge, practically significant and genuinely *complex* dynamic hierarchy of events filled with numerous links and conceptually specific phenomena is irreversibly lost in conventional theory in exchange for a deceptive 'unreasonable efficiency' of the perturbative unitary imitation. Its actual *in*efficiency appears through the basic fact of nonunitarity of any real micro-machine operation, including the unavoidable dynamical randomness of any single event of a real interaction process. Whereas this fundamentally substantiated conclusion completely devalues the conventional, unitary science of 'quantum computing' (as well as equally unitary imitations of 'quantum chaos', 'randomness' and 'complexity', see also Chapter 6), it opens up much *wider* prospects for micro-machine creation as explicitly chaotic, *dynamically multivalued* devices with qualitatively larger possibilities, which are *already realised* in all natural 'micro-machines' (or 'nanostructures') determining operation and development of natural physical, chemical and living systems, from the simplest molecules and surface structures to viruses and genetic code dynamics. The evidently 'hybrid', 'quantum-and-classical' origin and chaotic dynamics of those natural micro-machines provide an important general agreement with the obtained rigorous solution of (quantum or classical) many-body problem, its dynamical properties and universality, whereas the empirically well-known results of natural micro-machine dynamics confirm the practically unlimited possibilities of such unreduced complex dynamics. The properties of unreduced complexity, consistently obtained within the dynamic redundance paradigm [1-4,9-13] and briefly described above (Sections 4.1-4.5), such as dynamic adaptability, dynamic fractality (self-development) and intrinsic unification of regularity and randomness, actually explain the observed properties of real interaction processes, remaining fundamentally inaccessible to the unitary approach, which shows that the desired reproduction of all the possibilities of unreduced complexity in artificial (and modified natural) micro-machines can be a *realistic* task, but *only* if one uses the unreduced, dynamically multivalued description of all the interactions involved and the corresponding conceptual extension of the approach applied as compared to conventional, unitary thinking dominated by the purely 'calculative', one-dimensional logic (see also Chapters 7-9).

    An additional and very important confirmation of validity of the proposed extended approach comes from the fact that it provides the unique, totally consistent, realistic and *intrinsically unified* picture of the whole micro- and macro-world dynamics involving rigorous and physically transparent solutions of the known fundamental problems [1-4,9-13] mentioned above (Section 4.6) and remaining unsolvable within conventional, unitary science, despite all the 'post-modern' plays of words in its official sources. This property of unreduced, complex-dynamical (multivalued) description of the world reflects its *intrinsically cosmological* character: every entity, law, or property can only be explicitly and causally *derived* from lower-level entities, in exact correspondence with their *real emergence* (creation) processes, as opposed to positivistic, purely mechanistic 'registration' of

empirical observation results in conventional science and its version of 'cosmology' containing nothing but effectively one-dimensional, unrealistic 'photograph' of already existing, basically unchanged and fatally simplified entities and their properties (including the formally postulated dependence on abstract time-parameter, cf. Sections 4.3, 7.1). It is important that the complex-dynamical, explicitly creative cosmology of the universal science of complexity comprises within its unified picture the dynamically multivalued processes of emergence of *all* existing entities (including their unreduced properties and evolution laws), starting from the elementary field-particles, physical space and time themselves, up to structures and phenomena from the highest perceived complexity levels, such as human consciousness and all products of its activity (classified as 'inexact', explicitly subjective knowledge within conventional, unitary science).

In that way, the characteristic 'unsolvable' problems from the fundamental levels of conventional cosmology, such as the problems of time, wavefunction of the universe, classical world emergence and evolution, structure creation and the law of energy degradation, are naturally solved from the beginning [1-4,11-13], together with similar problems from higher complexity levels, such as life emergence and evolution. The first stages of real, physical universe evolution are naturally obtained thus as explicit emergence of the lowest complexity levels of the world, where the purely 'geometric' accent and totally abstract, zero-complexity picture of conventional 'general relativity' and related 'Big-Bang' cosmology (including all its 'modified', 'post-modern' versions) look as explicitly wrong, really over-simplified imitations of reality.[12] The link between the fundamental world structure (quantum cosmology), quantum computation and other related applications of quantum mechanics has been extensively (and speculatively) exploited within the unitary theory of quantum information tending to present the world as a gigantic unitary computer (see e. g. [62,63,74,75,76] and Chapter 7 for more details). However, the obtained causally complete results of quantum field mechanics permit us to state now that the universe is rather the process of unceasing *creation* distributed in the permanently evolving space and time and qualitatively different from the unitary 'calculation' schemes, at all levels of complexity (cf. Chapter 9). The unreduced, physically real information represented by the extended action and generalised potential energy is *qualitatively transformed*, in the course of those creative interaction processes, into the dual complexity form, generalised dynamic entropy (Section 7.1) [1], as opposed to mechanistic 'reading' of a 'ready-made', given 'world programme' expressed in terms of purely abstract, dimensionless 'bits' in the conventional theory. The real, irreducibly multivalued micro-machine operation, as well as any computation process, represent therefore particular manifestations of intrinsically unified, universally expressed and uniquely directed *symmetry of complexity*, this conclusion and the underlying causally complete picture having important *practical* consequences for further technological, social and intellectual development (Chapters 7-9).

---

[12] It is not difficult to see, in particular, that the recently re-established 'brane-world' schemes from the official field theory and cosmology represent but a unitary, purely abstract and formally imposed imitation of the physically real and unified system of two interacting protofields from the quantum field mechanics. This system is the simplest possible and therefore inevitable case of 'world-wide' interaction that does not contain a priori any observed structure. The latter, starting from the elementary field-particles is consistently and progressively *derived*, in full compliance with observations (including realistic resolution of the canonical 'quantum mysteries', causal, intrinsic relativity, dynamically unified and explained fundamental interactions, etc.). By contrast, the same structures, in their strongly reduced, abstract and contradictory version (including the 'necessary' number of pre-existing, abstract 'dimensions'), are artificially inserted (postulated) in any field-theoretical imitation of reality, together with a long series of 'fundamental principles' and other abstract rules describing their behaviour. Such are the conventional, microscopic 'strings' and 'branes', whose redundant number of versions, separation from reality and between them and inconsistent interference with, or mechanistic 'extension' to, the *macroscopic* 'world brane' only confirm the fundamental, irreducible limits of the unitary, dynamically single-valued imitation of reality.

# 5. Dynamically multivalued computation by real quantum and hybrid machines

## 5.1. The myth of unitary quantum computation: its formal origin and impossibility of realisation

The above rigorous analysis and solution of the generalised many-body problem (Chapters 3, 4) has revealed the fundamental origin of irreducible dynamic randomness and consistently defined complexity in any real system (interaction process) thus confirming the basic deficiency of unitary quantum computation, as it was outlined in Chapter 2 in terms of more general, qualitative (but still fundamental) arguments. Now we can provide a more specific and complete application of the obtained unreduced, complex-dynamical interaction description to the problem of computation by sufficiently simple structures containing essentially quantum elements or dynamic regimes. What we mean here by 'computation' is the real micro-system dynamics and its results, specified in our approach as a set of dynamically interconnected system realisations (Section 3.3), rather than formal manipulation with abstract symbols or mathematical 'bits' artificially attached to a simplified unitary 'model'. A more general theory of real computation, including its fundamental definition and related criteria, is given in Chapter 7.

We begin with a summary of the exact, rigorously specified meaning and origin of unitarity of conventional quantum computation, including its impossibility in real systems and related illusive, 'magically high' efficiency. As rigorously shown above (see especially Sections 4.5-7), any system evolution, determined by the unreduced development of the driving interaction within the system, contains the intrinsic, irreducible and omnipresent randomness of purely dynamic origin designated as dynamic redundance phenomenon, i. e. permanent system jumps between its different, but equally real states, or realisations, taken by the system in a causally random, chaotic order. It is clear that this genuine and inevitable randomness, occurring within any real, even elementary, interaction process, excludes any unitarity in principle. In particular, the dynamically chaotic change of realisations provides the intrinsic source of irreversibility and the unceasing, irreversible flow of the physically specified time [1,11-13] (see also Sections 4.3, 7.1), which agrees with the observed reality properties and diverges from the intrinsic reversibility of unitary evolution. Moreover, our analysis reveals the exact origin and meaning of the unitary model of natural processes, constituting the basis of conventional fundamental physics: it corresponds to incorrect, perturbative reduction of system dynamics to *only one*, arbitrarily 'averaged' or empirically adjusted, realisation, whereas all other realisations, extremely numerous and diverse in any real case, as well as chaotic transitions between them, are totally neglected or inconsistently imitated by 'stochastic' effects of external 'noisy influences' upon the single realisation, which does not change the basic, *dynamic* single-valuedness, and thus unitarity, of those reduced, 'exact' solutions of conventional approach.

The transient motion of a real system in each of its realisation is indeed approximately unitary (if one neglects other levels of the fractally structured complexity within each realisation, cf. Section 4.4), but it is always interrupted soon enough by a chaotic jump into another, randomly chosen realisation occurring, in addition, through a rapid transition to a qualitatively different, delocalised state of the intermediate realisation constituting the generalised, causal system wavefunction, or distribution function (Section 4.2). The system evolution in the whole, representing also the universal internal structure of any interaction process, consists therefore from such small 'pieces' of 'under-developed' (approximate) unitarity separated by explicitly nonunitary (but dynamically continuous) jumps, so that the quasi-unitary evolution within any single realisation can barely have enough time to fully develop (stabilise) itself before it is interrupted by a next irregular jump.[13] It is important that the highly nonuniform, essentially nonlinear jumps, involving system transformation between 'localised' and 'delocalised' states (generalised dynamical 'collapse'/entanglement and extension/disentanglement), are consistently derived as natural results of the *same* interaction that forms each pseudo-unitary (regular) realisation structure and therefore they cannot be neglected with respect to those quasi-unitary components, as it is done in the conventional approach by inconsistent extension of its perturbation theory approximation.

---

[13] This picture provides the causally complete, dynamically specified and universally applicable version of the popular idea of 'punctuated equilibrium' obtained in various forms and fields of science as formal generalisation of empirical observations clearly demonstrating the qualitatively uneven character of natural development that remains fundamentally mysterious within any conventional, unitary interpretation.

The unitary model imitation of reality may seem to be closer to observations for one limiting case of dynamically multivalued behaviour, the generalised, multivalued SOC (self-organised criticality), where the system realisations closely resemble each other and therefore their (relatively frequent) change can remain unnoticed (Section 4.5.1). However, as pointed out in the above Sections, this regime has little chance of appearance just within the essentially quantum interaction dynamics, which can be causally understood (Section 4.6) rather as the opposite limiting case of complex behaviour, the uniform chaos regime (Section 4.5.2), characterised by relatively big difference between realisations and visible, large manifestations of 'quantum' jumps between them. The standard, analytically derived properties of dynamically multivalued behaviour in the uniform, or 'global', chaos regime provide the causally complete explanation for the specific, 'quantum' features of elementary micro-system dynamics, such as 'quantum uncertainty', 'indeterminacy' and discreteness (quantisation) universally determined by Planck's constant [1-4,11-13] (see also Section 5.3), which are invariably endowed with a mystical air of 'veiled reality' [16] (or 'shadows of the mind' [15]) in the unitary quantum theory based upon the same, dynamically single-valued and necessarily abstract model of the world (it imposes indeed an *artificial* limitation, or 'veil', upon the genuine, dynamically multivalued, interactional and living world content, consistently 'unveiled' in the quantum field mechanics, once the evident simplification of the perturbative approach is abandoned). Even when a particular, delocalised version of self-organised multivalued structure, such as physically real de Broglie wave [1,2,11-13], appears in quantum dynamics, giving rise to the causally explained property of quantum coherence (Section 5.3), it involves a very small dynamic proportion of regularity that *cannot* be actually controlled by interaction with other objects, since it arises from the very *first* level of interaction between the primordial protofields giving rise to *every* perceivable entity of this world.

It is only at a higher complexity level of classical behaviour, starting approximately from the elementary bound state formation (such as atoms), that the (externally) more ordered behaviour and distinct, *actually* controllable structures of the SOC regime first appear in interaction between essentially quantum, uniformly chaotic components (Section 4.7). Therefore the conventional, unitary theory of quantum computation, including all its 'quantum control' schemes and mathematical fight against fictitious 'decoherence', ignores both the real, complex-dynamical (multivalued) content of essentially quantum behaviour and the inevitable emergence and utility of classical elements within the inseparably *unified*, self-developing microsystem dynamics (as it happens in all natural 'nanomachines', starting from functional molecules). The dynamical 'phase transition' from the clearly step-wise, 'yes-or-no' type of essentially quantum behaviour to more continuous, 'fine-grained' motions of classical type, allowing of their efficient (precise) control, is a necessary and common element of any practically useful micro-machine operation.

The finite irreducible increments of time, $\Delta t$, and space, $\Delta x$, in a regime of quantum, explicitly irregular and uneven dynamics are expressed through the fundamental, universally fixed increment of action, $\Delta \mathcal{A}$, known as Planck's constant, $|\Delta \mathcal{A}| = h$ [1-4,12,13]. The quantised action element describes, in the extended interpretation of the universal science of complexity (see also Sections 7.1, 7.2), the internally nonlinear, step-wise qualitative transformation of the unreduced dynamic (integral) complexity during one elementary cycle of system evolution including its reduction to the current realisation and the following transition to the next realisation. The lowest, properly 'quantum' levels of complexity differ from higher levels by the universally fixed value of this complexity-action quantum (negative by sign), $\Delta \mathcal{A} = -h$ (whereas at higher levels the complexity transformation quanta $|\Delta \mathcal{A}| \gg h$ are not strictly fixed and can vary, usually around one or several characteristic values of action). The elementary complexity change appears as *emerging* space structure element and therefore it is proportional to both space and time increments with the coefficients known as energy, $E$, (with the negative sign) and momentum, $p$:

$$\Delta \mathcal{A} = -E\Delta t + p\Delta x \ . \qquad (36a)$$

This relation is known from classical mechanics, but it acquires a universal and physically meaningful, essentially nonlinear interpretation in the unreduced science of complexity (see also Section 7.1).

Now, if we are interested only in the full temporal increment of action for a quantum system (so that $\Delta x = 0$ in eq. (36a)), this fundamental relation, reflecting the underlying complex dynamics, takes a familiar form, simply postulated in conventional quantum mechanics, starting from the pioneering work of Max Planck on the black-body radiation (see [3] for the detailed story and references):

$$\Delta \varepsilon = h\nu , \qquad (36b)$$

where $\Delta \varepsilon$ is the corresponding energy change (equal e. g. to the discrete energy-level separation), $h = -\Delta \mathcal{A}$ is the mentioned universal increment of complexity-action and $\nu = 1/\Delta t$ is the frequency of the process in question (equal e. g. to the frequency of radiation exciting a transition between the neighbouring energy levels, cf. the resonance condition and mechanism of global chaos, eq. (35)). It is important to understand that because of the complex-dynamical discreteness of the underlying interaction process (Section 4.3) the elementary 'step' of quantum system evolution described by eqs. (36) cannot be further subdivided into smaller parts or physically 'traced' in its continuous development, contrary to what is often done within mathematical theory of unitary quantum computation, conventional quantum chaos and related 'fine' experimentation schemes, in contradiction with even the standard quantisation postulate of conventional quantum mechanics.

The dynamically inhomogeneous evolution of any real quantum system is simulated, in the conventional theory and its applications to quantum computation, by the unitary 'evolution operator', $U$, having the standard, exponential form for the time-independent (conservative) system:

$$U = e^{-\frac{i}{\hbar}Ht} , \qquad (37)$$

where $H$ is the system Hamiltonian (operator), and in the general case of time-dependent Hamiltonian the term $Ht/\hbar$ in this expression should be replaced by a certain (unknown) Hermitian operator. Although it is supposed that $H$ has a generically discrete essential spectrum, according to the standard quantisation postulates the action of unitary evolution operator smoothly depending on time, eq. (37), on the system wavefunction (or 'density matrix', in the corresponding formalism) will contain a superposition of smoothly varying, regular functions that differs dramatically from the highly uneven, intermittent and permanently, chaotically changing mixture of transient quasi-unitary realisations and probabilistic jumps between them described above for the unreduced, real-system evolution. This comparison demonstrates the essential difference between the postulated, dy-namically single-valued projection of conventional theory, which is apparently 'successfully verified' only in the case of simplest, quasi-regular (separable, or integrable) problems and the underlying real, dynamically multivalued evolution that determines any quantum system behaviour, but shows especially pronounced, irreducible deviations from the unitary imitation for more complicated, 'nonintegrable' (or 'nonseparable') problems.

Indeed, the standard evolution operator expression of eq. (37) is obtained by the postulated extension of a result valid only for the exact eigenvalue problem solution, on which the operator is supposed to act. This is equivalent to the unjustified, and actually incorrect, extension of a perturbation theory result that can be obtained in the 'closed' form of 'exact' solution, to the arbitrary, 'nonintegrable' problem case, whereas the same conventional theory cannot say at all what should actually 'happen' to such 'nonseparable' system (it tends to assume that there should be 'some' solution, generally similar to the assumed perturbative one). Our unreduced interaction analysis shows that contrary to superficial hopes of canonical theory, something very special does happen to any 'nonseparable', and thus actually any real, system once one avoids any fatal reduction of perturbative approaches. The real 'evolution operator', even if it could have sense, cannot have the form of a smooth, exponential, or trigonometric, or any other 'analytic' function of time (like that of eq. (37) with an arbitrary anti-Hermitian operator in the exponent), since its unreduced action on any real-system state would immediately produce hierarchical system splitting into incompatible realisations, which automatically kills the assumed unitarity because of the related dynamical randomness, nonlinear jumps between realisations, etc.

Extensive applications of the basically incorrect construction of unitary evolution operator to quantum computation problem (and other cases of many-body problem) contain further incorrect approximations reduced to the same kind of unjustified extension of perturbative analysis results. Since for any nontrivial, nonintegrable problem the action of the formal evolution operator cannot be obtained in a 'closed', analytical form (which is equivalent to problem integration), applications using this formalism are forced to resort to a perturbation theory approximation, equivalent to the assumed smallness of the exponent determined by the characteristic action value expressed in units of Planck's constant. However, already the canon-ical 'quantisation postulate' (causally derived and extended in the quantum field mechanics as a standard property of the underlying complex dynamics [1-4,12,13]) shows that this quantity can take only integer values and cannot therefore be smaller than one by absolute value.

In the case of quantum computation theory, the 'necessary' perturbative smallness is often obtained by artificial subdivision of time and thus evolution into small enough intervals, after which the expansion of the assumed exponential function within each interval transforms the problem into an effectively integrable one, permitting various manipulations with the unitary evolution operator in the limit of infinitely small time intervals. The obtained results are said then to be valid for the unreduced system evolution. It is clear, already from a general point of view, that the problem decomposition into solvable subproblems cannot be as trivial as that, and the committed error is hidden in the assumed possibility of formal, arbitrary division of time (evolution), which is related also to the mechanistic, 'parametric' role of time in the unitary paradigm that does not see its genuine, dynamical origin just related to the complex-dynamical discreteness, or 'quantisation', of a real interaction process [1-4,11-13] (Section 4.3).

As we have seen above, in reality the system evolution cannot be represented as mechanistic addition (sequence) of practically independent small increments. It is composed of naturally emerging, inhomogeneous cycles, where each cycle is relatively big in the case of essentially quantum dynamics and combines system transient motion in a current realisation with its chaotic transition from a previous (or to the next) realisation, thus determining both causal, dynamic quantisation of motion and relatively big one-step advance of real, physical time of this level actually counted just by those cyclical events of dynamic reduction to and extension from a current realisation. In terms of formal time division into increasingly small intervals, this means that the evolution within each subsequent interval is not independent of previous intervals: consecutive (artificial) intervals gradually 'accumulate' potentialities for the system chaotic transition to another realisation and at certain 'moment' (within certain interval) the system starts leaving its current realisation and taking the next, randomly chosen one, in a 'catastrophically' uneven, self-amplifying way, which is a qualitative violation of the assumed unitarity, since it cannot be described by smooth, unitary 'small increments' even within the chosen 'small interval' of time. Therefore the formal time division into small intervals, introducing a perturbation theory version, does not make sense for real, intrinsically non-perturbative quantum system dynamics: any smallness will be broken and overridden by self-amplifying, essentially nonlinear reduction-extension cycles of dynamically multivalued interaction development.

The only type of time intervals pertinent to real quantum dynamics is given by the unreduced, essentially nonlinear version of that dynamics itself, in the form of duration of *unbroken* realisation change cycles obtained from the causally derived version of 'uncertainty relation' [1], $E\Delta t = h$, just describing the cycle result, eqs. (36). Since this time interval corresponds to modulus 1 of the exponential function argument in the expression for the unitary evolution operator, it becomes clear that any its perturbative, linear expansion and all related conclusions are irrelevant to real quantum dynamics. Moreover, as noted above, the exponential evolution operator, eq. (37), represents itself a basically incorrect imitation of reality justified by inconsistent extension of its integrable models. The real temporal change of a system during one complex-dynamical cycle of realisation switch is determined by an *externally* linear relation for its complexity-action, $\Delta \mathcal{A} = -E\Delta t$ (see eq. (36a)), or

$$\mathcal{A} = \mathcal{A}_0 - E(t - t_0) = \mathcal{A}_0 \left(1 - \frac{Et}{\mathcal{A}_0}\right) \qquad (38)$$

(for $t_0 = 0$), expressing the result of the underlying essentially nonlinear, chaotic system 'jump' between realisations. Similar linear dependences can be obtained for other quantities characterising one quantum evolution step, including the wavefunction [1,4,12,13] (see also Section 7.1). The expression in brackets in eq. (38) could be considered, within a perturbative approach, as a linear approximation to exponential (or trigonometric) dependence of eq. (37) that could be conveniently used for imitation of any system development or 'magic' calculative power (see below), as it is actually done in the unitary theory. The truth, however, is that there is no any exponential function at all behind the step-wise, pseudo-linear system advances described by eqs. (36), (38) or similar expressions for other quantities. Once the argument of assumed exponential function attains a value around unity (at $E\Delta t = h$), after which the exponential function differs essentially from a linear dependence and just 'becomes itself', the system performs an abrupt 'jump' to the next realisation and the pseudo-linear stage of eq. (38) restarts with a new system configuration (localisation centre).

The unitary quantum theory performs thus a doubly incorrect manipulation when it extends its perturbative, effectively linear models to the exponential form of the general evolution operator and then again approximates the latter with perturbative, linear expansions for arbitrary conditions. In reality, there is only the pseudo-linear time dependence (generalised to other power-law dependences for a sequence of many chaotic jumps between realisations), and it is valid only for the dynamically quantised, unbroken cycles of multivalued evolution. A more rapid system change does occur, but within a qualitatively different part of its interaction development, the chaotic jumps between realisations separating the above pseudo-linear steps, where the effective system evolution is even much *more* rapid than predicted by exponential dependence of eq. (37). In the case of isolated elementary field-particle or its simple enough interaction the pseudo-linear advances correspond to the localised, corpuscular state (and very short 'trajectory') of the particle, while the delocalised chaotic jumps form the physically real, dynamically probabilistic wave field of the causally understood wavefunction $\Psi$, thus providing a demystified explanation of the wave-particle duality (Section 4.6.1) [1-4,11-13]. The very uneven alternation between these two qualitatively different parts of any real interaction process is imitated by the 'smoothing', 'coherent' exponential dependence of the unitary evolution operator (similar to other appearances of the false exponential dependence [1], see also Chapter 6), so that the false exponential, quicker than linear (real) growth during one evolutionary stage (within each realisation) is supposed to be 'compensated' by its slower than real rate of change during system jumps between realisations. However, this extremely superficial, arbitrarily assumed 'compensation' cannot be satisfactory and leads to strong qualitative and quantitative deficiencies, such as the total loss of omnipresent, truly dynamical chaos (including the fundamental quantum indeterminacy, 'quantum chaos' itself and probabilistic 'quantum measurement').

It is clear now that the real (quantum) system evolution can be represented (here only schematically) not by a simplified, analytical time dependence of the conventional evolution operator, eq. (37), but rather by a highly nonuniform sequence of two qualitatively different stages:

$$\ldots\left(1-\frac{E_1(t-t_1)}{\mathcal{A}_1}\right)\uparrow\Psi\downarrow\left(1-\frac{E_2(t-t_2)}{\mathcal{A}_2}\right)\uparrow\Psi\downarrow\ldots\left(1-\frac{E_n(t-t_n)}{\mathcal{A}_n}\right)\uparrow\Psi\downarrow\ldots,$$
(39)

where $\Psi$ designates schematically the 'wavefunctional' stage of chaotic jumps between the 'regular' realisations, the arrows around it express the qualitative change occurring in the system during the extension/reduction process of each jump and the values of $E_n$, $\mathcal{A}_n$ can, in general, be equal or different for different pseudo-linear stages. The evolution in the whole can be described by the generalised Schrödinger equation corresponding to the canonical form for the lowest levels of quantum dynamics (without any external, artificially introduced 'stochasticity'), but it should be provided with the unreduced, dynamically multivalued solution, eqs. (20)-(31), just giving the nonunitary system evolution expressed by eq. (39) and leading to higher complexity levels. Note that the characteristic action change, $\Delta\mathcal{A}$, during one complete evolutionary cycle, determining the related increments of time and space, equals to $-h$ for the lowest complexity sublevels (so that $\mathcal{A}_n = nh$ for any $n$ in eq. (39)) and remains close to it for any 'essentially quantum' (delocalised) dynamics, but then starts gradually varying with further complexity growth ('semiclassical' dynamics) until a more serious change occurs with the first classical configuration appearance (Section 4.7), after which the direct relation between $\Delta\mathcal{A}$ and $h$ is lost, as well as universality (single-valuedness) of the elementary action increment. However, the discreteness of $\Delta\mathcal{A}$, its well-specified dynamical origin and 'evolutionary' consequences described above are preserved at any complexity level.

Finally, the difference between the exponential, unitary imitation of quantum system evolution, eq. (37), and its real, dynamically multivalued content, eq. (39), demonstrates the origin of the expected, but actually illusive, 'exponentially high' efficiency of conventional quantum computation, constituting the basis of increased popularity of quantum information processing in the last years. Indeed, the inevitable moments of 'quantum measurement', arbitrarily (and inconsistently) 'incorporated' into the usual scheme of unitary quantum computation correspond to anti-Hermitian parts (imaginary effective potential) of the Hamiltonian in the argument of the exponential evolutionary operator, eq. (37). Inserting the explicit Hamilto-nian decomposition into Hermitian and anti-Hermitian parts, $H = \text{Re}(H) + i\text{Im}(H)$, into eq. (37), we obtain the exponentially growing or decaying factor within the evolutionary operator (actually appearing in measurements through its observable matrix elements):

$$U = e^{\frac{\text{Im}(H)t}{\hbar}} e^{-\frac{i}{\hbar}\text{Re}(H)t}. \tag{40}$$

While $\text{Im}(H) < 0$ for ordinary dissipation effects (see e. g. [121]), which corresponds to the normal decay of coherent process intensity, the coherent computation dynamics can contain also some spurious, effectively amplified components with $\text{Im}(H) > 0$, giving eventually the exponentially growing computation efficiency. This effective amplification is due to the self-amplifying growth of the dynamical system 'volume', i. e. essential openness of computation dynamics, where ever larger number of system eigenstates is involved in the computation process (whereas any given dynamical volume can produce only decaying computation intensity). The exponential growth within the unitary system evolution reflects the self-amplifying propagation of its development by the conventional mechanism of unitary 'quantum entanglement' of system components, 'pinpointed' by the 'quantum measurement' elements (which realise, in particular, the quantum computer 'input' and 'output'). Any quantitative estimate of quantum computer efficiency will be proportional to matrix elements of the 'measured' unitary evolution operator, eq. (40), containing the exponentially large factors of the above origin, which demonstrates once again the nature of the 'magic', exponentially large efficiency of quantum information processing that constitutes its main expected advantage.

    It is important that the real, nonunitary quantum system evolution following the random, inhomogeneous sequence of stages of eq. (39) is incompatible with such coherent growth mechanism of exponentially high efficiency. Indeed, even replacement of chaotic jump phases by simple multiplication of linear stages in eq. (39) gives a power-law, rather than exponential, dependence. The 'magic', exponentially high efficiency can indeed be obtained in the real, dynamically multivalued system, but with the help of a quite different, incoherent mechanism involving dynamical randomness and self-developing fractality of the unreduced complex dynamics (Chapters 7, 8). The alleged exponentially large efficiency of uni-

tary quantum computation seems to be directly related thus to the (false) exponential dependence in the unitary evolution operator expression.

Based on the found qualitative difference of real quantum system dynamics from its unitary imitation, we can understand now why this strangely gratuitous advantages of quantum computation cannot be realised in principle, even in the ideally perfect system, devoid of any hypothetical, noise-induced 'decoherence'. The real evolution of a compound quantum system, even in the absence of any external 'noise', consists in a truly *irregular*, highly nonuniform sequence of jumps between incompatible system realisations, eq. (39), which *cannot*, therefore, be coherent or produce anything like the unitary, smooth time dependence of eq. (37). Deviations from coherent, ordered development are especially high just in the case of essentially quantum dynamics (Section 4.6), whereas the emergence of semiclassical and classical elements in the interaction development is described by *partially* coherent/ordered, but *always* dynamically *probabilistic* sequence of realisations (see also Section 5.3). It is important also that in any full-scale, practically useful quantum machine, realising a large number of elementary interactions with irreducible dynamic uncertainty within *each* of them, one will actually have quickly accumulating, *exponentially* growing randomness, which will be *added* to the fundamental 'quantum uncertainty' from the lowest sublevel of quantum complexity (Section 4.6) and would reduce to zero any 'exponential efficiency' of *unitary (regular)* computation, even if the latter could exist. In this sense, one should interpret exponentially growing factor in eq. (40) for a real, full-scale quantum computer rather as an *unpredictably* growing, fundamentally uncontrollable *uncertainty*, incompatible with any useful application of such *unitary* imitation of real, multivalued quantum computation dynamics.

The real quantum system dynamics is much closer to a random-walk process that exhibits, in average, a power-law dependence on time, with occasional very rapid transitions to higher complexity (sub)levels that form a quite different kind of 'magic' efficiency resembling a 'sudden' idea or unifying comprehension breakthrough in the process of human thinking. This another kind of efficiency of natural 'micro-machines' and 'computers' of neural network type (generalised 'brains') is based on the fundamental property of *creativity* of real, multivalued interaction dynamics (Sections 4.3, 4.4), so that the integrated result of the multi-component, probabilistically fractal hierarchy of natural interaction development emerges 'suddenly', in a single self-amplifying jump, thus largely compensating in one 'moment of truth' all visible 'losses' of the chaotic search process. By contrast, the unitary 'evolution operator', eq. (37), similar to any dynamically single-valued, perturbative imitation of reality *cannot* create anything new at all, including any expected result of a computation process. Whereas in many existing applications of conventional quantum theory to relatively simple systems interaction results are known (measured) empirically and are then somehow adjusted within an artificially composed, postulated 'model', the full-scale *computation* process, as well as other explicitly creative cases of more complicated quantum interactions with basically unpredictable results, cannot be treated in that way in principle, thus clearly showing the limits of the unitary, dynamically single-valued paradigm of conventional science.

The 'exponentially rapid' growth of unitary evolution result explaining its expected 'magic' efficiency, eq. (40), clearly appears now as a basically incorrect imitation of the natural, nonunitary (uneven and unpredictable in detail) process of dynamical fractal growth (Section 4.4). On the other hand, the relatively efficient control of usual, classical computer dynamics, reducing it to a highly ordered (but never totally regular!) case of multivalued SOC type of behaviour opposed to any such kind of 'extended' growth, cannot be applied to the inevitably coarse-grained structure of essentially quantum dynamics, as it is incorrectly, mechanistically done within the conventional quantum computation theory. In any case, the quasi-regular classical computer dynamics appears to be absolutely inefficient with respect to any higher-complexity, e. g. biological, system dynamics or brain 'computation' process, which gives rise to the idea of neural-network kind of macroscopic computer. What our unreduced interaction analysis shows is that for essentially quantum system dynamics of relatively low complexity levels, including partially classical (hybrid) configurations, one can have *only* such kind of explicitly chaotic, irregularly 'ordered' dynamics and dynamically chaotic (multivalued) type of 'computation', fundamentally different from any unitary, dynamically single-valued model (see also Section 7.3 and Chapter 8).[14] Hence, any real micro-machine dynamics

---

[14] Note, in particular, that any kind of conventional 'chaos' and 'complexity', giving rise to the corresponding idea of 'chaotic', or 'probabilistic', or 'stochastic' computation, can neither ade-

cannot be understood, even approximately, within the conventional, dynamically single-valued theory and should be described within the unreduced, dynamically multivalued interaction analysis [1-4, 9-13] (see Chapters 3, 4 for the general theory with selected applications and Section 5.2 for more detailed results for quantum computing systems).

We finish this Section with a summary of characteristic, most important deficiencies of the conventional, dynamically single-valued (unitary) concept of quantum information processing, outlined in Chapter 2 only with the help of fundamental principles of unitary science itself and now confirmed and specified by comparison with the unreduced, dynamically multivalued interaction analysis and the resulting causally complete many-body problem solution (Chapters 3, 4) [1-4,9-13]:

• The conventional concept and theory of quantum computation actually uses only the simplified, linear part of canonical quantum mechanics, leaving aside the evidently nonlinear effects of the underlying detailed dynamics, which are taken into account by certain standard postulates about probability involvement, quantum measurement and apparent wavefunction 'reduction'. Whereas the simplest quantum systems with explicitly observed interaction results can be (approximately) described using those 'mysterious' postulates and a number of adjustable parameters, more involved behaviour of 'computing' quantum systems with dynamically changing configuration and unknown (in principle, arbitrary) result needs the well-specified, causally complete understanding of the detailed interaction development in the many-body quantum system performing information processing. Any arbitrary manipulation with linear compositions of abstract 'state vectors' in the conventional theory, where the interaction result emergence is 'described' by a symbolical 'arrow' with a formally attributed, postulated output or by an equally poorly substantiated equation for an abstract density matrix (distribution function), cannot reproduce in principle the unreduced interaction dynamics that includes qualitatively specific effects of dynamically random (multivalued), physically real en-

---

quately describe the real micro-machine dynamics, since those conventional theories provide themselves only dynamically single-valued, mechanistic imitation of chaoticity/complexity trying to reproduce its given results, but not the unreduced, dynamical mechanism of their emergence in progressive interaction process development (which forms the genuine, universally defined content of the first-principles analysis). See Chapters 6-8 below for further development of this important difference of our reality-based dynamic complexity from its abstract, unitary imitations within the conventional science paradigm.

tanglement between the interacting degrees of freedom. Any 'stochastic' effects of external 'noise' and related abstract 'decoherence' or false 'quantum chaos' of conventional mathematical 'models' of quantum computing systems have nothing to do with the internal dynamical randomness of any real, even totally isolated, system with interaction, completely devaluing its unitary imitation and the derived results.

• Any real computation is a structure creation process necessarily involving essential, intrinsic irreversibility, uncertainty and nonunitarity that cannot be properly taken into account by conventional, basically unitary (dynamically single-valued) theory. The latter cannot consistently describe just the most important part of computation dynamics, its desired result emergence, represented by a new, dynamically created system configuration. Any reference to a poorly understood 'quantum measurement' from the evidently incomplete standard quantum mechanics cannot solve the problem, since any real measurement process constitutes an integral part of interaction development within the 'computing' quantum system. The unreduced interaction analysis shows that it is the causally complete picture of quantum measurement and classical behaviour emergence that can be obtained within the universally nonperturbative description of a quantum computer (or any machine) dynamics, giving simultaneously the dynamically complex (multivalued) result of 'computation'. It follows, in particular, that any useful 'quantum' machine should contain dynamically emerging classical elements (configurations) making it rather a 'hybrid', quantum-and-classical system, in agreement with the natural micro-machine operation. An indispensable part of a computing micro-machine dynamics is its memory whose useful operation can only be based on a strongly nonunitary, irreversible, and actually classical, behaviour, as opposed to the qualitatively incorrect imitations of memory by unitary dynamics within the conventional theory of quantum computation.

• The realistic, causally complete theory of quantum measurement within the dynamic redundance paradigm presents it as a particular case of the unreduced, dynamically multivalued interaction process in a quantum system, possessing a small dissipativity (openness to the exterior world). Therefore the notorious quantum uncertainty and discreteness of quantum measurement, as well as the accompanying wavefunction (quantum system) 'reduction' or 'collapse', simply postulated and remaining 'mysteri-ous' (highly contradictory) within the conventional theory, emerge as standard manifestations of dynamically chaotic (multivalued) behaviour at the quantum (lowest) complexity levels. The true quantum chaos, involving the genuine, well-defined randomness, is obtained as essentially the same phenomenon, but observed in the case of nondissipative, Hamiltonian dynamics. Both dissipative (quantum measurement) and nondissipative (Hamiltonian quantum chaos) manifestations of intrinsic dynamic randomness at the lowest (quantum) complexity levels provide the unified, realistic clarification and causal extension of obscure conventional theory ideas around unpredictability, undecidability and non-computability in quantum mechanics (and beyond). Considered against this background, the idea of unitary quantum computation (including unitary memory function) shows explicitly its fundamental deficiency: quantum computation, as well as any real interaction process in a quantum system, is always a particular case of unreduced quantum chaos, involving genuine randomness (nonunitarity) of purely dynamic (internal) origin.

• Because of the dynamic redundance phenomenon inherent in any real interaction, even in the total absence of noise, any real system cannot produce the regular, 100-percent-probability result. Schemes of 'chaos control' within the computing system, involving additional interactions, can decrease the dynamic uncertainty to a sufficiently low level only for macroscopic computers, operating at higher complexity levels. At the lowest, essentially quantum levels of system complexity, any control scheme cannot be efficient, as it is clearly implied already by the universal, irreducible discreteness (and related uncertainty) of quantum dynamics expressed by universal Planck's constant, which determines the unchangeable and relatively big action increment in any quantum process. Therefore the arithmetically 'exact', conventional computation type cannot be realised by an essentially quantum computer. On the other hand, the analogue 'simulation' of one quantum system by another cannot be useful in view of reproduction of the same problems within the computing system, including the irreducible and relatively great dynamical randomness. Such simulation of real interaction dynamics, contrary to its incorrect unitary imitation, cannot be 'universal', since the unreduced dynamic complexity always produces an 'individually specific' result. Instead of laborious production of an approximate quantum copy of each simulated system, it seems to be much more reasonable to use the universally applicable, causally complete analytic description of unreduced system dynamics in the universal science of complexity assisted by calculations on a classical computer, where maximum efficiency is ensured by hierarchical organisation of complex-dynamical problem solution. In any case, according to the universal complexity conservation law, a computer can properly simulate the dynamics of a system with the unreduced dynamic complexity not exceeding that of its own dynamics (which is the statement of the 'complexity correspondence principle'). In particular, any essentially quantum computation cannot correctly reproduce any classical behaviour (including its characteristic property of permanent system localisation), since the latter dynamically emerges from quantum interaction development as the next, higher complexity level.

• In summary, the widely acclaimed 'universality' of unitary quantum computation, as well as its related 'magic' (cost-free) efficiency, enter in fundamental contradiction with the demand for its 'physical realisation', after which the growing development of 'purely theoretic' aspect of quantum computation becomes completely senseless, especially taking into account its intrinsic subordination to the standard quantum theory, with its officially permitted, unitary 'interpretations' reduced inevitably to mere speculation, as opposed to the causally complete, explicit solution of persisting 'quantum mysteries'. The proposed 'small-scale' experimental 'realisations' of unitary quantum computation (see e. g. [38,97,107,108]) represent in reality but imperfect imitations of the expected properties, which cannot be extended to the full scale realisation in principle, as the long enough development of the field clearly confirms. The reason is the intrinsic, fundamental dynamic multivaluedness (randomness) of any unreduced interaction process and not some practical, 'technical' type of difficulties with micro-device fabrication or protection from imitative 'decoherence' induced by undesirable, external influences. The purely speculative, deceitful origin of conventional 'quantum computation' becomes evident here: one can always figuratively describe many of involved quantum processes with excitation and de-excitation of a complicated system energy levels as a discretely structured 'computation process' with unlimited, in principle, sophistication, where, for example, an excited system state corresponds to 1 and the ground state to 0, thus forming a 'quantum bit', but what does it actually change in our understanding of either quantum system dynamics or efficient computation processes? A similar kind of physically meaningless, trivial 'calculation' can be found within attempts to find the 'ultimate computing power' of a piece of matter, or even the universe as a whole, by simply dividing its total, unstructured energy by Planck's constant or its total equilibrium entropy by the Boltzmann constant [150,151], which is evidently nothing more than a mere play of words around trivially (and incorrectly) identified combinations of abstract symbols (see also Section 7.1). Quantum effects as such appear inevitably with decreasing size of real computer elements and will result in irreducible deterioration of conventional, 'digital' (regular) computation mode, with possible subsequent transition to the qualitatively different, dynamically multivalued type of behaviour, which can be useful, but necessitates the unreduced, complex-dynamical interaction description and understanding that involves new, essentially different computation purpose and criteria of result validity based on the causally complete concept of dynamic complexity (Chapters 7, 8).

 • In general, instead of 'calculation' and 'simulation' of one micro-system dynamics with the help of another one, it appears to be much more useful to create *productive* micro-systems with the desired properties that can be properly adjusted with the help of unreduced, dynamically multivalued approach of the universal science of complexity. These real micro-machines, similar to the existing natural ones controlling, in particular, the phenomenon of life, will combine and entangle in one dynamically multivalued, unreduced complexity development process the features of both (extended) 'computation' and 'production', which are mechanistically, artificially separated only within the basically limited unitary approach of the existing mode of science and technology. This well-substantiated new concept of micro-machine design includes serious clarification and modification of the growing direction of 'nanotechnology' (Chapter 8), also essentially based until now on the fundamental limitations of the mechanistic, unitary approach.

 • The conventional, unitary concept of quantum computation, as well as its evident and strangely persisting deficiencies, can be properly understood as a part of the much larger paradigm of the whole canonical science based on the same dynamically single-valued (unitary), perturbative approach closely related to the fundamentally abstract, imitative and 'symbolical', character of resulting unitary knowledge, as it is most completely ex-pressed in the idea of 'mathematical physics' born of the irreducible 'mysteries' of the 'new physics' at the beginning of the twentieth century. While the dynamically single-valued imitation of reality shows, in various fields of science, multiple and evident deviations from the observed total picture, it provides, like any low-dimensional projection, a superficial mechanistic 'simplicity' opening the way for an illusive, but quick 'success' in certain, simplest applications that admit straightforward 'postulation' of the (few) remaining 'mysteries' without their solution and actually give rise to the popular myth of "unreasonable efficiency of mathematics in natural sciences". However, as technology advances in its essentially empirical development, at a certain moment, actually just attained today in various fields of knowledge, it touches and starts essentially involving the full scale of unreduced dynamic complexity of respective systems, after which the mathematically 'guessed' and 'postulated', abstract kind of one-dimensional 'truth' becomes 'suddenly' absolutely inefficient and even unreasonably dangerous, so that any progressive, creative development of science and its applications is not possible any more without a decisive transition to the totally realistic, truly 'exact', causally complete understanding that actually involves the conceptually big, 'revolutionary' knowledge extension to the *unreduced*, dynamically multivalued and intrinsically unified image of reality at all its levels.

## 5.2. Dynamically chaotic (multivalued) operation of real micro-machines: general principle, universal properties and particular features

### 5.2.1. Time-periodic perturbation of bound quantum motion: Complex dynamics of elementary interaction act

We devote this Section to application of the general effective potential (EP) formalism of Chapter 3 to the particular case of a microscopic, 'quantum' element interaction with a time-dependent, periodic perturbation that constitutes a generic elementary stage of interaction process within a multi-component quantum machine of arbitrary final purpose. The perturbation, treated here as a non-small one, i. e. without any perturbation theory simplifications, can describe radiation coming from the outside (and thus realising the system 'input'/'output') or a signal transmitted from other ma-chine elements, which can eventually produce the quantised change of the element state constituting the smallest indivisible step in the machine operation. Note from the beginning that in view of the absolute universality of the basic formalism of unreduced science of complexity (Section 3.1) [1] this particular interaction and results of its development obtained below have a rather general meaning that includes, for example, any 'quantum-like' (e. g. optical) system operation [44], coordinate-dependent (rather than time-dependent), as well as non-periodic, perturbation, etc.

The quantum element dynamics modified by the time-dependent perturbation is described by the time-dependent Schrödinger equation for the system wavefunction $\Psi$, which is a temporal version of the simplest particular case of eq. (1) given by eq. (5b) (Section 3.1), with the unperturbed Hamiltonian $V_0(\mathbf{r})$ of the kind of eq. (2) corresponding to particle (e. g. electron) motion in a potential well:

$$i\hbar \frac{\partial \Psi}{\partial t} = -\frac{\hbar^2}{2m} \frac{\partial^2 \Psi}{\partial \mathbf{r}^2} + [V_0(\mathbf{r}) + V(\mathbf{r},t)]\Psi(\mathbf{r},t) , \qquad (41)$$

where $\mathbf{r}$ is the three-, two-, or one-dimensional particle position vector (depending on the effective system dimensionality), $V_0(\mathbf{r})$ is a separable (regular) motion potential of the binding type (i. e. a deep enough potential well) and $V(\mathbf{r},t)$ is the time-periodic perturbation potential.[15] The real quantum machine element described by eq. (41) can be represented by electron bound in an atom, or molecule, or so-called 'quantum dot', or other kind of natural or artificial binding potential $V_0(\mathbf{r})$ formed by the prefabricated element structure. The function $V_0(\mathbf{r})$ can also be periodic and describe a space-periodic potential, i. e. a lattice of potential wells, such as crystal lattice or various artificial periodic structures ('superlattices'). In this case the energy eigenvalues for $V_0(\mathbf{r})$ used below are split into 'energy zones', but this modification does not introduce any essential change in our description (cf. refs. [8,9]).

---

[15] A more general case, where the unperturbed motion in the potential well $V_0(\mathbf{r})$ is a nonseparable (chaotic) one, can also be considered within our universal approach. In this case we first choose an integrable (e. g. one-dimensional or centrosymmetric) component of $V_0(\mathbf{r})$ as the zero-order, regular-motion potential and obtain the dynamically multivalued solution for the unreduced potential well $V_0(\mathbf{r})$ in terms of our general EP (Chapter 3). In this sense, the 'unperturbed' potential $V_0(\mathbf{r})$ in eq. (41) should generally be considered as already an effective, nonlocal and multivalued, potential accounting for the chaotic motion in the stationary potential well. In the general case, therefore, $V_0(\mathbf{r})$ is not necessarily a regular-motion potential, but a potential for which we know the complete solution (maybe a multivalued, chaotic one) of the corresponding (stationary) Schrödinger equation, which is the essential information used below. However, we shall not explicitly introduce such additional complication here, in order to concentrate on the unreduced influence of the main, time-dependent perturbation $V_0(\mathbf{r},t)$.

The total interaction potential in eq. (41) can be considered as a time-periodic, binding potential that can be represented as a Fourier series:

$$V_0(\mathbf{r}) + V(\mathbf{r},t) = \sum_{n=-\infty}^{n=\infty} V_n(\mathbf{r})\exp(i\omega_\pi n t) = V_0(\mathbf{r}) + \sum_{n \neq 0} V_n(\mathbf{r})\exp(i\omega_\pi n t) ,$$
(42)

where $\omega_\pi$ is the perturbation frequency, $n$ takes only integer values, and we have actually included, according to the accepted convention, the time-averaged component of the perturbation potential (i. e. its zero-order Fourier component) into the unperturbed potential $V_0(\mathbf{r})$ (which does not limit the generality of our analysis and corresponds to a coordinate-independent value of the time-averaged perturbation potential or other its configuration that preserves integrability of the problem with $V_0(\mathbf{r})$). Since the Fourier series harmonics, $\exp(i\omega_\pi n t)$, play the role of known eigenfunctions of the 'decoupled' perturbing radiation dynamics (so that the time variable $t$ effectively replaces variable $q$ in our general analysis, cf. eqs. (5b), (6)), we expand the system wavefunction into a similar Fourier series (cf. eq. (7)):

$$\Psi(\mathbf{r},t) = \exp(-iEt/\hbar)\left[\psi_0(\mathbf{r}) + \sum_n \psi_n(\mathbf{r})\exp(i\omega_\pi n t)\right] ,$$
(43)

where $E$ is the energy eigenvalue to be found and $n$ takes only non-zero integer values (also everywhere below). Substituting expressions (42), (43) into eq. (41), multiplying it by $\exp(i\omega_\pi n t)$ and integrating over $t$ within one perturbation period ($T_\pi = 2\pi/\omega_\pi$), we get a system of equations for $\{\psi_0(\mathbf{r}), \psi_n(\mathbf{r})\}$:

$$H_0(\mathbf{r})\psi_0(\mathbf{r}) + \sum_n V_{-n}(\mathbf{r})\psi_n(\mathbf{r}) = \eta\psi_0(\mathbf{r}) ,$$
(44a)

$$H_0(\mathbf{r})\psi_n(\mathbf{r}) + \sum_{n' \neq n} V_{n-n'}(\mathbf{r})\psi_{n'}(\mathbf{r}) = \eta_n\psi_n(\mathbf{r}) - V_n(\mathbf{r})\psi_0(\mathbf{r}) ,$$
(44b)

where $n, n' \neq 0$,

$$\eta_n = E - \hbar\omega_\pi n, \quad \eta \equiv \eta_0 = E ,$$
(45)

$$H_0(\mathbf{r}) = -\frac{\hbar^2}{2m}\frac{\partial^2}{\partial \mathbf{r}^2} + V_0(\mathbf{r}).$$
(46)

It is clear that the system of equations (44) coincides with its universal prototype, eqs. (16) (Section 3.2), where the variable $\xi$ is replaced by $\mathbf{r}$, $\varepsilon_n$ (eq. (9)) by $\hbar\omega_\pi n$, $H_n(\mathbf{r}) = H_0(\mathbf{r})$ (because $V_{nn}(\mathbf{r}) \equiv V_0(\mathbf{r})$) and $h_0(\mathbf{r}) = -\frac{\hbar^2}{2m}\frac{\partial^2}{\partial \mathbf{r}^2}$ (see eq. (13)). This coincidence demonstrates the important general similarity, mentioned before, between the complicated (arbitrary) many-body interaction and the elementary interaction act (provided that both of them are considered in the unreduced version), as well as the equivalence between the time-dependent and time-independent problem formulations (Section 3.1). It clearly shows also that any other particular case or problem formulation (see e. g. eqs. (3)-(5) in Section 3.1) is reduced to the same its universal expression in terms of known eigen-solutions for the lower-level dynamics of non-interacting system components, after which we can follow the general way of its integration revealing the dynamic redundance phenomenon and inherent complexity (Sections 3, 4).

We can, therefore, apply the same method of 'generalised separation of variables by substitution' that was used in our universal analysis (Section 3.2) and has led to the 'effective' expression of the unreduced problem, eqs. (20)-(23). We do not reproduce here the detailed derivation (Section 3.2) and limit ourselves to the final results which can also be directly obtained by the above change of notations. The main system of equations, eqs. (44), and thus the unreduced problem itself, eq. (41), is equivalent to the following effective Schrödinger equation for $\psi_0(\mathbf{r})$:

$$-\frac{\hbar^2}{2m}\frac{\partial^2 \psi_0}{\partial \mathbf{r}^2} + V_{\text{eff}}(\mathbf{r};\eta)\psi_0(\mathbf{r}) = \eta\psi_0(\mathbf{r}) , \qquad (47)$$

where the operator of the effective (interaction) potential (EP), $V_{\text{eff}}(\mathbf{r};\eta)$, is given by

$$V_{\text{eff}}(\mathbf{r};\eta) = V_0(\mathbf{r}) + \hat{V}(\mathbf{r};\eta), \ \hat{V}(\mathbf{r};\eta)\psi_0(\mathbf{r}) = \int_{\Omega_\mathbf{r}} d\mathbf{r}' V(\mathbf{r},\mathbf{r}';\eta)\psi_0(\mathbf{r}'), \quad (48a)$$

$$V(\mathbf{r},\mathbf{r}';\eta) = \sum_{n,i} \frac{V_{-n}(\mathbf{r})\psi_{ni}^0(\mathbf{r})V_n(\mathbf{r}')\psi_{ni}^{0*}(\mathbf{r}')}{\eta - \eta_{ni}^0 - \hbar\omega_\pi n} , \qquad (48b)$$

and $\{\psi_{ni}^0(\mathbf{r}), \eta_{ni}^0\}$ is the complete set of eigenfunctions and eigenvalues of the auxiliary, truncated system of equations (cf. eqs. (17)):

$$H_0(\mathbf{r})\psi_n(\mathbf{r}) + \sum_{n'\neq n} V_{n-n'}(\mathbf{r})\psi_{n'}(\mathbf{r}) = \eta_n \psi_n(\mathbf{r}) \ . \qquad (49)$$

Thus, we have formally 'excluded' time from the problem formulation and replaced the time-dependent perturbation potential in the initial Schrödinger equation, eq. (41), by the 'effective', time-independent potential which, however, incorporates the unreduced perturbation influence entering through its frequency-dependent, dynamically entangled components and related nonlinear dependence on the eigen-solutions to be found. In order to complete problem solution, one should now find the eigenfunctions, $\{\psi_{0i}(\mathbf{r})\}$, and eigenvalues, $\{\eta_i\}$, of the effective Schrödinger equation, eqs. (47)-(48), which is possible to do due to integrability of the time-independent problem and existence of sufficiently good approximations for the auxiliary problem solutions (see Section 4.4). After expressing other wavefunction components, $\psi_n(\mathbf{r})$, through the found solutions for $\psi_0(\mathbf{r})$ (cf. eq. (22)), the total system wavefunction is obtained, according to eq. (43), in the following form:

$$\Psi(\mathbf{r},t) = \sum_i c_i \exp(-iE_i t/\hbar)\left[1 + \sum_n \exp(i\omega_\pi n t)\hat{g}_{ni}(\mathbf{r})\right]\psi_{0i}(\mathbf{r}), \quad (50)$$

where according to eq. (45) $E_i \equiv \eta_i$ and the action of the integral operator $\hat{g}_{ni}(\mathbf{r})$ is given by (cf. eq. (22)):

$$\psi_{ni}(\mathbf{r}) = \hat{g}_{ni}(\mathbf{r})\psi_{0i}(\mathbf{r}) = \int_{\Omega_\mathbf{r}} d\mathbf{r}' g_{ni}(\mathbf{r},\mathbf{r}')\psi_{0i}(\mathbf{r}') \ ,$$

$$g_{ni}(\mathbf{r},\mathbf{r}') = V_{n0}(\mathbf{r}')\sum_{i'} \frac{\psi^0_{ni'}(\mathbf{r})\psi^{0*}_{ni'}(\mathbf{r}')}{\eta_i - \eta^0_{ni'} - \hbar\omega_\pi n} \ . \qquad (51)$$

Similar to the general case (Section 3.3), the most significant feature of the obtained problem solution is the redundant number of eigen-solutions of the effective Schrödinger equation with the unreduced EP, eqs. (47)-(48), which is mathematically due to the EP dependence on the eigenvalues to be found, $\eta \equiv E$, and leads to the main property of dynamic multivaluedness related to the dynamic entanglement, intrinsic chaoticity and dynamic complexity. The detailed argumentation is given in Chapters 3 and 4 and remains practically without change for this particular case, which ac-tually describes also the 'standard' situation for appearance of the phenomenon of (genuine) *quantum chaos* [1,8,9]. The result can be summarised by the expression for the measured system density, $\rho(\mathbf{r},t) = |\Psi(\mathbf{r},t)|^2$, giving the general problem solution in the form of *dynamically probabilistic* sum over the explicitly obtained, incompatible system realisations numbered by index *r* (cf. eq. (24)):

$$\rho(\mathbf{r},t) = \sum_{r=1}^{N_{\Re}} {}^{\oplus} \rho_r(\mathbf{r},t) \,, \quad (52)$$

where $N_{\Re}$ is the total realisation number (actually determined by the number of essential Fourier components of the perturbation potential, cf. Section 3.3) and the individual realisation densities, $\rho_r(\mathbf{r},t)$, are obtained from the above wavefunction and EP expressions with the eigenvalues forming respective realisations (see eqs. (25)). Realisation probabilities and expectation value of system density are given by eqs. (26) and (27) with the evident change of notations.

The dynamically probabilistic sum of eq. (52) expresses the main property of the considered system distinguishing it from unitary imitations used in conventional quantum mechanics in general and in quantum computation theory in particular. Namely, the unreduced, dynamically multivalued solution shows that the simple quantum system of a bound particle subjected to the time-periodic perturbation follows the explicitly nonunitary evolution (in the absence of any noise) consisting in the permanent change of its redundant states/realisations in the truly chaotic, dynamically probabilistic order, each of them corresponding to its own EP that can differ considerably from the unperturbed system potential, $V_0(\mathbf{r})$. The system does spend a finite time, of the order of a period of realisation formation, in any its current realisation having a roughly unitary dynamics (at the current level of complexity), which is expressed by the obtained solution for the wavefunction, eqs. (50)-(51), with the eigenvalues corresponding to this particular realisation. However, this pseudo-unitary evolution is quickly interrupted by irregular system 'jump' into another, generally similar, realisation but with different EP parameters, eigenvalues and other measured characteristics. It is important that those jumps between chaotically chosen realisations, passing by the special 'intermediate realisation' qualitatively different by its configuration from the ordinary, 'regular' realisations (Sec-tion 4.2), can never stop, since they are driven exclusively by the *same* interaction that determines the structure of each individual realisation (i. e. the observed system configuration), which also demonstrates the fundamental difference of our results from the unitary imitations of quantum (and classical) chaos, uncertainty and classicality emergence with the help of postulated 'decoherence' by artificially introduced external 'noise' (though any real external perturbation can, of course, *additionally* influence the system when it is sufficiently strong).

This consistently derived result means, practically, that the full-scale quantum computation in the unitary, regular mode is impossible *in principle*, being fatally destroyed already at the level of each elementary, noiseless interaction act in a computing system, and any noise-reduction and dynamical 'chaos-control' measures cannot eliminate the underlying, purely dynamic uncertainty (which is similar in its origin to the fundamental quantum indeterminacy at the level of single particle dynamics [1-4,9-13]). The real, complex-dynamical realisation change by the system, driven by the main, totally regular interaction, provides also the causally complete extension of abstract uncertainty imitation by the conventional density matrix concept and similar postulated constructions extensively used, in particular, in the unitary quantum computation theory. The unreduced interaction description gives instead the dynamically probabilistic realisation sum of eq. (52) and the corresponding intermediate (main) realisation playing the role of the generalised wavefunction (or distribution function) that describes realisation probability distribution at various emerging levels of dynamics (Section 4.2) and obeys the generalised Schrödinger equation [1,4] (see also Section 7.1).

The described causally complete picture of irreducibly chaotic (multivalued) behaviour of a simple quantum system under the influence of a regular external perturbation is quite universal and provides the detailed understanding of the mechanism and 'internal dynamics' of 'quantum transitions' in general, described in conventional quantum mechanics with the help of standard 'postulates' that only fix the formal, empirically 'guessed' rules for possible initial and final states of a system and the related 'transition probabilities' while leaving the underlying system dynamics within the 'officially permitted' quantum 'mystery'. The obtained causally complete problem solution, eqs. (47)-(52) (eqs. (20)-(31) in the general case), shows that even the simplest and regular interaction produces deep qualitative changes in system configuration and evolution. Instead of straightforward, internally smooth unitary 'propagation' in an abstract 'state space' of the same initial configuration, as it is prescribed by the canonical quantum (and classical) 'postulates', we have a dramatic change of the potential and system configuration themselves splitted into multiple components with different parameters and fractally structured, unceasing system transitions between them, during which the system is transiently 'decomposed' and then again 'recomposed' into its new configuration explicitly created by the interaction process (through the 'dynamic entanglement' of system components, Section 4.2). If a group of system realisations thus formed constitutes a compact entity with relatively small difference between realisations within the group (with respect to other realisations and their groups), then the new, 'excited' system state and configuration emerges at a finite, 'quantised' distance from the initial, 'ground' state and other possible states. It is important that the detailed, fractally structured and dynamically probabilistic configuration of such 'quantum transitions' (occurring also at any higher, 'classical' levels of complexity [1]) can be completely described and understood, without any 'mysteries', within the proposed unreduced formalism (see Sections 4.1-4, while we leave for future work further refinement of these results for particular systems).

In a similar fashion, another quantum phenomenon known as 'quantum tunneling' and also remaining 'mysterious' within the conventional unitary scheme, obtains its causally complete interpretation as fractally structured, dynamically probabilistic (adaptable) 'wandering' of emerging (many-body) system configurations within the dynamical, EP 'barrier' to motion, occurring through its transient, fractally structured decrease for certain realisations [1,9] (which describes actually the self-consistent, dynamically multivalued 'percolation' of the system 'within itself', in the real space of its emerging configurations, differing qualitatively from its various unitary imitations by 'chaotic tunnelling' in the 'phase space', or another abstract 'space', in the conventional chaos theory, see e. g. [152]-[154] and references therein). Here again, the obtained universal description within the dynamic redundance paradigm provides the necessary basis and fundamental mechanism for the causally complete understanding of complex-dynamical reality behind the official 'mystery' of the quantum tunnelling phenomenon, whereas further refinement of details should constitute a part of the intrinsically complete, dynamically multivalued 'remake' of conventional quantum mechanics.

These explicitly obtained manifestations of the unreduced, multivalued dynamics of real interaction processes reveal their true character as a hierarchically structured sequence of unceasing 'quantum transitions' of various scales with probabilistically determined results that form the average, observed system 'tunneling' (understood in the most general sense) in the universally defined direction of growth of the unreduced dynamic complexity-entropy at the expense of diminishing complexity-information [1] (see also Sections 4.4, 4. 7 and 7.1). Any internal system 'wiring', exchange of 'commands', 'memorisation' and 'calculation' are mainly produced dynamically and intermixed among them into the single, dynamically multivalued (and thus everywhere probabilistic) interaction development process guided 'automatically' by the unreduced complexity development. External 'control' of the process is possible, but only as modification of informational dynamic complexity (encoded in interaction 'potential energy'), which is performed at certain, 'critical' points and moments of time and have nothing to do with the conventional idea of 'regularising', basically unitary, and therefore unrealistic, control that would correspond to (impossibly) decreasing complexity-entropy (see also Section 2(iii), Chapters 7 and 8). The dynamically multivalued results of even the simplest, externally regular interaction process considered in this Section clearly show that in the case of a realistically involved quantum micro-machine, starting already from a single multi-atom molecule, it is the *detailed*, dynamically emerging configurations, their probabilities and unreduced development, which are important for the adequate, practically efficient understanding and control of quantum machine operation (let alone construction). This means that the canonical, 'postulated' and 'mysterious' quantum mechanics is absolutely insufficient for this more complicated application (since it cannot predict even possible system states, let alone their probabilities and internal system dynamics) and the unreduced, conceptually extended analysis of the universal science of complexity becomes indispensable.

It would also be interesting to specify, for the simplest particular case of this Section, manifestations of the two universal limiting regimes of multivalued dynamics, uniform (global) chaos and multivalued SOC (self-organisation) as they were described in Section 4.5. Since in the simplest case we consider here the role of the variable $q$ is played by time $t$ and the corresponding 'interacting entity' (or system 'degree of freedom') is reduced to the time-periodic perturbation (e. g. a radiation field), the 'internal', or 'structure-independent' system frequency $\omega_q$ is given by the perturbation frequency $\omega_\pi$, while the 'structure-dependent' frequency $\omega_\xi$ corresponds to characteristic energy-level separation, $\Delta E$, of the unperturbed potential well $V_0(\mathbf{r})$: $\omega_\xi = \Delta E/\hbar$. The 'resonance' condition for the uniform (global) chaos onset, eq. (35), corresponds then, as it should be expected, to the resonance between the perturbation and system-structure potential:

$$\kappa \equiv \frac{\Delta \eta_i}{\Delta \eta_n} = \frac{\Delta E}{\hbar \omega_\pi} \cong 1 \;, \qquad (53)$$

The opposite regime of multivalued SOC is obtained at small values of the chaoticity parameter, $\kappa \ll 1$, or $\omega_\pi \gg \Delta E/\hbar$. The (generalised) 'phase transition' from this externally regular regime to the explicitly irregular case of uniform chaos occurs, with growing $\kappa$ (decreasing $\omega_\pi$), around the main resonance condition, $\omega_\pi = \Delta E/\hbar$, being preceded by a series of less dramatic, partial transitions to chaos around higher-order resonance conditions, $\omega_\pi = n\Delta E/\hbar$, $n = 2, 3, ...$ (Section 4.5.2) [1]. At very low frequencies, $\omega_\pi \ll \Delta E/\hbar$ ($\kappa \gg 1$), we obtain another case of SOC which is, however, less interesting in this context (see Section 4.5.2).

Physically, the obtained results mean that at very high perturbation frequencies ($\omega_\pi \gg \Delta E/\hbar$) the system can perform only relatively small, high-frequency deviations from its unperturbed motion, even though they are internally chaotic (dynamically multivalued), contrary to predictions of conventional self-organisation/SOC [129-131] or 'motion in a rapidly oscillating field' [138] (see also Section 4.5.1). The EP, eq. (48), is approximately reduced in this case to a local potential (cf. eq. (33)):

$$\begin{aligned} V_{\text{eff}}(\mathbf{r};\eta) &= V_0(\mathbf{r}) + \sum_n \frac{|V_n(\mathbf{r})|^2}{\eta - \eta_{ni}^0 - \hbar\omega_\pi n} \cong \\ &\cong V_0(\mathbf{r}) + \frac{1}{(\hbar\omega_\pi)^2} \sum_{n>0} \frac{|V_n(\mathbf{r})|^2 (\eta_{ni}^0 + \eta_{-ni}^0 - 2\eta)}{n^2} \;, \end{aligned} \qquad (54)$$

where the last (least exact) approximation corresponds to the ordinary, dynamically single-valued analysis (perturbation theory). However, if the perturbation frequency (or other related system parameter) approaches the resonance condition, $\omega_\pi = \Delta E/\hbar$, the deviations are quickly amplified to relatively large magnitudes and so is their internal chaoticity that becomes thus explicitly evident, giving the regime of 'global' (uniform) chaos. Similar, but less important, local growth of chaotic fluctuations is observed around higher-order resonance conditions, $\omega_\pi = n\Delta E/\hbar$, $n \geq 2$, while the decay of strong chaoticity at the other side of the main resonance, at $\omega_\pi < \Delta E/\hbar$, should also have more or less distinct step-wise character, with abrupt behaviour changes around higher-order resonances at $\omega_\pi = n\Delta E/\hbar$, $n \geq 2$.

This transition from global regularity to global chaos is well known from the classical chaos theory, but could not be properly obtained in the conventional quantum chaos theory [139-145,155], simply because there can be no true dynamical randomness in the unitary (single-valued) quantum dynamics. Now we see how this transition, as well as the unreduced dynamical chaos in general, naturally emerges in our dynamically multivalued description of the essentially quantum problem. It is not difficult to see that the obtained point of this transition satisfies the correspondence principle between quantum and classical dynamics[16] [1,8,9] (i. e. it gives the respective transition between regularity and chaos for the corresponding classical system under the straightforward limit $\hbar \to 0$), resolving the main problem of the canonical chaos theory, where this quantum-classical correspondence is absent, together with any true randomness (see Chapter 6 for more details). These properties of genuine, dynamically multivalued quantum chaos disclose the abuse of various existing imitations of dynamic randomness in quantum systems in general and quantum machines in particular, reduced most often to replacement of the intrinsic, explicitly derived system randomness by its postulated 'penetration' into the system from an indefinite outside source (see also below, Section 5.2.2).

---

[16] Indeed, since $\Delta E/\hbar$ tends, at $\hbar \to 0$, to the classical oscillation frequency in the unperturbed potential, $\omega_\xi$, it is evident that the criterion of eq. (53) can be presented in an explicitly classical form, $\omega_\pi \cong \omega_\xi$, which already shows a basic agreement with the correspondence principle. Moreover, the detailed comparison with 'standard' models from the classical chaos theory shows [1,8,9] that the 'classical' chaos criterion of eq. (53), derived quantum-mechanically, coincides with the corresponding results obtained in the framework of classical mechanics. Even more, the classical chaos description within the conventional, also dynamically single-valued, theory contains itself essential deficiencies (including the absence of true, internal system randomness) and should be replaced by the dynamically multivalued analysis at the level of classical mechanics, which is a version of our universal formalism (Chapters 3, 4) [1].

As far as quantum machine operation is involved, the regime of uniform chaos in the dynamics of its elementary step is especially important as the standard, dominating regime of essentially quantum dynamics. Indeed, if we have considerably different characteristic frequencies/energies, as it is necessary for the regime of multivalued SOC, we deal with at least semiclassical dynamics that will tend to the fully classical (predominantly localised) behaviour with growing frequency/energy ratio. The latter case corresponds just to the qualitatively more distinct, 'classical' structure emergence, where slowly changing, 'self-organised' external form contains within it many slightly different, incompatible system configurations that quickly replace each other in a random order (Sections 4.5.1, 4.7). By contrast, the essentially quantum, fully 'dualistic' behaviour is nothing else than the uniform (global) chaos regime whose existence already in the unperturbed, as well as perturbed, dynamics of individual quantum machine element, eq. (41), is determined by the universal criterion of eq. (53) (cf. Section 4.6.1). The latter actually describes the causally complete, complex-dynamical internal mechanism of the well-known main property of quantum behaviour, simply postulated in the conventional theory, i. e. *quantisation* itself of any quantum system dynamics determined by the *universal* action quantum $h$ [1,3], so that $\omega_\pi = \Delta E/\hbar$ in our case (recall the corresponding delta-function occurrence in the conventional expressions for transition probabilities), $\kappa = 1$, and chaoticity is global (or coarse-grained, or uniform). Correspondingly, any 'quantum control of quantum dynamics' cannot eliminate this strong chaoticity without being transformed into classical, SOC type of dynamics already during every elementary interaction act, occurring even in the most 'pure', totally 'coherent' conditions, which demonstrates once more the basic deficiency of respective results of unitary quantum theory [90-107] and its understanding of randomness/uncertainty. Any real system dynamics permanently 'decoheres' by itself, in any its single step, due to the very essence of the unreduced "*inter*-action" process; the truly quantum behaviour involves, by definition, a relatively large degree of this internal randomness.

A much more distinct regime of self-organised, 'classical' behaviour in principle allows of efficient control, but then one deals with classical machines, which do not possess any 'magic', cost-free efficiency of the unitary type. One can also express this result in terms of characteristic 'distance between (elementary) realisations', or (primary) 'realisation spacing', measured by the corresponding EP differences and expressing the chaotic 'uncertainty' (smearing) of system configuration (in terms of characteristic energy parameters) [1,9]. It is easy to see from eqs. (21), (33), (48), (54) that this realisation spacing, $\Delta V_{\text{eff}}$, is generally of the order of $(\Delta E)^2 / \hbar \omega_\pi \ll \Delta E$ in the multivalued SOC regime and of the order of $\Delta E$ (energy level spacing in the absence of interaction) in the uniform chaos (essentially quantum) regime (i. e. $\Delta V_{\text{eff}} \approx \kappa \Delta E$ in every case), which is another expression of the above conclusions.

## 5.2.2. Fundamental properties and causally derived general principle of real quantum machine operation

Based on the obtained causally complete description of the elementary interaction act of the quantum machine operation, we can now summarise the ensuing fundamental properties and related principles of real quantum machine dynamics, which reproduce and specify the universal results for the general case of complex (multivalued) dynamics of many-body interaction system (Chapters 3, 4). The major property is, of course, the *unreduced dynamic complexity*, or *chaoticity*, itself, understood as unceasing, internally inhomogeneous and causally probabilistic process of system realisation formation and change that naturally occurs, as we have seen in the previous Section (see eqs. (47)-(52)), already in any elementary act of quantum-mechanical interaction, eq. (41), within a larger micro-machine dynamics (Sections 3, 4). The explicit derivation of multiple, incompatible results (realisations) of unreduced interaction process in our approach provides the consistent, dynamically justified definition of randomness/probability and related notions (see also below), which should be clearly distinguished from various postulated, a priori defined versions of the same notions in conventional science, giving rise to what we designate as 'stochastic' (or 'statistical') version of unitary theory widely used, in particular, for introduction of 'probabilistic' and 'chaotic' elements in the dynamically single-valued description of quantum computation. It is important to emphasize that the universally defined, unreduced complexity, based on the physically real, dynamically derived phenomena of *dynamic redundance and entanglement*, naturally unifies, causally (realistically) extends and consistently specifies various related concepts, which are often intuitively (and ambiguously) introduced within the non-dynamical, unitary and abstract theories around quantum behaviour speculatively endowed with informational/computational 'interpretation'.

One of such 'information-theoretical' and 'quantum-informational' properties is *non-computability* attributed, in various ways, to any 'interesting' (and correspondingly 'sophisticated') dynamics/behaviour, including for example [15] the unified origin of 'mysterious' quantum postulates, consciousness and (quantum) gravity. The conventional non-computability concept tries to formalise mathematically the vague idea of ultimately random, 'unpredictable', 'undecidable', or 'inimitable' sequence of patterns (states) in system behaviour, but fails (and does not actually try) to explicitly, consistently derive these properties from a simple initial configuration of a system, where these properties would not already be present/inserted in an explicit, trivial fashion. By contrast, our results for both simplest and general interaction cases (Sections 5.2.1 and 3.3/4.1 respectively) just explicitly show the detailed dynamic origin and mechanism of 'non-computability', which simultaneously clarify the meaning of this notion itself appearing to be yet another expression of our unreduced dynamic complexity and chaoticity. Indeed, the 'ultimate' uncertainty and 'undecidability' in the redundant realisation choice, now *dynamically derived* in the unreduced EP method [1-4,8-13] (Chapters 3, 4) and expressed by the *dynamically probabilistic* realisation sum of the general system solution, eqs. (24)-(27), (47)-(52), just causally specify the implied meaning of 'non-computability', with all its possible nuances and aspects (including randomness). It is easy to see that all the ambiguities, abstraction and incompleteness of conventional imitations are eliminated in this dynamically derived property of the unreduced interaction dynamics. As far as quantum computation is involved, we show, in terms of this well-specified, dynamic non-computability, that any quantum machine dynamics (Chapter 3), and even any its elementary step (Section 5.2.1), is always non-computable itself, because of omnipresent, interaction-driven system splitting into many incompatible realisations, which is another expression of impossibility of ordinary, digital and unitary, modes of computation by quantum devices (cf. e. g. [27,78,117] for the opposite conclusion of the unitary theory).

Since natural interaction processes develop discrete, progressively emerging levels of complex behaviour, starting from the multivalued quantum dynamics [1] (Sections 4.4-4.7), the fundamental dynamic non-computability and its mechanism will be independently reproduced at each level (instead of being only directly 'inherited' from lower levels). One obtains thus the multilevel hierarchy of non-computable, or complex, or chaotic (in a well-defined sense) world dynamics, which explains the 'non-computable character of conscious thought' at the highest complexity levels [1,4], as well as non-computability of any other level of world dynamics, without its direct reduction to manifestations of ambiguous 'quantum weirdness' (cf. [15,68-71]). Instead, it is this 'quantum non-computability' and 'quantum mysteries' (intrinsically unified with the dynamic origin of gravity) which obtain their causally complete, and actually universal, explanation within the dynamic redundance paradigm. In particular, the most fundamental aspects of quantum mechanics, related to the dualistic behaviour of elementary particles and their very origin, naturally unified with the origin of gravitation, space, time, mass and other 'intrinsic' particle properties, are consistently explained as the lowest sublevel of 'non-computable', dynamically multivalued interaction between two primordial, effectively structureless protofields [1-4,11-13], the sublevel of complexity that precedes the one of interacting quantum entities within a quantum machine considered here. These dynamic relations between neighbouring levels of non-computable (multivalued) behaviour produce just the necessary, neither mechanistically reduced nor inconsistently mystified, hierarchy of world dynamics explaining its nontrivial, inimitable, undecidable, self-developing and sometimes externally 'mysterious' character.

Talking about the well-defined, dynamically multivalued complexity, chaoticity, entanglement, etc. of real quantum machine behaviour, we should emphasize the basic distinction of these unreduced, dynamically derived concepts from their dynamically single-valued, unitary imitations in conventional theory that do not reveal the essence of the unreduced interaction complexity, but only speculatively, artificially muddle the situation, which is already rather ambiguous because of the persisting 'ordinary' quantum mystification. Such are unitary imitations of (quantum) 'complexity', 'chaos', 'randomness', entropy, information (as well as related 'foundational' approaches in quantum mechanics), which are extensively used in 'quantum information theory' (see e. g. [14,53,59-67,72,73,81,110,118, 156-174], in addition to general reviews on quantum information theory cited in Chapter 2). It is not necessary to enter into technical details of each particular imitation in order to see their general, fundamental deficiency: all those 'quantum' and 'dynamical' versions of randomness, chaoticity, complexity, entropy and information are not obtained as a result of unreduced interaction development, but are postulated instead in the abstract form of 'ready-made', already existing and intuitively 'plausible' manifestation, or 'sign', of the corresponding property/phenomenon, which is actually equivalent to the unitary, dynamically single-valued, perturbative description leading to evident contradictions. Thus, 'randomness' and 'complexity' are often defined by formally introduced 'correlations' among elements of a given set, where minimal correlations are identified with maximum unpredictability and thus randomness. The unreduced, dynamic origin and the ensuing most important relations between elements remain, however, undiscovered within such mechanistic definition. The interconnected, and most often confused, values of complexity, entropy and information are defined in the unitary approach by analogy with the postulates of classical statistical mechanics, where the classical (probability) distribution function is replaced by the quantum density matrix, being itself a postulated and basically deficient construction (see also Sections 3.3, 4.2). In that way one can obtain arbitrary values of 'quantum complexity' or 'quantum entropy' with ill-defined, purely abstract meaning, depending on the particular approximation used. A non-zero 'complexity' is thus typically attributed to the effectively zero-dimensional (dynamically single-valued, or unique) problem solution, whereas the reality-based, unreduced complexity always implies a dynamically multivalued and thus truly chaotic solution (Section 4.1). Indeed, as noted in many places above (see e. g. Section 2(i)), the very problem of conventional, unitary quantum computation is that its true dynamic complexity is zero, whereas any real computational process needs positive, and actually quite large, dynamic complexity (see also Chapter 7 for more details). That's why the extensive speculations around 'quantum complexity' (or 'entropy', or 'information') within conventional 'quantum information theory' provide especially transparent examples of the absolutely irrelevant, qualitatively deficient content of unitary, dynamically single-valued and abstract imitations of the dynamically multivalued reality in conventional theory.

The same kind of deficiency is inherent in the dynamically single-valued imitations of 'quantum chaos' in quantum machines [14,65-67,156-166]. The unreduced, dynamically multivalued interaction description clearly shows why the dynamically single-valued theory cannot find any true source of randomness in principle and corresponds to the zero value of genuine dynamic complexity (Section 4.1). At the lowest, essentially quantum levels of world dynamics this absence of any truly dynamic randomness in conventional quantum mechanics becomes especially evident (see e. g. [155]), since here it cannot be so easily hidden behind imitations like 'exponentially diverging trajectories' used at higher, 'classical' (macroscopic) levels of dynamics, where 'randomness' can be 'quietly' inserted as a lower-level 'noise' or 'uncertainty in initial conditions' (see also Chapter 6). This difficulty does not stop, however, the most pertinacious imitators and they do find 'genuine' and 'dynamical' quantum chaos where other single-valued approaches have discovered only proven regularity. In reality, randomness is introduced of course only mechanistically, through a 'noise' within the system that can be somehow 'amplified' by an obscure, actually postulated mechanism and give chaos where it is expected to exist by correspondence to the respective classical system. As a result, one obtains a very peculiar mixture of contradictions, where the conclusion about possibility of stable (unitary) quantum computation also chaotically changes from one article to another (probably depending on 'noisy influences' at much higher levels of interaction) and where the 'post-modern' play of words, so characteristic of today's official science, attains one of its highest peaks, so that, for example, 'ergodicity in energy level distribution' is arbitrarily mixed with ergodicity and randomness in real, configurational system behaviour itself, etc. The 'rigour' of this 'ironic' science is 'confirmed' by 'pictures' produced in elaborate computer simulations within arbitrarily simplified, purely abstract 'models' with numerous adjustable 'parameters' (so that in practice it is rather useless to verify their correctness). The totally consistent results of our dynamically multivalued analysis of real system dynamics reveal the origin of ambiguities of conventional 'quantum chaos' theories (i.e. their perturbative single-valuedness) and thus show why there is no sense to plunge into more details of those fundamentally erroneous imitations of reality.

In connection to this difference between genuine and imitative complexity/chaoticity, one can mention the qualitative difference between our objections against unitary dynamics of real computing (quantum) systems and other existing doubts about conventional quantum computation [79-88]. The latter are based rather on 'practical' impossibility of realisation of a possible 'in principle' theoretical solution (obtained always within the same, unconditionally accepted unitary paradigm), which is attributed most often to the destroying influence of (external) noise supposed to be relatively strong for 'sensitive' quantum dynamics of 'very small' objects. Whereas this 'technical' kind of difficulty can always be eliminated, in one way or another (indeed, due to the fundamental discreteness of quantum dynamics the omnipresent noise does not prevent the observed existence of truly coherent quantum dynamics), we show explicitly that even the most ideal case of strictly 'Hamiltonian' (non-dissipative) dynamics is characterised by the universal, purely dynamical and therefore irreducible source of true randomness (Section 3.3), appearing already at every elementary interaction act within quantum machine dynamics (Section 5.2.1).

Moreover, it becomes evident that being the ultimate source of randomness as such, the intrinsic chaoticity of real system dynamics also governs the behaviour of any 'open', 'noisy' and 'dissipative' system (including e. g. the situation of 'quantum measurement' [1,10]) and provides both the ultimate origin of any 'noise' and dynamically random result of its further development/amplification at higher complexity levels. This dynamic origin of any 'noise', which otherwise needs to be directly and inconsistently 'postulated' in the conventional theory, is especially important and irreducible just for the essentially quantum dynamics, since it is confined to a few lowest sublevels of complexity, where no 'other' influence can come 'from below', from an indefinite, fine-grained reservoir of 'practically' non-computable dynamics (in other words, the smallest possible 'grain' of real world dynamics is always given by Planck's constant, whether it describes a conventionally defined 'noise' or the 'main' system dynamics). This is another demonstration of the profound deficiency of conventional, 'stochastic' approaches to randomness and chaoticity in quantum systems becoming evident already by correct application of standard quantum postulates.

The genuine dynamic randomness of real quantum system behaviour is closely related, by its very origin, to another manifestation of unreduced dynamic complexity, the dynamically multivalued, interaction-driven, probabilistic entanglement, or simply *dynamic entanglement*, between the system components. One should emphasize the essential difference of this physically real, fractally structured and permanently changing, probabilistic 'mixture' between the interacting entities from its mechanistic, purely abstract and especially actively mystified imitation known as 'quantum entanglement' in conventional quantum theory and unitary quantum computation (see e. g. [83,110,118,172-176], as well as general references on quantum computation from Chapter 2), where it is even used as the key argument in favour of the expected 'fantastically high' efficiency of quantum computation process. As it is clearly seen from our explicitly obtained general solution of the unreduced interaction (or 'many-body') problem for both general and particular cases, eqs. (20)-(25) and (47)-(52) respectively, the real entanglement in a quantum system has a directly visible dynamical, interactional origin, so that the explicit 'mixture' of the component wavefunctions for different entities and eigen-states in a single expression for the total wavefunction is always obtained as their interaction-driven, physically real and *unceasing* recomposition, assisted by the corresponding EP structure formation. Thus, the dynamic entanglement between interacting degrees of freedom $Q$ and $\xi$ in the general case, eq. (25a), contains the initial interaction potential and eigenvalues (eq. (25b)) determined self-consistently by solution of the Schrödinger equation with EP (eqs. (20), (25c)) that includes the whole process of unreduced interaction development and in particular real dynamical collapse/reduction to the current realisation configuration (see also Sections 4.2, 4.3).

As follows already from evident qualitative arguments confirmed by the rigorous analysis (Section 3.3), this real dynamical mixture of interacting components can only emerge in a large number of incompatible, but equally possible versions (realisations), which leads to their unceasing change in the dynamically random order, eqs. (24), (52), giving rise in particular to the true quantum chaos and probabilistic quantum measurement [1,9,10]. Each system realisation corresponds therefore to a particular, causally determined version of dynamic entanglement between its components, which are forced by the main, driving interaction to permanently entangle, disentangle and re-entangle into unceasingly emerging and replaced realisations, this intense internal *life* of a system constituting the essence of its existence as such, as opposed to 'general' definitions and mechanistic imitations of conventional science or 'systems theory'. Correspondingly, the real 'quantum entanglement' also represents the intense, physically real 'internal life' of a quantum system remaining completely beyond the 'very averaged' and purely abstract conventional description and consisting in unceasing sequence of interaction-driven system collapses to probabilistically chosen possible states (realisations) alternating with transient disentanglement phases during system transitions between realisations. It is this causally probabilistic internal structure of (quantum) entanglement, or *interaction*, process (originating from the fundamental protofield coupling) that explains the real, demystified, but complex-dynamical nature of 'quantum entanglement' and 'linear superposition' postulates of standard quantum mechanics (see also Section 5.3) and reveals the illusive basis of 'magic' efficiency of unitary quantum computation as being due to the fundamentally incorrect, dynamically single-valued (perturbative) imitation of real interaction processes. The necessity for the system to take many incompatible, 'incoherent' realisations will immediately destroy the false exponential 'inflation' of the single realisation in the unitary projection of reality (see also Section 5.1), but the same phenomenon of irreducible dynamical chaos can play a constructive role leading indeed to the greatly increased efficiency, though for a qualitatively different, multivalued (chaotic) kind of micro-machine dynamics, actually realised in all natural, living systems operation (Section 7.3, Chapter 8).

An inherent aspect of the dynamic entanglement concept is represented by the property of autonomous *creativity* of complex-dynamical interaction process (Sections 4.3-4) and related solution of the 'problem of configuration space' in quantum mechanics. Namely, it is by way of interaction-driven, physically real entanglement between the interacting quantum system components that the new entities, constituting the computation/interaction process result, autonomously, dynamically *emerge* in the essentially nonlinear process of system reduction/squeeze (Section 4.3) towards its consecutive realisations that actually form physically real *config-* *urations* of (quantum) system (see eqs. (22)-(23), (25a,b), (50)-(51)) and thus the basis of physically real *space* structure of the corresponding complexity level. These resulting system configurations, especially important for understanding/control of real quantum machine dynamics, cannot be obtained in conventional theory using only their postulated, imitative forms, which are artificially inserted into formal expressions describing 'quantum entanglement' and other abstract 'elements' of unitary quantum evolution. The dynamically emerging 'mixture' of the interacting degrees of freedom, such as $Q$ and $\xi$ in eqs. (25) for the general interaction case, actually describe the tangible 'texture' of the emerging interaction product, which determines its specific 'quality' as that of a 'new' and 'holistic' entity. This physically real creativity of complex-dynamical quantum mechanics is inseparably related, as we have seen, to the irreducible, dynamic *unpredictability* of creation/entanglement results that can, however, be causally described in terms of *dynamically determined* probabilities of emerging system configurations/realisations (Sections 3.3, 4.4, 5.2.1).

It is clear that chaotically changing configurations of a real quantum machine, obtained through the described process of dynamically multivalued entanglement, possess the intrinsic *inimitability*, leading to the general *non-universality*, or 'non-fungibility', of both the detailed structure and total 'level' of quantum computation dynamics. This property of real quantum behaviour agrees well with standard quantum postulates and corrects the corresponding erroneous expectations of unitary quantum computation theory, often referred to as its universality or 'fungibility' (see e. g. [21,25,32,46,103,177-181]) and implying the permanent, exact self-reproduction, (quasi-) regularity of unitary quantum evolution. Note the essential difference between this one-dimensional 'universality' of the unitary imitation of reality and unrestricted universality of the general principles and formalism of the unreduced complex dynamics (Chapters 3,4) [1]: the truly universal description of unreduced system complexity and well-defined direction of its general evolution (growth of complexity-entropy, Chapter 7) reveals the fundamental origin of the omnipresent *non-universality* (non-repetitiveness) of the *detailed* real system dynamics. Contrary to the *unified diversity* of the unreduced complex dynamics, 'universality' of unitary evolution is physically senseless: it is close to the *mathematically* universal linear decomposition of a function into a series/integral over a (complete) set of 'standard' functions (such as the Fourier transform). The conventional theory of universal quantum computation is reduced thus to infinite play with various versions of such abstract, linear expansion of mathematical functions over sets of other functions.[17]

The difference between that mechanistic universality of abstract unitary expansions and unified diversity of real quantum dynamics is conceptually close to the difference between the purely 'digital' mode of computing (including 'simulation') that encodes everything what happens in (almost) exact numbers and the 'analogue', or 'dynamical' (interactional), mode of natural micro-machine operation, where the resulting configurations are simply obtained and compared directly, such as they are (including their inevitable irregularities). It is evident that conventional, unitary quantum computation, despite any occurring 'general' speculation about 'complexity' and 'probability', tends actually and inevitably to the digital, 'exact-number' mode and the inherent 'linear' (sequential and abstract) logic (see e. g. [27,117]), which is closely related to the effectively zero-dimensional logic of 'exact solutions' of the underlying dynamic single-valuedness paradigm (unitarity). By contrast, the *dynamically* unpredictable behaviour of real quantum, classical and hybrid systems with interaction, where the emerging entangled configurations and their probabilities are directly determined by the interaction details, can be described and efficiently used only as analogue, 'creative', rather than simply 'calculative', kind of operation.

The self-amplifying character of a real interaction process (equivalent to its universal dynamic instability) leads to the *dynamically discrete*, or *quantised*, structure of interaction dynamics, which provides the causally complete, totally realistic explanation for the origin of 'quantisation', 'wavefunction collapse', Planck's constant and its universality at the lowest sublevels of complex world dynamics (Sections 4.3, 4.6) [1,3,12,13]. Since the internal structure of the quantised interaction 'steps' is formed by the dynamically probabilistic (multivalued) entanglement of interacting entities, it becomes clear why causal quantisation is inseparably related to the true randomness/probability: each quantised system 'jump' can be performed in many equally real, but incompatible ways (directions), which determines the fundamental unpredictability and 'undecidability' of the actual system choice (equivalent to 'non-computability' of quantum dynamics, as we have seen above in this Section). If one deals with 'essentially quantum' system dynamics, confined to the lowest complexity sublevels, then one has, by definition, the situation of 'coarse-grained' and strongly (or globally, or uniformly) chaotic case of complex dynamics (Section 4.5.2): at the lowest level of complex world dynamics, the discrete transitions between system realisations cannot be further subdivided into smaller steps by any real experiment within this world, while the 'distance' between realisations, or the size of a jump between them, is relatively big (comparable with the average realisation size itself). This general property of quantum dynamics was also confirmed for the particular case of typical elementary interaction act within quantum machine dynamics (Section 5.2.1).

---

[17] In a broader sense, the same refers to the whole 'new', or 'mathematical', physics and modern 'exact' science in general [1]: in accord with Bergson's sentence, it does not reveal any really new, qualitatively extended phenomena, properties, or entities, but provides instead a series of technically new reformulations, or 'interpretations', of the same, ultimately reduced, effectively zero/one-dimensional (zero-complexity) 'model' of reality, which correspond to various point-like or line-like projections of a multi-dimensional structure observed from different aspects, but always reduced *completely* to its current zero-complexity projection (see also Chapter 9). A simple, mechanical translation from one ordinary language to another, revealing no new meaning by definition, provides certainly an example of much more complicated and sensible work of that kind.

These causally derived, totally realistic and demystified properties of unreduced quantum interaction mean that the abstract 'quantum bit' of the conventional theory of quantum computation is determined in reality by the same universal quantum of action, Planck's constant $h$, that underlies other basic features of quantum behaviour and that any practical incarnation of such causally extended, physically real 'qubit' will always contain the irreducible dynamic uncertainty ('quantum indeterminacy') remaining relatively large for the essentially quantum dynamics. Both these properties of unreduced quantum interactions, quantisation and randomness, determine the essential, qualitative difference of real quantum machine operation from its unitary imitation (cf. e. g. [182]), which can explain why they are so 'strangely', but conveniently 'overlooked' by the conventional theory, in the evident contradiction to the standard quantum postulates themselves (see also Section 2(ii)). Indeed, although the real, complex-dynamical 'quantum bit' thus obtained possesses the fixed 'information content', equal to $h$ (see also Chapter 7), it appears to be extremely 'volatile' by its particular configuration (e. g. spatial location), which evidently puts an end to any hope for practical, full-scale realisation of any unitary, regular (re-versible) quantum computer, as well as to the very idea of such kind of operation of a real quantum machine.[18]

The physical reality and transparency of origin of this complex-dynamic elementary 'bit', and actual step, of quantum machine operation, always determined by *h*, shows also that the existing vague hopes of unitary theory to obtain a cost-free, reversible quantum computation scheme by means of some ill-defined 'non-destructive measurement' or a ghostly 'nonlocality' ('quantum teleportation' etc.) using especially some 'eluding', massless entities, like photons, are absolutely vain, irrespective of details determining only the exact manifestation of the unitary imitation deficiency. The dynamically discrete and probabilistic nature of real structure ('quantum bit') emergence will always appear in irreducible interaction processes that determine the quantum machine operation. It is not difficult to see also that the relatively strong manifestation of dynamical randomness and discreteness in the essentially quantum dynamics (which is a particular case of the uniform chaos regime, Section 4.5.2), underlies the correspondingly strong, pronounced character of specific features of complex quantum machine dynamics considered above, such as its intrinsic creativity, irreversible direction and non-universality.

Finally, with growing interaction scale/complexity the dynamic formation of *classical* (bound) configurations within quantum machine becomes inevitable (and necessary for a useful device); such states emerge as more regular, SOC type of dynamic entities/regimes (Section 4.5.1) possessing the internal, experimentally detectable fine-grained structure with the smallest grain size of the order of *h* (Section 4.7). In this sense, any real, useful 'quantum' machine is rather a *hybrid*, 'quantum-and-classical' device; its characterisation as 'quantum' object means that it contains *not only* classical, but also essentially quantum, strongly chaotic elements in its functional, operationally important dynamics.

---

[18] The high dynamic uncertainty of the real 'qubit' can be expressed as the problem of 'irreducibility of *n*-ary quantum information' [183], where already the standard quantum scheme implies that any '*n*-dimensional quantum state' (emerging in a generic interaction process) cannot be separated into binary components, i. e. 'qubits', without 'undesirable features' (such as the irreducible dynamic/quantum uncertainty). As our analysis shows, the quantum interaction uncertainty is present in its results in any case, even before any additional 'quantum measurement' is performed and for the full dimensionality of the emerging system configuration, which means in particular that any purely mathematical transition to 'n-dimensional' unitary description [183] cannot solve the problem. The same argumentation is valid with respect to a similar problem of separability of (generic) quantum entanglement (see also Section 5.3).

The outlined universal properties of real quantum machine operation are unified by the underlying concept of unreduced dynamic complexity (Section 4.1) and can therefore be considered as particular manifestations of a *general principle* governing the emergence and evolution of any real (complex) structure, the *universal law of conservation and transformation of complexity, or (dynamical) symmetry of complexity* [1] (see also Chapter 7 for more details). Indeed, since the unreduced quantum interaction dynamics is always represented by a globally chaotic regime and thus cannot be regularly 'controlled' (i. e. transformed into a globally 'self-organised' regime), it should possess its own, 'incorporated' guiding rule that determines the observed operation of natural atomic, molecular and living systems involving structure creation and development (where the eventual transition from quantum to classical type of dynamics also occurs 'from the inside', through the natural system development). The universal symmetry of complexity, constituting an integral part of the unreduced concept of complexity [1], just expresses that intrinsic 'programme' (and its progressive realisation) within any real system, actually provided at the beginning of system existence in the form of its main, driving interactions represented by their initial, 'potential-energy' configuration (it does enter in our starting 'existence equations', eqs. (1)-(5), (41)). This initial, 'hidden', 'latent', or 'folded', form of dynamic complexity, called *dynamic information*, undergoes the unceasing process of transformation into the complementary, 'explicit', 'apparent', or 'unfolded', form of dynamic complexity, called *dynamic entropy* (it extends the ordinary, equilibrium entropy to any kind of process or phenomenon), which constitutes the true, complex-dynamical essence of any system *evolution*. The total, strictly positive (and usually large) quantity of dynamic complexity does not change during its permanent transformation from the hidden form of (ever diminishing) dynamic information into the explicit form of (ever growing) dynamic entropy, which expresses the universal dynamical symmetry (conservation and transformation) of complexity extending and *unifying all* the conventional conservation laws and other postulated fundamental 'principles' of canonical science (including the generalised first and second laws of thermodynamics, i. e. the principles of energy conservation and degradation) [1].

Since this complexity development process constitutes the sense of any interaction (and thus any system existence), exactly expressed by pro-gressive emergence of changing realisations and their levels ('levels of complexity'), we see that the symmetry of complexity, understood as its conserving transformation from conceived (potential) information into tangible (realised) entropy, expresses the universal guiding line, or irreversible *direction* (orientation), of natural system evolution. And since any real system has a finite initial stock of dynamic information, its essential existence, or generalised complex-dynamical *life*, can continue only until the complete transformation of information into entropy, after which the system cannot change any more while remaining itself and thus cannot exist as such and falls into the state of generalised *dynamic equilibrium*, or complex-dynamical *death*, characterised by a local maximum of relatively uniform, highly irregular kind of chaoticity.[19]

Being applicable to any system dynamics, this unified law of evolution is especially useful, however, in the case of such 'indistinct' and 'uncontrollable' systems as quantum micro-machines, including their eventual dynamical transition to classical configurations of growing scale. Indeed, in that case the driving interactions, in their initial, 'potential' form, make the only possible 'programme' of quantum machine operation, with its dynamically probabilistic behaviour at any single step, while the global system evolution does have a well-defined general direction, the one of growing complexity-entropy and diminishing complexity-information, which ensures the conservation of their sum, the total dynamic complexity. Therefore any unitary, regular-step, one-dimensional (sequential) programming (and theory) of quantum machines has no sense at all (other than well-paid, but basically incorrect speculations and mathematical exercises), whereas the unreduced interaction analysis by the universally nonperturbative EP method just gives us the dynamically determined, really emerging system realisations and their respective probabilities (Chapters 3, 4, Section 5.2.1). The step-wise progressive emergence of complexity levels, including the important transition to classical behaviour, realises the universal direction of system development towards the growing complexity-entropy, which is the unique, and practically meaningful, guiding line for creation and control of 'truly small', quantum machines, or 'nanomachines', used for various purposes (Chapters 7, 8). This 'dynamically parallel' programming (and information processing) is also applicable, of course, to any systems with explicitly chaotic dynamics and differs substantially from its unitary imitation by the conventional 'parallel computation', which is nothing more than a practically convenient version of sequential computation split up artificially into simultaneously performed, but totally disrupted pieces.

---

[19] For example, the stage of maximal, 'developed' democracy in *any* civilisation evolution represents but a particular case of this kind of generalised system equilibrium, or death, after which the system (civilisation in this case) can only either totally disappear as such (by the naturally emerging processes of self-destruction) or perform a 'revolutionary' (quick and 'global' enough) transition to a superior, qualitatively different level of complexity, where the transformation of information into entropy will restart again, in a new way and with a new force ('élan vital'). The straightforward application of these results, totally confirmed by all previous civilisations history, to the modern stage of global civilisation development is of vital practical importance [1] and should quickly and definitely replace the dominating interplay of selfish, extremely short-sighted group 'interests' implemented by the ruling unitary system 'priests' in the form of low-level, political and apologetic, 'games' (manipulation) with the dangerously paralysed 'mass consciousness', which tend to present the 'democratic' civilisation decay and evident, fundamentally inevitable impasse as its desirable, 'progressive' evolution and the 'best possible' way of further development.

Note also the persisting and inevitable confusions of conventional theory around the notions of complexity, entropy and information related to its dynamically single-valued reduction of real interactions. Thus the conventional 'information' concept referring to operation of real computing devices corresponds rather to our generalised entropy, while the 'true', dynamic information is an equally tangible quantity, but related rather to conventional 'potential energy' of the driving interaction, as opposed to purely mathematical expressions for an arbitrary mixture of 'complexity-entropy-information' in terms of abstract-space elements (such as quantum 'state vectors') or empirically 'counted' states in conventional theories (see Chapter 7 for more details). The unreduced, complex-dynamical information and entropy, changing in any elementary act of real interaction, without any artificially imposed 'environmental' influences, are basically dimensional quantities (which reflects their physically real origin) directly related to (extended) action or its derivatives (such as energy and momentum) and naturally measured in the corresponding units [1]. This explains why the universal unit of information/entropy content in the unreduced 'quantum bit' (appearing in the essentially quantum part of computing system dynamics) is given by Planck's constant, $h$, contrary to artificial units and ambiguous realisation of conventional qubits in the unitary imitation of quantum interaction processes leading to the value of (abstract) information content in one quantum bit that varies as a function of changing results of system interaction with its unknown 'environment' [182].

The ambiguous 'environment' appears inevitably in the unitary theory as a (forced) imitation of irreversibility (entropy growth) in the otherwise reversible unitary dynamics of the whole system (quantum computer + environment). However, the 'irreversibility' (and related entropy, information, complexity, etc.) thus inserted from the outside have a fictitious, tricky character close to the corresponding introduction of the concept of entropy in classical (and quantum) thermodynamics, where the necessary 'uncertainty' in system behaviour can only be obtained as a practical 'lack/impossibility of knowledge' about 'too complicated' configuration of the 'environment' (playing the role of thermodynamical 'heat bath', or 'thermostat'), rather than any true, unreduced randomness that should have only purely dynamic origin, without any 'help' from the external 'noise'. Such true randomness, originally and causally derived within the dynamic redundance paradigm, is the only way to avoid contradictions inevitably appearing e. g. in 'quantum thermodynamics' relying on imitative randomness, or 'stochasticity' (cf. [184-186]). That's why the artificially inserted 'redundancy' of dynamically single-valued interaction between the 'system' and the 'environment' [182] is but a unitary, inconsistent imitation of our dynamic redundance (the latter being derived for the isolated, nondissipative computing system itself) and simply describes different hypothetical (and quite mutually compatible) excitations/changes in the 'environment' produced by its interaction with the system, which does not reveal any intrinsic randomness. The imitative character of conventional 'quantum entropy' (e. g. 'von Neumann entropy' expressed through the density matrix) is in the fact that it grows only for the system (coupled to the environment), but remains unchanged for the combined 'meta-system' including both the system and the environment, which leads to the mentioned contradictions with apparently universal thermodynamical principles [184-186] resembling those inherent in conventional interpretations of emergence of living systems and other 'strongly nonequilibrium' structures at macroscopically levels of reality (cf. [1]).[20] The universally derived dynamic multivalued-ness of any real interaction at any complexity level (Chapters 3, 4) provides the missing source of purely dynamic randomness within any kind of structure and thus the totally consistent, and actually unique, solution to all the thermodynamical and dynamical problems (including also that of quantum chaos, Sections 4.6.2, 5.2.1, Chapter 6).

---

[20] Many other references to 'quantum thermodynamics paradoxes' within the unitary science paradigm can be found in the materials of international conference dedicated to the problem (First International Conference on Quantum Limits to the Second Law, San Diego, California, 2002), http://www.ipmt-hpm.ac.ru/SecondLaw/index.en.html. Note that as follows already from the conference title, the underlying conventional science results seem to imply a violation of the ordinary second law for quantum systems, which is not surprising taking into account the intrinsic deficiency of the unitary, dynamically single-valued description used that actually cannot

Another practically important consequence of the universal symmetry of complexity is the *principle of complexity correspondence* which states, in its application to computational processes, that any closed computing system cannot correctly simulate the behaviour of systems with higher than its own values of total dynamic complexity. In other words, the correct simulation is possible only if the full dynamic complexity of the simulating system exceeds that of the simulated system/behaviour. In view of the above general theory of computation in terms of complexity transformation, the complexity correspondence principle emerges by the direct application of the complexity conservation law to a computing system that evidently cannot attain complexity levels higher than its internal complexity determined by all driving interactions (and thus it cannot reproduce any 'finer' structural/dynamical details of the simulated system with higher complexity). The correspondence of complexity is especially important for quantum computing systems, since they are confined to the *lowest* complexity levels and therefore cannot correctly simulate higher-complexity systems that constitute the majority of practically important phenomena.

Since classical behaviour emerges from essentially quantum behaviour as a definitely higher complexity level (Section 4.7) [1,12,13], it follows that any purely quantum system cannot correctly simulate any classical, quasi-regular or explicitly chaotic, system, contrary to the results from a great variety of papers on conventional quantum computation 'showing' the opposite. As elementary classical states are usually formed starting from the simplest bound states (like atoms), it appears that even the really small, but practically important, realm of atoms and molecules is already

---

explain *any* 'thermodynamical' behaviour in principle, including its application at the level of 'usual', macroscopic systems with classical dynamics (they simply provide a possibility of easier manipulations using their much higher complexity and its diverse manifestations, while the problem remains basically unsolved). In reality, as we have seen, it is the unitary quantum (as well as classical) mechanics that should be considerably extended to its unreduced, dynamically multivalued version, after which all the 'thermodynamical' (as well as 'quantum' and 'relativistic') 'paradoxes' disappear, being replaced by inevitable, natural manifestations of the unreduced dynamic complexity of any system dynamics.

beyond the possibilities of any quantum simulation. It is evident, in particular, that any essentially quantum dynamics cannot correctly reproduce the main property of permanent localisation of a classical system configuration without losing its quantum (delocalised) character.

This our result actually demonstrates once more the hopeless illusiveness of the major underlying motivation of conventional quantum computation theory developed as a 'qualitatively more efficient' way to calculate a large majority of various real, mainly macroscopic and classical, regimes of dynamical system behaviour. Indeed, the 'digital', 'exact-number' regime of a quantum computer is impossible because of the strong, 'global' chaoticity of essentially quantum dynamics (Sections 4.5.2, 4.6.2, 5.2.1), while its 'analogue', direct-simulation possibilities are strictly limited to quantum systems with lower dynamic complexity. We obtain thus another, fundamentally substantiated objection against the announced and 'rigorously proven' property of universality of unitary quantum computation, completing the arguments developed above in this Section and showing that unitary quantum computation is simply impossible as such. Now we see that not only unreal, unitary, but also any real, irreducibly chaotic quantum dynamics cannot properly simulate any classical behaviour (thus, the main property of classical behaviour, its permanent localisation, can hardly be reproduced by the irreducibly delocalised dynamics of essentially quantum system). As noted above, any real, practically useful (i. e. complicated enough) 'quantum machine' will always eventually perform (local) dynamical transitions into classical type of state, and it is in that way that such real, hybrid micro-machines will be able to approach the features of classical world, as it actually happens to natural micro-machines, including those that determine 'simulation' processes in the brain (the complexity correspondence principle provides thus another fundamental objection against all theories of essentially quantum basis of high enough, 'global' levels of brain dynamics [68-71]).

The outlined general limitations of purely quantum computation imposed by the complexity correspondence principle can be specified for several particular situations. One of them concerns the problem of 'quantum memory' supposed to participate in operation of conventional quantum computer [47-49] (see also Sections 2(iv,v), 4.5.1, 4.7). It becomes clear now that even if quantum memory could exist, it could 'memorise' at max-imum structures from lower, essentially quantum complexity sublevels, which means, taking into account the globally chaotic type of essentially quantum behaviour, a direct reproduction of the corresponding uniform regime of dynamical chaos (Section 4.5.1), having apparently no practical sense. As any function of memory is a complex-dynamical (multivalued) effect in principle [1], the idea of *unitary* quantum memory is totally erroneous because of the zero value of unreduced dynamic complexity for any unitary dynamics. And since any *practically useful* memory should have the properties of the (multivalued) SOC type of state with 'distinct' enough dynamical configuration, it becomes clear that even real, nonunitary, but globally chaotic quantum dynamics cannot realise a useful memory and the latter can actually start with classical, bound state emergence in quantum interaction dynamics (Section 4.7). Such states appear inevitably in any realistically complicated micro-machine and therefore the *natural* development of real, complex-dynamical 'quantum' interaction within a properly structured machine leads to production of those more 'fixed' and localised, classical configurations playing actually the function of 'memory' (that can be explicitly specified as such or not). It is important to note the essential difference of such kind of 'distributed', 'dynamical' (emergent) memory, *inseparably mixed* with 'calculations' and most probably dominating in natural nervous systems, from the pseudo-unitary, totally regular and 'separated' image of memory borrowed from the architecture of usual modern computers (for those macroscopic devices, the inevitable memory chaoticity and dissipativity remain often hidden within large enough hierarchy of participating complexity levels).

Similar applications of the complexity correspondence principle concern various other intuitive 'projections' of higher-level notions and phenomena onto quantum dynamics and the reverse (see e. g. [52,55-57,62,63, 68-71,187-190]), which become increasingly popular in conventional theory after the full-scale advent of quantum information idea (indeed, if 'universal' quantum computers can simulate, in principle, any process, then why can't one use this simulation results as the corresponding 'quantum' version/explanation of the simulated higher-level process/phenomenon?). The complex-dynamical basis of any real interaction and its results shows, however, that the naturally emerging hierarchy of unreduced dynamic complexity cannot be 'inverted' so that higher complexity levels would be somehow 'reproduced' at its lower levels. In particular, *any* quantum process, real or simulated, cannot reproduce just the essential, key properties of higher-level phenomena (classical, biological, social, financial, etc.). As for the *unitary* quantum schemes invariably used in conventional theory for 'quantum' reproduction of higher-level dynamics, they look especially absurd, since the unitary, zero-complexity quantum mechanics cannot properly explain even the real, complex-dynamical quantum phenomena themselves, using instead empirically based 'postulates' for the mystified 'substantiation' of its dynamically single-valued imitation of reality. Note that despite the intense 'post-modern' plays of words in 'quantum generalisations' of higher-level notions, the problem cannot be reduced to the choice of terminology, since the difference between various complexity levels is now properly explained and rigorously specified in the unreduced science of complexity (see Sections 4.1, 4.7). We deal here with a deeper, physical meaning of the unreduced dynamic complexity: its values corresponding to a certain 'level of complexity' directly account for a specific, tangible 'quality' and 'diversity of properties' of entities from that particular complexity level, which are well determined by the dynamic entanglement and dynamic redundance phenomena just constituting the real entities emergence. Therefore formal projection of higher-level notions to lower-level dynamics (which is actually described, in addition, by its effectively zero-dimensional projection) is strongly, *qualitatively* incorrect in principle, irrespective of details. By contrast, it is not impossible, in principle, to find some manifestations of properties from lower levels (e. g. 'quantum-like' properties) in certain patterns of higher-level system behaviour (confined to that, higher-complexity level), but such analogy implies the use of the adequate formalism containing the causally derived (dynamically emerging) features of the corresponding complexity level (it should be provided, in particular, with the unreduced, dynamically multivalued interpretation and solutions), instead of the opposite, unjustified (postulated) 'lowering' of higher-level entities/phenomena down to quantum level (utterly simplified by its unitary description).

A number of 'hyper-simplified' schemes within the unitary quantum computation theory itself can be mentioned as more explicit demonstration of its basic deficiency. These include such hypothetical possibilities as 'ground state quantum computation' [45] and various 'nonlocal', 'instanta-neous' and 'interaction-free' applications of 'quantum teleportation' and other 'quantum mysteries' (e. g. [191]). Note that in a way these are the most 'consistent' realisations of the unitary quantum computation idea just pushing it to the extreme of its most 'pure' form and thus demonstrating its true essence. It is clear, for example, that if ever the unitary quantum computation could be realised, it should form a sort of (relative) 'ground state' with (almost) uniformly chaotic internal dynamics, which actually constitutes the physically real, causally substantiated structure of any ground state or the state of rest (as opposed to a more ordered state of motion) [1,9,12,13]. However, the problem is that such globally chaotic, highly irregular state cannot simulate any more complex, SOC type of structure in principle, so that practically it could simulate only itself and similar, structurally trivial states of matter. This consequence of the complexity correspondence principle is generally applicable also to other 'ultimate' cases of quantum computation, which are different from its 'generic' case only in their more explicit demonstration of existing fundamental deficiencies.

A similar kind of abstract over-simplification of reality within the unitary quantum computation initiative is demonstrated by estimates of the 'ultimate computing power' of the real, tangible universe or any its finite portion [150,151]. The maximum (but really attainable) speed of computation (in operations per second) by the actual universe, considered as a unitary quantum computer [62], is obtained, for example, by simple dimensional division of its total energy by Planck's constant, while the universe's total informational capacity, or 'memory space', is given by its total equilibrium-state entropy divided by Boltzmann's constant [150]. By analogy to the above 'ground-state computation' and in agreement with the complexity correspondence principle, it is clear, however, that the assumed equilibrium, or uniform chaos state (Section 4.5.2), which has, of course, nothing to do with the real, highly nonequilibrium universe structure, cannot reproduce anything more complex and structurally distinct than its own ultimately randomised configuration, irrespective of the detailed 'computation' dynamics (see also Section 7.1). It is true that by dividing e. g. the energy of a free electromagnetic wave by $\hbar$ one obtains a quantity with the dimension of frequency, but does it really mean that the freely propagating, linear wave can actually 'compute' something (or even 'everything', according to [150]), with the number of operations per second equal to its frequency?

Popular references to e. g. 'complex adaptive systems' dominating in the universe [62] shows that the adherents to this, indeed 'ultimate' imitation of reality can well understand the difference, actually more than obvious, between the true universe content and the globally chaotic, hypothetical state of its total equilibrium (which is practically never totally realised even in any limited part of a real, structure-producing universe). Taking into account the huge and evident, 'premeditated' contradictions of such 'unlimited', truly apocalyptic simplification of reality by the officially honoured unitary science, it is difficult not to acknowledge not only its totally 'ironic', 'post-modern', speculative content [5-7], but also practically fraudulent, explicitly parasitic practice, taking especially pronounced forms just in the case of 'quantum computation' and other similar 'applications' of conventional, the more and more abstract and mystified, quantum theory and 'new physics' in general (see also Chapter 9). In the meanwhile, the most powerful supercomputers, using the most advanced abstract theories and their most efficient algorithmic realisations produced by armies of carefully chosen scribes from hundreds of prestigious laboratories, cannot reproduce, during many hours of work, the shortest moment of existence of the smallest observable objects of the universe, its elementary particles, taken in their physically real, unreduced version. On the other hand, the totally realistic, complex-dynamical content of the electron and other elementary particles, causally explaining their physical nature and all the 'mystical' peculiarities of their unified, 'quantum' and 'relativistic' behaviour, can be reproduced even by a 'hand-made' analytical theory, if one uses the consistent, logically correct and physically transparent description [1-4,11-13], which is possible simply due to the non-perturbative, honest, causally complete analysis of a configurationally simple, but unreduced interaction process. The same dramatic, 'non-computable' difference between the unitary, dynamically single-valued, and unreduced, dynamically multivalued descriptions of natural entities continues for all higher levels of complexity, including quantum chaos and measurement [1,9,10] (= real 'quantum computation'), classical behaviour emergence, many-body problems, living system dynamics, consciousness and all aspects of civilisation dynamics [1-4]. The system of close inter-connections, permeating the whole hierarchy of complexity and adequately described by the universal science of complexity [1], becomes only more evident by the easily made comparison between the results of both approaches and their respective practical support (at the presently dominating level of consciousness).

The physically real version of 'quantum information', being liberated from artificially mystified deviations of its conventional, unitary version, appears thus as a chaotic, dynamically multivalued and naturally quantised evolution of the corresponding physical system, which is different from the ordinary, 'classical' (macroscopic) and regular realisation of information processing just by the strongly irregular, 'globally chaotic' dynamic regime of any essentially quantum machine. The irreducible dynamic randomness, as well as the related quantisation and 'nonlocality' (now causally explained), of essentially quantum dynamics are due eventually to the ultimately low (but strictly positive!) values of the unreduced dynamic complexity of purely quantum systems, which cannot therefore reproduce, by 'simulation' or 'calculation', any higher-complexity dynamics starting already from the elementary classical (permanently localised) systems like atoms. This specific regime of information processing, consistently understood and described only with the help of unreduced, dynamically multivalued theory, can be present also at the level of classical (both micro- and macro-) dynamics of special, 'chaotic' types of computing system, but for systems containing essentially quantum operational elements this kind of strongly irregular dynamics is inevitable, in full agreement with the well-known fundamental properties of quantum behaviour. The usefulness of such chaotic quantum, classical and hybrid machines is proved by their successfully working natural versions (determining and causally explaining e. g. the behaviour of all living systems and their elementary 'units'). However, the practical, detailed understanding and reliable control of natural complex-dynamical machines, as well as creation of their useful artificial versions, becoming critically important today, definitely necessitate the unreduced, non-simplified description provided actually by the dynamic redundance paradigm and qualitatively exceeding the results of effectively zero-dimensional imitations within all possible unitary approaches of conventional science.

Note in this connection that the appearing merely speculative, empirically based observations about the eventual 'quantum' origin of any natural, 'complex' machine, followed by the conventional type of dynamically single-valued, abstract, mystified and arbitrarily postulated symbolism (see

e. g. [62,192-195]) represent only decadent, inconsistently deformed versions of the basically unchanged unitary approach, with all its irreducible deficiencies. The vain desire to obtain 'something from nothing' by any means, a qualitatively new, superior property from just another, as if successfully 'guessed', but totally speculative 'postulate' or imposed abstract 'principle', has already rendered so many bad 'services' to science (with the characteristic case of quantum computation), and now that the unreduced dynamic complexity of the world is practically subjected to the critically deep, purely empirical and therefore largely destructive modification at all scales and levels, one should be especially careful with the illusive 'unreasonable efficiency' of extremely superficial, over-simplified 'calculations' of both purely mathematical and purely egoistic origin that tend 'strangely' (but not unexpectedly) to be unified within one kind of 'tricky' thinking becoming critically dangerous right now by its unjust and compromising domination of entire scientific knowledge (Chapter 9).

Finally, it is important to emphasize once more that the fundamentally substantiated, well-specified and practically important conclusions about operation of real quantum micro-machines summarised in this Section demonstrate the usefulness of the underlying causally complete extension of quantum mechanics called 'quantum field mechanics' and constituting itself an application of the universal science of complexity based on the dynamic redundance paradigm to systems from the lowest complexity levels [1-4,9-13]. The real quantum computation theory emerges, in particular, as application and development of the causally complete, dynamically multivalued theory of quantum chaos and quantum measurement [1,9,10] (see also Chapter 6) revealing the irreducible and omnipresent, purely dynamic origin of true randomness in any real process of quantum interaction. The obtained results not only show why the conventional, unitary theory of quantum information processing, quantum chaos and quantum mechanics in general is basically deficient, but provide the causally complete, complex-dynamical theory of any quantum machine operation that reveals qualitatively new directions of their practical creation, in accord with the obtained physically and mathematically consistent picture of already existing, natural system dynamics (see also Section 7.3, Chapter 8). These are also the properties of other realised and outlined applications of the quantum field mechanics (see e. g. [13] and Chapter 3 of ref. [3]) and the universal science of complexity in general [1,4] supporting the causal completeness of its intrinsically unified description of the world dynamics, in contrast to the persisting ruptures and contradictions of the simplified, dynamically single-valued projection of reality maintained by the conventional paradigm of 'mathematical physics' type that shows only further, dramatically growing and now critically large separation from the real world dynamics, starting already at its lowest levels.

## 5.3. Complex-dynamical reality behind the mystified abstractions of official unitarity

Application of the universal concept of complexity within the dynamic redundance paradigm to the lowest complexity levels, called here quantum field mechanics, provides the explicit, totally consistent and reality-based derivation of all the peculiar features of quantum behaviour from the unreduced interaction process analysis (Chapter 4) [1-4,9-13]. Since the conventional, dynamically single-valued theory continues to impose its artificially mystified abstractions, especially within the unitary theory of quantum information processing, we shall briefly summarise, in this Section, the true, totally realistic and causally complete meaning of the corresponding 'quantum' notions, based on their complex-dynamical origin and actually extending the realistic approach of Louis de Broglie, the true founder of quantum (or 'undular') mechanics (see ref. [2] for the details and references). Any quantum 'computing' system necessarily represents a sufficiently involved system of interacting entities, where not only the postulated 'averaged', externally observed, but the detailed (though maybe hidden and probabilistic) behaviour becomes *practically* important, contrary to the case of more simple, elementary quantum systems dominating in conventional theory applications, where the problem of genuine, reality-based understanding of formally postulated results could most often be transferred to the realm of 'purely theoretical', 'interpretational' or 'philosophical' studies (see e. g. [15,16]). However, despite the dominating superficial illusions, the resulting glaring inconsistencies of the official fundamental science could remain outside the truly important applications only until the deep enough advance of the empirically based technology into the structure of matter, and the modern story of quantum computation (and other 'microscopic' applications, see Section 7.3, Chapters 8, 9) only confirms the well-known fact that any obvious lie in basic issues, *especially* when it is 'generally accepted' and seems to be 'practically unimportant', hides and prepares a major future failure and essential loss of opportunities.

The real world structure starts, according to the quantum field mechanics [1-4,11-13], from the simplest possible system of interacting entities, the two physically real, initially homogeneous fields, or 'protofields', uniformly attracted to each other with a force which is sufficiently high to produce a large (local) deformation of at least one of the fields. One of the protofields has an electromagnetic (e/m) nature and gives rise, after being perturbed by the other protofield influence, to the directly observable e/m entities and effects, while the other field is called gravitational protofield, or medium, since it is responsible for the universal gravitation, even though its 'matter' cannot be perceived as directly as that of the e/m protofield (the gravitational protofield content should actually be represented by a viscous and dense enough 'quark matter', whose detailed structure is less important for the observed results of protofield interaction).

The unreduced interaction analysis of the universal science of complexity (reproduced in Chapters 3, 4 for the general case of many-body problem) shows that already such simple initial system configuration leads, for generic interaction parameters, to emergence of randomly distributed local structures, each of them having the form of essentially nonlinear, dynamically multivalued protofield pulsation (self-oscillation) called *quantum beat* and constituting the essence of a simple (massive) elementary particle, such as the electron [1-4,11-13]. Each massive elementary particle is represented thus by a complex-dynamical *process* of quantum beat in the system of two coupled protofields, consisting in unceasing periodic cycles of self-amplified (essentially nonlinear) protofield squeeze (or reduction, or collapse) and the reverse transient extension to a quasi-free protofield state, where the positions of consecutive centres of reduction appear in a dynamically random order from the total set of their equally possible values that form 'realisations' of this particular system and give rise to the living, physically real structure of fundamental, 'embedding' *space* of the universe. The unceasing *chaotic change* of the *emerging* inhomogeneous structures (reduction centres) constitutes the naturally irreversible 'flow' of *events* thus defined, or physically real, fundamental *time* of the world, fur-ther 'modulated' (together with space) at higher complexity levels by their respective realisation change processes (Sections 4.3, 7.1).

The elementary particles thus formed (Section 4.6.1) enter in the next level of interaction among them through the two protofields they share, which explains the nature and number of the two long-range fundamental interactions, the e/m and gravitational interactions.[21] This level of complex-dynamical interaction development gives rise to the phenomena of quantum chaos, quantum measurement and elementary bound system formation (like atoms), the latter actually constituting the simplest structures from a higher-complexity, classical (permanently localised) type of behaviour (Sections 4.6.2, 4.7). This universal hierarchy of complexity levels continues its natural development in the same fashion, up to the highest known levels (represented by consciousness and all its products) [1,4], but we shall now limit ourselves to the lowest levels of complexity (up to the simplest classical structures) that actually form the world of 'quantum' phenomena and see how the universal, standard properties of complex, multivalued dynamics give rise to the characteristic features of quantum behaviour, remaining unexplained (postulated) and 'mysterious' [15,16] in the framework of dynamically single-valued, unitary, zero-complexity world projection that constitutes conventional quantum mechanics (including all its modern 'interpretations' and 'modifications'). We shall see once more that after being realistically and causally explained as manifestations of dynamically multivalued behaviour, the 'magic' properties of quantum systems cannot be considered as the basis for the equally 'magic', cost-free gain in computer efficiency predicted by the unitary, totally regular imitation of quantum dynamics. The unitary theory, in any its version, is not suitable at all for description of a real quantum machine operation and provides its qualitatively erroneous picture (cf. Section 5.1), while the dynamically multivalued, causally complete analysis of quantum interaction processes explains a quite different, omnipresent kind of magic of natural micro-machine operation (Section 7.3, Chapter 8).

---

[21] This living, complex-dynamical and naturally unified world construction of the quantum field mechanics [1-4,11-13] is imitated in conventional, unitary science by various versions of recently appeared 'brane-world' model. Without going here into details of the latter, note however that it has all the characteristic limitations of a unitary projection, including the purely abstract and formally imposed (postulated) origin of arbitrary number of introduced entities (such as 'dimensions', fields, interactions, particles and formal 'rules' of their description), replacing the intrinsic creativity of dynamic redundance and entanglement in the unreduced interaction analysis, which leads to numerous contradictions, unlimited multiplicity of 'suitable' models and predicted (but never observed) new entities, and fundamentally unclear physical origin of the main structures, their emergence, evolution and properties. The whole set of such modern versions, or 'scenarios', of conventional 'field theory' and cosmology form a strange agglomeration of abstract symbols artificially subjected to arbitrary number of 'conveniently' adjusted abstract rules and supposed to 'explain' the real structure of real world, from which the abstract picture of canonical science obviously continues to deviate, both basically and practically, including already the most important, qualitative physical properties, such as explicit dynamic emergence of entities, wave-particle duality, dynamic indeterminacy and uncertainty at various levels of behaviour, omnipresent and strong 'violations' of all the predicted 'simple' symmetries, natural unification of different fundamental interactions and 'principles' within real objects, etc. The situation in the canonical fundamental physics is the same as it would be in the field of graphical arts if we had there *only* works of extremely abstract painting representing real life by simple sets of separated dots and lines, but *obligatory for buying* from their authors by the state in practically unlimited quantities and for the prices determined by the authors themselves and their interested agents.

(A) <u>Quantum coherence, 'decoherence', randomness, chaos and density matrix</u>. As shown in previous Sections, the abstract and artificially imposed 'quantum decoherence' of abstract 'state vectors' attributed in conventional theory to the influence of ill-defined external 'noise' is nothing but unitary imitation of the totally internal phenomenon of dynamic multivaluedness and related randomness in the (spatial) sequence of system realisations, appearing as absence of their 'coherence'. However, one should also understand the true, physically real origin of 'quantum (undular) *coherence*' of *particles* and their ensembles, showing itself in many observable effects of wave interference for simple enough quantum systems. It appears that both quantum coherence and dynamic 'decoherence' (randomness) are present, in a variable proportion, in every quantum system behaviour and can be interpreted as average (probabilistic) order in system realisation sequence and its complementary, purely irregular component, respectively [1-4,11-13]. The cases of dynamically multivalued SOC (Section 4.5.1) and uniform chaos (Section 4.5.2) represent the extreme limits of this general coexistence of order and randomness, tending respectively to almost total regularity and maximum disorder, where the former is characterised by narrow, δ-like and the latter by quasi-uniform, flat distribution of realisation probabilities.

While the general coexistence of order and randomness, or partial coherence, is inherent (Section 4.1) in the very nature of the unreduced dynamical chaos/complexity concept (contrary to conventional imitations of chaoticity), the mentioned limiting cases of dominating SOC (coherence) or global chaos (incoherence) tend to alternate hierarchically in the progressively emerging levels of complex-dynamical interaction processes and resulting entities. Thus, the canonical diffraction patterns produced by many individual quantum particles contain within their regular (coherent) undular shapes the irregular distribution of particle reduction centres that can be observed experimentally (see e. g. [196]), in full agreement with the dynamically multivalued content of the field-particle process (whereas the 'Bohmian mechanics' cannot consistently explain this wave-particle duality, including especially the unavoidable wave transformation into particle and back, as well as dynamically random distribution of particles within the wave field). The irreducible 'coherence' of an elementary particle 'with itself' is related also to the holistic, *indivisible* nature of that complex-dynamical object emerging at the lowest complexity level as a dynamically multivalued structure, unified by its *physically real* wavefunction (Sections 4.2, 4.6.1) that just provides the well-specified, causal incarnation of the field-particle coherence. At higher quantum sublevels of complexity, involving interactions between at least several particles, the number of possible system configurations (realisations) grow dramatically and maintenance of high coherence becomes difficult, which gives rise to the (genuine) quantum chaos phenomenon (Section 4.6.2, Chapter 6) and proves practical impossibility of unitary quantum computation that would need the ideal coherence of isolated particle behaviour. The degree of coherence (order) within any quantum system state/behaviour can be quantitatively characterised by any measure of inhomogeneity of the dynamically determined (spatial) distribution of system realisation probabilities, which is given, in particular, by the generalised system wavefunction (or its squared modulus) extending the conventional wavefunction, density matrix and distribution function concepts (Section 4.2). It is important to emphasize that real, and always existing, 'decoherence' of a quantum system has a totally internal, purely dynamic origin determined by the driving system interaction itself and takes the form of dynamically random (probabilistic) order of realisation emergence, whereas the system coherence, also always present in its dynamics, is an equally emergent property appearing in the form of a partial order/inhomogeneity in realisation probability distribution. The related system properties, its dynamic complexity, 'non-computability' of behaviour, etc. (Sections 4.1, 5.2.2) are always determined by the total set of incompatible, dynamically obtained and therefore *permanently* changing system realisations and their respective probabilities (a separate, detailed dis-cussion of exact physical meaning of *information* and *entropy* as two dual forms of unreduced dynamic complexity, as well as their imitations in conventional theory, can be found in Chapter 7).

This dynamically determined, interaction-driven coexistence of intrinsic randomness and probabilistic order (coherence) is the unique property of our unreduced, dynamically multivalued description of a configurationally regular interaction process and can only be very inconsistently simulated by formally introduced, external 'quantum randomness', such as that of Nelson's statistical interpretation [197] of quantum mechanics (see e. g. [198-206] for only a limited selection of papers devoted to this kind of approach). Apart from arbitrary insertion of entities and rules, usual for the unitary, 'mathematical' physics, this approach has its own internal inconsistencies actually reduced to another formulation of standard 'quantum mysteries', which originate exclusively from the dynamic single-valuedness that underlies all official theories and permitted interpretations. Similar to so many other versions of unitary approach (involving e. g. all 'zero-point field' theories, conventional 'decoherence', quantum gravity and field theories), randomness is introduced here 'stochastically', i. e. formally, from the outside, and not as a dynamical result of internal, a priori absolutely regular interaction development. Unfortunately, any 'hidden variables' kind of approach, already looking quite 'revolutionary' within the heavily reduced framework of conventional science, is always understood, in addition, in the mechanistically simplified sense of a direct influence upon quantum system of some 'hidden' medium which, according to its introduction, accumulates and conveniently 'buries forever' all the existing 'mysteries' of quantum behaviour. The most consistent (and almost never referred to) approach of this series is the original de Broglie's theory of 'internal particle thermodynamics' [207-209] constituting an integral part of his 'nonlinear wave mechanics' (see [1,2,11-13] for further references): while the origin of 'subquantum medium' and its randomness remain unclear, the quantum 'particle' itself is outlined as a physically real entity dynamically emerging from the accompanying wave.

Note also the essential difference of the consistently derived, totally realistic phenomenon of dynamic multivaluedness, explaining dualistic field-particle transformations and their randomness, from the conventional science substitute in the form of postulated and abstract many-worlds, or 'multiverse', interpretations (especially popular in unitary quantum computation theory): the detailed, essentially nonlinear mechanism of dynamic redundance (Sections 3.3, 4.2) shows how the system can change qualitatively its state and autonomously 'choose' between multiple, dynamically emerging possibilities for its current configuration without leaving the physically real space of the single, dynamically emerging and self-developing ('living') universe. It is easy to see that the underlying mathematically and physically complete (i. e. truly 'exact' and 'general') solution to the unreduced interaction problem resolves thus, simply due to its consistency crudely violated in the unitary theory, bundles of inter-related 'mysteries' from conventional quantum mechanics, relativity, field theory and cosmology [1-4,11-13].

(B) <u>Quantum entanglement, nonlocality/'correlations', 'teleportation', reduction/measurement and duality/complementarity</u>. The unreduced, interaction-driven, physically real entanglement of quantum system components results only from the interaction development itself and has the permanently changing, causally probabilistic internal structure determined by the dynamic redundance phenomenon (Section 4.2). This complex-dynamical entanglement is a totally realistic and universal phenomenon occurring within any unreduced interaction process; its peculiarities for quantum systems are determined by their lowest position in the unified hierarchy of complexity, so that one cannot observe the fine-grained, quasi-continuous structure of occurring processes, which leads, in particular, to the canonical mystification of quantum properties. The unambiguous interactional, complex-dynamical origin of quantum entanglement is demonstrated, in particular, by the explicit entanglement expression in the obtained general solution for the many-body system wavefunction (or generalised distribution function), eqs. (24)-(25), (50)-(52). We see from those expressions that the interacting degrees of freedom are indeed physically, directly entangled among them, but this real quantum entanglement has the explicit origin in the driving interaction, which leads also to its dynamically probabilistic (Section 3.3) and dynamically fractal (Section 4.4) structure. It is the dynamically probabilistic structure of real, always *multivalued* entanglement of interacting entities, rather than arbitrarily varying 'decohering' influence of the environment, that makes impossible *in principle* the real existence of any essentially quantum machine with unitary, or even approximately unitary, dynamics (Sections 5.1, 5.2), as well as any particular application (e. g. [210]) of unitary, fictitious 'entanglement' from the conventional theory.

Another, related expression of complex-dynamical, interactional origin of real quantum entanglement is provided by the accompanying formulas for EP realisations, eqs. (25c), (48), showing that during each cycle of permanently changing entanglement the system performs the real dynamical squeeze, collapse, or 'quantum reduction' to the internally entangled configuration of the forming current realisation. The resulting, entangled system existence is impossible, therefore, without the transient phase of disentanglement, alternating with entanglement-reduction, during which the quasi-free system components 'liberate' and rearrange their compound configuration to start the next entanglement-reduction phase directed towards the next, probabilistically 'selected' realisation. This peculiar transient state of quasi-free, undular existence of quantum system components, common for all its 'regular', entangled realisations, just forms the causally extended version of (generalised) wavefunction (Section 4.2) resolving all the 'mysteries' of its conventional version [1,4,11-13].

It becomes clear, in particular, that the conventional paradoxes of linear 'quantum superposition' of states and (interacting) objects are resolved by the internal complex dynamics within each such 'linear superposition', where the total quantum system described by the superposition wavefunction performs the unceasing cycles of reduction-extension towards the constituent (superposed) 'eigen-states' describing in reality system realisations, with the average frequencies of their appearance determined exactly by the respective, dynamically derived realisation probabilities (Section 3.3). The permanent, dynamically dualistic transitions between the quasi-linear, weak-interaction state of wavefunction and *essentially nonlinear*, but largely hidden phase of entanglement-squeeze (strong effective interaction) explain all the 'mysterious' manifestations of quantum duality, or 'complementarity' (see also the end of this item), including the peculiar coexistence of external linearity of Schrödinger equation and the strongly nonlinear, corpuscular and probabilistic, properties of real quantum objects, giving rise to the causally complete derivation and *dynamic* interpretation of quantisation rules and uncertainty relations [1,4,12,13] (the coexistence of physically real linear and nonlinear aspects of quantum dynamics was proposed and defended by Louis de Broglie, see [1,2]).

It is not difficult to see that our universal expression of dynamic entanglement, eqs. (25), represents the generalised form and dynamical origin of all its particular cases (such as general linear superposition, explicit component entanglement in a many-body system, etc.), whereas conventional entanglement expressions hide the dynamic details revealed by the unreduced description behind the corresponding general 'coefficients' and 'eigenfunctions' obtained for each particular case by formal (and strongly simplified) postulation of directly observable, irreducibly 'coarse-grained' quantum structures (see item (C) below for the case of macroscopic and classically amplified superpositions). The mystery of quantum entanglement phenomenon, first highlighted by Schrodinger [175], obtains thus its consistent, physically realistic and conceptually nontrivial solution, explaining also the long problem stagnation within the unitary approach. The extended, complex-dynamical version of entanglement is much deeper than its canonical version not only because the former has universal origin and manifestations appearing at any higher level of complex dynamics, but also because already at the quantum level the causal entanglement phenomenon describes real, essentially nonlinear dynamical structure of any system or emerging entity, rather than exclusively a postulated linear combination of many-particle states with interchanged individual particle eigenstates (e. g. positions) implied by the conventional version of quantum entanglement and forming a part of the official 'quantum mystery'.

The canonical case of quantum entanglement can be causally interpreted now as real and permanent system jumps between its different realisations with interchanged particle positions, similar to analogous causal interpretation of any other 'linear' superposition of quantum states describing only the external 'envelope' of unceasing, highly nonlinear, explicitly described quantum jumps between those states. Such complex-dynamical, interaction-driven interchanges permanently occur in any essentially quantum many-body system, involving the complete hierarchy of various possible versions of its configuration (see also Section (C)) and largely exceeding the simplest canonical spin-flip case in a two-particle entangled system, which explains, in particular, the origin (and insufficiency) of 'symmetrisation' or 'antisymmetrisation' procedures postulated for such system wavefunctions by the conventional theory.

The same unified picture of unreduced complex dynamics and its formal expressions contain the causally clarified properties of 'quantum nonlocality' and 'quantum correlations', which appear to be more directly related to the mentioned 'wavefunctional' phase of disentanglement in quantum system dynamics. Any quantum (and actually any real, complex) system is always partially 'delocalised' because (i) it spends a part of its life in the delocalised state of wavefunction (or 'intermediate realisation') and (ii) due to the property of dynamic redundance, the system performs chaotic wandering in the localised, 'corpuscular' state of regular realisation, these two qualitatively different kinds of state (and manifestations of nonlocality) being related by the causally extended 'Born's probability rule' [1-4, 10-13] (the system tends to collapse towards the realisation localised around higher wavefunction magnitude and the reverse).

The situation of 'quantum correlations' usually refers to a particular case of quantum superposition in a many-body system, where different superposed states correspond to an 'interchange' of eigenvalues, such as positions or spin directions, between different system components (usually elementary particles or atoms). While the system components remain in direct, close enough contact/interaction among them, the 'correlations' between the properties of their observed configurations in each superposed state follow from the mentioned cyclic transitions of the whole system from one its configuration to another (generalised 'quantum beat' process) permanently occurring, it should be emphasised, within the system dynamics itself, apart from any 'measurement' from the outside that can only 'catch' a current phase of that internal system dynamics (one can also say that the unreduced interaction process permanently 'measures itself', passing unceasingly by all its possible, dynamically quantised and probabilistically selected states, or realisations). However, even after an arbitrary large separation between individual system components (*always* preceded by a stage of direct contact in that kind of setting) correlations between their configurations will persist due to the persistence of unstoppable quantum-beat pulsation within each component, which is driven by the fundamental, unchangeable protofield interaction and tends therefore to preserve its temporal phase just at the lowest, quantum level of complexity, where any spontaneous, or 'dissipative', modification of a (quasi-free) system is impossible [1]. The latter property is generally lost at higher complexity lev-els permitting system modification through lower-level dynamics (dissipativity), which explains 'mysterious' peculiarity of quantum behaviour, where any measurement of separated components gives 'mysterious' correlation between their properties (one should only take into account the quantum beat process, playing the role of internal particle clock, 'synchronised' by initial interactional entanglement with those of other participating particles). We shall not give here formal expressions of the described properties, easily obtained as particular cases of our general expressions, eqs. (25).

It is clear that such causally explained quantum correlations and nonlocality need not involve any 'infinitely fast signal transmission' and similar mystifications originating from total ignorance by the conventional theory of the underlying complex interaction dynamics. Further theoretical and experimental refinement of the proposed picture is possible, including e. g. experimental tracing and control of the outlined processes of random walk of a system among its realisations, where even faster-than-light jumps between individual realisations are not excluded, in principle (though they remain an exotic assumption), since they do not automatically lead to the superluminal propagation of the observed averaged, random walk of the system. Note, however, that such kind of experimentation at the ultimately low complexity levels will always involve much ambiguity, simply because all our 'instruments' and measurement processes also make part of this world and are therefore fundamentally limited from below by dynamics of the same, quantum levels of complexity. This basic limitation, usually neglected by conventional science because of its formal and abstract character, refers to any kind of 'fine' experimentation with elementary quantum systems consuming lots of resources and oriented to 'experimental confirmation' (or rejection) of fundamental quantum theory 'interpretations'. Specific manifestations of dynamic multivaluedness and entanglement at the level of highest quantum and lowest classical sublevels, determining real quantum machine dynamics and predicted by the quantum field mechanics (Chapters 3-5), can be better suited for 'experimental verification' and will certainly be much more useful practically, since they will constitute an integral part of real quantum machine development and control.

We have seen above that the internal, driving quantum system interaction leads to permanent 'self-measurement' of the system appearing as unceasing events of its localisation towards the constituent 'eigen-states' (alternating with reverse delocalisation events). The 'quantum measurement' process is the next higher level of such complex-dynamical interaction, this time between the whole system and another, also quantum, system constituting eventually a part of a larger, macroscopic and classical device. The two levels of interaction are closely related, so that quantum measurement, being generally a weaker interaction process, is reduced to 'catching' the measured (and 'self-measuring') quantum system in one of its internal, permanently probabilistically changing eigen-states. The specific feature of quantum measurement [1,10], distinguishing it from a similar process of (genuine) quantum chaos that occurs around the same sublevel of complexity (Section 4.6.2, Chapter 6) [1,9], is a relatively small, but finite dissipativity of interaction between the measured system and the instrument elements, which serves both to open the way to further amplification of measurement result towards higher (macroscopic) complexity levels (item (C)) and to choose the relevant basis of actually measured eigen-states among many other possible ones (this preferred basis, or 'representation', of the measured system wavefunction is often reduced to possible values of coordinate/position of the system localisation centre, which forms thus the actually measured quantity). Transient dynamic localisation of the externally measured quantum system is similar by its mechanism to the lower-sublevel localisation of self-measured system in the process of its generalised quantum beat (which always continues 'within' any quantum measurement event), but is quantitatively larger in its spatial and temporal extension and resembles actually a transient formation of a classical, permanently localised kind of state (Section 4.7, item (C)). This causally specified, physically real 'wavefunction reduction' during 'quantum measurement' emerges as a temporary 'bound state' of the measured quantum system and the participating (quantum) element of the measuring instrument, confined to its characteristic size (which explains the involvement of postulated 'instrument classicality' in the conventional theory of quantum measurement), but then quickly decays into the normal, 'extended' process of internal quantum beat of the measured system (in general, with the new wavefunction parameters) and strongly, irreversibly changed state of the corresponding instrument (detector) element(s), with its classical 'indicator' degrees of freedom showing the measured system eigenvalue (if the instrument is properly tuned).

The obtained physically transparent picture of quantum nonlocality, its manifestations and limitation by localisation (measurement) processes permits one also to understand both the real basis and inconsistent speculations behind various related, often deliberately 'mystified' notions of conventional theory, which appear with growing intensity especially within the unitary fantasy of quantum information processing. One of the 'central' ideas of this kind is the notorious 'quantum teleportation' (of quantum states), which not only constitutes the key element in many proposed operation schemes of quantum circuitry, but embodies the core of the whole quantum computation idea, with its magic and cost-free power to transfer real information and then why not real system states. Many other 'quantum miracles' born in the unitary science delirium, such as momentary quantum 'action at a distance', 'nondestructive measurements', underground 'correlations' and easily added 'waves of consciousness', have certainly much to do with the sweet dream of quantum teleportation. In the causally complete, complex-dynamical picture of quantum interaction dynamics a sort of small-scale 'teleportation' (one could call it so) happens actually during each real 'quantum jump' of the system, i. e. its transition from one localised realisation to the next one, usually very close spatially to the previous realisation. In the special case of spatially extended, 'macroscopic' quantum state (see item (C) below) consisting from many entangled individual elements, a collective 'teleportation' of simple enough structures at higher distances can be 'automatically' realised by the same mechanism, including many small-scale, coherent 'quantum jumps'. Finally, similar to the nonlocal 'quantum correlations' transmitted, in principle, over arbitrary large distances (in 'clean' enough conditions), a simple quantum 'state' understood only as its main parameter(s) (e. g. a given spin direction of particle of a certain species), but *not* as the *individual* carrier of those quantities, can evidently be 'transmitted' *in the form* of 'correlations' themselves. Therefore 'quantum teleportation' is a mere play of words, another way to describe the same, well-known 'quantum mysteries' remaining unexplained within conventional theory and acquiring causally complete interpretation in the quantum field mechanics. It is important that the realistically interpreted quantum 'nonlocality' and 'teleportation', for any their particular cases and 'schemes', always involve at least simplest, but real interaction processes and are therefore subject to their dynamic randomness (Sections 3.3, 5.2.1), which means that there can be no 'pure', unitary, totally coherent 'teleportation', but only *probabilistically* determined, *partially* regular 'transmission', coherence, 'correlations' and quantum state reproduction at a distance, even in the absence of any 'noise' beyond the main, driving interaction processes (this conclusion agrees, contrary to unitary 'miracles', with the standard quantum postulates, cf. Chapter 2 and refs. [86,87]).

Note also the related inconsistent confusion, within the conventional theory, between essentially different cases of massless (photons) and massive (electrons) particle dynamics, or those of quasi-free propagation, elastic interaction and dissipative (de)excitation processes. This characteristic and tricky 'entanglement' within the scholar theory itself, based on its total and 'officially permitted' ignorance of real, physical origin of elementary particles and their detailed dynamics, is often used for ambiguous 'experimental demonstrations' of 'really performed' elements of unitary quantum computation, where the undesirable 'small deviations' and theoretical difficulties, conveniently left 'for future studies', describe in reality the irreducible obstacle to realisation of any full-scale version of unitary quantum computation.

Note finally that the inseparable and irreducible mixture of nonlocal (undular) and local (corpuscular) system properties in its complex (multivalued) internal dynamics provides the causally complete extension of the famous feature of wave-particle duality in the essentially quantum type of behaviour. We see now that quantum systems behave 'dualistically' not because they are inexplicably 'weird' in principle, but rather because they actually and permanently change their current state between (generalised) localisation around different regular realisation configurations and extended state of the common intermediate (or main, or transitional) realisation constituting the physically real version of wavefunction. This complex-dynamical, interaction-driven duality of a real quantum system has the intrinsically probabilistic, nonunitary (and essentially nonlinear) character, which makes impossible its use in practical realisation of conventional, unitary schemes of quantum devices. The universal character of the complex-dynamical mechanism of quantum duality provides its direct extension to any level of world dynamics [1-4,11-13], where 'corpuscular' states are generalised as respective object 'shape' and trajectory ('localised' system configurations) and the 'undular' state becomes (extended) 'distribution function' (formed by delocalised system transitions between the quasi-regular configurations/trajectories). This universal, physically real duality of complex dynamics of any interaction process, having many particular manifestations [1], causally extends and explains the origin of the idea of 'complementarity' of Niels Bohr, even though his intuitively directed attempts to endow complementarity with the status of a universal principle of nature could inevitably have only ill-defined, obscure basis within the framework of unitary (single-valued) science paradigm.

(C) <u>Classicality, macroscopic quantum states and complex solid state dynamics</u>. As shown before (Section 4.7) [1-4,12,13], the classical, permanently localised type of behaviour emerges from the dualistic, quantum behaviour as the next higher level of developing interaction complexity represented by the simplest bound states, such as the hydrogen atom, which means that classical state emergence can be described as a generalised, complex-dynamical, internally driven 'phase transition' in a few-component (quantum) system with interaction. It is important that quasi-permanent localisation of bound states of quantum particles originates not in any 'decoherence' processes initiated from the outside, but in the unreduced, complex dynamics of the bound state itself and specifically in the dynamically random (multivalued) character of individual quantum beat processes within each bound system component. Since each component tries to perform its unceasing quantum jumps in random directions, the resulting mutual 'hits' of the bound particles almost compensate each other and the probability of a large enough series of independent random jumps of the components in one direction, necessary for the 'quantum nonlocality' of the whole system, is extremely (exponentially) low. This probability can become much larger, however, if such bound system, classically localised in its isolated state, enters in a suitably chosen interaction process because in this case the quantum beat oscillations of system elements, giving rise to their quantum jumps, can form a spatially (loosely) ordered, or 'coherent', structure, so that the correlated jumps of system components in one direction are greatly facilitated by the global-motion, regular and often resonant correlations between the individual quantum beat processes. It is this kind of quantum beat dynamics that gives rise to the causally complete version of 'quantum entanglement/superposition' situation in many-body systems (see item (B)), where the neighbouring particles exchange their positions in resonance with their (spatially coherent) quantum beat pulsation. As a result, a spatially extended, macroscopic quantum state can form as a partially spatially ordered system of many-particle quantum beat, which can be described as an *essentially nonlinear*, many-particle 'standing wave' in the system of two interacting protofields, forming the 'quantum condensate' of individual particles that 'freely' and coherently jump and exchange their places in resonance with each other.

It is this kind of dynamically chaotic, essentially nonlinear, partially spatially ordered (coherent) and internally entangled system of quantum beat processes that constitutes the physically real structure of all 'macroscopic quantum states' involving Bose-Einstein condensation, such as superconducting, superfluid states and (e. g. atomic) Bose-Einstein condensates (cf. [211]). In the case of atomic Bose-Einstein condensates this essentially nonlinear standing wave can also be called 'gaseous solid' because it unifies certain spatial order with the properties of a (quantum) gas, where atoms quasi-freely wander on a rather loose 'lattice'. We obtain thus a more specific justification for classification of this popular, Nobel prize-winning system [212] as a 'novel phase of matter', where we can see now (contrary to conventional description) the exact, complex-dynamical origin of the particular new order (symmetry) that appears in this qualitatively new phase of interacting atoms. We can also clearly understand the physical, dynamic origin of that 'mysterious' quantum affinity between identical Bose particles which is simply postulated in conventional science and then described within a formal statistical theory: the latter is but an external, averaged expression of the underlying *spatial coherence* (SOC) of dynamically multivalued quantum beat processes for *individual* particles. Conventional, unitary theories of many-particle quantum states always provide such 'collectively averaged' (or 'statistical'), ultimately simplified and abstract, empirically 'guessed' kind of description that does not take into account the detailed complex dynamics of *interacting*, coherent quantum beat processes of *individual* particles and their partially, *probabilistically* ordered spatial arrangement (thus, interaction is basically excluded from the conventional, purely statistical/averaged and formal description of Bose-Einstein condensation or sometimes tentatively included in terms of 'mysterious' quantum effects and arbitrarily postulated equations for an 'averaged' system wavefunction).

That is the reason why conventional theories can approximately account only for the simplest cases and properties, but in cases of more involved many-body interactions, such as the high-temperature superconductivity, they 'suddenly' become inefficient and give contradictory, ambiguous results, despite really huge quantities of efforts applied (see e. g. [213]). The unreduced, dynamically multivalued interaction analysis provides the unique possibility of correct description of those more explicit manifestations of the underlying complex behaviour (so that the hierarchically structured and totally consistent 'theory of everything' [214] is still possible, but outside the limits of conventional dynamic single-valuedness, and actually takes the form of the universal science of complexity [1]).[22] To realise this possibility in full detail, one needs only to apply the universal many-body problem solution (Chapter 3) to each particular case. While the detailed analysis of each application of the universal complex-dynamical solution deserves a separate investigation, it becomes clear why no essential progress in the description of those more involved, 'nonseparable' problems in the modern many-body/solid-state theory can ever be achieved within the conventional, dynamically single-valued (unitary) approach, irrespective of its particular version details and the quantity of efforts applied. The necessity of the ensuing conceptually big, well-substantiated change in the whole solid-state physics development represents an important result in itself, naturally giving rise to a large diversity of results for particular systems and applications.

As noted above, the situation of full-scale quantum micro-machine (including quantum computer) actually represents a generalisation of quasi-macroscopic quantum system, where the distributed dynamic emergence of a classical kind of state (and thus definite interruption of quantum coherence) is necessary for the useful machine operation. This shows, in particular, that practical applications of atomic Bose-Einstein condensates, often 'generally' implied behind their peculiar properties, are fundamentally limited, besides their inevitable fragility, by extremely low dynamic complexity of any essentially quantum behaviour excluding any involved enough (localised) structure emergence necessary for any useful quantum machine operation, such as computation process (cf. [215]). In this sense, the externally 'impressive' quantum mysteriology of official science, in both its theoretical and experimental aspects, is definitely and inevitably condemned to failure, as far as sensible systems and useful practical applications are involved, while any qualitative future advance can only be performed, for the same reason, with the dynamically multivalued, nonunitary, hybrid micro-machines already extensively realised in nature (see also Chapter 8).

---

[22] The novelty of the unreduced, complex-dynamical (multivalued) solutions to solid-state problems with respect to their solutions in the framework of conventional theory can be illustrated by the extended description of 'quasi-particles' in the dynamically multivalued theory: they are represented by an internally chaotic regime of multivalued SOC (Section 4.5.1), which means that any real 'quasi-particle' has its 'private', probabilistically determined 'life' in the form of a dynamically fractal, self-developing hierarchy of unceasingly changed realisations and their common transient state, or (generalised) wavefunction. Such unreduced, complex-dynamical quasi-particles can only be obtained beyond the limits of perturbation theory (invariably used in the canonical theory), which leads to the consistent and reality-based description of their emergence and any 'strong' interaction effects giving rise to new phenomena and entities from a higher sublevel of complexity. All the main limitations of conventional unitarity, becoming now explicitly evident for those more involved applications of solid-state theory, are due to the fact that it always deals with the perturbatively simplified, effectively one-dimensional and often directly postulated 'envelopes' of real, dynamically multivalued structures, such as particles, quasi-particles and their arbitrary agglomerates (see also Chapter 8).

Note also that real macroscopic, many-body quantum states can exhibit many global realisations (even for each particular system), which differ among them by the degree of probabilistic spatial order that can contain various (in general discretely structured, quantised) proportions of local 'defects' (like voids of various size) in its roughly regular spatial structure. 'Quantum transitions' between such various degrees of order/randomness in the microscopic spatial structure of the corresponding 'lattice' of interacting quantum beat processes should have a peculiar, rather 'abrupt' (catastrophic) character (compared with order-disorder transitions in classical, incoherent many-body systems), which can be properly described only within the unreduced, dynamically multivalued interaction analysis (cf. [216]). The important underlying property of the unreduced structure of a large-scale quantum state is the nonvanishing, dynamically meaningful interaction between its components that just gives rise to its peculiar features, such as 'revival' of quantum properties for 'condensates' of relatively heavy, individually classical entities (atoms and molecules), whereas the conventional theory is based on inconsistently *postulated* quantum properties of such extended states (including 'Bose-Einstein condensation') consisting of allegedly non-interacting, quasi-free components, which inevitably leads to another series of 'unsolvable' problems at the attempt to take into account the internal condensate interactions [213,214].

Even without the detailed analysis, it is clear that the proposed exact, dynamically derived picture of macroscopic quantum states can explain many observed properties of atomic Bose condensates, complex superconductors and other similar systems, starting from the true, irreducibly complex-dynamical (multivalued) nature of such 'peculiar' states of many-body quantum system. The experimentally observed, very diverse manifestations of that 'peculiarity', remaining obscure and disrupted within the conventional theory, obtain now a *unified* and consistent explanation as *explicitly* appearing, but *standard* properties of the *unreduced dynamic complexity*, which are present, though maybe in a more hidden form, in any many-body system, despite their complete omission in the unitary approach, leading it to the obvious impasse. One example is provided by the observed 'quantum entanglement' between two macroscopically large, many-atom states [217], which can now be understood microscopically, in detail revealing, in particular, the origin of macroscopically ordered (coherent) entanglement of atom-scale quantum beat processes in the system of individually classical, 'heavy' atoms (while, being considered within the conventional, unitary theory, the same situation contains a whole hierarchy of 'mysteries'). Similar examples are provided by various macroscopic quantum states of the 'Schrödinger cat' (or 'quantum cat') type, meaning that they contain a linear, coherent superposition of several macroscopic 'eigen-states' with explicitly different, macroscopic measured characteristics (see e. g. [218,219]). Whereas the conventional theory 'mysteries' are only amplified in proportion to the quantum state size (in particular, all 'decoherence'-based approaches fail by their nature), the above picture of quantum properties revival for macroscopic ensembles of interacting particles clearly explains, similar to the case of microscopic quantum superpositions, the complex-*dynamical*, interactional origin of such macroscopic linear combination as unceasing system transitions between the constituent 'resonant' realisations, which can be traced down to the corresponding transitions in the local motions of neighbouring constituent elements (atoms, molecules, quasi-particles, etc.).

Note once more that the dynamically random, globally chaotic character of internal structure of any such essentially quantum macroscopic state can produce only very simple, quasi-uniform kind of spatial structure (like an imperfect lattice) and is not compatible with any unitary or even chaotic, but totally quantum, scheme of (practically useful) computation/creation process. This fundamentally substantiated conclusion of the causally complete theory of many-body quantum interaction shows that the vague and general promises of a vast variety of 'magic' future applications, accompanying recent experimental boom in the field, are largely and evidently exaggerated, in contrast to the elementary limits imposed already by the well-established principles of standard theory (see Chapter 2). What those 'fine' and 'quantum' experiments actually demonstrate is the high fragility and esoteric conditions of existence of the 'promising' states involved (especially when they are obtained from heavier particles like atoms), which only confirms their intrinsic dynamic instability predicted by the dynamically multivalued description of their real physical origin.

As concerns the original Schrödinger cat paradox itself [175] and the related quantum measurement process, it contains essentially the causal complex-dynamic mechanism of classical or 'reduced' state emergence (Section 4.7, items (B) and (C) above), but in its simple, irreversible version, i. e. without the following return of the measured/classical element to quantum type of behaviour that takes place in macroscopically large quantum systems (including 'Schrödinger-cat states') due to a specially arranged, low-noise interaction between elements, as described above. Any slightly dissipative interaction process between two (or more) *essentially quantum* systems, actually constituting the 'quantum measurement' situation, leads to formation of a transiently localised (dynamically squeezed, or 'reduced'), pseudo-classical configuration of 'bound' quantum beats of interacting systems. Therefore the delocalised and 'coherent', or 'essentially quantum', 'linear' mixture of elementary constituent state-realisations of the 'measured' system is destroyed already at this first stage of the real measurement process that does not need any *postulated* and ill-defined 'classicality' ($\approx$ macroscopic size) of the other, 'measuring' quantum system. The *transiently* localised, 'reduced' configuration of the interacting couple of measured system(s) and *quantum* instrument element(s) quickly decays and the measured quantum system takes its 'normal', delocalised configuration (though generally changed in details), but certain essential moments of extended wave interaction (leading to diffraction in a double-slit experiment, etc.) are lost during the 'reduction' stage, while the irreducibly changed, excited state of the measuring instrument element can be fur-ther amplified during its avalanche-like transfer to higher complexity levels, occurring due to the mentioned small dissipativity of the element (i. e. its slightly open, leaky dynamics appearing as ability to be actually excited and transmit this small initial excitation through a hierarchy of further interactions).[23] It is that well-defined, dynamical, small, but finite 'leak' of interaction development to higher (macroscopic) complexity levels that appears to be necessary for the actual, experimentally feasible measurement of quantum system characteristics by the *real* measuring instrument, which needs thus to be fully classical in those its *final*, output structures (such as macroscopic, truly classical 'indicator'/'pointer'), as opposed to its sensory, input elements that should necessarily participate in a *purely quantum* interaction process.

Note the essential difference of this dissipative, but well structured aspect of classicality emergence in the causally explained quantum measurement process from any ill-defined and abstract 'decoherence' of the (purely mathematical) wavefunction that still needs postulation of unexplained instrument 'classicality' as its 'big mass' introducing 'much noise' and thus somehow eventually destroying 'quantum coherence' (cf. [220]). In this latter interpretation the 'cat' from Schrödinger's paradox needs simply to be macroscopically big in order to reduce the probability of 'quantum superposition' of several states to negligibly small levels (e. g. [219]), but this explanation cannot consistently account for existence of various macroscopic quantum effects and states (including coherent superpositions of the Schrödinger cat type), on one hand, and readily observed emergence of classicality in microscopic, atom-scale interactions, on the other hand. In reality, as we have seen above, the quantum wave 'reduction' (physically real dynamic localisation) and related transient classicality emerge already at the first, 'essentially quantum' stage of quantum measurement, which momentarily 'kills' not the 'cat' itself, but the possibility for it (or even for any its small part) to be in quantum 'coexistence' of several states. After that the obtained, already essentially classical (quasi-single-valued) result of 'eigenvalue' measurement of the essentially quan-tum measured system is simply stabilised by its amplification to higher complexity levels permitting one to actually register its irreversibly determined value ('cat dead' or 'cat alive'), while the measured quantum system returns to its 'normal', delocalised state of quantum beat between its eigenvalues. In a similar way, the simplest classical, permanently localised states appear already at a very small scale, together with the elementary bound systems, e. g. atoms (which does not prevent them from possibility of quantum properties revival in a suitably chosen interaction process involving quantum beat resonances, see items (B), (C) above and Section 4.7).

---

[23] Note that avalanche-like amplification of the primal quantum measurement result is necessary only in the case of real registration of the measured system characteristics ('eigenvalues'), involving macroscopic final indicator, while 'quantum measurement process' in general refers to the first, elementary stage of wavefunction collapse that commonly occurs in real micro-systems and accounts, in particular, for interference pattern destruction in 'quantum diffraction' experiments [1,10].

It is also important to emphasize the essential difference of our dynamically derived, truly first-principles theory of quantum system collapse and measurement from any kind of *postulated*, artificially imposed stochasticity of quantum systems in the unitary theory, existing in a large variety of formulations and often also called 'dynamical' collapse, (continuous) measurement, etc., which gives an illusion of a dynamically derived effect (like e. g. various versions of 'intrinsic decoherence' [221-225]). The problem of any such description is in the inevitably ensuing modification of the standard, extensively confirmed quantum formalism even in its properly dynamical, a priori linear part, whereas in our approach we *explicitly derive* the Schrödinger equation, in its usual, *externally* linear form, starting from the *reality-based* description of dynamically multivalued (chaotic) quantum beat process [1,4,12,13] whose *essential, though hidden, nonlinearity* provides the *causally complete* explanation for the 'mysterious' postulated 'additions' to the linear standard scheme, and we continue in the same fashion at higher levels of quantum/classical complexity (Sections 4.6, 4. 7, 7.1), where the dynamic randomness (redundance) and the related generalised Schrodinger equation are *independently* derived at each level within basically the same, universal mechanism, instead of their postulated, artificial 'extension' from other levels in the canonical, stochastic unitarity.[24]

---

[24] The essential, interaction-driven nonlinearity emerges dynamically at the sublevels of complexity both below and above the formally linear Schrödinger dynamics. The lower-sublevel nonlinearity behind the usual Schrödinger equation is hidden in the phenomena like wave-particle duality (i. e. corpuscular behaviour of the same 'wave') and its irreducibly probabilistic manifestation ('Born's probability rule'), which remain 'mysterious' within the conventional scheme and are formally imposed by the standard 'quantum postulates'. The higher-sublevel nonlinearity appears in the form of true quantum chaos (Chapter 6) [1,9] and causal quantum measurement (Section 4.6.2) [1,10], which demonstrate the unreduced, complex-dynamical development of externally 'linear' interaction in a configurationally simple quantum system described by the standard Schrödinger equation (which is provided now with the unreduced, universally nonperturbative solution).

Being artificially imposed upon quantum system dynamics, the 'intrinsic' decoherence/stochasticity of the unitary approach will inevitably violate the standard, extensively confirmed (though unexplained) rules and will always need introduction of its (postulated) source, thus simply displacing the same, unsolved problem to hypothetical (and always ambiguous) 'deeper' levels of reality (this kind of trickery is a typical and inherent manifestation of unitary science deficiency [1]).

Note that the true quantum chaos (see Section 4.6.2, Chapter 6) is close by its level of emergence and dynamically multivalued mechanism to quantum measurement process, but happens in the actual absence of (suitable) dissipativity, so that the interacting quantum systems, instead of being transiently (and irreversibly) localised around one of possible centres of reduction (realisations), take successively and in a dynamically random order all possible (usually spatially delocalised) realisations. Combination of genuine quantum chaos with quantum measurement (dynamically 'embedded' in it) is possible and naturally occurs in real quantum machine dynamics, where each of the interaction processes and their combination are described by the universal EP formalism (Chapter 3, Section 5.2.1) [1,9,10]. Note also the fundamental difference between the unreduced dynamic complexity of real, dynamically multivalued interaction processes, used here for explanation of quantum system behaviour, and the appearing imitations of quantum (and classical) complexity (e. g. [225-229]) actually reduced, despite the intense terminology of 'novelty', to the conventional, dynamically single-valued 'science of complexity' that replaces dynamic randomness by mechanistically inserted 'stochasticity' etc. (see also [4]).

The basically incomplete (unitary) and totally abstract interpretation in scholar quantum mechanics of the above realistic, complex-dynamical picture of unreduced quantum system behaviour (items (A)-(C)) shows especially clearly its inevitable deficiency just in the quickly growing field of quantum theory applications to more complicated, man-made, modified and controlled micro-systems, where the fundamental limitations of the canonical single-valued imitation of reality are 'suddenly' transformed from the 'unreasonable efficiency' of the evidently contradictory 'postulates' to the absolutely senseless and openly fraudulent system of arbitrary guesses and purely mathematical fantasies. One characteristic sign of this 'ironic' [5] and show-business kind of science, directly related to the present dis-cussion of the essence of 'quantum' behaviour, is the infinitely 'inventive' play of 'puzzling', 'quantum' words without sense intermixed with equally vain speculations of 'post-modern' kind of 'philosophy' (see also Chapter 9). It is impossible to miss the avalanche-like flux of all those quantum 'games', 'entanglements', 'distillations', 'nonlocalities', 'teleportations', 'tomographies', 'internets', 'contexts', logics, 'impossible' states and 'magic' properties, arbitrary endowed with a desired fantastic meaning, penetrating into all the 'solid' printed sources of official science, its most prestigious departments and popular branches, but contradicting already the standard postulates (cf. Chapter 2) and the common sense itself (the latter fact is considered, apparently, as a decisive *advantage* of applied 'quantum mysteriology', permitting its swift-handed promoters and the 'friendly' scientific bureaucracy to scrounge more money from the unaware, hypnotised 'public'). This latest, 'information-oriented' generation of unitary quantum speculation is added to and combined with previous, equally unlimited lies of abstract quantum-mechanical 'interpretations', including various versions of quantum 'histories', 'many worlds' (or 'multiverse'), 'decoherence' ('superselection of pointer states' and 'predictability sieve'), etc. (see also Chapter 9), which leads to a completely intractable and misleading mixture of useless, abstract, esoteric 'narratives' only obscuring still more the naturally 'veiled' reality and suppressing, by their artificially amplified noise, any attempt of truly consistent and realistic problem solution. Being often overcharged with technically sophisticated, tricky mathematical symbolism, the pseudo-scientific post-modern abstraction rarely deals with the unreduced system dynamics, often replacing its most essential, 'nonintegrable' parts with formal 'arrows' pointing to a 'guessed'/postulated result that expresses usually the evidently incorrect extension of a perturbation theory approximation. We could disprove and mention here only the most frequent, obvious and dangerously noisy verbal exercises (deceitful 'narratives') of conventional unitarity, with the hope that the proposed causally complete picture of the unreduced, multivalued dynamics behind 'quantum mysteries' can clearly show the universal way out of the science-killing impasse of the blind math-physical cabbala. It is evident that the latter cannot actually go any further than the canonical 'interpretations' of the standard quantum formalism proposed already at the time of its creation, but tries simply to 'redescribe' it in a philosophically large variety of fic-tions, which can only repel potential and serious science participants and supporters (and actually does so), but never lead to real problem solution. The inflating mega-joke of unitary quantum information and other officially promised 'miracles' of unitary science becomes now too banal even for a banal 'post-modern' show.

In conclusion of this brief summary of our causal description of the main 'peculiarities' of quantum behaviour as manifestations of respective universal properties of the unreduced complex (multivalued) dynamics, items (A)-(C), note that one general feature underlying their proposed, intrinsically unified and totally realistic description is the *physically unified* initial world configuration in the form of two physically real ('material'), interacting protofields [1-4,11-13] that eventually gives rise, after consistent analysis of the interaction process development, to such features of quantum behaviour as quantum coherence, nonlocality, dynamic duality (complementarity), classicality emergence, etc., in addition to the causally complete explanation, within the same picture, of intrinsic, universal properties of quantum entities, such as mass (in the dynamic unity of its inertial, relativistic and gravitational manifestations), electric charge, spin, number and origin of fundamental interaction forces, etc. This totally realistic and naturally, dynamically unified picture of all the main observed properties of micro-world behaviour, additionally amplified by their consistent macroscopic extension [1], provides in itself a strong argument in favour of quantum field mechanics and universal science of complexity in general, which can be compared with the persisting 'quantum mysteries' of conventional science able, at best, to arrange for a limited number of separated and artificially adjusted 'experimental confirmations'. It becomes clear, therefore, that further practical work with micro-machine control and design can be successful only within the unreduced, dynamically multivalued description of reality.

# 6. Genuine quantum chaos and its consistent transition to the true dynamical randomness in classical mechanics

Since any quantum machine dynamics is essentially based on non-dissipative interaction of machine elements among them and with external systems, it is clear that one deals here with the same general situation as in the quantum chaos problem considering similar cases of arbitrary, nonintegrable quantum interactions, even though a real quantum machine can include eventually many integrated systems with simple enough configurations considered within usual quantum chaos studies. The results of our analysis of both general (Chapters 3, 4) and particular (Section 5.2.1) cases of interaction within a generic quantum machine confirm this similarity with the quantum chaos problem because they actually provide its general solution, eqs. (20)-(27), (47)-(52), that reveals the intrinsic, purely dynamic source of true randomness in any real quantum system in the form of dynamically redundant number of mutually incompatible elementary solutions, or system realisations, eqs. (24), (52). Whereas in terms of quantum machine theory the obtained solution demonstrates fundamental deficiency of usual, unitary description of a quantum system and reveals the principles of unreduced, dynamically multivalued operation of any real machine, in the case of quantum chaos problem we obtain the source, and the meaning itself, of genuine, purely dynamic randomness in a Hamiltonian quantum (and eventually classical) system, which is absent in the canonical quantum chaos theory, in contradiction with the correspondence principle and dynamical, logical necessity for the true randomness existence at the very basis of complex, self-developing world structure. The most transparent relation to quantum chaos is provided by our analysis of an elementary case of quantum interaction, represented by a time-periodic perturbation of a bound motion (Section 5.2.1), which is also one of the few standard formulations of the quantum chaos problem considered previously within the same approach [1,8,9]. The obtained quantum chaos criterion, eq. (53), has a universal meaning and ensures the conceptually important correspondence with the respective criterion of classical chaos (see below).

The extended work on quantum chaos description within the conventional quantum theory [127,144,155,156,161-164,230-294] results mainly in the fundamentally substantiated absence of any true, intrinsic randomness in a (closed) quantum system dynamics that would be definitely chaotic for its classical analogue [155,230-257], which constitutes a major contradiction, but is not really surprising in view of the basic unitarity (dynamic single-valuedness) of the underlying conventional quantum theory. Correspondingly, the study of specifically 'involved', but fundamentally regular, zero-complexity quantum dynamics of such 'apparently chaotic' quantum Hamiltonians (any real system belongs to this class) leads to the concept of 'quantum chaology' [235] or 'pseudo-chaos' [236,247-250] including actually *all* conventional quantum (and eventually classical) 'chaos' and represented by some 'sophisticated', but *regular*, quantum (and classical) dynamics showing certain, more or less universal, 'signatures' (or 'signs') of that 'involved regularity' or related peculiarities of tricky mathematical objects like zeta-functions [231,233,234,237-240,254-257] (these manifestations of regular 'chaoticity' actually make part of the general phenomenon of external 'signatures' of basically absent 'chaos', and 'complexity' in the whole, in the conventional, dynamically single-valued 'science of complexity', see [1]).

The absence of any true randomness in 'chaotic' quantum systems, confirmed by 'rigorous' unitary analysis and apparently by the fundamental quantum postulates themselves, leaves a clear impression of confusion, which can only amplify the existing doubts about consistency (or 'completeness') of standard quantum mechanics (see e. g. [230]). As we have seen above (Section 4.6), it is the latter that should indeed be extended to the dynamically multivalued (nonunitary) theory [1,9,10-13] which naturally includes the irreducible and universal source of true, purely dynamic randomness without introducing any change in the basic Schrödinger formalism or 'borrowing' randomness from the environment. The canonical science does not dare to change its mechanistic, zero-complexity 'foundation' and tries therefore to resolve the evident contradiction around the absent true chaoticity in a way usual for the 'mathematical' physics, i. e. by a number of mathematical tricks replacing the missing unreduced, dynamically multivalued solution of 'nonintegrable' equations and 'showing' either that chaos (and actually any irregularity) *means*, also in classical phys-ics, something else than straightforward randomness (e. g. a sort of *externally* sophisticated regularity) [230-261,281-287], or that real chaoticity still exists, also in quantum systems, if one looks at them 'at a different angle' (but always within the same, unitary paradigm) [156,262-280,288-294]. None of such imitative 'solutions' and their existing arbitrary 'mixtures' can, however, provide true consistency and finally the whole problem is transformed into infinite series of mathematical games, verbal definitions and pseudo-philosophical speculations, with the harmful practical consequences for such critical applications as quantum computers and nanotechnology (Chapters 7, 8). The artificial, useless intricacy of unitary imitations of complexity attains a particularly high degree in such field as quantum chaos becoming an 'unsolvable', obscure problem in itself (it is objectively difficult to simply get through the misleading and ever growing agglomeration of vain technical exercises), which demonstrates the direct harmful influence of the conceptually dead unitary paradigm upon the scientific knowledge development in the whole (Chapter 9).

The interpretation of quantum and classical chaos in terms of 'involved regularity' starts already in the main quantum chaos theory and the ensuing concepts of quantum chaology and pseudochaos [155,235,236,245-251]. The fundamental absence of quantum chaos is matched here with the presence of its classical analogue by playing with various time scales at which the behaviour of a Hamiltonian system 'appears' to be 'irregular' (whereas randomness as such remains without any unambiguous definition and source). It is stated [247-250] that at short times of quantum, quasi-classical system evolution, $t < t_\mathrm{r} \sim \lambda^{-1}\ln(I/\hbar)$ (where $\lambda$ is the Lyapunov exponent of chaotic dynamics of the corresponding classical system, $I$ is the characteristic action value, and $t_\mathrm{r}$ is called the random, or Ehrenfest, time scale), the quantum chaotic system will manifest the 'apparent' motion instability due to the pseudo-classical effect of 'exponentially diverging trajectories' (or wave packets, in the quantum system) that can be distinguished only until the critical spread of quantum wave packets makes them unresolvable at $t \sim t_\mathrm{r}$.[25] In the limit of classical system the ratio $I/\hbar$

---

[25] Note that already this formal use of results obtained for classical chaotic systems in the description of purely quantum, though quasi-classical, evolution does not seem rigorous. This 'method' of silent, formal extension of classical chaos results to quantum dynamics in order to

goes to infinity and with it $t_r$, which is interpreted as the correct quantum-classical correspondence, since we obtain in this limit the trajectories that 'diverge exponentially' at any time, giving apparently a chaotic classical system. It is easy to see, however, that one obtains in that way not a truly chaotic, but only 'pseudo-chaotic', basically regular classical system without any true randomness. Indeed, for any value of $I/\hbar$, one finds a regular, quasiperiodic quantum motion at $t \gg t_r$. In other words, one obtains an imitation of chaoticity by a long-period, but actually regular motion that 'practically' never repeats itself and still shows no true randomness.

It is important to note that the conventional concept of classical chaos itself, used here for quantum chaos interpretation and considered to be endowed with a source of true dynamical randomness (see e. g. [139-145]), is reduced in reality to just that kind of simulated chaoticity taking the form of 'intricate' regularity (as it is especially explicitly seen in the famous 'period doubling scenario' containing in its 'chaotic' region only the 'infinitely long period' of regular motion, but not any direct, local motion randomness). In this sense one obtains indeed a 'quantum-classical correspondence' in the conventional chaos theory, namely the correspondence between unitary, incorrect imitations of complexity/chaoticity by respective, dynamically single-valued reductions of quantum and classical dynamics. The falsification is inherent in the divergent-trajectories criterion itself: randomness as such is actually introduced through 'unknown' deviations in initial conditions (*without* indication of any its *dynamical* source and meaning), whereas the exponential law of neighbouring trajectories/states divergence represents a totally regular function, supposed to 'amplify' the externally introduced, ill-defined and *postulated* randomness.

Now, this 'exponential amplification' itself is but an obvious mathematical trickery, in both official quantum and classical chaos theories. The main description of local motion instability, giving the notorious 'Lyapunov exponents' that permeate the whole conventional 'science of complexity', is based on perturbative, local *linearisation* of the allegedly 'nonlinear'

---

'derive' then its 'chaoticity' and 'correspondence' to classical dynamics has become common in the conventional quantum chaos theory (see below), whereas the same, 'exact' science cannot even properly explain what the essentially quantum and classical types of behaviour actually are, how the latter emerges from the former and vice versa.

system dynamics, which is valid only within a small spatio-temporal domain of its standard application limited just by the value unity of the obtained exponential function argument (i. e. at $\lambda t \ll 1$ in the case of temporal evolution). However, the main 'result' of classical chaos theory is obtained, starting from this perturbative expansion, just far outside the region of its validity, where (e. g. at $\lambda t \gg 1$) the exponential function can only 'become itself' and show clear deviations from a power-law behaviour that would not give the necessary, sufficiently high 'randomness amplification'.

In the above 'quantum-classical correspondence' of conventional chaos theory a tricky 'logic' is used even more than once: first, by inconsistent introduction of classical, false exponential divergence and Lyapunov exponent into purely quantum evolution and, second, by extension of the false result, obtained in the limit $t_r \to \infty$, to 'any time' $t$ (whereas in reality at $t > t_r$ one obtains the unitary quantum regularity). The latter circumstance is especially evident and leads to the contradictory conclusion that "the phenomenon of the "true" (classical-like) dynamical chaos, strictly speaking, does not exist in nature" [248]. As we have seen above (Sections 3.3, 4.1), the true chaos does exist, both in nature and its unreduced, realistic description within the dynamically multivalued interaction analysis, whereas its pathological absence in the unitary imitation of conventional science is the inevitable consequence of its dynamically single-valued reduction of reality [1,8,9]. In general, it is evident, even in the framework of conventional science, that a unitary system evolution, ultimately smoothened and non-creative as it is, cannot contain any source of true, dynamical randomness in principle, and this is the single and ultimate origin of all those 'irresolvable difficulties', 'fundamental contradictions' and 'mysteries' of conventional quantum chaos, quantum mechanics and other fields and applications of unitary science where the omnipresent manifestations of unreduced dynamic complexity simply become more explicit.

Actually the same kind of improper extension of local perturbative expansion is used in many other 'basic' cases of conventional science for production of *false exponential dependence* [1] (e. g. in conventional cosmology or unitary evolution operator expression in quantum and classical mechanics), which appears to be very convenient, because of its very 'extendible' variation, for unitary imitation of absent change/emergence and randomness provided in reality by the 'abrupt', faster-than-exponential, dynamically probabilistic creation and destruction of incompatible system realisations (see Sections 3.3, 4.1). This universal 'method of mathematical physics' starts with a quite general equation of the form $df/dt = L(f,t,...)$ (where $L$ can be any function of its arguments), then produces formal perturbative expansion of the function on the right into powers of $f$ and, taking into account *only* the term with the first power of $f$, incorrectly extends the ensuing 'exponential' solution beyond the range of the starting expansion validity (where the assumed 'exponential' dependence remains in reality but a linear or other power-law function). It is no coincidence that one and the same 'method' of false exponential dependence thus defined is used in the unitary science first to derive its incorrect, exponential-function expression for the 'evolution operator' and then to justify the related 'magic' properties and 'strange' results, such as exponential efficiency of (unitary) quantum computers (see Section 5.1) and the conventional chaos concept based on 'exponentially high sensitivity to perturbations' (see also below). This means, in particular, that the highly publicised 'butterfly effect' in the conventional chaos paradigm and its 'rigorous' expression in terms of Lyapunov exponents [141-145] provide a basically wrong image of the origin and character of real motion instability. This is not really surprising, since we deal here with the situation where the results of *linear, local* instability analysis are boldly applied to characterisation of *global* behaviour of *nonlinear* systems, without any serious justification of that 'spontaneous' extension. As our universally nonperturbative analysis of the unreduced interaction process shows, any real system evolution is strongly, permanently and globally unstable and consists in very abrupt, 'faster-than-exponential' and frequent enough change of system realisations, where both realisation dynamics and realisation change process are driven by the same, 'main' interaction between the system components, so that the exponential dependence cannot develop either within individual realisations, or between them, or even in the average system change determined by the observed power-law dependences [1] (see also Section 5.1).

Returning to quantum pseudo-chaology and its play with temporal scales serving to imitate the correspondence between quantum and classical 'chaos', one should mention another time scale, $t_R \sim \omega^{-1}(I/\hbar)^\alpha$ (where $\omega$ is a characteristic motion frequency and $\alpha$ is a model-dependent exponent of the order unity), called relaxation, or diffusive, or Heisenberg time scale [247-250]. It is a period during which (i. e. at $t < t_R$) the quantum system evolution shows a diffusion-like behaviour coinciding, but only *in average*, i. e. by its statistical properties, with the corresponding classical chaos. Since usually $t_R \gg t_r$, this statistically averaged imitation of classical chaos persists much longer than the above detailed quantum replica of classical instability at $t < t_r$. However, such apparent quantum chaotic 'diffusion' ends up, or 'localises', after $t \sim t_R$ and the inevitable regularity of unitary quantum evolution appears explicitly in the 'chaotic' system dynamics. The ambiguity of such 'pseudo-chaotic', but in reality absolutely regular 'diffusion' demonstrates once more the imitative 'power' of the conventional theory of dynamical systems. Transition to the classical limit, $I/\hbar \to \infty$, reproduces the situation with the same transition for the random time scale $t_r$: though formally $t_R \to \infty$ at $\hbar \to 0$, for each finite $I/\hbar$ there are times, $t > t_R$ (though maybe very large), for which quantum and classical behaviour differ not only in detail (since $t > t_R \gg t_r$), but also statistically (in average), while even at $t < t_r$ the quasi-classical quantum dynamics tends rather to a regular imitation of 'diffusive' manifestations of dynamical chaos (which may also be the case for the purely classical, but always dynamically single-valued, theory of dynamical 'chaoticity').

As the above analysis of the unreduced system evolution shows, in reality the system will tend to perform dynamically random transitions between its incompatible, but equally feasible realisations (in the real, 'coordinate', or configuration, space) with the average time period, $\Delta t$, of the order of the characteristic period of motion within each realisation, $\Delta t \sim \omega^{-1}$, so that $\Delta t \ll t_r \ll t_R$ and the true quantum (and classical) chaos is established as quickly as the system dynamics itself, without waiting long enough for a sufficiently large 'divergence of trajectories', asymptotic regime establishment, etc., as it should always be the case for any fundamental system property, such as chaoticity,[26] and we obtain another clear demonstration of the qualitative difference between the real world dynam-ics, adequately represented by the dynamically multivalued EP formalism, and its unitary imitation (including the canonical theories of 'ergodicity', 'mixing', 'entropy', etc. in classical mechanics and statistical physics).

---

[26] In the semiclassical and classical systems the difference between realisations can be relatively small in the SOC regime of multivalued dynamics, but this does not eliminate its intrinsic chaoticity, even if the latter does not appear explicitly in observations (cf. Sections 4.5-4.7).

In summary, the most pertinent, purely dynamic, 'direct' approach to evolution of Hamiltonian quantum systems with few interacting entities that show (imitative) 'chaotic' behaviour in their classical versions reveals, within its unitary, perturbatively reduced analysis, no true, or even imitative, chaoticity (dynamical randomness) in quantum systems, but predicts instead persisting deviations of their behaviour from the corresponding classical system, even in the quasi-classical range of parameters [231-257], thus confirming most honest (and actually rather pessimistic) estimates by the conventional quantum theory of its own completeness [230] (see also [1]). This fundamental inconsistency of conventional quantum mechanics is accompanied by a large series of 'smaller', but equally irreducible defects of its application to 'chaotic' dynamics. Thus, it remains unclear why one should look for dynamic randomness in quantum behaviour, or even its pseudo-chaotic 'signatures', only in the limiting, semiclassical regime of quantum dynamics. Whereas the latter could, in principle, provide some necessary 'hints' on quantum chaoticity by analogy with its classical version, the *quantum* chaos phenomenon in the whole, for any its content, should refer to arbitrary, but *especially* fully quantum (undular) regimes of Hamiltonian chaos, which is not the case of conventional theory. In the attempt to reveal the non-existing unitary quantum chaoticity and its correspondence with classical chaos, the standard theory performs the more and more serious deviations from consistency and wishfully inserts results of the classical phase-space analysis (often doubtful in itself) into quantum chaos description (see e. g. [152-154,258-264]), where the statistical properties of dynamical system behaviour ('ergodicity' etc.) become tacitly and strangely intermixed with those of its spectral characteristics (in contradiction with the main results of the same quantum chaology). Another serious drawback of conventional theory appears as inevitable involvement of time, and in particular infinitely large time, in the definition of chaoticity (and related notions, such as 'ergodicity', 'mixing', etc.), both in quantum and classical dynamics (and the correspondence between the two, as we have seen above). Similar to the previous (and actually any other) limitation of conventional chaos description, this one is the inevitable result of the fundamental absence of any true randomness in the dynamically single-valued imitation of real interaction processes, which leads to invention of various substitutes with the help of mathematical trickery of unattainable or 'complicated' limits, etc.

In addition, practically all results in the conventional theory of quantum (as well as classical) chaos are obtained for extremely simplified, schematic 'models' of real dynamical systems, represented most often by so called 'maps', which are purely abstract recurrence relations only imitating dynamical system evolution by an over-simplified mathematical 'game' that does not correspond at all, even approximately, to any real system behaviour. In addition to their artificially reduced dimensionality (often it is just one 'spatial' dimension plus time) and extremely simplified interaction models used, chaotic maps show serious deviation from reality by their formal, non-dynamical discretisation, which is especially important in the case of quantum dynamics, where the natural, complex-dynamical quantisation has a relatively large and well-defined 'step' size determined by Planck's constant (Sections 4.3, 4.6, 7.1). The apparent 'chaoticity' in such kind of totally artificial, ultimately simplified and non-realistic 'dynamical systems' is often estimated or 'confirmed' with the (essential) help of computer simulations (i.e. 'numerical experiments'), being finally reduced to an involved regularity of purely mathematical objects, like canonical fractals or particular number distributions (see also below), which has nothing to do with real, interaction-driven chaoticity and fractality of a generic dynamical system (Sections 4.1, 4.4). When such, already ultimately deformed 'model' of reality is analysed also in the framework of strongly reduced unitary scheme of quantum evolution (representing but another example of the above 'method' of false exponential functions, see also Section 5.1), then one can only ask what is the arbitrary deviation from reality and consistency that is not used in such imitation readily subjected, nevertheless, to 'successful' experimental verification (this is another demonstration of the 'unlimited' general possibilities of this 'decisive' argument of official science). By contrast, our results for quantum chaos are obtained as the unreduced (dynamically multivalued) solution of the exact (Schrödinger) dynamic equation describing a real, non-simplified interaction process (Chapter 3, Section 5.2.1), which leads to revelation of the explicit, omnipresent source of true, purely dynamic chaoticity in essentially quantum systems, naturally passing to its unreduced classical counterpart in the ordinary semiclassical limit ($\hbar \to 0$) [1,8,9].

Setting aside the helpless task of finding any true chaoticity in the unitary wavefunction dynamics, some approaches try to evade the unpleasant situation by using other, 'secondary' versions of quantum formalism that deal with directly observable quantities, such as various 'distribution functions' [265-269] (which should give one the impression of being 'closer' to classical mechanics with its apparently 'well established' chaoticity). The basically incomplete character of any such formulation of quantum mechanics with respect to not only realistic approach, but even the conventional Schrödinger scheme, makes it even easier to mechanically insert the missing randomness, thus replacing its dynamic origin with an artificial stochasticity (as a result of 'coarse-grained' distribution, 'irreducible density matrix representation' and other mathematically decorated tricks). In any case, the obtained quantum 'chaos' and its classical 'correspondence' enter in the evident contradiction with the opposite result ('pseudo-chaoticity') of Schrödinger evolution analysis in quantum chaology [230-264], which does not lead, however, to any noticeable confusion and thus falls within a permitted degree of 'consistency' of official theory.

The same actually refers to another, related way to find a 'true' quantum chaoticity by analysing sensitivity of quantum evolution to small perturbations of the Hamiltonian or similar effects of 'quantum decoherence' [164,270-280]. It is argued, within this approach, that if the absence of trajectories in quantum mechanics makes impossible the analysis of 'sensitivity to initial conditions' of classical chaos theory, it is still possible to consider quantum dynamics sensitivity to Hamiltonian (interaction) variation, after which the 'exponentially high' sensitivity of quantum chaotic systems to Hamiltonian perturbations is readily obtained as a hint on the 'true' quantum chaos and its (previously absent) correspondence with the classical chaos. However, the formerly established and actually undeniable absence of chaoticity in the unitary quantum evolution, including any its 'perturbed' version, should not depend on the way of system analysis. Indeed, it is not difficult to see that *any* attempt of motion instability introduction through 'sensitivity to perturbations', in both quantum and classical chaos theory, is reduced eventually to the evident fraud of 'false exponential dependence' described above (i. e. incorrect extension of perturbative expan-sion validity domain), while in reality the unitary (dynamically single-valued) system evolution always remains stable in the whole (regular), as should be expected. The omnipresent dynamic instability and true randomness of unreduced, dynamically multivalued interaction processes are incompatible with exponential or any other smooth 'sensitivity to perturbations', since they originate in the fundamentally nonunitary (qualitatively nonuniform) character of the main, internal system dynamics itself (Chapters 3-5), which cannot be even approximately imitated by artificial introduction of non-existing 'exponential divergence' and other similar 'tools' of unitary theory.

The inevitable regularity reappearance in conventional quantum chaology naturally degrades towards ultimately mechanistic, non-dynamical reduction of 'chaotic' quantum (and classical) behaviour to a 'visibly', externally 'intricate', but internally totally regular behaviour of purely mathematical entities [170,281-287], like various number distributions in their natural sets (irrational numbers, prime numbers, etc.) or very 'uneven' (but actually regular) sequence of steps in various simple recurrence relations (including the widely used 'random number generators'). This tendency towards 'chaos without chaos', 'predictable randomness' and 'regular complexity', obtained as an 'exact solution', can be clearly traced in the very basis of conventional, dynamically single-valued science and in particular its approach to quantum and classical chaoticity. It shows also a close correlation with a much wider tradition of official science mechanicism, taking the form of 'mathematical physics' (cf. [114]) and trying to reduce reality to purely mathematical, exact and fixed ('eternal') laws, remaining 'ideal' and simple, devoid of any imperfect 'flesh and blood', with their uncontrollable and living complexity. If the observed inimitability of natural structures is eventually determined by an 'arithmetical chaos', then there is no surprise that such 'arithmetical world' of 'exact solutions' not only can be universally simulated on a unitary quantum computer [21,24,25,46], but represents itself such a computer and results of its operation [62,63,150]. Everything in such mechanistic world acquires a special, cabbalistic meaning determined by a regular sequence of digits (or other abstract symbols), including space and time, 'chaos', 'complexity' and thus eventually consciousness and intelligence itself (cf. Chapter 9).

A tacit mixture of arithmetic 'chaoticity' and the above 'sensitivity-to-perturbation' kind of imitation is applied to generate a 'new' notion of quantum chaoticity in larger, many-body quantum systems [156,288-290] and it is this kind of approach which is extensively used in the analysis of unitary quantum computer dynamics and assumes, of course, the existence of special, 'decohering' interactions as the necessary source of (ill-defined) randomness [14,157-160]. In that way, the most complicated case of dynamical chaos in the quantum many-body system, directly related to the expected practical application, is considered in the mainstream research by combining the maximum number of erroneous concepts and ambiguous wordplays. The real many-body system is replaced by an ultimately simplified 'model' that appears to be totally misleading even for a small number of interacting entities in a non-dissipative system. Since even such over-simplified problem cannot be solved within the dynamically single-valued approach, computer simulations inevitably involving uncontrollable additional simplifications are heavily applied, which makes the situation yet more obscure and actually helps to hide the evident deficiency of the underlying imitation (cf. [158]). The false 'chaoticity' is actually introduced from the outside, through ambiguous 'random' influences, with the 'original' conclusion that a strong enough external noise can destroy the system dynamics when the noise magnitude is comparable to that of characteristic parameters of intrinsic system dynamics. This kind of 'dynamical chaos' is actually indistinguishable from the postulated randomness of the artificially imposed noise, while the statistical properties of spectral characteristics are arbitrary confused with the signs of randomness in the dynamical system behaviour. Moreover, the underlying imitative concept of 'hypersensitivity to perturbations', actually represented by 'residual' interactions in a quantum computing system, is evidently inconsistent from the beginning: one can thus consider *any* part of a useful, 'computing' interaction within the machine as perturbation and arrive at the conclusion that it will generically show 'exponential (strong enough) growth' and destroy the very computation process it is supposed to create (provided it is 'rigorously shown' that a generally comparable, or smaller, 'residual' interaction can do it).

As a result, the conclusions of this 'quantum chaos' analysis concerning the feasibility of unitary quantum computation vary chaotically themselves [14,157-160,164] from reaffirmation of 'exponentially high' efficiency of unitary quantum computers to prediction of very serious practical difficulties of their realisation.[27] We deal here with multiple, superimposed imitations of unitary theory: the non-existing 'chaos' of a quantum computing system is 'adjusted' so as to demonstrate its ability to simulate the equally false quantum and classical conventional 'chaoticity' (where the supposed 'universal' simulation of classical chaos contradicts also the fundamental 'complexity correspondence principle', see Sections 5.2.2 and 7.1-2). The possibility of 'inverting time arrow in macroscopic systems' with the help of unitary quantum computer [158], revealed in that way with the help of simplistic model simulation on an ordinary computer, provides a concentrated demonstration of genuine 'power' of officially dominating unitary imitations of chaoticity in any kind of system (as well as the corresponding quality of the 'well-established' and 'peer-reviewed' truth of conventional science in general).

The ultimate inconsistency of fundamental and applied theory of many-body quantum systems within the dynamically single-valued approach is the unavoidable consequence of its basic limitations. As the unreduced interaction analysis shows (Chapter 3), the intrinsic, dynamical randomness emerges in any real, even totally noiseless (non-dissipative) system of interacting elements (Section 3.3) and even within a single elementary interaction act (Section 5.2.1), which not only makes impossible practical realisation of a useful (large-scale) unitary quantum computation, but also shows that *any* real computation/production process cannot have, and be described by, a unitary, single-valued dynamics/evolution in principle (see also Chapter 7). It appears also that a positive issue from this impasse exists (contrary to the difficulties predicted by conventional theories of quantum chaos/decoherence) and consists in development of explicitly dynamically multivalued (complex-dynamical, truly chaotic) quantum machines (Sections 5.2.2, 7.3, Chapter 8) that will necessarily involve, however, a combination of dynamically emerging quantum and classical (localised) states/elements and can be described, even approximately, only within the universally nonperturbative, dynamically multivalued analysis of the universal science of complexity [1-4,9,10], which makes any conventional, unitary theory of quantum computation/production absolutely useless and practically harmful (as it is clearly demonstrated by the conventional theory of quantum chaos and its application to quantum computation analysis).

---

[27] That specific kind of 'rigour' and 'consistency' of results is actually characteristic of the entire quantum information theory and modern fundamental science in general, which is directly related to the fatal limitation of the underlying, artificially imposed unitary approach and the corresponding perverted organisation of official science (see Chapter 9).

Note finally that some other versions of the dynamically single-valued theory of quantum (and classical) chaoticity exist and continue to appear (see e. g. [291-294]), but as the above analyses and examples clearly show, they cannot step out of the basic limits of unitary theory and reveal any true, dynamical randomness, which is just situated definitely beyond those limits and takes the universal form of unreduced, dynamically multivalued evolution of any real, noiseless interaction process. This conclusion involves also various computer estimates and simulations in the field of quantum chaos, since they use themselves simplified, unitary models of real interaction processes, while in order to reproduce adequately the dynamically multivalued character of real system dynamics, one should explicitly take it into account in the detailed scheme of computer simulation (which can also increase dramatically its efficiency [1], even without illusive 'miracles' of unitary quantum computation).

Returning now to the unreduced, dynamically multivalued theory of quantum chaos (Section 5.2.1) and its universal criterion, eq. (53), we note first the different, *frequency resonance* character of this criterion, as compared to conventional quantum (and classical) chaos criteria, always essentially involving the direct *magnitude* of the driving interaction [139-145, 231,245-250]. In our approach the omnipresent chaoticity attains its maximum simply at resonance between essential components of system dynamics, where chaos becomes 'global', or uniform (see eq. (35) for the general case), irrespective of interaction amplitude, while the conventional theory emphasises the role of the latter, i. e. in the unitary approach chaoticity 'naturally' appears to be stronger for stronger 'perturbation' of a regular background motion. It is this qualitative feature of our theory that simplifies correspondence between quantum and classical chaoticity: the quantum ratio of quantised energy increments, eqs. (35), (53), naturally passes to classical frequency ratio as $\hbar \to 0$, and also physically quantum resonance condition of global chaos has the same meaning as the respective classical condition (at the corresponding, higher complexity level), which again demonstrates *universality* of our description of chaoticity/complexity.

The resonance criterion of maximal chaoticity has a transparent meaning in terms of our qualitative interpretation of dynamic multivaluedness (Section 3.3): the large number of incompatible entangled combinations of 'everything with everything else' (system realisations) within the unreduced interaction dynamics giving rise to its chaoticity is present always, but the dynamically random realisation change process has the most pronounced, explicit form just at resonance between the interacting eigenmodes, as it gives the maximum number of equally important and sufficiently (equally) different system configurations (realisations). By contrast, the association of classical and quantum parameters in the conventional chaoticity criteria cannot be put into such transparent 'proportionality' and universality just because of the improper occurrence of interaction magnitude, which is due, in its turn, to the basic deficiency of the unitary concept of chaoticity itself, where (false) randomness is introduced artificially and silently in a 'small' form of 'noise' and should therefore be inflated, or amplified, by the direct interaction force to the scale of global chaos.

Note also the general physical transparency and technical simplicity of our resonance criterion of global chaoticity, eqs. (35), (53), as compared to much more involved interpretations of the unitary chaos theory (including e. g. the ambiguous condition of 'overlapping resonances' in the classical chaos theory [139-142]; it is clear from the above that at the global chaos onset the 'resonance width' can only coincide with the 'separation between resonances'). We can clearly see now that the 'well-known' and 'elementary' phenomenon of resonance itself, with its canonical, 'catastrophically amplified' behaviour, forms the natural, universal basis for emergence of global chaos regime and the phenomenon of true, dynamical chaos in general (including the complementary regime of multivalued SOC structures, if the resonance is 'out of tune'), provided the unreduced, dynamically multivalued analysis of the driving interaction is performed (which is not the case for the conventional theory of resonance phenomena and various speculations around the 'role of resonance' in the official, unitary 'science of complexity').

The interaction magnitude still can indirectly appear in a reformulation of our resonance chaos criterion, if we notice that practically any real part (element) of a system with interaction has a range of many 'eigenfrequencies' and it is an 'average' or 'effective' frequencies (or level spac-ings) of interacting parts that enter the global chaos condition, eqs. (35), (53). The existence of more than one eigen-mode frequency for an element is a result of its nonlinearity (here understood in the ordinary, 'non-dynamical' sense), i. e. the difference of the internal element potential from the quadratic coordinate dependence of the harmonic oscillator potential (which gives the single oscillation frequency). The real element nonlinearity usually leads to a decrease of its eigen-frequency (energy level spacing) for higher (excited) energy levels with respect to the lowest (ground state) value. One can characterise this zero-order nonlinearity (of the system element) by replacing the single element frequency, $\omega_\xi$, or level spacing $\Delta E = \hbar \omega_\xi$, in the elementary (quantum) chaos criterion, eq. (53), by the effective frequency, $\omega_\xi^{\text{eff}} = \omega_\xi / l_0$ (or effective level spacing $\Delta E_{\text{eff}} = \Delta E / l_0$), where $l_0 > 1$ is a measure of the zero-order potential nonlinearity (the uneven 'steepness' of the potential well) and $\omega_\xi$ is the harmonical-approximation frequency, obtained for the lowest bound states. It is clear that $l_0$ can also be expressed through the effective value, $V_0^{\text{eff}}$, of the zero-order potential, $l_0 = V_0^{\text{eff}} / V_0^0$, where $V_0^0$ is close to the ground-state energy. In a similar fashion, the frequency, $\omega_\pi$, of the (nonharmonic) driving interaction should be replaced by the effective frequency $\omega_\pi^{\text{eff}} = \omega_\pi / l_1$, where $l_1$ can be expressed through the effective value, $V_1^{\text{eff}}$, of the driving interaction potential, characterising nonlinearity of its temporal dependence: $l_1 = V_1^{\text{eff}} / V_1^0$. The global chaos criterion of eq. (53) can now be rewritten as

$$\kappa = \frac{\Delta E_{\text{eff}}}{\hbar \omega_\pi^{\text{eff}}} = \frac{\Delta E l_1}{\hbar \omega_\pi l_0} = \frac{\omega_\xi l_1}{\omega_\pi l_0} = \frac{\omega_\xi}{\omega_\pi} \frac{V_1^{\text{eff}}}{V_0^{\text{eff}}} \frac{V_0^0}{V_1^0} \cong 1 \ . \tag{55}$$

Note that the appearance of potential values here does not change the basically resonance, force-independent character of our main chaos criterion, eqs. (35), (53), providing instead just another its mathematical expression. In this form, however, the rigorously derived criterion of global quantum chaos, eq. (55), practically coincides, up to notations, with the *classical* chaos criterion from the conventional theory [139-142] (our resonance parameter $\kappa$ is equivalent to the 'resonance overlap parameter' from the classical chaos theory, whereas a related parameter $K = \kappa^2$ is also used in the quantum and classical chaos description, see also [1,9]). This natural quantum-classical correspondence is the intrinsic property of our universal analysis and its general criterion of uniform chaoticity as a characteristic limiting regime of complex (multivalued) dynamics (Section 4.5.2).[28]

By contrast, the conventional criterion of classical chaos [139-142] is obtained for a very special, unrealistically simplified 'model' (the 'standard map'), under a number of additional approximations and unproved assumptions, which can be 'convincingly confirmed' only by (largely adjustable) computer simulations. Most important is the fact that a consistent criterion and the very notion of dynamic randomness cannot be unambiguously introduced in the conventional theory because of its single-valuedness and are finally replaced by a number of imitations of the 'arithmetical chaoticity' type described above. As for the quantum chaos phenomenon and criterion, they both are actually absent in the conventional theory, since classical imitations of chaoticity cannot be extended to quantum mechanics even as imitations because of a more explicit appearance of the unreduced, dynamically multivalued complexity at those lowest levels of world dynamics. At the same time, it is the *single* criterion of eqs. (35), (53), (55) that gives, within the unreduced interaction analysis, both physically transparent (resonance) criterion of global chaos onset in *any quantum* , including *essentially quantum*, system dynamics and its classical version (directly or in the straightforward semiclassical limit).

Moreover, a related criterion, $\kappa \ll 1$ (and $\kappa \gg 1$), determines, within the same theory, emergence of the complementary limiting regime of multivalued 'self-organisation', or SOC, in quantum, classical, or any arbitrary system (Sections 4.5.1, 4.6, 4.7, 5.2.1), thus unifying all the separated and often mutually opposed imitations of conventional theory of 'complexity' ('chaos', 'self-organisation', 'self-organised criticality', 'adaptability', 'phase transitions', etc.) within a single, extended and well-defined concept of universal dynamic complexity [1] (Chapters 3, 4), where every structure is explicitly obtained (emerges) as a 'confined chaoticity', while the omnipresent and true dynamical randomness is always 'entangled' with a degree of regularity (appearing through the structure of individual realisations and their probability distribution). The resulting crucial extension of conventional theory demonstrates convincingly the advantages of the unreduced, dynamically multivalued description of reality over its single-valued, perturbative projections of canonical science.

---

[28] If one actually considers the corresponding, periodically perturbed classical system, then the criterion of its global chaoticity can and should, of course, be obtained from the classical equations of motion. However, due to the unrestricted universality of our analysis within the unreduced EP method (Chapter 3), one can be sure that the result, given by eq. (35) for the general case, will be the same as the one obtained by the direct semiclassical transition in the purely quantum-mechanical solution, demonstrated by eqs. (53), (55). We see here two general aspects of universality, the 'absolute', direct universality of the main results, such as the dynamic multivaluedness/entanglement phenomenon and the related criteria of global chaos/SOC regimes (Section 4.5), and their detailed correspondence between the neighbouring levels of complexity (such as quantum and classical dynamics), explicitly realising the unbroken 'reductionist' programme (i. e. consistent derivation of a higher-level phenomenon/property from its lower-level components). The latter is possible in the universal science of complexity [1] due to its unreduced interaction analysis describing explicit emergence of (dynamically multivalued) new entities (Sections 3.3, 4.1-4.3, 4.7, 5.2.2) and impossible in conventional science because its dynamically single-valued imitation can never reproduce the creative interaction (emergence) process, which shows how the well-known, but unexplained failure of the 'principle of reduction' in conventional science is clarified and repaired in its complex-dynamical (multivalued) extension.

Other known features of chaotic behaviour, such as 'stochastic layer' in the regime of global regularity, fractal structure of system dynamics with its properties and 'signatures of (quantum) chaos', can be obtained [1,9] within the same method of unreduced EP (Chapters 3, 4, Section 5.2.1) and demonstrate the same kind of 'natural' quantum-classical correspondence (at $\hbar \to 0$) as the one demonstrated above for the global chaos criterion.

The existence of true chaoticity of the same, universal origin (dynamic multivaluedness) in any, even quasi-regular (SOC) regime of system dynamics follows directly and naturally from our theory, in the form of the above well-specified coexistence of randomness and regularity. It is the remnant true chaoticity in the regime of multivalued self-organisation, at $\kappa \ll 1$ or $\kappa \gg 1$ (Section 4.5), that appears in the form of 'stochastic layers' in the conventional theory of classical chaos [142]. However, 'stochasticity' in the latter is reduced inevitably to ambiguous, perturbative imitation of 'exponentially amplified' external randomness or 'involved regularity' (e. g. 'mixing'), completely separated from explicit emergence of structures (incompatible with the perturbative analysis used) or even from its imitation in usual 'science of self-organisation' (synergetics, self-organised criticality), which forms a separate group of approaches (see e. g. [129-131]) without any natural place for randomness [132-134]. The conventional theory of stochastic layers (or 'weak chaoticity') is necessarily limited also to classical systems, whereas our universal analysis reveals the same phenomenon in quantum (though usually semiclassical) systems, with its natural transition to classical stochastic layer at $\hbar \to 0$. Without going here into the detailed theory of chaotic SOC dynamics, note that the graphical analysis of the unreduced EP equations, eqs. (20)-(21), (47)-(48), shows [1] that in the limit of low-frequency perturbation ($\kappa \gg 1$) the magnitude of chaotic fluctuations of changing realisation properties, or 'stochastic layer width', is directly determined by the adiabatically induced shift of realisation parameters and therefore proportional to a small power of $1/\kappa$, while for the high-frequency perturbation ($\kappa \ll 1$) the differences between chaotically changing realisations decay exponentially as a function of $1/\kappa$, in agreement with the corresponding results of classical chaos theory [142] (but obtaining now a qualitatively extended meaning mentioned above).

The remnant stochasticity of multivalued SOC regimes can be considered as a particular manifestation of the unreduced, dynamically multivalued fractality (Section 4.4) that has also many other features (note again that the conventional, non-dynamic fractality [122-128], with its mechanistic 'scale invariance', is totally separated from both conventional self-organisation/SOC and classical 'stochastic layer' studies, despite its evident general relation to the latter). All the properties of quantum (and classical) systems with various degrees of dynamical randomness, including the observed eigenvalues and density distributions, should possess dynamically fractal structure, which implies, according to the unreduced EP analysis (Section 4.4), the unceasing, dynamically probabilistic system change at each level of fractal hierarchy, leading to the autonomous dynamic adaptability of real interaction processes [1]. Contrary to this naturally appearing, intrinsic fractality of unreduced quantum interaction, the conventional theory of quantum chaos tends to arbitrarily confuse fractal structure of a quantum system with that of its classical limit (see e. g. [154,260]), where fractality is also inserted artificially and in the reduced, dynamically single-valued version, excluding permanent probabilistic variation.

A major manifestation of the unreduced dynamic fractality, especially important just for quantum systems, is that it serves as a universal causal, complex-dynamical mechanism of the phenomenon of quantum tunneling [1] (that can now be generalised to any higher complexity level). Indeed, already the general structure of the universal EP solution to the many-body interaction problem shows that *any* real potential barrier has actually an 'ef-fective', dynamical origin, and its observed smooth, average shape hides within it permanent chaotic and fractally structured potential variations that lead to finite barrier 'transparency' with respect to a confined entity. Contrary to the 'mysterious' origin of tunneling in the standard quantum mechanics, this complex-dynamical penetration 'within' the barrier occurs by quite real motions of gradual chaotic 'percolation' realising a particular case of the universal dynamic adaptability mechanism mentioned above and introduced in Section 4.4 (it remains more hidden in the dynamically single-valued approach just at the microscopic, quantum levels of dynamics). One should also distinguish this dynamically multivalued mechanism of tunneling in real, configurational space (where possible system configurations/realisations are clearly determined themselves within the same theory) from the so-called 'chaotic tunneling' or 'chaos-assisted tunneling' of conventional, dynamically single-valued theory of chaos [152-154], which is analysed in abstract (phase) spaces and represents either a play of words or another unitary imitation of unreduced, dynamically probabilistic tunneling in natural processes.

Finally, the 'signatures of quantum chaos', such as specific statistical properties of chaotic system energy spectra or 'quantum scars' in the system density distribution, replacing the absent true chaoticity in conventional 'quantum chaology' [155,231-235], can also be provided with possible complex-dynamical interpretation [1]. Thus, the tendency to 'energy level repulsion' in the presence of chaos can be related, in terms of the unreduced EP approach, to the appearance of a finite, perturbation-induced energy distance in the system spectrum, described by terms like $\varepsilon_{n0}$ and $\hbar\omega_\pi n$ in the denominators of EP expressions, eqs. (21b) and (48b) respectively. Quantum 'scars' in the probability density distribution are likely to be due to close and 'entangled' separations between energy levels of different realisations, appearing especially around resonances of various orders, which are conveniently described by the geometry of dynamically multivalued 'tori' of the 'extended KAM picture' (Section 4.5.2) and the graphical version of EP equation solution [1].

Note that being a technically simple, basically analytical and physically transparent theory, our approach to quantum chaos analysis readily finds applications to real, unreduced quantum systems and processes repre-sented in the theory as they are, contrary to over-simplified models (like 'maps') from the unitary theory. In addition to the quantum computation dynamics considered in this work, we can mention a group of proposed applications to energetic particle scattering in crystals [8], where a number of interesting and easily observable effects is predicted, in compliance with the already existing experimental data, for various cases of swift particle scattering, including channeling and electron microscopy.

In this Chapter we have emphasised that the unreduced, dynamically multivalued solution to the problem of interaction in any real quantum system provides the universal source of genuine, purely dynamic randomness, or (extended) quantum chaos [1,8,9], naturally passing to the true classical chaos in the ordinary quasi-classical limit and absent in the conventional, dynamically single-valued (unitary) theory of quantum chaos. This fundamental difference is especially important in theory applications to description of more complicated cases of quantum interaction, such as quantum information processing (quantum computers) and quantum machines in general. It becomes clear, in particular, that because of irreducible quantum randomness within any real interaction process, the conventional, unitary scheme of quantum computation cannot be realised in principle (in a practically useful version), even in the total absence of 'decohering' interactions or noise and irrespective of 'error correcting' tools or other 'protective' mechanisms applied. The realistic, dynamically multivalued version of quantum machinery needs the unreduced description (provided by the universal EP formalism, Chapters 3, 4, Section 5.2.1) and the ensuing, qualitatively new purposes, criteria and strategy (Sections 5.2.2, 7.3, Chapter 8).

# 7. Computation as a complexity development process: The physical information theory

## 7.1. Dynamic information as a form of dynamic complexity, its unceasing transformation into dynamic entropy and the universal symmetry of complexity

The universally nonperturbative analysis of arbitrary, many-body system (interaction process) within the unreduced EP method of the universal science of complexity (Chapters 3, 4) [1] describes explicit emergence of qualitatively new entities made up by multiple, permanently and probabilistically changing, quantised versions of entanglement of the interacting system components, or system 'realisations'. One and the same system is described, on one hand, by the starting dynamic equations expressing 'problem conditions' at the beginning of interaction process and generalised as the 'system existence equation', eqs. (1)-(5) (Section 3.1), and, on the other hand, by the resulting dynamically probabilistic sum over system realisations, eqs. (24)-(25) (Section 3.3), provided with the dynamically, a priori determined values of realisation probabilities, eqs. (26), and expressing the unceasing process of realisation change in a dynamically random order, as a result of the same interaction process. The qualitative difference between the initial and final system configurations is evident and provides the explicit expression of creativity of any real interaction process leading to new entities emergence and absent in the dynamically single-valued (unitary) imitations of conventional science, for which there is no qualitative, explicitly obtained difference between the 'beginning' and the 'end' and thus no unreduced event, no real change (the latter can only be postulated and described by a mechanistically simplified imitation). Now having explicitly obtained the events of change which clearly demonstrate system development from the beginning to the result of the driving interaction process, how can we conveniently, universally describe the course and direction of that real, qualitative development?

The plurality of explicitly obtained, dynamically related and probabilistically changing system realisations expresses the universally defined dynamic complexity of the final system state (Section 4.1): a consistent measure of the unreduced system complexity, $C$, is provided by any growing, positively defined function of the total number, $N_\Re$, of its (observed) realisations, or their rate of change, equal to zero for the (unrealistic) case of only one realisation (for example, $C = C_0 \ln N_\Re$ or $C = C_0(N_\Re - 1)$). Since the number of system realisations (as well as the related distribution of their probabilities) is the dynamically, a priori determined quantity (the maximum realisation number is determined, in particular, by the number of interacting modes/degrees of freedom, see Section 3.3), it is clear that the total system complexity, *C*, cannot *quantitatively* change and *should remain constant* during the driving interaction development from the initial to the final system configuration. What evidently changes is not the quantity, but the *qualitative form* of dynamic complexity: whereas in the final system state we have explicitly emerging, fully created (and changing) system realisations, at the beginning the interacting system components possess only a *potential* for realisation creation, even though this potentiality takes an equally *real*, 'material' form of (generalised) 'potential energy'.

Due to the explicitly specified system change in the course of its driving interaction development, we have the right to introduce the corresponding two qualitatively different forms of complexity, one of them describing the form of the *quantitatively* permanent *total* system complexity, *C*, at the start of interaction process and called *dynamic information* (or simply information), *I*, and another one referring to the same system complexity at the end of the driving interaction development, describing its results and called *dynamic entropy* (or simply entropy), *S*. Both dynamic information and entropy (as well as the total dynamic complexity) thus introduced have absolutely universal meaning and application, determined by the unreduced, multivalued dynamics of a real interaction process or system and therefore qualitatively different from the corresponding imitations of conventional (dynamically single-valued) science, including various branches of the scholar 'science of complexity'. In particular, all the three forms of unreduced complexity are basically positive, $C, I, S > 0$, and the conserved total complexity equals to the sum of information and entropy:

$$C = I + S = \text{const} > 0 \ . \tag{56a}$$

Since total complexity remains unchanged (for a closed system), we have the following expression of this *universal law of complexity conservation* (applicable to all its manifestations and measures in particular phenomena):

$$\Delta S = -\Delta I > 0 \ , \tag{56b}$$

where the increments of information and entropy refer to any given stage (period) of system evolution. The last condition in eq. (56b) means that the permanency of the sum in eq. (56a) is actually maintained by the *unceasing* and unavoidable *transformation of complexity* from the permanently diminishing information $I$ to the increasing entropy $S$, driven by the main interaction itself.

It is extremely important that one obtains, in that way, the universal expression of the direction and meaning/content of any system evolution (interaction development). At the beginning of interaction development one has $S = 0$ and $C = I$, so that the total (positive) system complexity is present in the 'folded', or 'latent' ('hidden') form of dynamic information (or generalised potential energy). The system (interaction process) development consists in the unceasing, interaction-driven decrease of $I$ that should be compensated by the equal increase of $S$, in accord with the conservation, or *symmetry*, of complexity. This interaction development process, or generalised system 'life' [1], is finite and continues until (effective) information stock drops practically to zero, while entropy attains its (local) maximum, so that $C = S$. This means that the system complexity $C$, remaining unchanged in quantity, is now totally transformed into its explicit, or unfolded, form of dynamic entropy. One can also describe this transformation as the development (or unfolding, or extension) of the system (interaction) potentiality, represented e. g. by its potential energy and providing the causally specified version of Bergsonian 'élan vital', into its final *(spatial) structure*. It is this final stage of system development that is described by the dynamically probabilistic change of system realisations, eqs. (24)-(25), together with their dynamically random fractal structure (Section 4.4). It is also the *generalised state of equilibrium*, or system 'death': the system still possesses the full hierarchy of realisation change processes, but this hierarchy cannot develop its new levels any more and will therefore (usually) degrade into processes with more chaotic and less distinct structure.

One obtains thus an essential extension of the conventional 'entropy growth law' (known also, in a reduced formulation, as the 'second law of thermodynamics'), eliminating all its difficulties and naturally unifying it with the 'first law of thermodynamics', or 'energy conservation law' (where energy is now extended to any measure of dynamic complexity): it is the universal *conservation, or symmetry*, of (unreduced) dynamic com-plexity that determines any system (and world) evolution/development, while this symmetry can be realised only through the permanent, omnipresent, interaction-driven and therefore unavoidable transformation of the qualitative form of conserved complexity from (diminishing) information to (growing) entropy. It is very important that the omnipresent growth of the extended, dynamic entropy accompanies not only processes of evident decay/disordering, but also *creation* of any *inhomogeneous*, externally *ordered* structure, since every such structure, according to our results for the multivalued SOC regime of complex dynamics (Section 4.5.1), is actually represented by the *dynamically random* transitions between similar enough, but different realisations, densely packed within the observed regular, or even static, external shape (representing the average system realisation). Hence, in the unreduced science of complexity the (extended) 'law of entropy growth', 'law of information decrease' and 'law of complexity conservation' suppose and involve each other, since they describe one and the same development and the *unique way of existence* of any real system, process, or phenomenon, which can be described as the *universal (dynamical) symmetry of complexity*.

It can also be shown [1] that this universal symmetry of complexity is the *unified* and 'ultimate', totally realistic extension of *all* the postulated 'conservation laws' (such as conservation of energy, momentum, or electrical charge) and 'principles' (such as all 'variational' principles or 'principle of relativity') of conventional science and is therefore uniquely confirmed by the *whole* body of observations supporting *all* the (correct) fundamental laws known in conventional science, but remaining irreducibly separated within its invariable unitary approach. Moreover, the universal symmetry of complexity, in its unreduced form, can be provided, contrary to the *postulated* conservation laws of unitary science, with a solid theoretical foundation using the fact that violation of complexity conservation could only be compatible with existence of highly irregular, *non-dynamically* chaotic structures, where unpredictable, 'virtual' emergence of an ephemeral inhomogeneity does not originate in any underlying interaction, but appears without any meaningful, causal reason, just 'by itself' (the conventional science idea of 'vacuum fluctuations' in the form of 'virtual (massive) particles' is close to that kind of randomness and therefore clearly contradicts the complexity conservation law [1]). It is evident, therefore, that *any kind* *of existence* of anything 'real' in a widest possible sense, i. e. perceptible and interacting (producing sensible consequences) at least 'in principle', cannot be realised without strict validity of the unreduced complexity conservation law (its 'small' violation will inevitably propagate to the whole system dynamics and lead to the same result as a 'big' one, similar to the 'sensitivity' of chaotic dynamics or gas escape through a small leak in a large container).

It is not surprising that the universal symmetry of complexity constitutes simultaneously the *unified evolution law* obtained in the form of *universal Hamilton-Lagrange-Schrödinger formalism* (two related equations reproduced below in this Section) [1] that not only generalises all possible, already known (correct) and yet unknown dynamic equations for particular systems from any complexity level, but also considerably extends each of them by providing its causally complete, dynamically multivalued (and eventually fractal) solution, eqs. (24)-(31), instead of usual 'unique', dynamically single-valued (perturbative and effectively zero-dimensional) *imitation* of solution basically different and separated from the unreduced reality it is supposed to describe 'rigorously' and 'exactly' (let alone the absence in the unitary science of objective, repeatable and clearly expressed laws describing *higher-complexity*, 'non-physical' system behaviour).

A particular corollary of complexity conservation law is especially important for the theory of complex-dynamical, including quantum and computing, machines. It is called *complexity correspondence law* (also rule, or principle) and states that a system with a given total complexity cannot correctly reproduce, 'compute' or 'simulate' in any sense, a dynamic system behaviour with greater unreduced complexity (cf. Section 5.2.2). Indeed, if the unreduced complexity is characterised by the total number of incompatible, different realisations inevitably taken by the system (or the equivalent rate of their change), then a greater system/behaviour complexity will necessarily contain some really new elements, which are absent in its lower-complexity imitation. Since complexity usually emerges in the form of holistic 'levels', each of them containing 'exponentially many' new realisations that provide a 'qualitatively new' behaviour, a lower-level system will typically miss something 'essential' in its imitation of higher-complexity behaviour. The limiting case of such deficiency is provided by nothing else than the whole body of dynamically single-valued knowledge of conventional, unitary science (including its versions of 'complexity') that tries to imitate the extremely high multivaluedness of the real world dynamics (where one of the simplest basic blocks, the free electron, contains already around 100 permanently changing realisations [1]) with the help of a single, unchanged realisation equivalent to exactly zero value of unreduced dynamic complexity (Section 4.1), including all the imitations of conventional 'science of complexity' (see [1] for more details).

Since the zero-dimensional projection of conventional science does not see the true, irreducible character of the difference between systems from different complexity levels (they all are reduced to artificially, smoothly 'moving' and dimensionless 'points' from abstract, postulated 'spaces'), it is not surprising that the scholar theory tries to inconsistently imitate and arbitrarily mix dynamic behaviours from absolutely different complexity levels: quantum behaviour is supposed, for example, to reproduce classical dynamics [158,189,190], including that of the whole universe [62,150], or economical system behaviour [57,187], or (generalised) biological evolution [56,74], or even conscious brain operation [15,68-71]. It is interesting to note that all such qualitatively incorrect, eclectic mixtures of different complexity levels in conventional science constitute the recently appeared and very intensely promoted, 'advanced-study' concept of unitary 'interdisciplinarity' (or 'cross-disciplinarity', or 'trans-disciplinarity').

In the theory of (unitary) quantum computation this evident violation of the complexity correspondence principle (and the directly related symmetry of complexity) takes the form of the concept of 'universality' of quantum computation pretending to be able to reproduce any process/phenomenon, but especially those more complicated cases, where usual, classical computers fail [21,25,46,58,64,67,81,103,157,158,177-181]. As shown above (Chapter 5), conventional, *unitary* quantum computers cannot correctly reproduce any, even quantum system dynamics (quite similar to the underlying unitary paradigm of conventional quantum mechanics). But even real, dynamically multivalued (chaotic) quantum systems could at best correctly reproduce the behaviour of other quantum systems with the same or lower complexity, but never any higher-level, classical behaviour, starting already from the simplest bound systems, like atoms (Section 4.7). Moreover, it appears that such purely quantum simulation of purely quan-tum behaviour can hardly be realised in practice and in addition does not seem to have any sense with respect to more realistic, hybrid (and explicitly chaotic) systems with the dynamically emergent classical behaviour (Sections 5.2.2, 7.2-3, Chapter 8).[29]

Before outlining the exact dynamical expression of the universal evolution law, it would be not out of place to emphasize the essential differences of the universal concepts of dynamic information and entropy introduced above from their multiple imitations in conventional science especially abundant and confusing within its 'quantum information theory' and related mystified 'interpretations' of 'quantum weirdness'. Note, first of all, that what is usually meant by 'information' in conventional science (including the official 'science of complexity' and 'quantum information processing') corresponds rather to a simplified, unitary version of our dynamic *entropy*, whereas our dynamic information has no analogue in conventional science, including its 'theory of information', and corresponds rather to extended version of 'potential (interaction) energy', being actually represented by the (generalised) mechanical action and its derivatives (see below). Indeed, any *real* representation of information 'bits', either 'classical' or 'quantum', deals with the *fully developed* structure of material carriers of those abstract bits. Therefore if one thinks of the *physical* theory of information, one should not forget that any really registered/processed 'quantity of information' is actually represented by *entropy* (in its generalised version of dynamic complexity-entropy). Conventional theory cannot discern between information and entropy because of its invariably single-valued, one-state structure that does not accept any real change, transformation between qualitatively different states of the same system. Since the existence of such changes in many dynamical processes is empirically evident, con-ventional science starts an infinite series of imitations around entropy, complexity, information and randomness arbitrarily confusing them with one another (for example, by defining information or complexity as entropy with the negative sign, or 'negentropy'), inserting randomness, its arbitrary measures and related entropy/information/complexity in an artificial, external and formal way, etc. [53,59-62,72,73,81,110,115,129,130,150,169,170, 174,182,183,270,272,274,295-336].

---

[29] Note that the complexity correspondence principle does not exclude certain, limited resemblance between patterns of behaviour from different complexity levels. Moreover, such very general 'repetition' of dynamical patterns follows from the 'holographic' property of the fractal structure of the world complexity meaning that universality of the dynamic complexity mechanism (Chapters 3, 4) leads to a general (but not detailed!) 'resemblance' between emerging structures and their dynamic regimes, which irregularly and roughly 'repeat' themselves in the sequence of emerging complexity levels (Sections 4.5-7) [1]. In any case, lower-complexity behaviour can at best only roughly reproduce the features of higher-complexity dynamics, similar to a small portion of a hologram that reproduces the whole image, but with a poor quality. It is probably this, holographic property of the unreduced dynamic complexity that is actually implied behind the arbitrary 'interdisciplinary' generalisations within the zero-complexity world projection of unitary science that leads to incorrectly 'extended', presumably complete similarity between different complexity levels starting from a few 'generally similar' features.

Moreover, if we take the conventional version of each of those concepts, even apart from its ambiguous relations to (actually absent) other forms of complexity, we find fundamental differences with respect to the corresponding unreduced, dynamically derived versions of the universal science of complexity. In our description we first *explicitly obtain* the multiple, *incompatible*, but *dynamically related* system realisations, which are forced to permanently replace each other by the action of the *same*, driving interaction that *creates* the explicitly obtained structure of each individual realisation and then transiently disentangles it into the *common* state of the *wavefunction*, or intermediate realisation (Section 4.2), during system transition to the next realisation. Only after having explicitly derived this dynamically entangled, unified, but permanently internally changing, 'living' construction of system dynamics, can one 'count' the *explicitly emerging, individually* specified and *a priori* defined system realisations and be sure that *all* the relevant 'units', in their unreduced configurations, and *only* them, are taken into account. Instead of all this involved internal structure of unreduced complexity, constituting the main sense of system dynamics and its development, conventional science simply postulates various 'reasonable', purely formal (mathematical) expressions for probability, entropy, information, or complexity based on counting arbitrary, ill defined, *abstract* 'units', 'states', or 'events' expressed by *identical* 'bits' and supposed to be eventually accessible through experimental observations (which become unrealistically complicated, including their computer versions, starting already from the case of several interacting bodies). Such 'arithmetical' complexity, entropy, information and probability are quite close conceptually to the 'arithmetical' kind of conventional 'dynamical' chaos [281-286] mentioned above (Chapter 6) and are not really more sensible than the underlying simple enumeration of observation results with the help of natural number sequence (1, 2, 3, ...) or any other formal, al-ways effectively *one-dimensional* set of simplified, homogeneous symbols and artificial rules applied to them.

It is not surprising, therefore, that conventional definitions of information (arbitrarily confused with entropy and complexity) are completely devoid of the main quality one would expect from a reasonable notion of information, its irreducible *meaning*, or *information content*. They just compute and recompute different mathematical quantities defined in a formal way on fundamentally senseless, one-dimensional mathematical strings (see e. g. [169,170, 295,299,300,305,309-312,314,331,334-336]), but cannot consistently define, in terms of information, the elementary difference between any two simplest, but meaningful notions, or 'qualities', from the 'ordinary life', which constitutes another 'mystery' of the 'exact' conventional science, the mystery of (real, meaningful) information, or knowledge content (see below), that persists despite all the technically 'elaborated' (and highly speculative) studies within the conventional, dynamically single-valued paradigm around 'physical' theory of (classical and quantum) information (see e. g. [37,80,113,150,170,300-306,310,321, 330-336]).[30]

It is clear from the above that the irreducible, and actually causally complete, content, or meaning, of real, implemented information (represented in practice by complexity in the form of entropy) is described just by the unreduced dynamics of multivalued entanglement-disentanglement process, constituting the true essence of any real interaction (Section 4.2). When we count the explicitly obtained realisations we know, in our approach, the rigorously defined *qualitative characteristics* of the units we count, including the detailed internal structure of realisations accounting for their 'material' quality, or texture, in the form of dynamically entangled system components, and the dynamical, probabilistically fluctuating fractal structure, both within and outside the realisations. It is that tangible, individually specific (inimitable), permanently changing, but intrinsically unified, externally irregular/asymmetric, and at the same time highly dynamically ordered, fractally tender construction that determines the unreduced structure of the real, physical and human, 'information (knowledge) con-tent', including the finest details of its always somewhat uncertain and changing meaning, within both elementary notions or 'messages' and the entire body of knowledge. All those properties are rigorously derived within our mathematical description of the unreduced interaction results (Chapters 3, 4) constituting a particular level of complexity-entropy.

---

[30] Conventional science tries to look for a formal solution of the mystery of meaningful information by defining various forms of 'conditional', 'mutual', 'joint' entropy/information, etc. (see e. g. [300,301,309,310,335]), but it is evident that such kind of 'solution' cannot advance further than just a reformulation of the same problem, remaining always 'unsolvable' within the unitary paradigm.

In that way one obtains also the rigorous definition of knowledge itself, without any reference to its 'carrier', type, or particular way of production: *knowledge* is represented by the *unreduced dynamic complexity* of the cognising system, in the form of its tangible *dynamic entropy* (as opposed to one-dimensional and ambiguous conventional 'information' expressed in abstract, dimensionless 'bits'), including its detailed, dynamically multivalued (multi-dimensional), fractal structure and dynamic uncertainty, as they are explicitly derived in our approach (it becomes clear that a real cognising system should possess a high enough minimum complexity) [1]. As concerns the *process* of cognition (acquisition of knowledge), it is actually represented by the above universal transformation of dynamic information (hidden complexity) into the explicit (unfolded) form of entropy, or knowledge, of cognising (learning) systems, driven by their interaction with the 'objects' of cognition that involves also their internal interaction processes.[31] It follows that knowledge is represented by a tangible, 'material', spatial (and fractal) dynamical structure subjected to the unceasing realisation change of unreduced dynamic complexity, rather than a fixed set of abstract symbols and rules, as it is often imposed by conven-tional science, in agreement with its mechanistic, dynamically single-valued paradigm. The unreduced, complex-dynamical knowledge concept possesses the unlimited universality and exact definition of completeness, so that being applied at the level of human civilisation complexity, it naturally comprises and unifies all kind of knowledge, such as all 'exact', 'natural', 'humanitarian' sciences, 'arts', spirituality and various practical skills, whereas they remain irreducibly separated into small unrelated pieces within the conventional, unitary and ill-defined, kind of knowledge (with its unitary concept of information). On the other hand, the dynamic information, being a qualitatively different (but equally real) form of complexity, as compared to entropy/knowledge, plays rather the role of a necessary 'élan', or 'desire'/predisposition, for acquisition of knowledge, often imitated within various inexact, 'philosophical' or 'spiritual', ideas around 'teleology' and 'vitalism' [1]. The true, dynamic information is therefore something that one 'does not know yet', but it is a definite, real *potential* to know always involving, however, an irreducible dynamic uncertainty of the result, whereas the dynamic entropy-knowledge describes just a particular result of this intrinsic potential realisation (possessing its own internal uncertainties). This dualistic relation between the two fundamental forms of unreduced complexity, information and entropy, is the same at every level of complexity, but one can meaningfully describe it in terms of knowledge and its acquisition process for high enough levels of complexity containing autonomous systems with sufficiently stable memory functions etc. [1].

---

[31] Already this rather general, but fundamentally substantiated, concept of knowledge and cognition leads to practically important conclusions and implies, in particular, that the maximum attainable knowledge for an individual, a civilisation (group, society), or any other cognising system is predetermined by the complexity (in the form of information in this case) of the world being cognised, cognising subject and intensity/depth of interaction between the two (depending critically on their internal interactions, especially those between the elements of the cognising subject), even *before* the process of cognition has started. If, however, the complexity of the object of cognition (e. g. the world in the whole) and intensity of its interaction with the cognising subject remain approximately fixed (as it often happens in reality) then the key quantity determining the finally attainable quantity of knowledge and the related system ability to induce the world complexity-entropy growth, or system *intelligence* [1], is the total *cognising system complexity* initially present in the hidden form of dynamic information and not subjected to change during various system interactions. It is therefore not the mechanistic sum of various acquired informations, or 'erudition', and neither the abilities to increase it that determine the practically important intelligence of the cognising system, but rather its total, unreduced dynamic complexity mainly given at the system birth and then only changing its form from the hidden information to the explicit entropy in the process of cognition. These empirically confirmed conclusions are now rigorously derived within a fundamentally substantiated and mathematically exact theory, which is not limited in its further detailed applications to intelligent systems of any complexity and origin (see also Section 7.3).

It is clear from the above analysis that everything one observes, measures and 'processes', including 'information', is actually represented by a form of our universal dynamic entropy, as it should be expected taking into account the physically 'realistic', tangible and 'unfolded' character of this form of complexity equivalent to the generalised space structure (whereas its folded form, the dynamic information, is involved in the internal dynamics of emergence of those observable structures and relates more closely to the entity of time, see below). In this connection it is important to understand the advantages of the extended entropy concept within the dynamic redundance paradigm with respect to its imitation within conventional, dynamically single-valued (unitary) science. Thus, the extended entropy, together with the unified dynamic equation it satisfies [1] (see below), is explicitly *derived* from the unreduced *interaction dynamics* of any,

'nonequilibrium' state (including the case of causally understood quantum systems), upon which the physical meaning itself of 'nonequilibrium' and 'equilibrium' states, as well as the unceasing system evolution from the former to the latter, is causally specified in the form of the above universal law of conservation and transformation of complexity. In particular, the absolute (total) *equilibrium*, or generalised complex-dynamical *death*, of any system (even the one consisting from just a few interacting bodies) is universally *derived* as the state of totally developed complexity, where the dynamic information (generalised potential energy) is totally transformed into entropy (generalised kinetic/thermal energy) attaining thus its absolute maximum for this system. In a similar way, a *partial* equilibrium state is attained by the system at the end of a given complexity level development, after which a new highly nonequilibrium state can emerge in the form of the next complexity level (provided the system is not at the end of its complexity development, where the last, most complex partial equilibrium coincides with the absolute equilibrium).

This causally substantiated definition of equilibrium in terms of system dynamics itself, without evoking any ambiguous exchange with the environment ('thermal bath') and related abstract postulates, is important especially in cases where conventional 'interaction with the environment' becomes explicitly contradictory, for example in cosmology, where the 'system' is the whole universe with unknown 'boundaries' and ill-defined 'exterior world', or in quantum mechanics, where any system interactions lead to either no change at all or a relatively big, step-wise change. One can compare the totally consistent and universal concept of dynamic entropy from the unreduced science of complexity [1] with the situation in conventional science where even the simplest notion of equilibrium is simply postulated in purely abstract terms, so that the fundamental origin of the 'thermodynamical arrow of time' (the irreversible tendency towards equilibrium) remains unclear, whereas generic, nonequilibrium states, necessary for any process ever to occur, are inconsistently described in terms of (postulated) 'local' or 'pseudo-' equilibrium dynamics, again not revealing the fundamental, physical origin of nonequilibrium state and its tendency towards equilibrium (note that these issues of the 'old', classical physics are not clarified by the latest results of the scholar 'science of complexity', despite the evident involvement of dynamic complexity in these phenomena).

Another major advantage of the extended concept of (dynamical) entropy with respect to its unitary imitation is related to the intrinsic, probabilistic entanglement between order/regularity and disorder/randomness within a generic system state of the multivalued SOC type (Section 4.5.1), which means that the emergence/existence of *any*, even quite externally 'regular' structure or dynamic behaviour is expressed by a definite *growth* of entropy, rather than its decrease postulated in conventional science that inconsistently attributes emergence of structures to 'virtual', local violations of the entropy growth law or to 'openness' of the structure-forming system supposed, following Schrödinger [337], to 'consume the negative entropy' from the 'environment' (whereas an elementary estimate easily reveals the evident quantitative inconsistency of this assumption).[32] Not only the universality and *true* origin of the entropy growth law are thus established within the unreduced concept of complexity, but this 'asymmetrical' principle of growth/maximum appears now rather as intrinsic *mechanism* of maintenance of the universal dynamic *symmetry* of complexity, due to the existence of the dual, 'opposite' complexity form of dynamic information (absent in the unitary science) whose decrease compensates exactly the increase of entropy, thus leaving their sum, the total dynamic complexity, always constant (if all interactions and complexity sources are properly taken into account). Therefore the universal symmetry of complexity, obtained as the unique principle of existence/development of the world and any its part, can actually be maintained only by permanent, unavoidable transformation of the qualitative form of complexity, from the folded (latent) form of dynamic information (potential energy) to the unfolded (explicit) form of dynamic entropy (spatial structure).

---

[32] Thus, in the canonical case considered by Schrödinger a living organism acquires its dynamical order, or negentropy, by consuming the ordered environment (mostly as food) or, equivalently, renders its 'excessive' entropy to the environment (basically quite similar to a man-made machine like refrigerator). It is evident, however, that the amount of complexity, or 'dynamical order', of the organism is much greater than that of the food (or any other resources) it consumes. This fact constitutes even one of the main principles of living matter existence: fabricate complex structures from much simpler components. Most evident manifestations of this principle include production of very complex, living forms from non-organic substances by plants and generation of higher brain activities, such as conscious thought, by sufficiently developed animal species that physically consume only pieces of biologic, but much less complex, matter. On the other hand, the negentropy consumption concept refers only to 'wide open' processes, while many systems producing very complex order within them are practically *closed* to any noticeable influx of a suitable 'food' (e. g. a planetary system evolving up to life emergence on one of the planets), which clearly designates the limits of the unitary imitation of reality within conventional science. In fact, the main part, or even the totality, of the developing complex structure of both open and closed systems does not come from the outside, but is encoded from the beginning in the internal dynamic information, or generalised 'potential energy' of internal system's interactions (which just determine its specific nature). The dynamic information realisation in the case of living organism is evident: it is the *system of interactions* of the organism's *genome* that determines almost entirely the resulting organism complexity. Note the essential difference of this result from the corresponding postulates of conventional, unitary genetics considering genome only as a one-dimensional, weakly interacting, quasi-regular 'list of instructions' that closely resembles an ordinary computer programme, in contradiction with the evident and huge gap in complexity and autonomy between such computers and living systems.

One may also note the fact that in the case of our unified evolution law its formulations as conservation *or* symmetry of complexity mean the *same*, whereas every classical *conservation* law of a particular quantity (always being a measure of unreduced complexity) is *different*, and usually *derived*, from the corresponding postulated, formal *symmetry* of an underlying abstract entity (this transition from symmetry to conservation is known as Noether's theorem). Thus, the momentum and energy conservation laws do not result, in our description, from system invariance with respect to spatial and temporal shifts respectively (as it is the case in the canonical science), but rather both momentum (energy) conservation *and* space (time) uniformity are manifestations of the *symmetry/conservation of dynamic complexity* of the considered system, which reflects the specific, 'emergent' character of the universal science of complexity (absent in conventional science), i. e. the fact that every entity, including space, time, momentum and energy, is explicitly obtained, in accord with its actual emergence, as a result of unreduced interaction development expressed by that unified (evolution) law, the symmetry of complexity.

The obtained intrinsic, dynamic 'entanglement' between causally derived randomness of unceasing realisation change and order of their average envelope, expressed by the extended entropy growth law, confirms once more that 'complexity' and 'chaoticity' mean actually the same, within their unreduced, dynamically multivalued description (Section 4.1), and more complicated, involved, elaborated spatial structures, associated with growing complexity (emergence), contain in them also proportionally *more* of chaoticity, randomness, dynamic disorder (expressed by dynamical entropy growth), contrary to the opposite ideas of conventional, dynamically single-valued science, leading to so many irresolvable 'paradoxes' in thermodynamics, understanding of temporal change/evolution, quantum mechanics, life sciences, etc. (see also Section 7.3, Chapter 8). As concerns particular quantitative measures of complexity, entropy and information, expressing the universal symmetry/transformation of complexity in various its particular manifestations, they can take correspondingly diverse forms, but usually differing between them only by a constant factor. Thus, if integral measures of complexity characterise its development through the dynamics of the total (effective) number of realisations, or (generalised) action $\mathcal{A}$, its differential measures describe the same processes in terms of temporal rate of realisation change, or (relativistic) energy $E$, and spatial rate of realisation change, or (generalised) momentum $p$, equivalent (actually proportional) to the integral measures [1] (more details below).

The notion of dynamic information is characterised by still greater differences than entropy with respect to any conventional imitations already because, as explained above, the canonical 'information' is always reduced in reality to dynamic entropy, whereas our unreduced, dynamic information cannot exist in the unitary theory with its basic absence of qualitative change/transformation and can be considered as the complex-dynamical, *essentially nonlinear* extension of conventional 'potential energy' and 'action' (see below) which, however, are never recognised in conventional theory as forms of clearly defined information. It is not surprising that our *unreduced information* is directly related to the universal dynamic origin of *randomness* revealed by the unreduced concept of complexity: it is the causal, interaction-driven *development* of the a priori totally regular and relatively poorly structured initial system configuration, containing dynamic information usually in the form of 'potential energy of interaction' (see e. g. eqs. (1)-(5)), that gives the dynamically multivalued/redundant general solution necessarily involving causally random realisation change (eqs. (24)-(25)), which just expresses the result of information transformation into entropy. It is clear that any postulated, mechanistically inserted kind of randomness, or 'stochasticity', of unitary quantum information theory (e. g. [109,338]) has actually nothing to do with that fundamental connection between the unreduced interaction development and inevitable dynamic randomness of any real information implementation in the form of dynamic entropy. In a similar way, it becomes clear that various unitary-theory speculations about the '(thermodynamical) cost' of 'real'/'physical' information (e. g. [80,113,300-304]) cannot actually reflect real system operation already because the difference between unreduced information and entropy cannot be seen within the single-valued imitation of dynamics, whereas application of conventional 'equilibrium' thermodynamics and the related mechanistic uncertainty of 'noise' have little to do with the true, dynamical origin of randomness in the essentially nonequilibrium process of information transformation into entropy. Contrary to some very selective and limited 'principles' of the unitary computation theory around 'thermodynamical cost', we have shown in our unreduced interaction process description (Chapters 3-6) that any real, practically useful 'operation' within a 'computation process' has the strictly positive cost of increasing dynamic uncertainty that can be expressed, as we have seen above in this Section, as a growth of dynamic entropy at the expense of corresponding decrease of dynamic information, ensuring the exact conservation (symmetry) of the total dynamic complexity (= information + entropy).

The dynamically multivalued concept of information and its transformation into entropy in the course of any interaction, process, motion (all of them realising universal complexity development) is therefore much larger than the conventional idea about (abstract) information 'bits' starting from an ill-defined 'incomplete knowledge' whose mechanistic, meaningless 'completion' independent of the content is described by an acquired quantity of thus formally defined (postulated) 'information'. The origin of this conventional imitation of information becomes clear now: in the absence of the true, dynamically defined change in its dynamically single-valued (unitary) projection of reality, conventional science tries to imitate the actually 'felt' process of novelty acquisition in a system by a formal division of already existing, unchanged elements into 'already known' and 'yet unknown' ones, while the postulated, mechanistic growth of the number of the former at the expense of the latter is presented as 'information growth'. Since nothing really new (rather than existing, but 'previously ignored') can appear in system dynamics, according to the unitary science results, it tends, instead of acknowledgement of this evident deficiency (cf. [339]), to its presentation as fundamental and universal 'law of conservation of information', especially in relation to recent propagation of ideas around information and quantum computation (see e. g. [327,333,335]). In reality, as shown in our unreduced, dynamically multivalued interaction analysis, a qualitative change is omnipresent and described just by inevitable, unceasing and universal information *decrease* (rather than its acquisition/growth or 'conservation') compensated by the equal *growth* of dynam-ic *entropy* (e. g. in real computing systems), even for the processes of external *order (structure) emergence* in a *closed* system (such processes are, in fact, denied by conventional theory, but permanently happen in nature). As concerns the subjectively felt growth of intuitively defined 'knowledge' about previously existing, but subjectively 'unknown' entities (reduced in conventional theory to abstract 'bits' of one-dimensional information) it is also correctly described by the dynamic *entropy* growth of the cognising system, which is usually smaller, however, than the entropy growth associated with those entities emergence itself. Therefore we deal here not with a simple difference in verbal definitions (terminology) around information/entropy, but with a fundamental, qualitatively big difference between the real, dynamically multivalued and *creative* world dynamics (adequately described in the universal science of complexity) and its reduced, mechanistically fixed imitation within the dynamically single-valued paradigm of the whole conventional science (including usual 'science of complexity', 'theory of systems', 'cybernetics' and 'information theory') that does not contain any real, qualitative change in principle.

It is important to note also that finer *fractal parts* of the unreduced interaction results (Section 4.4) represent the *physically real*, causally specified, dynamically multivalued *potential energy* and thus dynamic information for the *next (higher) level(s)* of complexity development (provided there was enough of dynamic information in the initial complexity stock of the system in question), which shows how the two dual forms of dynamic complexity alternate their dominance of system behaviour at the consecutive levels of its developing complexity. In fact, it is this alternating, interaction-driven emergence of *fractally* structured dynamical, or 'physical', information (or causally specified 'potential energy') from 'nonintegrable' parts of the *current-level* spatial structure (dynamical entropy) *and* the 'reverse' transformation of this physical, tangible information into the *next-level* entropy (spatial structure) that provides the causally complete understanding of the fundamental property of *autonomous creation (self-development)* of system structure from the lower-level, previously existing dynamic information, which is directly related to the problem 'nonintegrability'/'nonseparability' (see Section 4.4) and absent in conventional theory in principle (being replaced with various 'proto-scientific' speculations and plays of words, such as 'becoming', 'self-organisation', 'emergence of meaning', 'autonomous agents', 'adjacent possible', 'autopoïesis', etc.). That is the unreduced essence of the notion of *physical* information, which indeed reveals itself as being a tangible, 'material' entity, *a part of the developing system*, its 'living body', represented by the probabilistic dynamical fractal and possessing intrinsic, causally defined (dynamical) 'fertility' with respect to 'unfolded' (spatial) structure production by its interaction-driven transformation into entropy. Needless to say, the underlying explicit, causally complete, dynamically multivalued, fractally structured problem solution (Sections 3.2-3, 4.4) [1] is very far from the abstract guesses, 'arithmetical' imitations and related vain (but extremely intense) speculations of conventional science about information, its 'physical' content ("information is physical") and role in both classical and quantum applications (e. g. [61,62,72,73-75,80,81,110,113,115,150,169,170,174,182,183,272, 274,299-311,316,319,322-336]).

It would be especially important to emphasize that within this new, unreduced understanding of the universal science of complexity the dynamic information (including its lowest, quantum levels) is obtained as a well-specified, integral and tangible *part of the system* and its dynamic complexity and constitutes thus an intrinsic, major system property showing its (permanently changing) *potential for creative development* (thus also generalising usual, formally introduced 'potential energy' from classical mechanics). This feature of unreduced information should be compared to conventional, mechanistic information as if consisting of a special, 'mathematical matter' of 'pure bits', which is basically detached from, and exists independent of, *any* tangible system structure (contrary, for example, to usual energy) and is often endowed therefore, directly or indirectly, with the mystical quality of an 'informational aether' distributed throughout space as a sort of 'primal', *supernatural* 'tissue', or 'programme', of the world that determines the observed, tangible structures by way of mysterious (and most certainly 'quantum'!) influences. Such 'informal' attitudes of official, 'objective' and 'exact', science are quite clearly implied behind the famous 'it from bit' hypothesis [316] and its further development intending to say that 'it', or 'material world/structure', somehow emerges, and is therefore to be derived, from 'bit', or 'pure information' (exclusively coherent, of course, with a 'pure thought' of a 'chosen' community of 'mathematical' sages). In reality, as we have seen, the dynamic information (generalised potential energy) is a quite 'material', tangible entity (which does not prevent the existence of other, much higher, actually imperceptible levels of both information and entropy [1]) and it is due to its well-specified physical origin that the unreduced information is *directly* transformed into a more 'visible', 'unfolded', actually observed structure of space, or dynamic entropy of a higher emerging level of complexity, which gives rise, in its turn, to the next level of dynamic information (potential energy), and so on. The *purely dynamic uncertainty* is intrinsically involved in the dynamic information structure, the process of its transformation into entropy and the emerging results, which form, in addition, the hierarchy of *qualitatively* different levels of complexity and therefore the regular, identical and abstract 'bits' of conventional information concept, extensively used in modern (or rather 'post-modern') physics, provide but a very rough approximation of reality, which is not more exact than its description through enumeration of existing entities by integer number series (which is a good illustration of the true, 'ultimate' motivation of canonical, 'mathematical' physics: reduce the world to numbers and science to counting). The essential difference of real interaction processes from their unitary imitations is that the former cannot perform merely an abstract 'calculation' process, as it is implied behind the conventional 'computation' idea, but always involve also irreducible, explicit creation, or 'production', of new entities (less visible, but present, in the case of usual computers), which changes completely the sense and strategy of practical applications of quantum (and explicitly chaotic classical) computation processes (Sections 7.2, 7.3, Chapter 8).

Whereas our universal notion of dynamic information is applicable at both classical and quantum levels of complexity, the conventional mysteries of information, arbitrarily confused with equally inconsistent imitations of entropy and complexity, become increasingly obscure within its applications to quantum systems, especially popular in relation to unitary quantum computation (see e. g. [61-63,72,73-75,110,150,167,169,174,182,183,301, 316-333]). The ill-defined 'informational' properties are attributed here to everything including the already highly mystified and misunderstood conventional notions of wavefunction, density matrix, 'quantum entanglement', 'teleportation', 'nonlocality' ('Bell inequalities'), 'quantum bit', 'subquantum matter', etc. (see also Section 5.3). The extreme simplifica-tion of the unitary, effectively zero-dimensional projection of real, multi-valued dynamics and its evident contradiction to observations inevitably lead thus to another extremity, that of unlimited 'fantasy' around 'quantum information'. As becomes clear from our unreduced interaction process analysis, the causally complete, physically real versions of *both* quantum system dynamics *and* its information/entropy/complexity result from the same, complex-dynamical (multivalued) picture of interaction development described by the universal symmetry of complexity (at its lowest, 'quantum' sublevels). The specific features of dynamic information (and entropy) at the quantum levels of complex world dynamics are reduced mainly to the well-defined, dynamically fixed quantum of dynamic information (action) change, equal to Planck's constant $h$, as will be specified below (whereas the same information at the classical levels of dynamics is transformed into entropy in various portions of action determined by characteristic 'periods' of a particular system behaviour), and also to very low stability of physical realisation of a bit within any essentially quantum dynamics representing a particular case of the uniform (global) chaos regime (Sections 4.5-7, Chapters 5, 6). This causally complete, truly realistic understanding of physical information transformation at the quantum levels of dynamics shows fundamental impossibility of unitary quantum machine realisation (related to the unitary imitation of information) and also puts an end to various tricky, artificially mystified 'narratives' of the 'post-modern' official abstraction around 'informational interpretation of quantum mechanics', representation of the whole world or its arbitrary parts as a (unitary) quantum computer (information processing machine), 'it-from-bit' speculations, 'quantum networks', 'spin networks', etc., creation of direct links between quantum behaviour, living cell dynamics, brain activity, consciousness, gravitation and ... actually any arbitrary, mystified imitation of reality by the 'top' official science, all of it being thoroughly maintained and published, despite the glaring inconsistency, in the best sources of 'exact' science with the help of quite real, 'classical' networks of educated swindlers.

A meaningful illustration of the difference between the unitary imitation of quantum and classical information and its unreduced, complex-dynamical version can be obtained from elementary analysis of unitary estimates of the 'computational capacity of the universe' or any its material part [150]. These estimates are based on the formal attribution of the capacity to compute (in bits per unit time) with the maximum speed of the order of $E/\hbar$ to any 'piece' of mass-energy $E$, whereas the ultimate capacity of matter to register/memorise is expressed through its equilibrium, conventional entropy (divided by the Boltzmann constant $k_\text{B}$ and $\ln 2$) in a state with "maximum entropy", when "all matter is converted into radiation". As follows from the above, the unreduced, dynamic entropy, determining the memory capacity of a real system, need not attain its absolute maximum in the equilibrium, highly disordered state (of 'uniform chaos') and, moreover, the system *should* be in a qualitatively different, highly *nonequilibrium* and well-structured state (of 'multivalued SOC') in order to be able to register the information with a reasonable efficiency/stability. The total system complexity-entropy (transformed from the complexity-information in the course of information registration) will be determined by the total number of realisations, which need not coincide with the number of fixed identical system 'units'.

Consider, for example, a small part of the 'universe capacity' represented by a single human brain. In the simplest approach, registering structures are composed from synaptic links and if there are of the order of $10^4$ links for each of $10^{10}$ neurones, then the total number of memory system elements, $N$, is not smaller than $10^{12}$ to $10^{14}$ (taking into account possible differences between neurones, but also the existence of other possible links and memory elements). Now, the key point is that the complex-dynamical, nonequilibrium interaction processes between $N$ elements can produce up to $N!$ different combinations of elements, or operative system 'realisations', where each combination can encode, or register, a 'bit' of information (let us accept here this reduced designation of our complex-dynamical realisation). Therefore the 'ultimate' brain memory capacity is of the order of $N! \sim \sqrt{2\pi N}(N/e)^N$, which for the lowest estimate of $N = 10^{12}$ gives the capacity of not less than $(10^{12})! \gg (3.4 \times 10^{11})^{10^{12}} \gg 10^{10^{13}}$. This number is so much greater than not only one-dimensional unitary estimate of the 'ultimate universe capacity' (of the order of $10^{90}$ bits [150]), but also any reasonable numbers concerning mechanistic, 'linear' universe structure, that it is clear that any possible uncertainty in the model or parameter values cannot change the qualitative result: a *single brain* memory capacity exceeds by an actual infinity anything ever imagined by the conventional unitarity for the *whole universe*, and the real brain operation is limited not by its maximum formal capacity, but rather by practical impossibility (and absence of necessity, at least today) of dealing with such huge volumes of information.[33]

It is important that the above 'miraculously' high memory capacity is obtained not by means of some obscure 'quantum magic', but simply due to the unreduced complexity of real, 'classical', but *multivalued* and *creative* system dynamics (considered in a simplified, minimal version for this quick estimate), and therefore the same kind of increase will be obtained, of course, for any other explicitly chaotic, dynamically multivalued 'computing' system (the maximum increase will always be much greater for systems with higher complexity and thus, for example, for classical systems it will be greater than for lower-level, quantum ones).[34] In addition, *the same*, causally complete and universal picture of multivalued interaction dynamics explains the canonical quantum 'miracles' themselves (Sections 4.6, 4.7) [1-4,11-13], but similar to higher complexity levels, the 'magic' efficiency of *real* quantum dynamics has a *real* cost in the form of its fundamental dynamic uncertainty (which is especially and irreducibly high at those lowest, quantum complexity levels).

We obtain thus indeed 'exponentially high efficiency' of *any* real, quantum or classical, explicitly multivalued (complex) system dynamics with respect to its unitary imitation simply due to the exponentially large growth of the number of state-realisations, actually (and autonomously) taken by the system. In the case of pseudo-regular, SOC type of dynamics of ordinary modern computers this multitude of practically indistinguishable (in that case) realisations is effectively 'hidden' within each system element, or physical 'bit', remaining thus sufficiently stable, but in the case of explicitly multivalued, considerably irregular 'computer', such as any quantum computer or classical brain, those multiple, hierarchically branching realisations are effectively 'deployed' by the driving system interactions up to their individual distinguishability, so that the total system element number $N$ 'jumps' to the argument position of the exponential function determining the total number of actually taken, distinct state-realisations. It is also clear why this dynamically multivalued 'information decompression' with 'exponentially high efficiency' is obtained not for nothing (as it seems to be the case in the unitary theory of quantum computation), but in exchange for the irreducible and essential dynamic uncertainty of system dynamics, which alters completely the purposes and principles of construction of such complex-dynamical 'computer' (see Section 7.3, Chapter 8). The latter kind of machine, realised in various natural systems with high enough dynamic complexity (including eventually the whole universe), uses the above multivalued, exponentially efficient and 'automatic' (purely dynamical) 'decompression' of its own information processing for 'adaptable' and common solution of conventional 'difficult' and 'unsolvable' problems, usually formalised under the unitary, effectively one-dimensional notions of 'NP-complete', 'non-computable', or 'non-decidable' problems, these notions themselves being a huge (one-dimensional) simplification, or 'compression', of the now clearly specified, dynamically multivalued origin of all the unitary 'difficulties'.

---

[33] Note in passing, however, that the obtained 'incredibly high' capacity of real, complex-dynamical brain operation shows also that the system possesses a new, superior *quality*, rather than simply quantitatively increased performance. This 'new quality' can be associated with *consciousness* obtained in the detailed, dynamically multivalued picture of brain dynamics as an *emergent* property accompanying the dynamic appearance of the corresponding, high enough level of the same, universally defined hierarchy of complexity [1,4] (see also Section 7.3).

[34] Consider, as additional confirmation, a slightly different estimate of the number of brain operational units that takes into account only combinations of directly connected links between neurones. If each of $n = 10^{10}$ neurones has $L = 10^4$ links to other neurones, then possible (different) link combinations form a series of connected hierarchical 'generations' of consecutive neurones, with the number of combinations in the last, *n*-th generation growing as $L^n$ (which is much greater than the total number of links, $N = nL \leq 10^{14}$). The total number of connected links combinations from all generations is then obtained as $LL^2L^3...L^n = L^{(1+2+3+...n)} = L^{n(n+1)/2} \gg 10^{2\times10^{20}}$ for the above parameter values. Even if there is an over-estimate because of inevitable links repetition, we basically obtain the same exponential growth of the number of operational states (or memory cells), with the exponent determined by the system element (neurone connection) number.

A similar, qualitative difference appears between the unitary and complex-dynamical estimates of the speed (power) of computation. Whereas the mechanistic, physically ill-justified estimate (conceptually reproduced also within the conventional science of neural networks) gives the number of $10^{120}$ of 'ultimate', 'quantum' operations for the whole universe during its whole history [150], the modest frequency of switch of each element state in the brain of the order of 10 or even 1 Hz gives the maximum number of elementary changes (or operations) of not less than $10^{10^{13}}$ per second (it is again clear that a real brain uses but a small fraction of this maximum capacity which considerably exceeds, nevertheless, the parameters of any existing, or even ever possible, computation device or scheme of conventional, unitary type). The detailed, purely *dynamical*, 'emergent', interaction-driven origin of the state-realisation plurality of a real system is especially important for the correct understanding of this exponentially high speed of complexity development, appearing as *complex-dynamical, or chaotic, parallelism*, or dynamic adaptability (Section 4.4), of the explicitly multivalued interaction process that occurs within the same, fixed number of 'processing units' due to their chaotic realisation change, as opposed to the conventional concept of 'parallel information processing' always reduced to mechanistic multiplication of operational, simultaneously working units. It is the unreduced, complex-dynamical parallelism of brain operation that explains its crucial differences from any conventional 'super-computer' and makes unnecessary any additional assumption about obscure 'quantum consciousness' mechanisms allegedly hidden at a deep sub-neurone level and explaining the 'miraculous' power of conscious brain operation (e. g. [15,68-71]).

Note also that though being quantitatively simplified, our estimate of real, multivalued complex dynamics refers to actual computing system, like the brain, while the unitary estimate of the computation speed by $E/\hbar$ does *not* reflect *any* feasible computation dynamics and actually crudely violates the law of entropy growth, complexity correspondence principle and thus the complexity conservation law in general (see above in this Section). Thus, the quantum beat process within every single electron produces indeed of the order of $10^{20}$ reduction-extension cycles and associated *chaotic* spatial jumps per second (an estimate for the quantum beat frequency, $v_0 = m_0 c^2/h$, where $m_0$ is the electron rest mass) [1-4,11-13], but considering seriously any such elementary free-electron life-cycle as an operation of any (let alone unitary) 'computation process' by the universe is but a misleading play of words (even if the associated 'information' could be used, the electron would need to be transformed into e/m radiation, which makes the estimate senseless). Moreover, when one passes from the indivisible natural entities like elementary particles to their various agglomerates, constituting the actual universe structure, then the total energy of every agglomerate and their ensemble reflects only the number (per unit time) of chaotic and chaotically added quantum beat jumps of all elements, or system inertial (and gravitational) mass, but *not* any computationally sensible changes, or 'operations', at all, even for any 'ultimate', imaginary computation process (the number of computationally useful operations is determined only by suitable, higher-level interactions within the system, preserving its basic structure). Therefore the tremendous *under-estimation* of real system computation capacities within the unitary imitation contains, in addition, a fundamentally incorrect *over-estimation* of mechanistic computation capacities of any 'piece of energy' formally divided by $\hbar$ and any 'piece of entropy' formally divided by $k_B$ (in both cases one deals actually with states close to the 'uniform chaos', or 'equilibrium', regime of essentially quantum dynamics and therefore possessing the lowest possible *actual* computation capacities because of their least distinct/stable structure for the given entities). The obtained qualitatively big difference between the unitary and complex-dynamical (multivalued) estimates of real system capacities is a good illustration of the advantages of development of the explicitly chaotic (dynamically multivalued) kind of micro-machines, as compared to their pseudo-unitary existing versions and illusive unitary quantum schemes (see Section 7.3, Chapter 8).

Now, in order to obtain the adequate mathematical expression of the above universal law of complexity conservation and transformation, eqs. (56), taking into account all the mentioned properties of the unreduced entropy and information, we need first to determine the suitable physical quantity for dynamic information and entropy representation (as discussed above, information expression in dimensionless 'bits', as well as entropy expression in usual 'thermal', equilibrium-state units, does not reflect the real interaction/computation processes). The physical transformation of information into entropy, appearing as interaction/computation process, takes the form of emergence and change of incompatible system realisations, where each realisation configuration is the causal expression of the generalised, but physically real, observed space 'point' of the corresponding complexity level, while the neighbouring realisation separation gives the dynamically discrete (quantised) element, or unit, or increment of that dynamically obtained, or 'emergent', space (Section 4.3) [1,11-13]. Specifically, the size, $\Delta x$, of these natural space elements, determining the dynamical structure and thus properties of really perceived, physical space (at each given level of complexity), is obtained in our universal analysis as the separation between the 'centres of reduction' of neighbouring realisations, determined, as can be seen from the explicit expressions for the realisation configurations, eqs. (25) (Section 3.3), by the difference, $\Delta \eta_i^r$, between the corresponding eigenvalues of the effective dynamic equation, eqs. (20), that can be estimated, in its turn, by the eigenvalue separation of the auxiliary equations, eqs. (17), approximated by an averaged, or 'mean-field', interaction, eqs. (28), (29) (Section 4.4). Keeping in mind this well-specified link between the quantised space structure and the underlying essentially nonlinear, dynamically multivalued interaction process, all that we need to retain for our present purpose is that a given complexity level emerges in the form of spatial structure determined by its element(s), $\Delta x$, which result, and can be causally derived in our approach, from the unreduced interaction dynamics. The second universal form of dynamic complexity appears as the time period, $\Delta t$, of realisation change process determining the (irreversible) flow of the emergent time (at a given complexity level), which characterises the 'intensity' (temporal rate) of emergence of space elements (Section 4.3) [1,11-13], so that the temporal rate of the emerging structure propagation, or velocity, $v$, of thus rigorously defined system *motion* (Section 4.6.2), is given (for the simplest motion case) by the ratio of the emerging space and time elements, $v = \Delta x/\Delta t$.

It is clear from the above analysis of complexity transformation process that the emerging spatial structure elements, $\Delta x$, and their averaged (global) period of emergence, $\Delta t$, characterise the unfolded, explicit form of the driving interaction complexity, or dynamic entropy $S$. Since the emergent space and time are quantised (Section 4.3) [1], the same refers to the complexity transformation and all its 'participants', i. e. dynamic entropy, information and total complexity. It becomes clear that, in agreement with eqs. (56), the interaction complexity is transformed from the 'hidden' (folded) form of information into the observed form of entropy (space structure) in discrete 'quanta', where the emergence of each element of space structure, $\Delta x$, during the time period of $\Delta t$ realises the dynamically discrete entropy increase by $\Delta S$ from the equal, but 'disappearing' quan-tum of dynamic information, $\Delta I = -\Delta S$, while the total complexity $C$ remains *quantitatively* unchanged, $\Delta C = 0$ (but it experiences the quantised *qualitative* transformation). Contrary to the general expression of complexity conservation law, eq. (56b), here the quantised changes of information and entropy are fixed by the interaction dynamics and directly related to the underlying quanta, $\Delta x$ and $\Delta t$, of the two main manifestations of emerging complexity. In view of the fundamental character of participating quantities, this relation can only take the form of proportionality between $\Delta I$ or $\Delta S$ and both $\Delta x$ and $\Delta t$, which means physically that the complexity transformation, $\Delta I \to \Delta S$, is realised and observed as nothing else than the time-dependent structure emergence. Since, however, the emerging complexity increments, $\Delta x$ and $\Delta t$, characterise *directly* the corresponding change of the dynamic entropy $\Delta S$, it is more pertinent to express the *dynamics* of complexity transformation/conservation (or interaction development) as a relation of proportionality between the dynamic *information* quantum $\Delta I$ and space and time quanta $\Delta x$ and $\Delta t$:

$$\Delta I = -E \Delta t + p \Delta x , \qquad (57a)$$

where $-E$ and $p$ are coefficients (the negative sign before the positive $E$ expresses the 'natural', but actually conventional, condition that time permanently grows with always decreasing information). This complex-dynamical relation is immediately recognised, however, as a similar expression for the increment of mechanical action, $\mathcal{A}$, known from classical mechanics (eventually extended to quantum mechanics), where our universal description should be valid.

We obtain thus an important generalisation of the quantity (function) of *action* and its meaning, which shows that action can be universally interpreted as the most natural *integral measure of dynamic information* determined by the underlying *essentially nonlinear*, dynamically multivalued and entangled process of interaction (taken at the very beginning of its development). This *actual*, realistically founded meaning of action cannot be revealed and remains totally unknown in conventional, unitary science, where action, similar to other quantities, is introduced as an abstract value by means of formal postulates. Now the basic expression of the disappearing complexity-information increment through the emerging complexity-entropy (space and time) increments, eq. (57a), takes the familiar, but universally extended form,

$$\Delta \mathcal{A} = -E\Delta t + p\Delta x \ , \qquad (57b)$$

where *E* and *p* have the meaning of generalised (total) *energy* and *momentum*, respectively, while all the increments are *finite* and determined by the complex interaction dynamics (cf. Section 4.3). The related expressions for energy and momentum generalise their well-known expressions as partial derivatives of action to the case of finite increments:

$$E = -\frac{\Delta \mathcal{A}}{\Delta t}\bigg|_{x=\text{const}} \ , \qquad (58)$$

$$p = \frac{\Delta \mathcal{A}}{\Delta x}\bigg|_{t=\text{const}} \ . \qquad (59)$$

Note that *x* here and below can generally be understood as vector, with the ensuing vectorial meaning of the obtained quantities and expressions. The last equations show that energy and momentum represent the universal *differential measures of dynamic complexity-entropy*, generally equivalent and directly related to action as integral complexity measure. The differential form of dynamic information transformation, eq. (57), can also be written as

$$L = pv - E \ , \qquad (60a)$$

where the system *Lagrangian*, *L*, is defined as the discrete analogue of the total time derivative of action:

$$L = \frac{\Delta \mathcal{A}}{\Delta t} \ . \qquad (61)$$

It is also easy to understand that the lowest, essentially quantum levels of world dynamics are characterised by the unique, fixed value of the dynamic information quantum equal to Planck's constant, $|\Delta \mathcal{A}| = h$, and transformed, in the course of interaction development, into the equal quantum of dynamic entropy (represented by the emerging space structure in the form of de Broglie wavelength, see below). This uniqueness and universality of the action quantum are due to the lowest position of the quantum complexity level in the hierarchy of world dynamics, so that the fundamental interaction between the primordial protofields, as well as its results, cannot be further subdivided within this world (contrary to higher complexity levels, where the action-information quanta are not uniquely fixed) [1,3,12,13]. On the other hand, we see that the extended, dynamical 'quan-tum information', contrary to its speculative unitary imitation, is *naturally and universally quantised* and measured in units of *h*, in accord with the underlying quantum dynamics, while preserving the totally *realistic* physical meaning of 'potential', always diminishing form of interaction complexity (equivalent, in the differential expression, to the generalised potential energy). The quantum versions of the basic relations of eqs. (58), (59) have the well-known form (not understood, however, in its real meaning, within conventional quantum mechanics):

$$E = h\nu \, , \tag{62}$$

$$p = \frac{h}{\lambda} = hk, \text{ or } \lambda = \frac{h}{p} \, , \tag{63}$$

where $\nu = 1/(\Delta t)|_{x=\text{const}}$ is the internal, quantum-beat frequency of transitions between realisations of essentially quantum system, $\lambda = (\Delta x)|_{t=\text{const}}$ is its de Broglie wavelength and $k = 1/\lambda$ is the corresponding 'wave vector' (in that way one obtains, in particular, the realistic interpretation of de Broglie wave as the emerging, complex-dynamical space structure of a quantum object) [1-4,12,13].

Both the simplest, quantum-mechanical and general manifestations of the complexity development law, eqs. (57)-(63), show that the interaction-driven transformation of dynamic information into entropy, constituting the physical essence of any 'information processing', is indistinguishable from a *structure emergence* process, where the appearing structures realise the final complexity form of (dynamical) entropy, but are usually taken in the unitary science paradigm as empirical expressions of dimensionless 'bits' of subjectively defined and arbitrarily measured 'information'. The universal expression of the complexity conservation and transformation law, determining both structure emergence and information processing, is obtained from the differential form of the universal symmetry of complexity, eq. (56b), and the above energy definition as a universal measure of dynamic complexity, eq. (58). Dividing eq. (56b), expressed in units of action, by $\Delta t$ at $x = \text{const}$, we get

$$\frac{\Delta \mathcal{A}}{\Delta t}\bigg|_{x=\text{const}} + H\left(x, \frac{\Delta \mathcal{A}}{\Delta x}\bigg|_{t=\text{const}}, t\right) = 0 \, , \tag{64}$$

where the *Hamiltonian*, $H = H(x,p,t)$, considered as a function of emerging space configuration (or 'generalised coordinate') *x*, momentum $p = (\Delta \mathcal{A}/\Delta x)|_{t=\text{const}}$ and time *t*, just expresses the implemented, entropy-like form of differential complexity, $H = (\Delta S/\Delta t)|_{x=\text{const}}$ (*S* is expressed in units of action here). In agreement with eq. (60a), the Hamiltonian can also be expressed through the system Lagrangian that depends canonically on coordinate, time and velocity (rather than momentum):

$$H = pv - L \ . \qquad (60b)$$

The obtained eq. (64) is the universal extension of the Hamilton-Jacobi equation known from classical mechanics (e. g. [138]), but now valid for any, always *complex-dynamical* process of structure creation and dynamic information transformation into entropy [1]. Among the essential differences of the extended version, eq. (64), from the standard Hamilton-Jacobi equation we emphasize the causally complete, complex-dynamical interpretation of action, based on the unreduced analysis of the underlying essentially nonlinear, dynamically multivalued interaction (Chapters 3, 4) and including the emergent quantisation of all participating quantities. It is important also that the solution of eq. (64) or any its equivalent form (see below) is to be found by the same unreduced, universal analysis, which leads inevitably to the dynamically multivalued entanglement at a new, emerging level of structure expressed by the causally probabilistic general solution, eq. (24) (Section 3.3), and describing fundamentally uncertain, but also permanently developing, creative process of complex-dynamical 'computation' (Section 7.3, Chapter 8).

As explained above, the universal conservation (symmetry) of complexity has two inseparable aspects. One of them is the permanence itself of the total, unreduced dynamic complexity that can only be attained through equal by modulus, but opposite in sign changes of dynamic information and dynamic entropy, realising thus the exact transformation between these two forms of complexity just directly expressed by eq. (64). The second aspect concerns the unreduced interaction mechanism underlying this universal evolution law, i. e. the dynamically multivalued entanglement between the interacting system components (Chapter 4), which consistently explains why the number of chaotically changing realisations, and thus complexity-entropy, can only grow in any real (interaction) process. One has here the causal, rigorously derived and physically transparent mecha-nism of the inevitable complexity development and its irreversible, well-defined direction from diminishing information to growing entropy that can be formally expressed by the following condition added to eq. (64):

$$\left.\frac{\Delta \mathcal{A}}{\Delta t}\right|_{x=\text{const}} < 0 , \qquad (65a)$$

or

$$E, H\left(x, \left.\frac{\Delta \mathcal{A}}{\Delta x}\right|_{t=\text{const}}, t\right) > 0 . \qquad (65b)$$

We see that the generalised entropy growth law can be universally expressed as positive definiteness of the generalised Hamiltonian or total energy-mass, eq. (65b), which provides a new insight into this 'well-established' (but poorly understood) law of fundamental physics. The 'natural' condition of positive value of the ordinary, 'relativistic' energy-mass of a particle or body obtains now a rigorous foundation that reveals the internally chaotic, *multivalued interaction dynamics* as the universal *origin of mass* [1-4,11-13] which, among other important consequences, makes unnecessary the resource-consuming search for (and interpretation of) an artificially invented, mechanistic source of mass in the form of additional entity ('Higgs particles/fields'). Moreover, both the causal definition of energy, intrinsically equivalent to mass, and the condition of its positivity as the universal expression of generalised entropy growth are automatically extended now to any system and complexity level. We emphasize that this latter condition, eqs. (65), is rigorously and universally justified by the multivalued interaction dynamics (Chapters 3, 4), as it is explained above. The resulting natural unification of the extended energy conservation and entropy growth laws within the universal symmetry of complexity obtains now its convenient formal expression in eqs. (64), (65) revealing the true, qualitatively extended meaning of the 'well-known' quantities and relations mechanistically postulated in conventional science.

    The dynamical expression of the universal entropy growth law by eq. (65a) can also be interpreted, in agreement with 'intuitive', empirically based postulates of conventional science, as unambiguous, consistently derived *direction of causal time flow*, or 'time arrow': time unceasingly grows, or any system inevitably evolves, only in the direction of decreasing dynamic information and increasing entropy, corresponding to *both* visible

structure degradation *and* creation, which means that complex-dynamical (multivalued) transformation of the folded, informational complexity form into its explicit form of entropy is the single possible, omnipresent *way of existence* of anything. It is remarkable also that this apparent *asymmetry* of the time flow, absolutely necessary for any real entity existence and development, constitutes itself a part of the encompassing universal *symmetry* of complexity, eq. (64), so that the true origin and purpose of the 'asymmetric' direction of time is but *realisation and preservation* of that global *symmetry* of complexity. The unified evolution law of eqs. (64), (65) implies that it is the Hamiltonian function that actually provides the exact 'entropy growth rate' expressed by the action time derivative with the negative sign and physically determined by the specifically, locally configured *balance* of the universal symmetry of complexity (the dynamic information 'wants' to be unfolded into new entropy, but the already existing structures 'resist'), rather than by any skewed 'principle of extremum' helplessly sought for by the unitary theory of complexity and dynamical systems (see also the end of this Section). The suitable expression for the Hamiltonian results from the same multivalued dynamics analysis, involving derivation of the generalised dispersion relation (see refs. [1,2,12,13] for details).

If the system is in a state of global motion ($p \neq 0$), then the condition of eq. (65b) can be further specified. Indeed, the system Lagrangian, equal to the total time derivative of action, eq. (61), expresses the temporal rate of action change 'in the moving frame' of the global system motion and should be negative, $L < 0$, as follows from the generalised entropy growth law. This means, according to eqs. (60), that

$$E, H\left(x, \frac{\Delta \mathcal{A}}{\Delta x}\bigg|_{t=\text{const}}, t\right) > p\upsilon > 0 \ . \tag{65c}$$

Since the energy $p\upsilon$ here expresses the energy of global system motion (see also [1-3,12,13]), the latter condition means simply that practically any real system (interaction process) possesses, in addition to this global motion part, a positive component of the total energy corresponding to the rest energy-mass and given by the temporal rate of action change (with the negative sign) in the process of irregular system wandering, or 'random walk', around its averaged, global motion. In other words this means that, irre-spective of its state of motion, the system preserves its minimum, intrinsic identity in the 'rest frame', expressed by the finite, positive complexity (multivalued chaoticity) in the form of its rest energy-mass. Since $E > E_0 = m_0 c^2$, where $E_0$ is the rest energy of the globally moving system with the total energy $E$ and $p = mv$ for a large enough class of motions [1-3,12,13], it becomes clear that the condition of eq. (65c), expressing the generalised entropy growth law, is equivalent to the basic relativistic limitation $v < c$, which reveals the deep complex-dynamical origin of relativistic effects [1,12,13] simply postulated in the form of abstract 'principles' and rules in the canonical interpretation.

The relation between the dynamic entropy growth law and causally extended relativity can be further developed, if we rewrite eq. (60a), using eqs. (58) and (61), as

$$\frac{\Delta \mathcal{A}}{\Delta t} = \frac{\Delta \mathcal{A}}{\Delta t}\bigg|_{x=\text{const}} + pv , \qquad (60c)$$

which is none other than the standard expression of the total derivative through partial derivatives (see also eq. (59)). Taking into account the fact that $\Delta \mathcal{A} = -|\Delta \mathcal{A}| < 0$, while $v_0 = 1/\Delta t$, $v = 1/(\Delta t)|_{x=\text{const}}$ are realisation change frequencies determining the 'velocity of time flow' in the 'moving frame' with respect to itself ('intrinsic' time flow) and in the 'rest frame' respectively, we see that the observed time flow *causally (dynamically)* slows down in the 'moving frame' with respect to the 'rest frame' (i. e. the intrinsic time of globally moving interaction processes goes slower than the one for processes globally at rest) because in a globally moving 'frame' (interaction process) a proportion of the purely random energy reservoir of the rest energy-mass growing with the global motion velocity is spent for the global system motion [1,2,12,13], *in addition* to its purely regular component ('one cannot obtain a useful result without some losses'):

$$v_0 = v - \frac{pv}{|\Delta \mathcal{A}|} = v - \frac{v}{\lambda} = v - v_{\text{gl}} , \quad v_0 < v , \qquad (65d)$$

where $v_{\text{gl}} = v/\lambda$ is the average frequency of system realisation change within the global (averaged) system motion and $\lambda = |\Delta \mathcal{A}|/p$ is the emerging characteristic length of the global motion. For the lowest, quantum levels of complexity, $|\Delta \mathcal{A}| = h$ is Planck's constant and $\lambda = h/p$ is the de Broglie wavelength (revealing thus its causal origin [1,2]). This relation can be further transformed [1,2,12,13] into the standard expression of the effect of 'relativistic time retardation' that now obtains, however, a considerably extended, causally substantiated (realistic) and complex-dynamical meaning, which naturally unifies not only quantum and relativistic effects, but also the dynamic entropy growth law within the universal symmetry of unreduced dynamic complexity (in particular, relativity effects, including those of extended general relativity, are obtained, as in eq. (65d), for arbitrary systems and levels of complexity, time and space [1,13]).

If the system is isolated, so that its energy $E = -(\Delta \mathcal{A}/\Delta t)|_{x=\text{const}}$ remains constant, then the universal complexity transformation law, eq. (64), takes a reduced form that does not contain explicit dependence on time:

$$H\left(x, \frac{\Delta \mathcal{A}}{\Delta x}\bigg|_{t=\text{const}}, t\right) = E \ . \qquad (66)$$

Note, however, that contrary to the conventional science version of this equation, the real physical time, determined by the unceasing change of explicitly obtained realisations, reappears in the general solution of eq. (66), in agreement with eq. (58) applied at the new sublevel of complexity, where it can be written as

$$\Delta t = -\frac{\Delta \mathcal{A}}{|\Delta V|} \ , \qquad (67a)$$

$|\Delta V|$ being the EP difference between the neighbouring realisations found from eq. (66) (see eq. (25c)) and $\Delta \mathcal{A}$ the corresponding action (complexity) increment. The total discrete time flow at the corresponding complexity level consists from the individual time increments $(\Delta t)_i$:

$$t - t_0 = \sum_i (\Delta t)_i = -\sum_i \frac{\Delta \mathcal{A}_i}{|\Delta V|_i} \ , \qquad (67b)$$

It can be seen from here that the fractal structure of realisations (Section 4.4) determines, through $|\Delta V|_i$ and $\Delta \mathcal{A}_i$, the fractal structure of time flow.

The universal evolution equation of a newly created structure, eq. (64), is formulated in terms of action-complexity change between 'localised' system states, each of them corresponding to a fully developed realisation. However, the system also spends a part of its time in transitions be-tween those regular realisations occurring through a specific, intermediate realisation-state of quasi-free, disentangled and delocalised system components, forming the physically real system 'wavefunction', or 'distribution function' (Section 4.2) [1,4,11-13]. Therefore the 'localised' type of description of eq. (64) should be completed by the related dual, 'delocalised' equation for the generalised wavefunction (distribution function) $\Psi(x,t)$ describing system behaviour in its qualitatively different state during transition between the regular (localised) realisations. The system in the state of wavefunction as if transiently returns to its state at the beginning of its interaction process, when the system complexity (at the current level) had the form of dynamic information, contrary to its transformation into the dynamic entropy in the developed-interaction state of regular realisation.

Since the integral complexity measures (number of system realisations) multiply for the adjacent sublevels, the total complexity of realisation change process, or 'wave action' $\mathcal{A}_\Psi$, can be presented as the product of its wavefunction (lower sublevel, or dynamic information) and action expressing actually the developed complexity form of well-structured realisations (higher sublevel, or dynamic entropy), $\mathcal{A}_\Psi = \mathcal{A}\Psi$ [1,4,11-13]. The total system complexity estimated by the logarithm of realisation number takes thus the proper form of the sum of information and entropy. It is advisable then to consider the change of the wave action during one cycle of reduction-extension (consecutive realisation change) of the considered interaction process. Since the total system complexity remains unchanged and physically the system returns, after a cycle, to its previous state, the wave action increment equals to zero:

$$\Delta \mathcal{A}_\Psi = \mathcal{A}\Delta\Psi + \Psi\Delta\mathcal{A} = 0 , \qquad (68a)$$

or

$$\Delta\mathcal{A} = -\mathcal{A}_0 \frac{\Delta\Psi}{\Psi} , \qquad (68b)$$

where $\mathcal{A}_0$ is the characteristic action value of the emerging realisation dynamics (it may include also a numerical coefficient accounting for the undular nature of the wavefunction, for certain complexity levels [1,4,12,13]).

The obtained relation, eq. (68b), is the generalised, causal *quantisation relation* for *any* complex (multivalued) dynamics taking the form of standard, but now *causally explained* 'quantisation rules' at the level of quantum-mechanical behaviour (consistently derived as quantum beat dynamics in our approach, Section 4.6.1). The realistic physical meaning of the quantisation relation of eqs. (68) is that it describes the cycle of real, qualitative system transformation from the 'extended', disentangled, 'informational' form of the generalised wavefunction to the localised, entangled, entropy-like (structural, or spatial) form of a regular realisation and back, which is the elementary cycle of consecutive realisation change in the multivalued interaction dynamics. This physically real system transformation during realisation change is replaced in conventional, dynamically single-valued quantum mechanics and field theory by *formally postulated* quantisation rules and related *abstract* 'operators of creation and annihilation' of postulated (and ill-defined) localised states, or 'particles', obtained as if from nothing, which is the unitary, linear imitation of dynamically multivalued entanglement and disentanglement of the unreduced, essentially nonlinear interaction process (Chapters 3, 4) [1], whereas the conventional 'science of complexity' or 'information theory', operating at higher complexity levels, cannot use this simplified construction of unitary quantum theory and therefore do not consider any explicit object appearance and disappearance in the course of interaction at all.

The dual, wavefunctional form of the universal complexity transformation law, eq. (64), is now obtained by substitution of causal quantisation rule, eq. (68b), into eq. (64), which gives the *generalised Schrödinger equation* (extending its causal derivation for quantum complexity levels to any higher level of complexity [1,4,12,13]):

$$\mathcal{A}_0 \frac{\Delta \Psi}{\Delta t}\Big|_{x=\text{const}} = \hat{H}\left(x, \frac{\Delta}{\Delta x}\Big|_{t=\text{const}}, t\right) \Psi(x,t) , \qquad (69)$$

where the 'operator' form, $\hat{H}$, of the system Hamiltonian is obtained from the ordinary Hamiltonian of eq. (64) with the help of the above quantisation rule, eq. (68b), followed by multiplication by $\Psi$. As a result, the operator Hamiltonian is a function of the discrete analogue, $(\Delta/\Delta x)|_{t=\text{const}}$, of the partial derivative operator from the standard quantum mechanics $(\partial/\partial x)$, causally originating now from the generalised momentum definition of eq. (59) in combination with the dynamic quantisation of eqs. (68). Similar to the conventional quantisation, higher powers of this operator correspond to higher derivatives (rather than to higher powers of the first derivative), which can be explained by the properties of the underlying reduction-extension (realisation change) process, together with the proper 'order' of operators (where every real 'annihilation' of an entity can only follow, but not precede, its creation) [1]. The usual, quasi-continuous form of the universal Schrödinger equation (and the corresponding form of the generalised Hamilton-Jacobi equation for action) can be justified in cases of sufficiently 'fine-grained' (quasi-uniform) structure of the emerging entities flux on the real (observed) scale of a problem:

$$\mathcal{A}_0 \frac{\partial \Psi}{\partial t} = \hat{H}\left(x, \frac{\partial}{\partial x}\bigg|_{t=\text{const}}, t\right)\Psi \ . \qquad (70)$$

The two dually related forms of the universal complex-dynamical evolution equation, eqs. (64), (68)-(70), describing (together with the accompanying relations of eqs. (58), (59), (65)-(67)) the way of existence and development of any system with interaction, can be called the *universal Hamilton-Schrödinger formalism* of the unreduced science of complexity, where the universal method of finding the causally complete, dynamically multivalued problem solution (Chapters 3, 4) at each naturally emerging complexity level is actually included in the problem formulation. In particular, we can see now that the accepted quite general form of the initial problem expression (see eqs. (1)-(5) from Section 3.1) also corresponds to the universal Hamilton-Schrödinger equation with the explicitly designated interaction potential, even though we did not make any such assumption at the beginning of our analysis.

As mentioned above, the universal symmetry of complexity, representing the generalised expression of the unreduced development of any interaction process, naturally constitutes the unified, causal extension of *all* the conventional 'fundamental principles' and 'conservation laws'. For this reason, the universal evolution equations, eqs. (64), (69), being the unified mathematical expression of the universal symmetry of complexity, should represent the *single dynamic equation* of the universal science of complexity, which is the unified, causally explained extension of *all correct equations* of conventional science (usually accepted by way of only empirically justified, formal postulation of an abstract mathematical relation). Whereas the detailed confirmation of this statement may come from various comparisons of our unified equations with the known basic constructions of usual science, the formal correspondence of the universal Hamilton-Schrödinger formalism with practically any correct, 'linear' or 'nonlinear', dynamic equation can be directly seen from eqs. (64) and (69). Thus, if we use the following expansion of the Hamiltonian in power series of its arguments:

$$H\left(x, \frac{\partial \mathcal{A}}{\partial x}, t\right) = \sum_{m,n} h_{mn}(x,t) [\mathcal{A}(x)]^m \left(\frac{\partial \mathcal{A}}{\partial x}\right)^n , \qquad (71)$$

where $h_{mn}(x,t)$ are the known expansion coefficients (and dependence on $\mathcal{A}(x)$ can appear from an effective interaction), then the generalised Hamilton-Jacobi and Schrödinger equations take the form

$$\frac{\partial \mathcal{A}}{\partial t} + \sum_{m,n} h_{mn}(x,t) [\mathcal{A}(x)]^m \left(\frac{\partial \mathcal{A}}{\partial x}\right)^n = 0 \qquad (72)$$

and

$$\frac{\partial \Psi}{\partial t} = \sum_{n>0, m} h_{mn}(x,t) [\Psi(x)]^m \frac{\partial^n \Psi}{\partial x^n} + \sum_m h_{m0}(x,t) [\Psi(x)]^{m+1} \qquad (73)$$

respectively (where the corresponding powers of $-\mathcal{A}_0$ are included in coefficients $h_{mn}(x,t)$ in eq. (73) for simplicity). It is clear that many model 'evolution equations', 'master equations', equations for 'order parameter', 'nonlinear', 'wave' equations and other *postulated* equations of usual theory (see e. g. [340]) remaining separated within it, can now be classified as particular cases of eqs. (72), (73), provided with a well-specified, realistic interpretation; others can be transformed into a suitable form.

An equivalent, 'Lagrangian' formulation of complex interaction dynamics, extending the corresponding Lagrange method from the classical (dynamically single-valued) mechanics [138], can be obtained by discrete integration of the Lagrangian definition of eq. (61), giving the expression for the integral action-complexity:

$$\mathcal{A}(x,t) = \sum_{\mathfrak{R}=\mathfrak{R}_{\text{in}}}^{\mathfrak{R}_{\text{curr}}(x,t)} L_\mathfrak{R} \Delta t_\mathfrak{R} , \qquad (74)$$

where the sum is taken over all consecutive realisation-trajectories (or *real* 'paths') $\mathfrak{R}$ between certain initial realisation $\mathfrak{R}_{\text{in}}$ and the current realisa-tion-trajectory $\mathfrak{R}_{\mathrm{curr}}(x,t)$ determining the current configuration, or generalised position, of the system. We obtain the causal substantiation and extension of the canonical 'least action principle', unifying it with the extended 'entropy growth law' and 'energy/mass conservation law' into the universal symmetry of complexity. Indeed, the causally extended action $\mathcal{A}(x,t)$ of eq. (74) permanently decreases in the course of system jumps between its realisations along averaged trajectory because action is the expression of dynamic information decreasing at the expense of growing dynamic entropy, in order to ensure conservation of their sum, the total dynamic complexity (represented, for example, by the total energy of a closed system). Moreover, system jumps along trajectory occur probabilistically and therefore the real trajectory is unceasingly and chaotically 'shifted' within its averaged, smeared trace, which provides causal, realistic extension for the idea of 'virtual' (imaginary) trajectories used in the formulation of the classical least action principle [138]. It is the *symmetry* (conservation) of complexity that determines *real, chaotic trajectory fluctuations* in the real interaction processes, as opposed to the usual formal, inexplicable *minimisation* of the *abstract* action function with respect to *imaginary, infinitesimal* trajectory variation around its allegedly 'real', but unrealistically smooth and 'infinitely thin', version.

Note that the extended version of Lagrangian approach is formally as universal as the above Hamilton-Schrödinger formalism and actually they constitute together related aspects/formulations of the unified description of complex (multivalued) system dynamics that can be called the *universal Hamilton-Lagrange-Schrödinger formalism*. However, the Lagrangian version of this unified formalism is oriented to systems with more localised, not too smeared trajectory/configuration (SOC type of complex dynamics, Section 4.5.1), while the Hamilton-Schrödinger description is more convenient in the case of strongly chaotic, 'distributed' kind of complex dynamics (uniform chaos regime, Section 4.5.2), which explains, in particular, why we do not develop here, in the study of essentially distributed system behaviour, the least action principle up to the differential form of dynamic equations (the so-called Lagrange equations, such as Newton's second law, which replace the Hamilton-Jacobi equation from the 'nonlocal' problem formulation). Here again, we obtain the causally complete, realistic explanation of the difference between the canonical approaches of Hamilton and Lagrange in terms of the underlying multivalued dynamics that remains hidden within usual, dynamically single-valued science approach.

The 'wavefunctional' version of Lagrangian approach also exists and takes the form of the modified and generalised, complex-dynamical version of conventional 'Feynman path integrals' [1]. It is obtained by using the causal quantisation rule, eq. (68b), in the above expression for the action-complexity, eq. (74), which gives the following relation between the generalised wavefunction (distribution function) as a distributed system feature and localised system jumps between realisations expressed by its Lagrangian and the intrinsic time increments:

$$\Psi(x,t) = -\frac{\Psi_0}{\mathcal{A}_0} \sum_{\Re=\Re_{\text{in}}}^{\Re_{\text{curr}}(x,t)} \Delta \mathcal{A}_\Re = -\frac{\Psi_0}{\mathcal{A}_0} \sum_{\Re=\Re_{\text{in}}}^{\Re_{\text{curr}}(x,t)} L_\Re \Delta t_\Re \ , \qquad (75)$$

where $\Psi_0$ and $\mathcal{A}_0$ are characteristic values. Note the essential difference from the exponential function under summation (integral) in the canonical path integrals and related methods of 'mathematical physics', which is a particular manifestation of the 'false exponential dependence' phenomenon in the unitary science (cf. Section 5.1, Chapter 6) [1], resulting from the incorrect extension of a perturbation theory approximation beyond the limits of its validity, whereas in reality one deals with a linear or close power-law dependence (which is none other than the integral version of the universal quantisation rule of the unreduced interaction dynamics, eqs. (68)). Other differences from the unitary version are the dynamically discrete (quantised), rather than quasi-continuous or formally discrete, structure and physically real, rather than 'virtual' and abstract, character of actual system wandering among different 'paths' (realisations). The unitary version refers, by definition, to a single realisation, which explains its deficiency.

We shall not further develop here the details of the extended Lagrange version, eqs. (74), (75), of the universal formalism of the unreduced science of complexity because in this work we consider the applications, such as essentially quantum machines, where system dynamics is basically nonlocal and uniformly chaotic, so that the Hamilton-Schrödinger version, eqs. (64)-(73), of the universal formalism usually appears to be more con-venient. Let us emphasize once more that the unity of all the related expressions of the universal complexity development law comes from the underlying symmetry of complexity, maintained dynamically within any system (interaction process) by the unceasing transformation of dynamic information into entropy, eqs. (56). Dynamic discreteness (quantisation, uncertainty) and causal randomness (indeterminacy) are universal and unavoidable features of any real system (interaction process), but their relative magnitude and character vary considerably for different complexity levels. At the lowest, quantum levels of dynamics we have universally determined quantisation (the complexity/information/entropy quantum $\Delta\mathcal{A}$ is *always* equal to $h$, Planck's constant) with relatively big quanta and strong (uniform/global) chaoticity, while most simple (externally 'regular') cases of classical behaviour correspond to non-universal and relatively small magnitudes of both quantisation and chaoticity.

The universal symmetry of complexity, substantiated and quantified in this Section, should be distinguished from various recent attempts, always performed within the unchanged, unitary science paradigm (postulated, dynamically single-valued and abstract imitations of multivalued reality), to guess a new, additional 'principle of nature' that could explain the behaviour of special, 'complex' or 'far-from-equilibrium' systems, desirably in terms of most popular substitutes, such as unitary (arithmetical) 'information' or 'complexity'. Apart from the false 'information conservation law' invented in that way (e. g. [327,333,335]) and already discussed above in this Section, we can mention various versions of equally illusive 'fourth law of thermodynamics' that vary arbitrarily between the 'law of *fastest* possible entropy (complexity) *growth*' in nonequilibrium systems, including fastest possible free-space invasion by greedy 'autonomous agents' [336], and the opposite laws of 'informational entropy'/complexity *decrease* as a result of system 'self-organisation' [299,341] or its '*slowest possible growth*' [342-344], or else obscure 'laws of chaos' [130,226,241,266] based on a 'plausible' philosophy, but actually formally postulated for abstract (and very limited) mathematical 'models', etc. (see also [345,346] as examples of conventional science limitations in the subject). As already noted above, in reality the system always *transforms qualitatively* (from dynamic information into entropy) its *quantitatively un-* *changed* complexity in an *optimal*, rather than extreme, way determined mathematically by the Hamiltonian function (see especially eqs. (64), (65)) and realising a sort of probabilistic *balance* (= symmetry!) between the generalised *potential* of dynamic information, as if trying to unfold 'as quickly as possible' under the driving interaction influence, and the generalised *inertia* of dynamic entropy of already existing/created structures, as if resisting 'as much as they can' to further novelties (the effects of finite compressibility, friction, 'dissipativity', etc. within the interaction process).

Contrary to the unitary guesses, always inconsistently 'skewed' and incomplete, the unreduced, really universal and totally realistic law of nature, the dynamic *symmetry* of complexity, eqs. (56)-(75), is consistently *derived* within the nonperturbative, multivalued interaction analysis (Chapters 3-5) and is actually *equivalent* to the unified, *unreduced* description of any system *dynamics*, i. e. *any* real world dynamics *is* the omnipresent symmetry and not only 'is described by', or follows from, a formal 'law of symmetry'. Instead of inserting artificial entities and new 'principles', actually only inconsistently modifying the well-established laws, the symmetry of complexity just takes into account the dynamic multivaluedness of really existing system realisations and thus provides the truly causal explanation for the 'established' laws of conventional science, but now liberated from the inevitable contradictions and defects of their unitary versions. In the following Sections we show in more detail how the obtained essential advance in understanding of fundamental evolution laws leads to qualitative change in the direction and efficiency of practical problem solution.

## 7.2. Computation as complexity conservation by transformation of information into entropy and its particular features at the level of micro-machine dynamics

Any computation or 'information processing' is physically realised as a series of interactions and can therefore be described in terms of the universal symmetry (conservation) of complexity maintained by transformation of dynamic information into entropy (Section 7.1). It is important to understand that it is the latter form of dynamic complexity, the dynamic entropy, that represents physical, always intrinsically random (dynamically multivalued) implementation of 'bits' of conventional, abstract 'infor-mation', whereas our dynamic information is the 'hidden', potential state of dynamic complexity at the very beginning of the corresponding interaction process, generalising usual 'potential energy'. This nontrivial internal structure of *any computation* reveals its meaning and character as unceasing series of hierarchically ordered *qualitative changes*, or *events* (cf. Section 2(i)), causally random in their spatial order and irreversible in time, in contrast to the conventional, unitary picture of computation (information processing) showing it as a reversible and nondissipative, in principle, manipulation with totally controlled (regular) carriers of abstract, pre-existing 'bits' of canonical, senseless 'information'. This unitary imitation of computation includes various 'physical' theories of information pretending to be especially close to reality (e. g. [37,80,113,150, 300-306,310,321,330-336]), but actually reduced to ambiguous speculations around the same, unitary and abstract theory involving various 'mind games', 'gedanken experiments' and unrealistic, meaningless assumptions (such as 'infinitely slow' information processing).

The unitary theory of information and computation is certainly inspired by operation of usual, classical and *apparently* regular, computers. In terms of unreduced, complex-dynamical (multivalued) theory of computation, this case corresponds to the self-organisation limiting regime of multivalued dynamics (Section 4.5.1), which explains the visible regularity of usual computers, but leaves the irreducible place for the internal randomness and irreversibility. These latter properties appear in the form of thermal dissipation (usually having a rather low, but finite, limited from below level) and occasional 'malfunction', such as unexpected 'halt' or 'cycling' (in addition to 'algorithmic' instabilities, such as catastrophic error accumulation processes). Note that it is the dynamically multivalued, composite structure of every physical 'bit' realisation in conventional computer that determines *both* its (small) intrinsic randomness *and* relative stability with respect to various small (e. g. thermal) perturbations and thus ensures the *almost* unitary (regular) mode of its operation.[35] Therefore the detailed analysis of conceptual and experimental structure of usual computer operation shows that its apparent 'unitarity' is basically illusive, even though this illusion does have a rational basis. Another essential part of this illusion is related to the fact that many explicitly complex (dynamically multivalued), creative actions indispensable for actual, useful operation of a usual computer, including its 'programming', control and elimination of dynamical halts at all levels, are 'tacitly' inserted by the effectively strong interaction with another, *explicitly chaotic* (multivalued) 'computing system', the human brain (see also below).

---

[35] In other words, if you want to obtain the maximum possible, *almost* total regularity, you *should* accept a small *intrinsic irregularity* constituting the very *mechanism* of, and the inevitable payment for, that almost complete regularity. This rather natural consequence of the entropy growth law is somehow 'lost' in the conventional theory of systems, computation and control which assumes a major possibility of totally 'coherent', regular processes, where one may deal only with *external* irregularities that can be reduced to any desired level by monitoring system interaction with the environment.

Each dimensionless 'bit' of abstract unitary 'information' is physically realised in the ordinary computer as a multivalued dynamic regime (interaction process) of the SOC type characterised by certain finite, classically large value (or 'quantum') of complexity-action, $\Delta \mathcal{A} \gg h$, needed to change, with a high enough probability, the *global* bit state (i. e. to change '0' to '1' or the contrary) with the help of an 'external', higher-level interaction. The classical bit state in the usual, pseudo-unitary computer is highly (but not infinitely!) stable due to its physical composition of many sufficiently independent, lower-level components (subsystems) with dynamically random behaviour, such as atoms, which reduces the probability of their *simultaneous* random ('spontaneous') 'switch'/motion in one direction to exponentially small values.[36] This mechanism of (relative) classical computer stability is quite similar, and even directly related to, the mechanism of dynamic formation of elementary classical, localised state in the form of a bound state of two quantum particles (Sections 4.7, 5.3(C)). If, however, the interaction between different classical bit states becomes comparable with the characteristic external influence necessary for the individual bit switch, then the quasi-uniform chaos regime (Section 4.5.2) reappears at the level of global system dynamics and one obtains a qualitatively different, explicitly chaotic (multivalued) type of 'computation' by a classical system, actually realised in neural networks (see below).

---

[36] Under the simplest assumption, the negative argument of the exponential function here is proportional to the number of subunits to the power around unity. A more complicated structure of interactions between the components, their association in a hierarchy of groups can lead to other particular forms of the exponential smallness in question. However, the main feature of interest remains unchanged: the negative argument of the exponential function is large by its absolute value for large number of components and drops to order unity (thus eliminating the exponent) when the number of components falls down to one.

In that way the intrinsic nonunitarity of any computation becomes more evident in cases of explicitly complex (multivalued) computing system dynamics, such as any essentially quantum dynamics (quantum computers) or explicitly chaotic classical dynamics (brain, neural networks). In particular, the specific properties of *quantum computation and information processing* are determined by the characteristic features of respective complexity levels (Chapter 5). As a result, one obtains such unique properties of quantum computation as the universally fixed and relatively large value of the dynamic information quantum (or real 'quantum bit'), equal to Planck's constant *h*, or the irreducible and large magnitude of dynamical randomness at any step of quantum system dynamics, which can be reduced and partially 'controlled' only after transition to the next, superior level of complexity known as classical (permanently localised) dynamics (Section 7.1), contrary to the basically wrong, mechanistically applied unitary-science idea of 'quantum control of quantum systems' [90-107].

It is evident that these particular features and limitations of real quantum system dynamics are related to its causally complete description as the *lowest* levels of complex world dynamics, so that the natural *diversity* of the intrinsically multivalued dynamics is reduced to its absolute *minimum* at the level of essentially quantum behaviour (which also explains its peculiar, apparently 'exotic' properties). This result can be extended to a universally applicable consequence of complexity conservation law called *complexity correspondence principle* (see also Sections 5.2.2, 7.1) [1]. It states that any system cannot correctly simulate/reproduce the behaviour of any other system whose unreduced dynamic complexity is greater than the same quantity for the simulating system, or in other words, the (effective) dynamic complexity of the (correct) simulator should be greater than the simulated system/behaviour complexity.[37] Any violation of this rule enters in direct contradiction with the universally confirmed and rigorously substantiated complexity conservation law (Section 7.1). The relative error of imperfect, low-complexity simulation of a higher-complexity behaviour is determined by the relative complexity difference between the two systems.

---

[37] Taking into account the real dynamics of computation process, one can advance even a stronger version of the same principle stating that in practice the effective complexity of a correct computation process should be at least one complexity level higher than the simulated system complexity. It is a consequence of the generalised entropy growth law stating that one can never avoid having a finite quantity of useless work and wasted resources. In the case of computation, one needs to have some 'auxiliary' actions (like energy dissipation or intermediate result memorisation) that do not directly contribute to the reproduction of simulated system properties, but are necessary for realisation of the computation process. This amplified version of complexity correspondence rule implies, for example, that one can simulate an essentially quantum behaviour starting only from quantum-and-classical (or hybrid), rather than purely quantum computation level (see also below).

The most evident manifestation of the complexity correspondence principle just refers to possibilities of real quantum computation in simulation of other system behaviour. Namely, it is rigorously proved now that any essentially quantum system/behaviour could properly simulate only other essentially quantum behaviour (with lower complexity), but definitely not any higher-level behaviour, starting already from elementary classical systems. According to the above estimate, the relative error in quantum simulation of classical behaviour will be close to unity, which has a transparent physical interpretation: the essentially quantum behaviour is qualitatively different from classical behaviour by definition, since quantum behaviour is irreducibly nonlocal and coarse-grained, while an elementary classical system is permanently localised (around its classical trajectory) and fine-grained. In other words, the irreducible quantum nonlocality prevents any reproduction of classical trajectory localisation (which is a manifestation of a considerably *higher* complexity of classical behaviour, contrary to what one could imagine within the unitary approach and its mechanistic 'intuition').[38]

This fundamentally substantiated and realistically explained result of the universal science of complexity shows that the opposite statement of the unitary theory of quantum computation, insisting upon 'universality' of (unitary) quantum computation (see e. g. [21,25,46,58,64,67,103,157,158, 177-181]), is strictly wrong. The evident falsification of reality by the conventional theory of quantum computation turns out to be especially grotesque taking into account its extremely simplified, *zero-complexity* (unitary) approach, which represents the fatal and artificial reduction to zero of already *relatively* low, but definitely *non-zero* and absolutely high complexity of real quantum dynamics. In addition, conventional quantum computers are supposed by their proponents to be uniquely efficient in solution of just particularly complicated, 'non-computable' problems, 'unbreakable' by ordinary, classical computers (which actually possess much *higher* complexity than any quantum computer).[39] As concerns the announced 'successful experimental realisations' of conventional quantum computation, their details clearly show that they can well be sufficiently 'inexact' to account for the irreducible quantum errors, but those unavoidable errors are incorrectly presented as acceptable, 'technical' deficiency of 'first', imperfect attempts that can be eliminated in subsequent, more involved versions. We have seen, however, that any quantum dynamics is irreducibly 'coarse-grained' and random, including any possible methods of its 'quantum control' (this conclusion agrees, by the way, with the standard quantum postulates, contrary to the conventional schemes of unitary quantum computation, see also Section 2(ii)).

---

[38] Note that any abstract, 'arithmetical' calculations can also be classified as 'classical' type of dynamics because of their intrinsic regularity (localisation).

It is evident that the same complexity correspondence principle refers to any kind of machine (not only 'computers') at any complexity level. If we apply it to arbitrary quantum machines, we immediately come to the conclusion that in order to be useful for our real, higher-complexity, macroscopic and 'classical' needs, quantum machines should necessarily include some essential classical elements, naturally entangled with their essentially quantum parts. This 'quantum-classical entanglement' takes the form of dynamic emergence of classical behaviour from purely quantum dynamics, being a particular case of higher complexity level emergence in the course of complexity development process (Section 7.1) [1] and usually occurring as formation of elementary bound state of several quantum systems (Sections 4.7, 5.3(C)). The intermediate, transient phases of this irreducibly complex-dynamical (multivalued) transformation constitute, where they can appear, the process of 'quantum measurement', now obtaining its causally complete interpretation [1,10]. The resulting system, consisting from at least two large neighbouring (groups of) complexity levels, can be called quantum-and-classical, or *hybrid machine*, irrespective of its detailed origin and function (natural, artificial, computer, generator, etc.). Taking into account the physical structure of real material objects, we can conclude now that *all* the existing structures, including living systems, operate, at their several lowest (microscopic) complexity levels, as such hybrid machines whose irreducibly complex-dynamical (multivalued) functioning, revealed within our unreduced interaction analysis, explains the observed unique properties of nature in general and living organisms in particular, including their quasi-autonomous evolution and the emergent property of intelligence [1]. We can see also the objective reason for the fundamental impossibility of causal understanding of natural micro-machine dynamics and any its property within the dynamically single-valued, perturbative approximation of conventional science (including the scholar solid-state physics, chemistry, biology, medicine, etc.).

---

[39] Note that already from a general point of view the idea that unitary quantum computers (or any other computers) can solve 'non-computable' problems only due to their 'exponentially high' power looks as a basically incorrect one. Indeed, all the obscure notions of conventional science around 'non-computability', 'nonintegrability' and related properties show in any case that these are manifestations of a *fundamentally* and *qualitatively* different character of problems in question, with respect to 'ordinary', much more simple, 'computable' or 'integrable' problems. It is difficult to expect therefore that such fundamental obstacle can be surmounted just by highly ('exponentially') increased computation power. In order to master a qualitatively more complicated range of problems, one should certainly introduce a qualitative, conceptual novelty in the approach to their solution (analytical or numerical), which should correspondingly describe qualitatively new, extended *properties of reality* actually accounting for the 'non-computability', 'nonintegrability', or 'non-decidability' of its *fundamentally simplified* representation in the non-extended approach. The dynamic multivaluedness paradigm confirms this general conclusion and specifies the conceptual novelty hidden behind the conventional 'non-computability' and other 'difficulties' as the *dynamically multivalued*, self-developing entanglement of interacting entities within any system, behaviour, or phenomenon (Chapters 3-5). It is the unreduced dynamic complexity of a problem (and the real system it describes) that constitutes the true, universal and causal, origin of all the 'irresolvable' difficulties of the unitary problem reduction. The effective complexity correspondence between a problem and the totality of tools applied for its solution (including the explicitly man-made, 'analytical' parts) also becomes thus universally and transparently justified by the ensuing complexity correspondence principle, which states that the computation process complexity should exceed that of the computed system/behaviour.

This *actually unavoidable* transition to classicality in the dynamics of any real, practically useful 'quantum' machine (including quantum computers) means that essentially quantum dynamics can exist *only* as an integral part of a real quantum machine that will necessarily contain also essential classical parts/stages, *naturally emerging* as various bound states of initially purely quantum (unbound) components by the complex-dynamical, totally realistic interaction mechanism of (generalised) 'quantum measurement' [1,10] (which has nothing to do with the artificially inserted, abstract 'decoherence' of conventional theory, see Sections 4.7, 5.3(C)). Therefore the real, natural or artificial, quantum machines are *essentially* hybrid devices in their *internal dynamics*, which is quite different from the conventional theory expectations with respect to 'natural' possibility of essentially quantum machines (also at higher, macroscopic levels like e. g. in the 'quantum brain/consciousness' hypothesis), where the emergence of classicality can be 'somehow' added as a result of 'quantum measurement' performed rather *outside* of the proper system dynamics and remaining characteristically 'mysterious' in detail.

It is important also to emphasize the essential difference between this clearly specified, dynamically multivalued character of irreducible complexity manifestations in micro-machine dynamics and various 'guesses' about complexity ghosts, or 'signs', within the unitary quantum theory (e. g. [192-195,345-352]), based rather on empirical intuition about natural 'quantum machines', but actually using unitary and highly speculative imitations of 'complexity' that remain fundamentally incompatible with any unreduced complexity manifestations in natural micro-systems (Section 7.1). All those unitary imitations of explicitly complex-dynamical, irregular and diverse micro-systems, including 'quantum neural networks' [192,193, 347-349], 'synthetic' and 'smart' quantum structures [194,195, 348], 'quantum biocomputers' [352,353] and unifying quantum 'biologic' principles [354], demonstrate especially grotesque, surrealistic contrast between the general promises and particular results of the unitary, 'exact' science paradigm, revealing the sheer contradiction with its own 'criteria of truth' and the elementary consistency demand.

## 7.3. Universal direction of system evolution, causal interpretation of intelligence and transition to creative computation/production processes

The described qualitatively new properties of real quantum machines, based on the explicitly chaotic (dynamically multivalued) character of the driving interaction processes, change completely the meaning itself of micro-machine operation and related concepts of their creation, use and control. As shown above, the unitary, regular control assumed in conventional theory is fundamentally impossible. On the other hand, the unreduced complex dynamics has its own, natural law and 'direction' of development corresponding, according to the universal symmetry of complexity, to the permanent growth of complexity-entropy at the expense of equal decrease of dynamic complexity-information (Section 7.1). The unavoidable emergence of classical behaviour within the essentially quantum dynamics (in the form of elementary bound states) is an example of such natural complexity development within the real micro-machine. The details of complexity development process depend on the system interactions (including outside influences), in agreement with the universal formalism of the unreduced science of complexity (Chapters 3, 4, Section 7.1). Therefore the regular 'programming' and control of unitary, conventional machines is replaced for any explicitly complex-dynamical (multivalued) machine, including all quantum (hybrid) machines, by the *universal criterion of optimal (or desired) complexity-entropy growth* realised through the proper modification of system interactions, which should be considered, however, only in their unreduced, dynamically multivalued (essentially nonunitary) version. One always obtains, for that kind of machine, the explicitly coarse-grained, dynamically random system evolution, unpredictable in details and irreversible. In return, the system possesses its own intrinsic *creativity* (the capacity for self-development, or dynamic adaptability) and related exponentially high efficiency of *real*, 'incoherent', dynamically multivalued evolution (see Section 7.1 for detailed estimates).

This superior efficiency of creation by unreduced complexity development with respect to its quasi-unitary reduction within the ordinary, regular machines is the evident result of the main feature of dynamic multivaluedness itself, explicitly appearing at the truly chaotic stages of natural interaction processes and adding many *supplementary*, real and dynamically created state-possibilities to the single realisation of the unitary operation scheme. Since the developing interaction complexity naturally forms the multitude of its dynamically related levels with many realisations at each of them, we obtain indeed *exponentially many* realisation-possibilities for the explicitly complex system instead of only one 'linear' realisation of quasi-regular machine: $N_{\text{tot}} = (N_1)^n$, where $N_{\text{tot}}$ is the total number of system realisation-states, $N_1$ is the average number ('geometrical mean') of realisations at each level and *n* is the number of levels. It is important that hierarchical complexity levels develop the fractal structure of their realisations *dynamically*, i. e. in correspondence, guaranteed by interactive multivaluedness, with already existing structures (this is the property of *interactive, or dynamic, adaptability*, Section 4.4), so that the system develops its structure in the direction of the 'right solution' by actually performing only a 'linearly' small part of all the exponentially large number of complex-dynamical 'operations' (probabilistic realisation switch), which provides the complex-dynamical extension of conventional, unitary 'parallelism' in the machine (especially computer) operation, necessarily based on propor-tional, mechanistic *addition*, or (linear) superposition, of individual unit powers. One should not forget also that deeper-level realisation switches occur at progressively (exponentially) decreasing time scales, which are determined by corresponding interaction splitting itself, eqs. (67), so that the whole exponentially large hierarchy of $(N_1)^n$ realisations is 'processed' by the system 'in the same time' (comparable with the duration of a higher level realisation switch), which reveals another aspect of extended, dynamically multivalued parallelism of unreduced dynamics. Due to the hierarchical, dynamically determined structure of unreduced interaction complexity, the 'computing' power contributions from whole separate *levels* (rather then single units of one level) are thus *multiplied* (rather than added), which gives the exponentially high efficiency of properly oriented multivalued dynamics for both quantum and classical degrees of freedom, without any 'quantum' or other 'miracles' from conventional theory (see also Section 7.1).

Correspondingly, the result of any explicitly chaotic machine operation, including that of real quantum (hybrid) machines and neural networks (brain), cannot be reduced to any unitary 'calculation' or 'information processing' expressed in purely abstract 'numbers'. It always contains really emerging, tangible and unpredictable in detail (asymmetric) structures consisting from permanently changing system realisations and constituting 'material products' of system operation, with their inimitable properties and internal complex-dynamical 'life' (Section 7.1). The underlying complexity development process can be described as an intrinsic, inseparable mixture of the essentially inexact 'calculation' of the necessary parameters of the 'adaptable production line' and the production (creation) itself subjected to the same uncertainty in details (but not in the general process direction, determined by irreversible complexity unfolding). Therefore the purpose and strategy of construction and complex-dynamical control of real micro-machines (and other explicitly chaotic machines and processes) is quite different from the mechanistic unitary 'instructions' directing the system along the regular, shortest trajectory towards the desired, exactly predetermined state. We do not need any more to mechanistically and senselessly separate the processes of unitary calculation with questionable efficiency and the following production stage, but can concentrate on the final purpose of production within a single process, leaving to the complex machine dynamics itself to choose the optimal way to the final goal in accord with the universal symmetry of complexity (usually this way will be extremely involved in detail, dynamically fractal, probabilistic and therefore 'non-computable' for the unitary machine and approach).[40] Not only despite the multiple 'errors' the explicitly chaotic computing system can correctly and efficiently unfold its hidden complexity and thus 'solve a difficult problem', but also *only due* to the intrinsically random, dynamically driven deviations can it possess the superior capacity of creation, inaccessible for any ordinary, regular machine and including such features as autonomous structure development according to the inherent 'purpose', 'desire', or 'élan' (the 'free will' property) and ability to understand (intelligence, consciousness).

The ordinary, unitary 'programming' of a fixed structure and regular instruction list for the conventional computer is replaced, for the explicitly multivalued (including quantum) machines, by (1) the proper choice of initial interaction configuration, determining the stock of system complexity, in the form of dynamic information, to be developed into the final form of dynamic entropy/structure (the complex-dynamical 'hardware', or 'device configuration') and (2) additional control (complexity introduction) in some key, 'turning' points of interaction development that can be realised with the help of *analytical* description and *causal* understanding of the universal science of complexity (Chapters 3-5, Section 7.1) assisted, where necessary, by numerical calculations on a usual, classical and quasi-unitary computer working within the unreduced complexity description (the complex-dynamical 'software', or 'user programming'). It is important that both the detailed process development and its general direction/purpose will be determined by the single, universal criterion of explicit complexity development (optimal growth of dynamic entropy). As follows from this criterion, complex-dynamical (multivalued) processes determine any machine operation, including usual, quasi-regular machines/computers. Indeed, programming of ordinary devices should always be done by programmers with explicitly complex, multivalued dynamics (actually represented by human brain operation), which remains 'in the background' in that case, but is indispensable just because it introduces the necessary complexity. Therefore the realisation of *both* (generalised) hardware *and* software for *any* real machine/computer is based on essentially complex-dynamical, multivalued interaction processes, but in the case of conventional, externally 'regular' machines the unreduced complexity manifestations remain more hidden within the detailed computation process or its human preparation dynamics. Note that the unitary approach in general and especially its application to computation process description are based on the opposite picture, where not only 'small' but inevitable deviations from regularity of the ordinary, quasi-unitary machines/systems are neglected, but also the relatively high and *obviously* irreducible irregularity of explicitly chaotic computation dynamics (both quantum and classical) is deliberately replaced by a unitary imitation, usually in the form of external, artificially inserted 'stochasticity' subjected to regular 'control' procedures.

---

[40] Note that even the operation of usual, classical and quasi-unitary computers can be considered as a limiting, degenerate case of multivalued dynamics, which approaches closely enough to the limiting case of SOC regime (Section 4.5.1) with very dense realisation distribution within each emerging quasi-regular structure. The high enough (though never complete) regularity and pre-determined parameters of macroscopic structures within the usual computer permit their interpretation in terms of abstract, dimensionless 'bits'. The situation changes, however, in cases of unavoidable 'halts' or 'non-computable'/'unsolvable' problems, where each conventional computer at each particular state behaves specifically and unpredictably, suddenly revealing the unreduced, though indeed largely hidden, complexity of its 'computation' dynamics.

The described complex-dynamical 'autonomy' of explicitly chaotic machine dynamics that chooses itself the 'best' way to the final goal of total complexity unfolding (with a finite risk of local 'impasses') has the property of elementary 'understanding' by the system of its own dynamics and purposes, *inherent* in the system, as opposed to the completely 'stupid', senseless and mechanistic character of any quasi-unitary, regular machine operation (and any unitary construction/scheme in general), which should be totally administered from the outside, through the imposed deterministic 'program'. This actually brings us to the causally complete, objective interpretation of the property of *intelligence* and *consciousness* in the universal science of complexity [1,4]. Intelligence can be consistently defined as the *total dynamic complexity* of a sufficiently autonomous system from a high enough level of complexity (entering thus in the category of intelligent, or *cognitive*, systems) [1], which gives automatically the universal *quantitative*, dynamically derived *measure of intelligence*. The universal symmetry of complexity shows immediately (see Section 7.1) that intelligence is a quantitatively conserved, but *internally, qualitatively developing* system property (during its finite lifetime), with its unceasingly and irreversibly *growing* part of *wisdom (knowledge)* equal to the system *dynamic entropy* and diminishing part of *cleverness (keenness, eagerness to know)* expressed quantitatively by *dynamic information*.

Note that in the usual case of a reasonably open cognitive system one can include the effective 'environment' into the system while applying the above definition of its intelligence. However, the system intelligence as such is always basically determined by its *internal* (total) complexity fixed at birth (system emergence), while the necessary interaction with the environment plays rather the role of a necessary 'trigger', or 'catalyst', of development that can introduce only small direct changes to the total system complexity (even though relative variations of the current level of dynamic *entropy-wisdom* at *first* stages of development can be greater). In fact, this property of cognitive system, i. e. relative independence of its intelligence from the environment, constitutes an integral part of its definition, specifying the demand for a 'high enough' level of its complexity [1]. These conclusions of the universal science of complexity enter in fundamental contradiction with a recently proposed concept of intelligence representing the best results of conventional, unitary science of complexity and based on the popular method of computer simulations with a group of competing 'agents' [355]. It is concluded that intelligence is 'inseparable' from the environment and totally determined by system (or 'agent') interaction with it, being defined by an 'operational', rather than analytical, criterion of 'victory' of a 'more intelligent' agent in a market-like competition, obtained as a result of its more successful cheating on its less 'advanced' colleagues and depending essentially on various, generally random external factors [355]. As a matter of fact, the contrast between the two approaches is not surprising at all and only reflects the corresponding difference between the unreduced (dynamically multivalued) and imitative (unitary) versions of complexity, thus confirming the former. Indeed, the effectively one-dimensional, zero-complexity, mechanistic 'agents' from any unitary version of 'complexity' are always totally dependent upon 'undirected', random environmental fluctuations just because of ultimately low, zero value of their true, unreduced complexity, this situation being quite similar to various mechanistic imitations of 'complexity' and 'chaoticity' in the unitary science (see e. g. Chapter 6). Note, however, that when such higher-level notions as intelligence intervene in the ordinary play of words of the unitary 'science of complexity', a deeper, 'human' preferences and qualities become more evident and come out of their usually hidden (but always existing) 'background' within a formally 'objective', but actually quite *intelligent* (in the *desired* sense) scientific study (see also Chapter 9).

It is not surprising that such environment-dependent cheating, 'minority games', or any other unitary imitation of intelligence cannot afford any reasonable definition of consciousness representing a superior level of intelligence. By contrast, the universal science of complexity is not limited 'from above' and therefore the causally complete concept of *consciousness* is naturally obtained as the same, *unreduced dynamic complexity* (of an autonomous, cognitive system), but starting from a still higher (than simple intelligence) level of complexity [1], encompassing system interactions with, and *classically fixed (localised)* knowledge about, a much larger, 'embedding' reality (the 'world', or 'universe'). Due to the hierarchical, 'emergent' structure of unreduced complexity (Section 7.1), this means that consciousness includes (developing) intelligence and that a system can be quite intelligent, but not really conscious (or 'human'), which only confirms, of course, empirical observations and certain intuitive expectations. In a similar way, the real, complex-dynamical consciousness (contrary to its unitary imitations) includes itself a hierarchy of smaller and bigger levels appearing through the actually accessible scale of unreduced reality comprehension, which is limited from below (by the maximal practical, globally chaotic, 'animal' intelligence), but need not be limited from above.

Consciousness realisation within the human brain (providing its unique known case) can be described in more detail in the universal science of complexity by application of its unified approach to the unreduced electro-chemical interactions between and within the brain cells, which leads to the generalised Schrödinger equation as a result of causal quantisation procedure (see eqs. (68)-(73) from Section 7.1), where the generalised wavefunction $\Psi$ describes now the real, fractally structured electro-chemical wave field of intelligence/consciousness, or the *brainfunction* [1,4]. The complex-dynamical reductions, or 'collapses', of the brainfunction, occurring by the universal mechanism of essential nonlinearity development (Sections 4.2-3) at the corresponding, macroscopic level of electro-chemical interaction in the brain, appear in the form of 'suddenly' emerging 'impressions', 'thoughts' and 'ideas'. As always, the dynamically multivalued, fractal and probabilistic entanglement of interaction components (eqs. (20)-(31)) constitutes the key feature of the complex-dynamical system description, distinguishing it qualitatively from any formal imitations of unitary science around fundamental understanding of brain operation 'complexity' and providing, in particular, the essential properties of conscious brain operation, which remain basically, causally 'inexplicable' within the unitary approach [7]. Consciousness emergence from the lower complexity level of non-conscious, 'animal' intelligence can be understood, in particular, as a higher-level analogue of transition from essentially quantum, delocalised to classical, permanently localised behaviour of elementary bound systems (Section 5.3(C)).

These results can evidently have a multitude of constructively oriented, creative applications in various branches of 'brain science', including especially the use of the obtained causally complete, unified understanding of real brain operation for creation of artificial intelligent and now even conscious systems. We also obtain a *rigorous* proof of *fundamental impossibility* of these problems solution within the conventional, unitary approach, which is confirmed by existing results (see e. g. [5,7]) and points to the necessity and well specified direction of qualitative change in strategic research orientation. In connection to quantum machine problem, it becomes evident, in particular, that any 'quantum' concept of consciousness (see e. g. [15,68-71]) is basically wrong, while any essentially quantum machine *cannot* be 'smart' (cf. [194,195,351]), i. e. possess properties from higher, classical levels of complexity. Both conclusions directly follow from the complexity correspondence principle, being itself a manifestation of the universal symmetry of complexity and stating that a (computing) system cannot simulate/reproduce any behaviour with the unreduced dynamic complexity higher than its own complexity (Sections 7.1, 7.2, 5.2.2). Since causally explained quantum behaviour is limited to lowest complexity levels (Section 4.6), any quantum dynamics could at best reproduce another quantum behaviour with the same or lower complexity.

The qualitative difference between explicitly complex-dynamical (multivalued) operation of intelligent/conscious systems and conventional, quasi-unitary 'computation' becomes thus clearly and rigorously specified within the dynamic redundance paradigm and the ensuing universal concept of complexity [1], which reveals, in particular, the deeply erroneous, totally unrealistic nature of conventional science illusions about 'quantum' and other unitary versions of artificial 'thinking machines' that could reproduce, at least partially, the essential properties of natural intelligence (see also Chapter 8). It is also important that the obtained guidelines for development of a qualitatively new, essentially chaotic (multivalued) kind of machines and technology have the universal meaning with possible applications to various levels of machine/system dynamics and different types of systems: quantum (hybrid) and classical, micro- and macro-, 'computing' (simulating) and producing, 'thinking' and 'practically realising', 'individual' and 'social', etc.

The main, essential feature of this new kind of machines, also present in natural systems, but thoroughly eliminated from conventional, unitary theory and technology, is the omnipresent, autonomous *creation* of new entities (structures and dynamic regimes), with the *unavoidable payment* for this capacity in the form of irreducible, and relatively large, *dynamic uncertainty* in the detailed parameters of the 'products' and their creation process. The created new entities form the emerging system *configurations* described in the universal science of complexity by system realisations, with their well-defined, entangled and fractal, internal structure and a priori probability distributions (see e. g. eqs. (24)-(27)). By contrast, unitary science, intrinsically opposed to the natural system creativity, cannot find the unreduced, nonperturbative problem solution and is limited therefore to semi-empirical postulation (guessing) of the emerging system configuration and properties, which cannot be efficient in any case of explicitly complex dynamics (cf. e. g. the problem of 'configurational space' and other canonical 'mysteries' in conventional quantum mechanics [1-4]).

The unavoidable dynamic randomness of explicitly complex dynamics appears not as a shortcoming, but rather as an advantage at this superior level of science and technology and can actually be regulated in a necessary way (but not eliminated!) with the help of universal formalism and general results of the new science of complexity (Chapters 3-5) [1]. It is interesting to note, in particular, that essentially quantum stages of real, chaotic micro-machines do not critically depend on various external, 'noisy' or parasitic influences, just because of their already present *genuine, purely internal* chaoticity (multivaluedness), as opposed to the fatal role of 'decoherence' for the unitary quantum computer dynamics (including unitary imitations of 'quantum chaos') related to the *absolute*, and therefore infinitely vulnerable, regularity of unitary dynamics (so that the assumed 'decohering' influences should eventually produce a 'chaotic', though ill-defined, state of unitary machine, in contradiction to its postulated principle of action, cf. refs. [14,157-160] and Chapter 6).

Another characteristic feature of the new old machinery with 'living' dynamics is that various levels, scales and regimes of dynamics, remaining separated within the unitary operation, tend now to be dynamically, 'naturally' *unified* within a single system that autonomously evolves and switches between them, realising the optimal complexity development (from dynamic information to entropy) in accord with the imposed initial system/interaction configuration. Thus, as we noted many times in previous Chapters, the lowest, essentially quantum levels of machine operation cannot solve any sensible, useful task without producing dynamically elementary classical (bound and localised) states that can be directly transmitted to user or further evolve, within the same system or 'dynamical factory', into states from still higher complexity levels (as it permanently happens within every living organism or ecosystem). Therefore the essentially quantum stages of a useful 'quantum' machine operation can actually exist only within a hybrid, 'quantum-and-classical' machine, where both quantum and classical stages, as well as their dynamical links are important, contrary to the case of 'purely' classical machines, where the underlying quantum dynamics remains hidden and unimportant because their essential operation (complexity development) starts from classical complexity levels.

Another unifying tendency joins inseparably, within a single complex-dynamical process, what is called 'computation' and 'production' in the unitary technology. Indeed, we have shown above (Sections 5.2.2, 7.1, 7.2) that the canonical, regular, unitary 'calculation' is impossible within any real quantum or other explicitly chaotic system, whereas its actual dynamics does *consist in* unceasing production, or creation, of new structures and dynamic regimes. The necessary 'computation' of their optimal sequence and parameter adjustment is naturally incorporated into the same complex, multi-level, fractal dynamics, which simultaneously constitutes the essence of autonomous, complex-dynamical, nonunitary (probabilistic) system '(self-)control'. Correspondingly, the conventional, unitary, 'straightforward' and regular control is never possible in its pure form, even for quasi-unitary (SOC) kind of dynamics, and becomes practically senseless in the case of explicitly chaotic machines. The omnipresent and evident examples of the 'mixed', 'computation-and-production' kind of operation are provided by many natural micro- and macro-machines, such as a single living cell or the global brain dynamics, the latter case clearly demonstrating the 'mixed' character of intelligence/consciousness that unifies such functions as 'calculation', 'memorisation', creation and rearrangement of results into one, inseparable interaction process, in sharp contrast to conventional, quasi-unitary computers.

It is important that the outlined qualitative specificity of explicitly complex-dynamical machines obtains a causally complete, reality-based explanation within the dynamic redundance paradigm, as opposed to purely empirical, intuitive guesses, pseudo-philosophical speculations and postmodern, 'fashionable' and 'advanced' plays of words by self-chosen 'priests' of unitary science, so abundantly appearing lately in the most 'solid', professional and popular science sources. Since the empirically based technology development has definitely brought civilisation, just at the present particular moment of its development, to direct practical exploration of those 'explicitly complex-dynamical' (chaotic) regimes of behaviour (such as 'quantum limit' in micro-electronics, or brain dynamics, or genetic manipulations, or the planetary ecology and development levels), we can state that a 'sustainable', definitely and provably progressive future development of science and technology (and thus life in the whole) can only be realised by passing, starting from now, through the profound transition to the described 'creative computation/production processes' in practically any kind of activity. In the next Chapter we specify the essential details of this qualitative transition concerning especially micro-machines and taking into account some recently advanced 'initiatives' of unitary science (such as 'nanotechnology'), which remains unchanged in the intrinsic limitations of its basic, dynamically single-valued and mechanistic approach, but often tries to dress them in the fashionable clothes of a 'new paradigm'.

# 8. Dynamically multivalued, not unitary or stochastic, micro-machines as the real basis for the next technological revolution

Although applications of conventional quantum mechanics to micro-system description have been concentrated lately on 'quantum computers' as a real quantum machine prototype, it is not difficult to see that other branches of applied science and technology quickly converge to similar types of ultimately small structures representing just another kind or aspect of micro-system composed of *essentially quantum* (i. e. explicitly multi-valued) elements. One can refer, for example, to the actively promoted and rather smeared group of concepts unified under the term of 'nanotechnology' and studying possibilities of creation of various 'nanomachines' and 'nanostructures' [356-359] based on the original concept by Richard Feynman [360]. Whereas many objects and structures with characteristic element size of nanometre scale ($\sim 10-100\,\text{Å}$), encompassing just a few typical interatomic distances in a condensed phase of matter, can now be really produced and studied in detail (up to one-atom manipulation), the central, widely announced and 'truly fantastic' ambition of nanotechnology concept [356,357] goes much farther and apparently encounters a strong barrier of practical realisation: it deals with creation of *active*, productive *machines* conceived as nanoscale copies of ordinary, macroscopic mechanisms with a similar spectrum of complicated, locally complete dynamical functions, as opposed to only passively used, static (and often quasi-regular) 'nano-structures'.

Yet another direction of 'spontaneous' quantum machine emergence today comes from molecular biology, including especially practical genetics and related micro-system analysis (see e. g. ref. [361] on this subject discussed here in its conceptual, rather than technical, aspects). This tendency towards natural, already existing nanomachine understanding and control could, of course, be expected and makes progressively its way among more artificial approaches (while retaining usual limitations of unitary science, see the end of Section 7.2 for discussion and references). It is important to emphasize that both artificial and natural micro-machine studies are performed inevitably within the same over-simplified, dynamically single-valued (unitary) approach of conventional science, meaning that 'machines' and systems of *any* origin and size are always basically interpreted in the extremely *mechanistic* sense of a totally regular, 'clock-work' kind of *effectively one-dimensional*, sequentially acting mechanism (including the artificially inserted, non-dynamical, *external* 'randomness', 'chaoticity' and 'multi-stability'/'parallelism'), exactly as it happens within the conventional version of a particular kind of micro-machines, quantum computers (see e. g. refs. [362-364] for general presentations of the field and Chapters 2, 5, 7 for detailed discussions and references). This conclusion and the underlying causally complete picture of complex (multivalued) dynamics of any real micro-system (Chapters 3-7) remain totally valid for and applicable to conventional *genetic* studies and related *living cell (organism) dynamics*, which generally follow the same reduced, unitary, sequential logic of usual, mechanistic, predetermined and non-creative 'programming' deprived of unreduced, strong-interaction effects, despite many 'general' speculations about genome 'interactions' and 'complexity' understood as a loosely interpreted 'intricacy' and the evident qualitative difference of living and biological 'machines' from any man-made device, including multiple, well-known 'enigma' around emergence, evolution and essential dynamics of life.

Since our approach and its results are intrinsically universal (Chapters 3, 4), we shall concentrate our present discussion of practical micro-technology strategy on the most popular and 'hot' idea of nanomachines (nanotechnology), while noting that all the main conclusions and their fundamental substantiation are equally well and directly applicable to other micro-technology directions (including its biological branches) and have already been outlined and practically specified in previous Chapters for the case of quantum computers.[41] It is important to emphasize, in this connection, that the unitary nanotechnology ideas also have their underlying, more 'fundamental' (though completely erroneous) basis in the conventional, unitary theory development, starting from the corner-stone formulation of the idea by Feynman [360] (it is actually only repeated with variations in the modern 'practical nanotechnology' approach [356-359]),[42] then passing by the quantum technology/revolution branch [362-365], up to post-modern, pseudo-philosophical speculations around 'quantum brain', 'quantum consciousness', 'quantum self', 'quantum society', etc. [366,367] (see also [15,68-71,350]). It is not difficult to see that those unlimited 'quantum' fantasies of unitary approach are closely related to a yet more general, deeply mechanistic way of destruction of real-world complexity in favour of 'spiritual machines' persistently promoted within certain direction of thinking, which is hardly distinguishable from one-dimensional imitations it tries to impose (see e. g. [368] and further discussion in Chapter 9).

---

[41] It may be worthy of recalling that the prefix 'micro-' ('machine', 'technology', etc.) in our present terminology actually includes all 'small' structures and scales (such as 'nanostructures') for which the unreduced dynamic complexity appears in explicitly nonunitary (dynamically multivalued) forms, including essentially quantum, hybrid and explicitly chaotic classical types of behaviour.

Returning to the situation in conventional nanotechnology [356-359], based on such 'revolutionary' ideas as 'nanorobots', 'molecular motors', or 'atomic assemblers', we note the relation with our general results (Chapters 3, 4, 7) applied above mainly to quantum computers (Chapters 5, 7), but actually valid for any kind of real micro-machine. It has been rigorously shown, in particular, that any essentially quantum behaviour (inevitably determining major stages of nanomachine dynamics) is characterised by causally derived and relatively large ('global') dynamical randomness and discreteness ('quantisation') due to the dynamically multivalued entanglement between the interacting system components. These irreducible manifestations of genuine quantum chaos (Chapter 6) destroy all the key as-sumptions of the conventional nanotechnology concept, such as unitary design, 'programming' and 'control' of a basically regular, sequential (i. e. unitary) machine function realised according to the desired pattern that should be predictable in any its essential detail. The real micro-machine dynamics has a *qualitatively different character* described by the hierarchical, *creative* and *unpredictable in detail* (explicitly multivalued) process of complexity development (Chapter 7), which can be represented mathematically by the dynamically probabilistic fractal [1] (Section 4.4), inaccessible to any unitary imitation. Note that this conclusion remains valid with respect to classical elementary stages of nanomachine dynamics because they are also characterised, due to the main concept of 'ultimate smallness', by an intrinsically coarse-grained structure, where the universal action quantum $\Delta \mathcal{A} = h$ of essentially quantum dynamics is replaced by a non-universal, but relatively large action quantum $\Delta \mathcal{A} = \Delta x \Delta p$, $\Delta x$ and $\Delta p$ being the characteristic minimal increments of element (e. g. atom) position and momentum.

---

[42] It is interesting to note that the conventional nanotechnology idea comes thus from the same, widely advertised 'genius of theoretical physics', R.P. Feynman, who was at the origin of the unitary quantum computer idea [18] evoking eventually all (post-) modern 'quantum miracles' (on paper) and quantum mystification (real). These two extraordinary 'successes' of unitary thinking are not occasional and just continue the series of previous 'breakthroughs' promoted with the invariably high intensity of praise. Thus, 'Feynman path integrals' originate from improper extension of perturbation theory results beyond the region of its validity (which leads to the 'false exponential dependence' [1], see also the end of Section 7.1), while they do not resolve any of the canonical 'quantum mysteries' and remain totally within its purely abstract formulation in terms of mathematical 'spaces' of 'state vectors' (the same is actually true for path integral applications beyond quantum theory). Finally, 'Feynman diagrams', probably the most popular 'achievement' from the same series, are nothing but 'figurative' and largely 'intuitive' (subjective) representation of perturbation theory results, preserving all its 'fatal' limitations, such as omnipresent divergence and inability to describe any truly 'interesting', qualitative novelty, whereas the 'intuitive' construction of such 'tools' as diagrams certainly opens much larger space for unjustified extension of result validity, play with 'free parameters' and vain, 'philosophical' speculations. The illusions of unitary quantum computation and deceptive abundance of "room at the bottom" [360] (today's 'nanotechnology' hype) represent therefore just a small (and let's hope final) part of a long-lasting flux of losses generated by only one, disproportionally boosted and unreasonably canonised 'high priest' of unitary thinking whose influence largely dominates until now over its well-organised troops remaining 'brainwashed by Feynman' [369], even more than they can imagine (see also Chapter 9).

One may note also that the naïve fantasies of conventional nanotechnology concept trying to directly, mechanistically endow micro-machine dynamics with the properties of ordinary, macroscopic mechanisms are fundamentally deficient because they represent *explicit violation* of the universal *complexity correspondence principle* and the underlying symmetry of complexity (Sections 5.2.2, 7.1, 7.2), which state that a lower-complexity, e. g. quantum or simplest classical, system cannot reproduce a higher-complexity, 'macroscopic' behaviour in principle, since the difference between the two just determines the essential difference between the conserved complexity levels. A loosely formulated version of complexity correspondence principle implies that the number of 'useful things', or actions, a mechanism can produce is roughly proportional to the height of its complexity level within the total hierarchy of complexity and thus also, in average, to the characteristic length scale at that level because those things or actions are represented by system realisations whose number determines the unreduced dynamic complexity of the system (Section 4.1). Therefore the ultimately small systems, nanostructures and nanomachines, can be useful only in producing the simplest, elementary operations, which usually need then to be directly utilised and amplified within the connected hierarchy of larger structures with growing complexity, as it actually occurs in all natural, biological microsystems exemplified by the hierarchy of cell structure and dynamics.

This means, in other words, that the ultimately small parts of a machine or device cannot in principle be as autonomous as usual, macroscopic constructions, contrary to the main 'motivating' ideas of conventional nanotechnology [356-360]. The underlying unitary approach statement [360] is therefore basically *wrong* and the 'room at the bottom' is *strictly limited* by the relatively low maximum complexity and thus diversity of functions it can contain, which is a typical limitation to illusive unitary 'miracles' imposed by the realistic, complex-dynamical (multivalued) analysis of unreduced interaction processes (see Chapters 5, 7 for similar conclusions for quantum computers). Indeed, complexity can be most consistently measured in the units of generalised action (Section 7.1), so that the characteristic complexity value is proportional to the characteristic system size and momentum (see above), where the latter is limited by the characteristic binding (interaction) energy in the system (i. e. by the condition for the system to remain intact as such).

One should note here that the comparison should be made between respective system dynamic regimes that can be closer either to the global chaos (Section 4.5.2), like it occurs in the case of living micro-system dynamics, or to the multivalued SOC (Section 4.5.1), as it is the case for the ordinary macroscopic machinery (including, of course, conventional 'microprocessors' and other formally 'small' units). The explicitly chaotic dynamics can produce complex enough effects already at the nanoscale size of elementary cell structures, but it absolutely needs the unreduced, *dynamically multivalued* analysis for its adequate description and control. The unitary approach of conventional nanotechnology refers to the pseudo-regular, SOC type of device and it is this kind of structure that needs much higher complexity level to show a sufficiently involved (and *externally regular*) behaviour.

Now one can clearly see the *objective, unavoidable origin* of the 'fatal' reality simplification within the conventional nanotechnology concept: it is the *dynamic single-valuedness* of the whole unitary science approach, which artificially neglects, within its invariably perturbative analysis, the possibility of any serious, qualitative change (i. e. *explicit emergence of a new entity*) as a *major* result of arbitrary, generic interaction process, nec-essarily involving the *irreducible, intrinsic randomness*. That's why our objection against the conventional nanotechnology idea exceeds considerably any particular, technical details or formal, postulated 'principles' of conventional, unitary science used as main arguments *both for and against* nanotechnological fantasies in the existing professional and popular discussions (see [356-359] and references therein).

However, much more important is the fact that our *causally complete* analysis provides the *universally* applicable, *objectively correct* (consistent and realistic) *problem solution* that should be used to replace the oversimplified mechanistic approach of unitary science and show explicitly the *qualitatively new direction* of nanotechnology development based on the totally causal, realistic *understanding* of detailed nanomachine dynamics. The artificial nanostructure dynamics is not fundamentally different, within this causally complete description, from the natural nanomachine operation: both can be objectively described as *living machines*, which means that the *objective purpose* and inevitable result of the underlying interaction development is the system complexity development, from the 'folded' form of dynamic information (generalised potential energy) to the explicit form of dynamic entropy (generalised kinetic or thermal energy). This complexity transformation describes *explicit, dynamically probabilistic emergence of new entities* and levels of complexity, impossible in any version of unitary approach in principle. Therefore the illusion of 'exact' unitary programming/design of conventional nanotechnology is replaced by the explicit *creation design* and control with the help of unreduced, dynamically multivalued analysis of the universal science of complexity [1] (Chapters 3-7), which implies a 'mild' control and 'approximate' kind of change introduced rather at certain, key points of intrinsically unstable interaction dynamics. This qualitatively new, complex-dynamical system construction and monitoring is governed by the universal criterion of optimal complexity development (Chapter 7) realised by hidden complexity introduction in the form of properly modified interaction potential (system configuration). By contrast, the unitary, 'totalitarian' control of conventional approach directly *contradicts* the property of system creativity and actually totally suppresses it (with the exception of rarely occurring and undesirable machine *failures*).

It is evident that the described general picture and detailed mechanism of real micro-machine dynamics change completely the strategic directions, purposes and accents in practical micro-technology development. One line of this development is based on the permanently growing miniaturisation of microstructure elements that approach today the 'quantum limit' (usually attained just within the nanometre size scale), where essentially quantum effects become unavoidable. Our results show (Chapters 5, 6) that the pseudo-unitary character of conventional, classical micro-machine (being in reality a multivalued SOC regime, Section 4.5.1), cannot be preserved in the quantum domain and will inevitably be replaced by the qualitatively different, essentially chaotic (explicitly multivalued) behaviour, contrary to conventional science illusions about a possibility of unitary quantum interaction dynamics. This means, practically, that one can preserve approximate unitarity (regularity) of usual machines only above the 'quantum limit' (i. e. without any essentially quantum element operation),[43] or else one should pass to a quite new, explicitly complex-dynamical (multivalued) type of micro-machine that can be properly described only within the unreduced, dynamically multivalued approach and formalism (Chapters 3, 4) and will contain *both* essentially quantum elements *and dynamically emerging* classical, chaotic or pseudo-regular, behaviour.

Note in this connection that although classical behaviour, contrary to essentially quantum interaction dynamics, can take the form of a pseudo-regular, 'self-organised' type of dynamics (Section 4.5.1), the tendency towards the true, intrinsic chaoticity increases with decreasing size of classical structures. This result follows directly from the physical origin of the unreduced SOC regime, where the 'regular', embedding shape is provided by a higher-level, generally much larger structure, which cannot be fitted within very small element/system construction. One can see also that the purpose of preserving maximum regularity (pseudo-unitarity) is not only fundamentally unattainable for ultimately small, nanoscale micromachines, but *should not* constitute the main goal of the future nanotechnology development, which can enter into the stage of practically unlimited creation just by stepping outside of unitary thinking limits underlying actually the whole conventional technology. We argue that it is this 'shift of paradigm' from unitary to explicitly multivalued, 'living' machines that determines the truly revolutionary status and wide perspectives of nanomachine creation and control within a clearly understood, intrinsic unification of more 'artificial' branches of nanotechnology and more 'natural' biotechnology fields.

---

[43] Note that it is this way of successful, but *empirically* based (technological) miniaturisation of working machine elements *above* the quanticity and chaoticity thresholds that constitutes the *actual* source of promotion of the conventional nanotechnology initiative, rather than any its 'futuristic', and actually wrong, fantasies. However, there is nothing conceptually or scientifically (fundamentally) new in that kind of 'initiative', which simply tends to say that all those already existing micro-structures have been really useful and therefore one should continue to develop them. One deals here with the invariably repeated way of the exhausted unitary paradigm to prove its importance and ask for huge extra support by making reference to purely empirical advance of technology that actually has nothing to do with abstract manipulations of conventional unitarity (see also Chapter 9).

Moreover, that kind of new progress does not need to be limited to microscopic structures and machines (where it becomes practically inevitable as we have seen), but can and should be extended to the total human technology in the widest possible sense, including not only 'machines' as such, but also living and thinking systems (Sections 7.1, 7.3), man's interaction with 'natural environment', social processes, 'psychological' levels of complexity, etc. It is not difficult to see that one obtains thus a *unique* opportunity for positive solution of the accumulating 'global', 'unsolvable' problems within a unified new way of the whole civilisation progress towards qualitatively higher complexity levels, which can only be based on the unreduced complexity development [1] and should replace the currently dominating 'industrial' level and related unitary thinking that quickly degrade to ultimate 'singularity' of over-simplified, 'machine-like', effectively one-dimensional living and thinking representing actually only inevitable and now close collapse of that kind of civilisation.

It is important to emphasize once more that the outlined bright perspectives of the unreduced, complex-dynamical version of nanotechnology (and eventually any other technology) are totally due to its fundamental difference from the conventional, unitary version and the existing 'objections' against it always remaining within the same, unrealistic, artificially reduced doctrine of unitary thinking (see e. g. [356-359]) and leaving no place for a realistic and consistent solution of arising problems. The same refers to various appearing substitutes for the unreduced dynamic complexity, causally complete quantum effects and other explicit manifestations of dynamic multivaluedness of real interaction processes, those substitutes pretending to bring 'true novelties' and 'new understanding', but actually remaining within the same, dynamically single-valued paradigm of conventional science and only playing with its unavoidable 'mysteries' as if permitting every poorly substantiated fantasy in their interpretation (see Chapters 5-7 for more detailed discussions and references). In particular, the *dynamically derived* and thus *causally defined* randomness of the dynamic redundance paradigm (see e. g. Section 4.1) is replaced in the unitary theory, including official 'science of complexity', by external, artificially inserted (postulated) 'stochasticity', 'probability', associated 'stochastic equations', etc. The ensuing concepts of 'chaos', 'complexity' or 'multistability' and related ideas about 'chaotic computers', 'control of chaos' in various kinds of machines and any other 'applications' of mechanistic 'stochasticity' can only introduce further confusion in the already quite 'mystified' conventional science (see also Chapter 6) and lead inevitably to technological impasses and real danger at the attempt of their serious introduction into practice.[44]

It is important therefore to develop the emerging new, indeed very powerful practical technologies exclusively on the basis of the unreduced, truly 'exact' understanding of real, unreduced interaction processes and use the ensuing qualitatively new concepts, such as the dynamic multivaluedness paradigm, as a guiding line for further empirical studies. Such totally consistent derivation of general direction and methods of applied, technological research is the true, practically indispensable role of fundamental science, dangerously missing today in its fruitless unitary version that becomes the more and more detached from reality and artificially mystified in its 'theoretical' branch, or else remains totally dependent upon chaotic and often dangerous jerks of the blind technological empiricism.

---

[44] It is sufficient to recall the failure of 'controlled' nuclear fusion and 'fifth generation computers' (or 'thinking machines'), 'unsolvable' and dangerously growing environmental and social problems, clearly felt, but actually unopposed dangers of biotechnological manipulations, among many other 'big' problems and related 'mega-projects' of conventional science, all of them being unified by the unreduced complexity manifestations and insurmountable, now clearly explained resistance to the unitary science approach, despite really fantastic quantities of efforts, publicity and financial resources being helplessly wasted for their accomplishment (see ref. [1] for more details). The characteristic lie-theft cycle of conventional *fundamental* science, starting from mega-money obtained with a reference to previous successes of *empirical* technology (which is a lie) and ending with a silently buried mega-failure (which is a theft), continue to reproduce itself (e. g. within the conventional nanotechnology initiative [356-360]) due to the deeply rooted deficiency of unitary thinking and organisation of that kind of knowledge, which in reality is much less irreplaceable, than it tends to state in its 'public relations' discussions (see also Chapter 9).

# 9. Human implications of quantum computation story: Unitary calculations and show-business kind of 'science' vs real problem solution within intrinsically creative knowledge

> *Beware of false prophets, which come to you in sheep's clothing, but inwardly they are ravening wolves. Ye shall know them by their fruits. Do men gather grapes of thorns or figs of thistles? Even so every good tree bringeth forth good fruit; but a corrupt tree bringeth forth evil fruit. A good tree cannot bring forth evil fruit, neither can a corrupt tree bring forth good fruit. Every tree that bringeth not forth good fruit is hewn down, and cast into fire. Wherefore by their fruits ye shall know them.*
>
> Matthew 7:15-20

We have shown, in previous Chapters, that the consistent analysis of the unreduced interaction processes in any quantum, classical and hybrid computer or other machine reveals a great deal of complexity that can now be rigorously and universally defined within the emerging new paradigm of dynamic multivaluedness (redundance) reflecting the totality of really occurring phenomena. It becomes clear also that the conventional, unitary theory is reduced, in any its particular version or application, to a huge, fatal simplification of real interaction processes, where the plurality of permanently changing, incompatible system realisations is replaced by only one, 'averaged' realisation corresponding to the zero value of unreduced dynamic complexity and leaving no place for any qualitative change, or 'event', of new entity emergence and thus for the related useful result expression. Nevertheless, the unitary theory of quantum computation, information and other quantum-mechanical applications, as well as dynamically single-valued description of 'complexity' and 'chaoticity', continue their prosperous development in practically all 'prestigious' and 'solid' establishments of official science, including teaching courses and research in 'top' universities, as well as powerful state-supported 'initiatives' with invariable 'deep hints' on the expected 'top secret' and 'strategic' applications (now often provided with a mystical aura of 'quantum' magic), whereas any attempt of realistic description of the same processes is thoroughly excluded from any 'peer-reviewed', i. e. subjectively controlled, publication of official science, let alone financial support from 'solvent' public or private sources.

The accumulated contradictions between real situation in science and its official practice are so great, evident and stagnating that their persistence cannot be explained only by the effectively totalitarian, 'plutocratic' kind of practical organisation of science (to be discussed below), but should also involve deeper, mind-related aspects, which become thus inseparably entangled with practical research development and results and actually step forward as the factor of main importance, exceeding now the usual dominance of 'practical needs', as it can be especially clearly seen just for the case of quantum computation and related subjects. Of course, the existence of a connection between the dominating 'state of mind' and progress in science and technology is not new or surprising: in a very general and loosely interpreted sense one can describe all human activity as 'machines producing other machines/structures', where the quality of products is certainly determined by that of producers. However, the modern critical stage of *empirical* technology and *industrial* society development capable, for the first time, to *practically*, essentially influence and transform the 'natural' world complexity within its *entire* range of scales, pushes this eternal relation to the boiling point of a 'generalised phase transition' [1], where the existing 'directions of thinking' and fundamental approaches are intensely 'self-organised', on a highly chaotic background, into much more distinct and very divergent, differently oriented tendencies, while the whole system instability is so high that objective or even occasional domination of one or another tendency can easily lead to extremely serious changes varying between fatal, catastrophic degradation and explosive development into a superior level of living.

Returning to the quantum computation story, we note that although the unreduced, complex-dynamical (multivalued) description of the underlying interaction processes involves important conceptual novelty (called here 'dynamic redundance paradigm' [1]), it does not involve extremely complicated mathematical methods and results and emerges *inevitably* and 'naturally', if only one *avoids* to fall into the evident sin of ultimate reality simplification within a version of 'perturbation theory' just killing all the essential, 'nonintegrable' and 'nonlinear', links of the considered interaction process. Correspondingly, the dynamic multivaluedness phenomenon, as well as its *universal* presence in any unreduced interaction process, has a physically transparent, qualitative interpretation (see Section 3.3) confirming its *inevitability* and *omnipresence* in terms of really simple explanation, which, however, continue to be strangely missed or rather deliberately ignored by conventional science. On the other hand, the idea of unitary quantum (or any other) machines enters in a direct and evident contradiction with so many basic, well-established and multiply confirmed laws of conventional science (see Chapter 2) that their strange coexistence and financially prosperous development within the same framework leave at least an 'ironic' impression (cf. [5-7]) of the officially maintained doctrine of 'rigorous', 'objective' and 'honest' character of such kind of knowledge (always evoked in justification of its large financial support from common sources).

The mental, content-related reason for those 'unlimited' contradictions within conventional science can be better understood just for the case of quantum, explicitly complex-dynamical computers, since their analysis inevitably involves higher-level concepts and leads to the causal theory of intelligence and consciousness (Chapter 7). It would be difficult to imagine such counter-productive, massive and evidently inconsistent obsession by the fundamentally wrong concept of unitary dynamics if that very unitarity was not inherent to a certain kind of intelligence and direction of thinking that became dominant in science soon after the 'new physics' breakthrough, which was *actually* performed at the beginning of the twentieth century by the last classical *realists* (see refs. [2-4]), immediately and very actively replaced, however, by abstract *formalists* and mystery-makers, or 'mathematical' physicists, for whom that unitary, 'calculative' thinking and closely related obscurantism were just as natural and exciting as the 'complex-dynamical', 'creative' thinking has always been and remains for the minority of 'physical' physicists (one should, of course, distinguish the genuine realism and creation, oriented towards objective, consistent truth, from their numerous imitations within abstract unitary 'models' and inconsistent fantasies). What the story of quantum computation clearly shows is that one should not (and actually cannot) avoid the adequate understanding of those objectively existing, practically important differences in thinking and levels of consciousness, since they are inseparably entangled with the resulting knowledge content and play the key role in its development, especially at the current critical stage of revolutionary change.

Another characteristic manifestation of intrinsically mechanistic, low-complexity, unitary tendency in thinking takes the form of 'computer physics' (or 'computer science' in general), practically dominating today's development of many most prosperous and ambitious fields of *fundamental* science and implying that the progress in reality *understanding* can be basically obtained by computer simulations alone, which use only simplest, conceptually trivial input data. This 'new kind of science' tends to have really universal ambitions (e. g. [311,370]) and even pretends to dominate unitary complex system studies [371], despite the evident contradiction between the mechanistic, basically trivial character of usual computer operation and the unreduced complexity of higher-level natural, e. g. living and intelligent, systems. That the 'strange' persistence of those contradictions in the official science should necessarily have deeply lying mental roots is confirmed by appearing insights into *that* kind of intelligence, performed within most developed approaches of the unitary science itself. Thus, a computer simulation of interaction processes in the medium of 'intelligent' agents with pronounced 'market' talents [355] arrives at a peculiar conclusion that their intelligence, and by a deeply rooted extrapolation *any* intelligence (natural or artificial), has a purely 'social', *external* origin without any essentially inbred, intrinsic root, since it can be attributed mainly to each agent interaction with the exterior environment, based in addition upon ultimately 'calculative', self-seeking, but also mechanistically randomised (stochastic) type of behaviour. It is evident that such specific, externally driven and thus inevitably parasitic kind of 'intelligence' is objectively opposed to intelligence interpretation in terms of unreduced (dynamically multivalued) complexity [1] (see Section 7.3), but shows, on the contrary, perfect correspondence with the dynamically single-valued, zero-complexity content of unitary science imitations.

> *L'essence des explications mécaniques est en effet de considérer l'avenir et le passé comme calculables en fonction du présent, et de prétendre ainsi que tout est donné. Dans cette hypothèse, passé, présent et avenir seraient visibles d'un seul coup pour une intelligence surhumaine, capable d'effectuer le calcul.*
>
> H. Bergson, *L'évolution créatrice* (1907) [339]

The unitary (dynamically single-valued) theory of quantum computation and the underlying concept of mathematical physics in general are thus evident manifestations of *unitary thinking*, which cannot be strictly unitary, of course, but still definitely tends to zero-complexity, simplified, mechanistic 'calculations', either in abstract mathematical constructions or in computer simulations that apply great speed of calculations to oversimplified 'models', or even in practical matters of research organisation and realisation. That tendency towards technically involved, but effectively zero-dimensional projection of dynamically multivalued reality explains the 'strange' persistence, during the last hundred years, of evident inconsistencies in the whole body of 'exact' modern science, as well as the resulting 'end of science' [5], of *that* kind of science, to be exact (indeed, a zero-complexity world has no place for any development in principle, cf. Section 7.1). The natural or acquired adherence to zero-complexity, point-like world 'model' favours its simplistic reduction to a sequence of 'pure' numbers (symbols, geometric constructions) detached from any other, 'subjective', 'qualitative', material content and equivalent to an 'ideal', regular and sequential (unitary) computer programme and results of its realisation qualified as 'information' (see Section 7.1). However, the unreduced, physical reality – ensuring, by the way, everyday existence of any, even most 'abstract-minded' mathematical physicists – does not want to enter into the narrow framework of mechanistic unitarity, which explains the evident inefficiency of unitary theory applications to phenomena involving explicit manifestations of complexity, such as essentially quantum behaviour or higher-complexity systems showing pronounced tendency towards involved structure creation. This leads the externally dominant, but internally bankrupt unitary thinking to inevitable payment for its ultimate destruction of complexity, in the form of vain and confusing *multiplication of artificial, abstract entities* (attempting to replace the *dynamic* plurality of real world structures) and related mutually incompatible 'theories' absolutely separated from reality (and from each other), as well as intentional, unjustified *mystification* of the *unexplained* and actually *missing* world dynamics (which is especially clearly seen in the case of notorious 'quantum mysteries', cf. Section 5.3, or else astronomical 'dark matter' enigmas [13]).

It is important to emphasize that conventional, 'mathematical' physics, and unitary science in general, is *not* more *mathematically* consistent, or 'exact', than other possible kinds of knowledge, contrary to what is implied by the conventional science paradigm itself and especially by the self-seeking, deceitful propaganda of unitary science merits. In fact, as it is clearly shown within the unreduced interaction analysis of the universal science of complexity [1] (Chapters 3-5), the dynamically single-valued imitation of multivalued reality in conventional science, including its most mathematically 'heavy' branches and 'complexity' versions, corresponds to the *maximum* possible inconsistency of *mathematical* presentation of reality: indeed, unitary science always deals with the *minimum* possible number of realisations by considering only one, averaged and actually arbitrarily deformed system realisation that replaces the multilevel, fractally structured and 'living' plurality of permanently changing realisations of any real system. The persisting fundamental problems of conventional science and its strange dependence on the officially accepted 'mysteries' result just from that ultimate *mathematical inconsistency* of its dynamically single-valued approximation of reality (one can recall the long series of 'rigorously' proved, but actually totally wrong, 'uniqueness theorems' of conventional science as a characteristic example of fundamental, evident inconsistency, now revealed and causally explained in the universal science of complexity [1], see also footnote 3 in Section 3.3).

We see that the difference between the conventional, dynamically single-valued and the unreduced descriptions of reality is not in the choice of mathematical tools used (*basic* tools remain the same and are quite simple), but rather in the *way* they are used in each case. The unitary science of the twentieth century has silently transformed mathematics from a simple, though indispensable, technical *tool*, evidently *blind* as such, into the unique and absolute *purpose* of any fundamental research. This quick, arti-ficial promotion of mathematics from a 'servant' to the 'queen' of sciences has not been useful, however, for either science or mathematics, since while science is artificially put into a state of conceptual stagnation, its technical tool, remaining inevitably blind and simple as it is by its nature, takes now a myriad of ambiguous, arbitrarily produced and perverted forms in the vain hope to guess by chance only the *external* shape of nature's real complexity. One obtains thus the well-known 'uncertainty' of modern mathematics reduced in practice to unlimited cabbalistic delirium.

As for the officially presumed 'great successes' and 'unreasonable efficiency' of the standard, 'mathematical' physics, they can always be simulated by introduction of the necessary number of artificial, abstract entities and rules and their adjustment to the 'experimental data', but *without* any consistent, causally complete *understanding* of real, *physical* world structure and explanation of the *inevitably* emerging para-scientific 'mysteries' and 'strange' postulates, which are arbitrarily *excluded* from the main criterion of truth of unitary science (correspondence between theory and observation). What *is* 'successful' in the unitary science farce is the tricky publicity and 'management' of its 'public relations', using real advances of empirical, intellectually blind technology (applied science) for hiding the disastrous failure of fundamental knowledge and justification of its ever growing, unconditional financial support.

Remaining restricted by its zero-dimensional image of reality, the official unitarity tends to arbitrarily 'guessed' and artificially *fixed* (postulated) *imitations*, or 'models', of natural system properties 'existing' only in *abstract* structures or 'spaces', where purely mathematical 'elements' and 'dimensions', absent in physical reality, serve as exclusively symbolical, conventional substitutes for real, empirically observed entities, so that 'correspondence between theory and experiment' in *separate*, subjectively chosen points is artificially arranged from the beginning, by the *postulated* choice of spuriously *redundant* (rather than *dynamically* redundant!) abstract entities. The growing number of 'justifiable' unitary 'models' also obtains its explanation: the number of single-valued models is determined by system realisation number (actually huge) and thus its unreduced *complexity* (which explains, in particular, the strong, *a priori* failure of unitary science methods at higher complexity levels [1]). The scholar fundamental science considers, however, such grotesque caricature of reality and its mechanistically adjusted, provocatively incomplete 'agreement with experiment' as 'perfect understanding of reality', often equipped with such epithets as 'first principles', 'causality', 'realism', or 'new paradigm' and provided correspondingly with all possible and impossible technical investments,[45] personal profits and honours (despite the stagnating 'unsolvable' problems and scandalous 'mysteries'), which only confirms a quite specific, internally *over-simplified*, but externally, practically infinitely tricky character of the underlying unitary 'intelligence'.

Moreover, being eventually poisoned by its own myth of a 'proven' success, the dominating unitarity becomes *really mad* and tends to claim that it is those grotesquely simplified, deadly fixed and disrupted constructions of conventional symbolism and its abstract 'spaces' with the 'necessary' elements and desired number of dimensions which constitute the fundamentally *genuine*, 'mathematical' reality, whereas all 'ordinary', observed and measured world entities are *only* a sort of its 'envelope', or a 'secondary', superficial reality somehow 'spanning', or mechanistically 'moving through', a 'manifold' of those unchangeable, 'ideal' (but now, in fact, arbitrarily deformed and perverted!) Platonic constructions (see e. g. [372-377]). Any interaction *development*, expressed by the mathematical procedure of consistent *equation solution* (or result *derivation*), let alone its dynamically random, probabilistic version, remains a much too complex, impossible action for that kind of unitary madness, actually just expressing the *true essence* of the whole conventional, dynamically single-valued the-ory with maximum transparency that avoids any artificial and irrelevant appeals to 'realism' performed sometimes within certain versions of the same unitary symbolism. Needless to say, it is just those extreme branches of the unlimited unitary cabbala that dominate absolutely in the departments, centres and laboratories of theoretical physics throughout the world. Insisting on the primacy of their abstract, 'pure' constructions, the masters of unitarity 'suddenly' become surprisingly practical and particularly attached to the rough, material world, demonstrating truly professional use of some of its most dirty practices, once the problem of financial support for their 'disinterested' research is involved. The grotesquely inflated, impertinent and crazy pretensions of unitary thinking champions hide only perverted *emptiness* of spirit and *mediocrity* of thought.

---

[45] Recall e. g. astronomical spending on the search for doubtful, and most probably nonexistent, gravitational waves and other 'cosmological' experimentation in favour of over-simplified, subjectively assumed 'models', as well as so many other 'exotic', unnecessary experiments in confirmation of blind assumptions of today's 'post-modern', purely speculative and occult version of unitary science, oriented to nowhere from the beginning, but selfishly promoted under general slogans like 'the search for truth' or using atavistic beliefs in simple, 'material' irrationality or 'supernatural' physics from the part of unitary system governors (recall the generously sponsored 'antigravity' research within the 'zero-point field' theory of the 'respectful' unitary science, among so many other similar 'initiatives', especially in the high-energy physics, actually nourished by 'secret' hopes for another source of a 'super-power' that could permit 'anyone' who controls it to conquer and rule the world without any progress of intellectual or spiritual capacities). While the latter hope seems, but only seems, to be supported by the superficial 'dominance' of the 'nuclear club' members in the world affairs, their evident inability to solve any of the serious real problems, so dangerously accumulating today (including those in the content and organisation of fundamental science) clearly demonstrates that the true human 'power', providing the unique source of real progress, can only be based on the *totally consistent understanding of reality*, situated far beyond the limits of blind semi-empirical 'tricks' of unitary thinking. The 'developed' unitary science of today can only use the *standard* features of the unreduced complexity, looking 'mysterious' *within its own, artificial limitations*, for self-seeking speculations and deceitful promotion of its 'magic' power.

> *Tandis que, par la force même des choses, s'appesantit sur la recherche et sur l'enseignement scientifiques le poids des structures administratives, des préoccupations financières et la lourde armature des réglementations et des planifications, il est plus indispensable que jamais de préserver la liberté de l'esprit scientifique, la libre initiative des chercheurs originaux parce qu'elles ont toujours été et seront sans doute toujours les sources les plus fécondes des grands progrès de la Science.*
>
> Louis de Broglie, *Nécessité de la liberté dans la recherche scientifique* (1962) [381]

The unique direction of escape from the fatal diseases of stagnating unitary science is clearly specified by the realistic world picture of the universal science of complexity [1]: one should definitely abandon the artificial, subjectively imposed limitations of unitary thinking and apply the mathematically and physically complete, unreduced description of natural phenomena inevitably involving a qualitative, *conceptual* knowledge extension, that of the omnipresent dynamic multivaluedness. On this way one immediately encounters, however, the rigid wall of specific *internal organisation* of unitary science, hidden behind its externally 'liberal' and 'open' façade and closely related to the above limitations of its content. Since the body of unitary knowledge is formed from various *equally* incorrect, zero-dimensional (zero-complexity) projections of dynamically multivalued reality, separated from both each other and the real world structure, its *actual* organisation follows the same simplification and degenerates into effectively totalitarian rule of competing clans of self-designated 'sages', forming veritable 'scientific' *mafias* and subjectively 'chosen' due to their 'best' internal organisation (realised through unlimited *favouritism*) and 'proximity' to the ruling powers and money sources (usually equally corrupt and forming the same kind of structure on the whole society scale, irrespective of external, 'embedding' regimes and ideologies). Any 'success' in that kind of science organisation is determined in reality not by the officially announced 'search for truth', i. e. totally consistent understanding of real world behaviour, but rather by agreement with a subjectively dominating, unitary imitation of reality usually originating from, and therefore exclusively supported by, local and global 'high priests' (mafia bosses) of unitary science.

The universal concept of complexity can be directly applied to science (and society) organisation itself and such application shows immediately that the unitary science content and its dominating, also quasi-unitary organisation type are *objectively unified* by the fundamental tendency of artificially imposed *simplification*, or *reduction*, of natural dynamic complexity. Such relation between form and content of unitary knowledge is not occasional and can be considered as manifestation of the general 'complexity correspondence principle' applied above (Sections 5.2.2, 7.1-2) to micro-machine operation and being itself a corollary of the universal symmetry of complexity. This causally derived principle shows also that any further progress in the knowledge content (towards the unreduced complexity of the real world), apparently indispensable for further existence of fundamental science as such and civilisation progress in general, needs a higher-complexity system of science organisation that should progressively replace the rigid totalitarian structures of unitary knowledge organisation [1] (see also the end of this Chapter).

It is clear from the above difference between unitary imitations of truth and its unreduced, dynamically multivalued content exactly reflecting the real world structure that the latter kind of knowledge, based on the causally complete, intrinsically consistent *understanding* of reality, just have the *minimal*, practically vanishing chances to be accepted, or even simply 'taken into account', by the corrupt unitary hierarchy that cannot, on the other hand, realise any true progress of knowledge, so badly needed today, and thus forms a deadly fixed *jam* on the route of progress. In the evident absence of truly consistent problem solutions and thus any real progress, the official science naturally degrades to thoroughly decorated, 'elegant' lies serving *exclusively* for *subjective*, unmerited ambitions satisfaction of their authors and having nothing to do with the officially proclaimed science goal of 'search for the (objective) truth'. In fact, the massively supported, dominating lie of the unitary imitation of reality acts right against the unreduced truth and any honest attempt to find it.

It is not surprising that arbitrary deviations in the unitary science content are accompanied by equally impressive spectrum of all possible versions of *bad practice* in its organisation and functioning, such as using the notorious 'peer-review' system for efficient and extremely strict prevention of support and publication in 'recognised' printed sources of any professional, but 'free' result, obtained beyond subjective interests of the governing scientific mafias, or 'elites', *especially* if it contains an evident 'grain of truth' able to disprove the 'officially accepted' doctrine and thus presents a potential menace to high positions of its *officially* 'prominent' authors. In that way a special, governing caste is formed and put into a position of uncontrolled and unjust dominance, where it has the exclusive 'right to kill' or to publish anything it wants, combined with the unlimited access to unpublished results of those who are actually deprived from any real possibility of publication in 'official' sources supported, by the way, by public, 'common' money. In this climate of scientific criminality and intellectual, if not financial, fraud the 'usual' moral, unofficially maintained rules and 'good manners' in research, such as reference to previously published, or even unpublished, work by other authors on the same subject, correct authorship, or proper respect for junior colleagues and their rights, become quite *un*usual and practically extinct, which often transforms the real situation within science into an *intellectual prison or concentration camp*. As a result of all those disgusting 'minority games', the real quality of 'solid', generously supported publications of official science is often indistinguishable from a parascientific delirium, including all kind of unitary imitations around occasionally appearing sound, realistic ideas which, however, cannot be accepted for publication themselves, in their original version by a 'rigorous' peer-review selection procedure.[46]

One obtains thus an 'ironic', post-modern science choice between the absence of any really useful, new results (in which case the paper can be published) and the absence of publications in 'recognised' sources (i. e. the true novelty is permitted only if it can be ignored). Although this situation is established now under the cover of 'developed', 'Western' democracy, it practically realises the purpose of the *openly* totalitarian version of unitary organisation that had existed, for example, in the communist Soviet Union, where the obligatory official permission for open publication, delivered and *personally signed* by an *openly* gathered peer-review panel of chief scientists, was concluded with a notorious phrase: "This paper can be published in an open source because it does *not* contain any essential, new result". One needs only to compare the modern content of any of the most prestigious physical journals, especially taken per unit volume, with that of the same journal but edited several tens of years before in order to see the striking resemblance between the purpose of 'totalitarian' and the practical result of 'democratic' peer-review version, i. e. the ever more dominating mediocrity of the essential content.

Although the true origin of the evil, the unitary way of thinking, is now clearly visible behind the official establishment cover, one cannot help wondering at the speed and degree of degradation of the entire 'enterprise' of conventional science, which takes especially large scale just in the most prosperous parts of modern 'technological paradise', where the scientific activity profits from quite comfortable conditions, élite social status and in-creased public care. Whereas much smaller deviations in economic life of the same countries are *still* treated as a serious crime or at least 'misconduct' and *usually* punished or at least 'denounced', the evident forgery and well-known moral debauchery in the official fundamental science can only be mildly criticised in special, 'intellectual' sources practically limited just to 'closed' circles of the 'averted' participants of the process, while they continue to profit from ever growing material support from *public* sources, most prestigious prizes and practically unlimited dissemination of their bankrupt imitations in the press and *all* educational establishments.[47] It follows that what is considered as a serious, unacceptable deviation within the 'classical' set of values becomes now a common, practically acceptable kind of behaviour in the system dominated by unitary thinking and its intrinsic adherents.

---

[46] In that way one obtains the 'sustainable' vicious circle of lie, theft and perversion of the 'developed' unitary system of knowledge and power, where a previous, 'well-established' lie is used for justification of a new theft and perversion producing the 'next generation' of officially supported lie and so on. And when finally the deviation from reality and harmful practical consequences become too big and create a catastrophic system *crisis*, those who have thoroughly developed the devastation present themselves rather as its 'innocent victims' and after having 'severely denounced' their own methods and (sometimes) visibly punished some formal 'top clerks' of the compromised system, enthusiastically give rise to its new version involving new vicious circles of lie. In science the start of the 'developed unitarity' domination, with its characteristic lie-theft cycle, can be clearly traced down to some 'new physics' interpretations at the beginning of the twentieth century, where e. g. an especially 'prodigious' amateur scientist quickly becomes an official 'genius' by producing and surprisingly easily publishing a conveniently formalised, but also fatally simplified compilation of the ideas of other, much deeper thinking scientists, without any reference to their work (the practice considered as unacceptable, severe deviation before that new 'liberation', which revealed quite soon its other aspects and consequences). Being initiated in that way, the 'new physics' development proceeded in the same direction and has inevitably attained the 'end-of-science' situation of today, including the over-simplified, trickily 'arranged' content and unlimited, criminal deviations in practical organisation.

The modern mafia-like mode of unitary science organisation is the inevitable consequence of its split, decaying content and therefore both have emerged, not occasionally, immediately after the definite 'crash' of the 'perfect', classical version of the same unitary science at the beginning of the twentieth century known for its destructive tendencies in all spheres of life. Whereas the true, fundamental origin of that scientific revolution was just the beginning of the end of visible, but actually superficial 'successes' of unitary approach of the classical epoch (known as 'Newtonian science') and the advent of unreduced, multivalued complexity manifestations (in the form of quantum behaviour, relativistic effects, chaoticity, etc., see [1-4]), the emerging 'new physics' doctrine had failed completely to reveal the underlying complex interaction dynamics and had actually reproduced the unitary, dynamically single-valued approach of classical physics, but now inevitably in its *explicitly* incomplete, contradictory and therefore mystified (irrational), abstract and 'paradoxical' (internally frustrated) version strangely resembling most occult branches of medieval mysticism (such as the notorious cabbala), but promoted 'paradoxically' under the brand name of 'mathematical physics' pretending to be especially 'exact' (i. e. 'objective') and 'rigorous' (i. e. 'consistent'). Contrary to the self-conceited statement of modern official science, the 'new' physics (e. g. [378]) represented by the canonical *postulated, abstract* and *irreducibly separated* theories of quantum behaviour, relativity and chaos ('statistical physics') *cannot* be considered as a natural *extension* of classical, Newtonian science, since it has *multiplied* (considerably) the number of *mysteries*, rather than consistent *explanations* (solutions) of old and new puzzles, and this is clearly because it remained completely within the *same qualitative concept* as the classical physics (now *causally specified* as dynamic single-valuedness, or unitary), but has been vainly applied to *explicitly* multi-valued, empirically obtained manifestations of the unreduced complexity of nature. As a result, in modern science, contrary to the classical science, *the more we know, the less we understand*.[48]

---

[47] This situation in science shows, by the way, that the so-called '(developed) democracy', often presented as 'at least a reasonable (or the best possible) guarantee' against *the worst*, the authoritarian, unconditional domination of rulers at power, not only does not provide any such guarantee, but actually shows itself as but a more elaborated, hidden and therefore much more dangerous version just of that absolute, 'infinitely' unfair and unreasonable domination of a group of tricky 'usurpers' of power, against which this same democracy should provide the 'best' cure. Indeed, any most 'democratic' system of *unitary* power, with all its official 'hierarchy' of absolutely dominating *establishment* constitutes evidently just the most 'developed', i. e. decadent and hypocritical, version of the same, unitary kind of social structure [1], where the explicit, 'honest' oppression rule of a 'naïve', 'non-democratic' unitarity is simply replaced by a more involved, but actually not less strict domination of 'bosses' with unconditionally high possibilities (though they may not directly coincide with the officially proclaimed, formally changeable 'governors', being hidden instead within an intricate enough system of monetary and political manipulation). It should not be surprising that, as both history and today's reality invariably confirm, the 'democratic' version of unitarity actually serves only as a transient, precursory stage for the inevitable following advent of its another, open, natural and therefore more stable, authoritarian version (one can compare the total historical duration and real creation results of 'democratic' and 'authoritarian' phases within all known, unitary civilisations). Transition to the qualitatively different, nonunitary, intrinsically creative (progressive) kind of social structure and way of development becomes possible and even irreplaceable today [1], but it has nothing to do with any version of conventional, unitary 'democracy' that provides only mechanistic, formal *imitations* of 'liberty', 'equality' and 'progress' used *actually* for suppression of the real, unreduced version emergence (including any superficial 'liberalisation' of the decadent unitary power, which always leads in reality to the opposite result, i. e. establishment of a particularly 'hard', totalitarian version of unitary rule becoming unavoidable in the atmosphere of omnipresent chaoticity that results inevitably from that *unitary* 'liberalisation').

---

[48] An important nuance should, however, be added to the snapshot of the 'revolutionary' epoch. It appears that the key discoveries of the new physics themselves, constituting its main content and remaining unchanged during a century, such as Planckian quantisation of radiation processes (Planck's constant), de Broglie wave of a massive particle, Schrödinger equation, Lorentz transformations of space-time and first insight into dynamical system complexity by Poincaré, have been performed by convinced *realists* strongly oriented towards 'objective', absolute and universal kind of truth (see e. g. [2-4] and further references therein). However, those solitary giants of the ending classical epoch were extremely quickly replaced, together with their 'old-fashioned' values and 'too long' search for the unreduced, consistent truth, by well-organised, collectively acting and thinking groups of swift-handed parvenus of the new, 'revolutionary' wave and characteristically 'prodigious' flavour who used the confusion of the transitional period and the justified 'thinking pause' of realistic scientists in face of the discovered qualitatively new phenomena in order to promote their unmerited domination by 'quick and easy' kind of answers to all 'big' questions, inevitably involving that grotesque mixture of abstract mathematics, postulated mysteries, unlimited relativism, shameless fraud and dirty politics which actually dominated the whole physics of the twentieth century, despite the long and vigorous opposition to it from the defeated 'giants' [2-4]. It is the modern 'end of science', in all its aspects, which realises the unavoidable payment for that kind of 'progress'.

The century of destruction, with its world wars, atomic horrors, ideological oppositions, mass media, mass production and mass extinction, created a quite prolific ground for the most prosperous growth and unjust domination of fruitless unitary substitutes, using the new, 'collective' modes of their own promotion within various political versions of unitarity deadly ill with the same disease, 'social AIDS', hitting and paralysing first just the central 'units of control' (it is not difficult to see that all those destructive social phenomena and related tendencies of moral degradation always originate themselves from the same simplified, unitary kind of thinking that further uses them for unfair promotion of its one-dimensional imitations, in science and elsewhere). The resulting 'feast of destruction' has, however, a naturally limited life cycle and can continue only until the destruction is complete and one is left with a total chaos of low-level dynamics of decay. At that point, just attained around the recently celebrated 'millennium border', the futility of conventional *fundamental* science becomes especially evident on the background of *purely empirical* technology and applied science successes, in spite of desperate, but less and less convincing efforts of the bankrupt unitary science to attribute the latter to its own (actually non-existing) advances. The difference between the two is clearly illustrated, for example, by the persisting contrast between quickly advancing *practical* development and applications of ordinary, classical and regular, computers and exclusively *academic* 'successes' of conventional quantum computers, despite all their promised 'miracles' and high quantity of efforts applied during the last twenty years. In the meanwhile, the 'liberated' and *technically* super-powerful, but intellectually blind technology modifies today (and inevitably *destroys*) the *full* range of natural complexity, without any idea about its true content and consequences of its change, becoming therefore *objectively dangerous on the global scale*, whereas the *unitary* science remains *enslaved* by the militant empiricism and *cannot* help in principle.

The ensuing justified and critically growing 'public distrust of science' is one of the worst practical consequences of unitary science separation from reality, even not so much because of the underlying intuitive (but actually appropriate) *fear* of unpredictable technology results, but because of the clearly perceived, real and persisting *absence of understanding* of those results, arbitrarily extended to 'science' in general, by not only any 'lay', but also quite highly educated people (including, in fact, professional scientists themselves). Whereas the usual 'political trick' of the dominating science elite consists in infinite (and unjustified) complaining about low quality/intensity of science teaching and popularisation, presented as the main origin of the dramatic fall of younger generation interest in it, we clearly see now that the true, *objectively specified* source of conventional science esotericism is hidden deeply in its *essential content* and is the *same* one as that which separates it from reality it pretends to describe, while making it instead obscure and abstract: this source of failure is the fatally reduced, *scientifically* incomplete, dynamically single-valued image of reality inherent in the conventional science approach. A 'lay' person (but also any 'professional' with a normally 'human', nonunitary kind of intelligence) can well understand and be interested in *both* the observed reality as it is *and* its unreduced, complex-dynamical and hence realistic picture within the causally complete kind of knowledge (exemplified by the universal science of complexity), but the same person (as well as *any* 'education system') cannot be responsible for the intrinsic limitations of the artificially reduced unitary science content just giving rise to all the 'bad consequences' and creating the obscure and abstract, really repulsive flavour of that *particular*, very *special* kind of knowledge (and associated way of thinking). In other words, 'unreduced complexity', 'realism' and 'understandability' (attractiveness) of knowledge *mean the same*, upon which we obtain the scientifically rigorous, exact interpretation of those 'vague' notions around social impact of science, as well as a clearly specified, fundamental reason for the growing and *thus* totally justified public 'distrust' of the *unitary* kind of science (which is still massively imposed as the *unique* possible kind of scientific knowledge, including its reduced, ugly and inefficient, versions of 'complexity').

Therefore, instead of spending lots of public money on the vain publicity campaigns for the *conceptually* dead unitary science, totally *enslaved* by the pure, blind *empiricism*, it is time for the whole 'establishment' to acknowledge the evident facts and start listening to those propositions of change which involve *true, explicitly demonstrated solutions* to fundamental and practical problems (including science organisation itself). It is not by low-level popularisation of a 'physics of beer froth', or fraudulent play of words around 'quantum teleportation', or with the help of artificially forced, massive promotion of 'elegant' lies of unitary imitations of reality[49] that one can hope to regain the motivated public interest in and trust of science, but rather by explicitly proposing the new, realistic, transparent and unified kind of knowledge whose existence is proven *de facto* within the dynamic redundance paradigm by actual demonstration of its fundamental basis and numerous applications to real problem solutions at various levels of complexity [1-4,13] (including the case of micro-machine dynamics analysed here in more detail).

In direct relation to 'public understanding of science' is the popular 'intellectual' talk about the 'two cultures', where the first, 'artistic' culture is represented by the humanities in the university courses and the second, 'technical' culture originates in the 'exact sciences'. The growing rupture between the two cultures was emphasised by C.P. Snow [379] and then permanently discussed, mostly from practical points of view, up to recent attempts to unify the splitted whole within a superficial 'third culture' (see e. g. [380]) always representing, however, only perverted variations of mechanistic 'interdisciplinarity' and cumulative 'erudition'. We can see now that what was intuitively characterised as 'technical' culture, whose understanding is difficult for people with the 'artistic' kind of thinking, corresponds to a rigorously specified *simplification* of reality by unitary thinking, i. e. *artificial* reduction of real, dynamically multivalued world complexity to the zero complexity value of only one, fixed realisation, starting already from the most fundamental, 'physical' entities, which not only are 'difficult to understand', but actually could not exist in the zero-complexity version attributed to them by the unitary science. It is quite natural, however, that the difference between the unitary reduction of reality and its observed manifestations generally grows with the unreduced dynamic complexity and becomes practically unacceptable starting from *high enough* complexity levels, expressing thus the *objective* meaning of conventional knowledge division into 'exact (fundamental) sciences' (lowest complexity levels), 'natural sciences' (medium to high complexity) and 'humanities', or 'liberal arts' (highest complexity levels) [1]. One should also take into account the fact that the first, 'artistic' culture deals with the unreduced dynamic complexity only until it remains at a purely empirical, 'subjective' level. Once it tries, however, to impose a 'science-like' order upon its objects of interest, i. e. to *understand* them, it immediately falls into the same deadly sin of unitary thinking, i. e. complexity reduction, appearing in the form of simplified 'classification schemes' and other fixed, abstract constructions. Therefore in reality there can be only one, universal kind of intrinsically complete and permanently developing knowledge and culture (or understanding) that inseparably unifies within it the qualitatively extended versions of conventional 'sciences' and 'arts' and where the 'non-quantifiable' notions of the latter can be explicitly and consistently quantified, while the 'mathematical' ideas from the former can be expressed in qualitative, physically transparent terms of really existing entities, without losing any 'scientific' rigour [1].

---

[49] Note, in particular, massive introduction of such 'interdisciplinary' subjects and even scientific degrees as 'public understanding of science' into study programmes of many 'leading' universities, which means that the farther unitary science is from real, fundamental problem solution, the more public money is spent by its decadent hierarchy for large-scale propaganda of its ever more obscure and abstract speculations. It is evident that such 'practical measures' as if 'in favour of science' can only give the opposite, negative result, since as recent experience convincingly shows, in reality one cannot deeply influence anybody's consciousness by force and the unaware 'general public' vigorously attacked by the militant unitary scientism can only externally yield to the official pressures, but intrinsically will be only more turned away from support and understanding of such kind of knowledge, dangerously associating it with science in general, with any kind of ordered, not purely empirical knowledge.

It is not surprising that in the persisting 'contradictory' situation in science the ongoing discussions of ontological and organisational problems of its unitary version (though silently and incorrectly considered as the *unique* general kind of scientific knowledge) acquire increasingly intense and public character (e. g. [5-7,381-410]) thus actually confirming the essential doubts [339] about the conventional science validity that appeared together with the 'new physics' (and rightfully rejected the latter as a really new paradigm). However, the situation does not seem to show any sign of positive change and it is not difficult to see why: any, even most vigorous criticism of the existing system of knowledge cannot induce a constructive change without a clear, realistic alternative to that kind of knowledge being designated at least in its main lines. The existing 'official' propositions involve only fruitless, ill-specified generalities (like the recent tendency for mechanistic 'interdisciplinarity'), or else amelioration of details or external aspects ('façade repainting') of existing system, without any real, deep and well-specified change of content and form of the fundamental knowledge system that can alone induce the necessary qualitative transition in practical, socially large role of knowledge. Indeed, almost all the 'officially accepted' criticism of unitary science is produced by its well-placed adepts and demonstrates nothing but various attempts to *save* the *existing* system by modifying only its details, including especially reinforcement of one's own position within it (which closely resembles the case of Soviet 'perestroyka' or else a modern 'enlightened' bureaucrat complaining about 'too much bureaucracy' in his work).[50] Moreover, criticism of conventional science without clear designation of its really different, extended version often produces a mechanistic opposition reaction of unitary 'priests' in power who take the role of 'shamans of scientism' [408] and tend to present that criticism as an attack against *any* kind of scientific, properly ordered knowledge, or 'anti-science' movement, favouring, according to them, the development of disordered, anarchical state of knowledge (see e. g. [409]), whereas in reality the observed 'chaos in science' is the direct result of artificial, huge limitation inherent in its very *special* (though unfairly dominating) version, the unitary science. This blind and deadlock opposition can only amplify the existing crisis, leading already to destructive 'science wars', which cannot produce any positive result in principle.

In the meanwhile, further decadence of content and organisation of unitary science gives rise to its ever more odious, grotesque forms which, however, continue to be mechanistically maintained, 'despite everything', within the officially adopted system. Thus, the self-seeking promotion of senseless and contradictory abstractions in professional and popular sources of information borrows increasingly its methods from the *show-*

---

[50] Note, in particular, the ambiguous role of various 'commissions on ethics in science', related discussions and 'subjects of study' inserted at the dawn of the new millennium at various levels of the official science establishment and supposed to provide 'ultimate', higher-order solutions to various particular problems of potentially dangerous science applications or incorrect practices within science itself. It remains unclear, however, how such kind of purely 'moralistic', non-professional intervention can resolve a difficulty originating from fundamental deficiency of the well-specified, professional science, without introducing a major qualitative change in that science content and organisation. The illusion that it can only amplifies the existing flaws and dangers of unitary science trying to preserve in that way its basic structure intact. Which 'ethical commission' would decide, for example, that the whole content of unitary quantum chaos, quantum computation and information theories, richly decorated with all kind of fictitious 'teleportations' and supported by many most 'prominent' scientists in 'prestigious' institutions for many years, is but a huge scientific fraud full of evident lies and elementary contradictions, such as direct, unexplained deviations from the fundamental, well-established and recognised laws of the same science (see e. g. Chapter 2)? And even if it would one day, then why shouldn't the same decision be taken with respect to exactly the same situation in such equally 'élite', but totally corrupt domains as modern 'quantum field theory' and related official 'cosmology' and 'gravity theory'? In the meanwhile, the false 'ethical' zeal around real dangers of unitary science is actually used by all kind of parasitic 'scientologists', having nothing to do with the scientific creation itself, for grabbing their share in the pie of public spending on science actually controlled by their equally blind and corrupt 'colleagues' from the 'political sector' of the same unitary system 'elite'.

*business* kind of activity of a corrupt monetary system, where what really matters is not the actual content of a result, but rather its 'appreciation' by easily manipulated crowds of intellectually paralysed 'lay public' and conformist 'professional critics' chosen, of course, from the same milieu of interested 'priests' and their 'clans' that governs the estimated 'research' and strongly defends its absolute right to subjectively decide who tells the truth worthy of support (see e. g. [411]). "How to *sell* science better" is the major problem and purpose of the unitary science 'paradigm' that actually and openly dominates in its most 'advanced' structures, determining their real activity.

Today's system of public money distribution for science support deliberately, and often even *officially* (by its rules), *suppresses* explicit solution of real, fundamental or applied, problems practically always resulting from the individual work, but favours instead creation of highly centralised, effectively totalitarian 'networks' of 'post-modern', speculating scientists around various unitary imitations, actually forming self-seeking clans, so that these 'networks' *themselves* are considered as the *main result* of 'scientific' activity [412] (including the generous payment for the pleasures of 'networking' process from 'public' money sources, which are actually alienated from the public and its true interests by the destructive, unitary system of governance and its selfish 'elites'). One always gets exactly what one really wants to obtain and therefore it is no wonder that after having spent many billions of dollars, one year after another, just for the development of parasitic, fruitless clans, one gets just the mafia-like structure of European science, which finds itself, however, 'strangely' behind other 'developed' (also corrupt) science structures in essential, properly scientific results involving real problem solutions that need actually only a small portion of the vainly wasted treasure for their detailed elaboration.

A 'successful', but in reality wrong and fruitless, theory, 'concept', or 'approach' thus imposed in the ugly arrangement of selfish 'political' interests between the clans is then intensely advertised and popularised *for sell* on the world-wide scale with the help of totally subordinate 'science media' and money sources that rely essentially on the virtual absence of any truly critical, motivated attitudes within a 'developed' industrial society, with its 'post-modern' (parasitic and indifferent) life style of a banal show that has been gradually invading its 'mass-consciousness' state with the help of the same, only formally 'free' media and really interested, but selfish governing structures. How many publicity noise, highest distinctions and prestigious prizes were associated only in more recent times with all those useless, inconsistent and imitative 'renormalisation groups', 'diagrams', 'path integrals', 'quantum field theories', '(quantum) cosmologies', 'cybernetics', 'artificial intelligence' promises, 'general systems theories', 'complexities', 'chaoticities', 'time arrows' and many other 'breakthrough concepts' and 'advanced study' paradigms (cf. [1,5-7]), but the fundamental problems involved remain unsolved and the related contradictions accumulate, while the announced 'prodigious' theories, being in reality but extremely superficial fakes and deliberately intricate plays with adjustable symbols, rules and parameters, simply replace one another and then silently disappear, after having vainly consumed the desired quantities of public money for selfish pleasures of the 'happy few' from self-designated 'scientific elites' and 'educated community'.

Now one can clearly "know them by their fruits", but using their unbalanced power, the "false prophets" of the unitary science continue to transform the temple of knowledge into "a den of thieves" and lead the whole society to a *catastrophic* version of the 'end of science', so that a deliberate, professional sabotage of science development could not be more efficient than the 'spontaneous' action of unitary approach promoters. The destroyed, ultimately perverted intellectual landscapes and critically high, *justified* public distrust of *that* kind of science are the main results of unitary thinking domination with unlimited 'glory and power' during the twentieth century. The same, well-known story helplessly and grotesquely repeats itself like a phrase from a spoiled record: again and again "a corrupt tree bringeth forth evil fruit" and the crazy, catastrophically growing obsession with the idea that the crude force of unitary system power and money 'can do everything' leads inevitably to the proportionally terrible disaster, as it was clearly demonstrated many times only during the last hundred years of modern history. There is no other choice *within* the unitary kind of thinking and organisation: being intrinsically limited to the cage of its ultimately low complexity, the unitary system inevitably and painfully alternates between the 'totalitarian', more or less open, domination by authoritarian 'clans' and 'democratic', more or less deceitful, but completely fruitless and auto-destructive state of 'global chaos'. However, contrary to its own, self-protective mythology, the unitary system, including *all* its versions and their *inevitably* destructive dogmata, is *not* the unique, and certainly not the 'best possible', way of real world existence and development (see the end of this Chapter).

The super-authoritarian, self-seeking structure of the official science operation, badly hidden now behind its 'democratic' discourse, can have only one, easily predictable orientation and result. Particular speculations and justifications of evidently false imitations of reality take, on the contrary, a practically infinite variety of most exotic forms and colours. It is enough to have a look at the main directions and examples of recent 'interpretations' of quantum mechanics or metamorphoses in the officially supported 'quantum field theory' and 'cosmology' to see that there indeed 'everything is possible' (on paper), every kind of deviation, inconsistency, or sudden change of opinions depending simply on the last cry of a purely subjective, 'post-modern' vogue promoted by a privileged 'priest' and his clan or recast of the haphazard distribution of 'experimentally measured' (but in reality subjectively interpreted and arbitrarily fitted) parameters of a postulated, but inconsistent 'model'.[51] The muddy flux of unitary imitations of reality of 'post-modern' age includes also permanently appearing pretentious versions of completely 'new kind of science', 'third culture', 'spiritu-ally upgraded' or 'ethically controlled', 'interdisciplinary' science and other 'edge' ideas from specially 'chosen' minds, 'most complex and sophisticated' by their own definition[52] and therefore supported by equally 'prodigious' market dealers, often claiming their 'strong' opposition to conventional knowledge, but proposing, in fact, just another set of postulated, abstract entities and simplified deviations from truth, quite falling within the same, unitary way of thinking and its practical consequences (this 'false prophet' phenomenon is characteristic of any apocalyptical 'epoch of change', or 'generalised phase transition', in science and society [1]).[53] In view of their evident inconsistency, the post-modern 'narratives' of decaying unitarity are often justified by the purpose of 'free', 'artistic' creation as if approaching science to a desirable 'interdisciplinary' state bordering on a futuristic art. However, when they fight for an effectively *unconditional* support for their 'art works' from the part of *unaware* public/sponsors, the unitary 'artists' forget about the 'beauty' of artistic creation and 'suddenly' change their decorative "sheep's clothing" into the real habits of "ravening wolves", thus confirming once again the ancient wisdom.

---

[51] In connection to quantum computation story one can recall, for example, an especially 'prodigious' version of 'multiverse' interpretation of quantum mechanics, where the multiple 'possibilities' of quantum behaviour are 'obtained' by postulation of an artificial, mechanistic dissection of the abstract Hilbert space of states, or 'multiverse', into 'slices' corresponding to those individual possibilities, or 'universes', the whole 'process' being arbitrarily equipped with a wordplay label of 'information flow' [74,75]. Another particularly 'prodigious' speculation considers the whole universe evolution as operation of a unitary (sequential and regular) computer [150], after a very intense talk about 'complexity' in quantum computation under the auspices of the most 'advanced' complexity studies in the Santa Fe Institute [62] (see also Section 7.1). The 'informational' [72,73,317,319,330], 'consistent-histories' [329,413-415], 'decoherence' [92,146-149,182,224,225,274,414,415] and various other 'interpretations' of quantum mechanics (e. g. [15,16,74,75,77,196-206,416-434]) contribute to infinite series of vain post-modern 'narratives' chaotically flickering around the same, persisting 'mysteries' of the standard theory [2-4,15,16,435,436] and involve also the expensive, 'super-fine' experiments, where plays of words and related logical 'loopholes' (discovered, of course, only after the 'seminal paper' publication) are accompanied by expensive plays with devices and numbers around basically deficient unitary 'models', while the results obtained never clarify anything in reality understanding (see e. g. [437-440]). The underlying pretensions of the 'advanced' unitary science to 'ultimate reality' understanding (e. g. [441]) only emphasize its imitative content and truly 'unlimited' organisational possibilities (where for example some main participants can vigorously deny the very idea of ultimate reality and unified understanding in favour of different 'discussion clubs' [411], while an existing consistent version of unified reality description cannot have any chance to be even considered as such without being subjectively 'promoted' by one of the leading 'clubs').

The ultimate subjectivity, abstraction and fruitlessness of content of modern official, unitary science, as well as 'Jesuitical' methods of its practical organisation, actually throw it back to the level of medieval, pre-Cartesian *scholasticism*, i. e. infinite and vain *interpretation* of the *officially fixed*, pre-established (and therefore *fantastically* incorrect) 'truth', which effectively cancels, rather than extends, all the really great achievements and values of realistic, classical physics based on the ideas of Renaissance. The unlimited, grotesquely disproportional *idolisation* and related *politicisation* of knowledge are the accompanying human, 'social' components of the 'developed', scholastically fixed unitarity, where what really and exclusively matters is not the announced higher-level search for an 'objective', independent truth showing *what* is right, but a low-level fight of selfish, 'personalised' interests aiming to show *who* is greater (since it is *his*, privately owned 'interpretation of (the fixed) truth' that will be recognised by the corrupt system as the *official*, currently absolute truth, irrespective of all those 'living natures' and 'unreduced realities'!). Vain, *artificially mystified symbolism* and intentional, tricky *obscurity* are the main, clearly visible properties of the most 'advanced' branches of 'exact' unitary science that closely resembles now, by both its content and 'methods' of intrinsic adherents, such forms of knowledge as medieval alchemy, occult cabbala, notorious astrology and other esoteric 'metaphysics'.[54] The effective, real *darkness* of the modern 'scientific' age is deeply lurking behind the external bright cover of illusive technological 'enlightenment', which is the inevitable final stage of the destructive 'new science' influence on knowledge, starting from its 'triumphant' appearance at the nineteenth century fall.

---

[52] For a particular illustration of the true content and 'flavour' of that kind of activity, see e. g. a short account by J. Mejias, "Which universe would you like?", originally published in Frankfurter Allgemeine Zeitung (28 August 2000, No. 199, p. 33) and reproduced in English translation at http://www.edge.org/documents/press/faz_8.28.02.e.html, as well as many other materials from the same web site (see also refs. [150,368,380]).

[53] Those particularly 'unlimited' fantasies of the unitary theory, clearly revealing its badly hidden internal dissatisfaction and evident contradictions, are often based on arbitrary, postulated 'mixture' between different complexity levels. Thus in many cases related to 'quantum information' concept and mentioned in this work, a 'quantum' type of behaviour and formalism are directly attributed to higher complexity levels or, on the contrary, higher complexity manifestations, such as consciousness, are used for a 'radical' explanation of the canonical quantum mysteries (see e. g. [15,68-71,77,187,189,190,192-195,350,354,366,367,419,421,422,442-450]). This kind of approach directly contradicts the complexity correspondence principle (Section 7.2) supported eventually by all well-established conservation laws and therefore could be easily avoided upon elementary application of the unreduced science of complexity [1-4].

---

[54] Referring to a deeply rooted link between the unitary kind of thinking and irreducible mystique in the official science base, we can cite a well-known statement by Albert Einstein, the father of the canonical 'new physics': "The most profound emotion we can experience is the sensation of the mystical. It is the sower of all true science." Indeed, although Einstein produced a famous 'attack' against the irrational 'enigmas' of quantum mechanics canonised by another great mysteriologist, Niels Bohr, those 'friendly' objections did not contain any hint on a positive, causal solution. Moreover, Einstein's main contribution to the 'new physics', his interpretation of the Lorentz relations between space-time and motion, or canonical 'special relativity', as well as gravity inclusion within his 'general relativity' scheme, rely totally on postulated formal 'principles' and 'adjusted' (guessed) mathematical relations, without any causal understanding of occurring physical processes within the tangible reality (such realistic understanding of 'relativistic' effects, intrinsically unified with the causally complete picture of 'quantum' behaviour, can be obtained only taking into account the unreduced complex dynamics of the underlying interaction processes [1-4,11-13]). As a result, one reaps in today's scholar science as one has sown, with that kind of sower, including relativity and gravity that remain causally unexplained (i. e. 'mysterious') and irreducibly separated from equally mysterious quantum mechanics (within their canonical, unitary versions). Fruitless and endless fantasies of modern, purely abstract field theory are also a direct result of this Einsteinian approach which he vainly tried to develop himself in his later years. It is clear, however, that *any consistent* world image can only be *totally realistic* (describe the physically real, rather than purely mathematical, constructions and their real change) and *intrinsically unified* from the beginning, irrespective of details (so that 'space', 'time', 'particles', 'fields', their properties, 'quantum behaviour' and 'relativity' are explicitly obtained *all together*, in exact correspondence with the real universe evolution [1-4,11-13]). Any other 'attempts', including handicapped formal adjustment between a system of mathematical postulates and specially chosen, point-like measurements (the intrinsic 'criterion of truth' within the 'mathematical physics' kind of fraud), should be rejected as definitely insufficient. It is the 'evident', 'natural' character of this attitude that is disputed by the unitary thinking deeply concentrated instead on *abstract* (rather than realistic) *fundamental origin* of things and their respective 'understanding', including 'mystical', frustrating separations in both reality and scientific knowledge about it.

Although 'human dimensions' of unitary thinking adherents do not seem to allow of any positive issue from the modern crisis of conventional fundamental science, it is the nonunitary, multivalued *dynamics of the real world itself* that puts rigid bounds to further domination of zero-complexity imitations of reality. The quickly growing difficulties are clearly visible now and range from the glaring rupture between canonical science abstractions and public interest to real practical dangers created by intellectually blind, but empirically powerful technology manipulations within the 'developed' unitary society based on profit. It is but another manifestation of our universal principle of complexity correspondence (Chapter 7) applied this time at the level of complex civilisation dynamics itself: a civilisation able to modify its full complexity scale empirically cannot afford the 'luxury' of remaining around the zero-complexity level intellectually, in its conscious, scientific understanding of the modified reality. Further persistence of the current way, as if 'violating' the symmetry of complexity, will inevitably and very soon result in a catastrophic (self-) destruction of the quantitatively 'powerful', but stupid species, as it happened many times in ancient and recent history. The importance of the dynamic redundance paradigm and related concept of complexity, applied in this work to description of real micro-machine dynamics, is in the fact that not only it specifies the fundamental, irreducible origin of conventional theory difficulties, but also shows the clear and fundamentally substantiated way out of the resulting impasse through the truly novel, *conceptual*, but also quite natural, *extension* of the dynamically single-valued (point-like), *unnatural* projection of conventional science to the dynamically multivalued, intrinsically complete picture of unreduced, living reality. It is also important that this crucial extension is obtained not by artificial postulation of another series of additional, abstract entities (the only possible kind of 'novelty' in the unitary science), but by the *truly consistent*, nonperturbative analysis of canonical, configurationally simple dynamic equations, which uses a causally extended, but technically straightforward version of the same mathematical tools that seem to be so much appreciated by the tricky adherents of unitarity (in reality, they roughly and inconsistently cut them at the most essential stages in favour of deceptive simplicity of the postulated mechanistic imitations and vain, pseudo-philosophical speculations). In other words, it is enough to be simply honest (consistent) while using the (elementary) mathematical tools in order to avoid conventional science cheating on public confidence.

Whereas the strategically important applications of the obtained micro-machine theory are specified above (Section 7.3, Chapter 8), it is also essential that the method used and the emerging causally complete world picture include any kind of system and level of complexity, which confirms the particular results obtained and opens the way for their multiple extension to other applications (see e. g. Chapters 6, 8, Sections 7.1, 7.3), up to the highest, man-related levels of complexity (intelligence, consciousness, society transformation) usually only empirically studied in the humanities, but now consistently derived from lower complexity levels and causally understood within the truly 'exact', intrinsically complete theory [1], realising thus the ultimate goal of science. The modern 'critical' state of the world shows that the intense use of these *already obtained* results of the unreduced science of complexity in order to realise the necessary positive changes at superior complexity levels is urgently needed and irreplaceable (see also below). The general concept and its applications, resolving bundles of fundamental problems, which otherwise helplessly stagnate within the unitary science approach, are described in publications made widely accessible through the physical preprint Archives (arXiv.org) [1-4,8-13][55] (see also *https://sites.google.com/site/unifiedcomplexity/*), so that their further ignorance or reduced, unitary imitation would reflect only the ultimately corrupt character of the existing system of knowledge.

---

[55] Note, for example, the consistent solution of quantum chaos [1,8,9] and quantum measurement [1,10] problems (see also Chapter 6), together with many other ones from the same group of fundamental levels of reality (see e. g. [13] and Chapter 3 in ref. [3]), as well as readily emerging solutions to particular problems from various fields (e. g. [8] and Section 5.3(C)).

> *Toutefois leur façon de philosopher est fort commode, pour ceux qui n'ont que des esprits fort mediocres: car l'obscurité des distinctions, et des principes dont ils se servent, est cause qu'ils peuvent parler de toutes choses aussy hardiment que s'ils les sçavoient, et soustenir tout ce qu'ils en disent contre les plus subtils et les plus habiles, sans qu'on ait moyen de les convaincre: En quoy ils me semblent pareils a un aveugle, qui pour se battre sans desavantage contre un qui voit, l'auroit fait venir dans les fonds de quelque cave fort obscure: Et je puis dire que ceux cy ont interest que je m'abstiene de publier les principes de la Philosophie dont je me sers, car estans tres simples et tres evidens, comme ils sont, je ferois quasi le mesme en les publiant, que si j'ouvrois quelques fenestres, et faisois entrer du jour dans cete cave où ils sont descendus pour se battre.*
>
> René Descartes, *Discours de la Méthode* (1637) [451]

A qualitative, deep change is necessary in science. Whereas this conclusion is often accepted as a 'general wish', we provide the *fundamentally substantiated* and *causally specified* version of that change avoiding thus any subjective illusions and random deviations. The proposed new, truly universal concept of complexity, really *derived* from the *objectively minimal* assumptions ('first principles'), shows *what exactly and why* is wrong in the conventional science content and organisation and how exactly and why should *that* form of knowledge be *extended* to the unreduced, totally realistic and intrinsically complete version. We rigorously *prove* thus and *practically confirm* by *explicitly obtained problem solutions* that the latter kind of knowledge exists and can be revealed, which is not evident at all within the perverted realm of unitary thinking. We prove not only the necessity, but also *reality* of science change involving *eventually* the equally necessary and now objectively specified transformation of the *entire society* and *way of living* [1], which is another indispensable aspect of any realistic change of knowledge system that modifies the leading *way of thinking*. It is the fundamental, qualitatively big and well-specified transition from the unitary science content and organisation to their ultimately extended versions of the unreduced science of complexity, which leads to the realistic, intrinsically progressive (or 'sustainable') version of 'society based on knowledge': the kind of 'knowledge' on which the future, prosperous and sustainable, society can only be based is *not* the unitary 'science', or 'culture', or 'religion' of today, nor their mechanistic, 'interdisciplinary' superposition, but rather dynamically multivalued, *intrinsically* unified and creative extension called here universal science of complexity (including its extended organisational structure).

One should be clearly aware of a qualitatively large scale of change, which *cannot* be smaller in principle, as it is practically confirmed by the fact that all smaller, quantitative, 'horizontal' changes have already been tried and the result *is* the continuing, catastrophically spiralling degradation of fundamental knowledge that cannot be hidden anymore behind the 'solid' façade of its bureaucratic governance and technically powerful equipment. Therefore any proposed 'amelioration' of the existing science structure, including pretentiously 'radical' switch between various 'modes' of its realisation (e. g. [397-399]), is simply irrelevant with respect to the emerging problem scale. In reality one deals today not with the choice among one or another version of a 'trade agreement' between an 'elitist' scientific community and the 'lay' society, actually reduced to finding more efficient schemes of cheating money out of the unaware 'public' for the useless or eventually dangerous 'research', but with the deepest possible change in the essential content of scientific knowledge itself, inevitably involving the corresponding, equally deep change in its organisation and understanding by the large public.

The new science *content* will be dominated by *causally complete solutions* to *real* problems, fundamental or practical, clearly specified as full sets of system realisations at all the essential complexity levels (Chapters 3, 4, Section 7.1), which is equivalent to establishment of the *absolutely consistent system of correlations* within the whole, internally entangled, 'theoretic-and-experimental' structure of knowledge (as opposed to never-ending 'search' for an ill-defined solution to unrealistically simplified, abstract 'models' of reality within the unitary science, accompanied with its equally endless and useless justification of persisting contradictions). This *genuine criterion of truth* expresses the universally realised *general purpose* of (new) science in the form of *unceasing* (though maybe uneven) *de-* *velopment of the unreduced dynamic complexity* of civilisation (including its *increasing* interaction with 'natural environment') by the optimal *growth* of its generalised complexity-entropy at the expense of complexity-information (Chapter 7), as opposed to the *zero-complexity content* of the now dominating unitary thinking and the resulting degradation (purely destructive end of complexity development). The ensuing *practical* usefulness of *fundamental* knowledge is *inseparable* from its *cognitive* role ('curiosity satisfaction') and centred now around its *steering role* in the civilisation complexity development (determined by levels of *individual consciousness*), where it can *provably* obtain the *whole* spectrum of existing possibilities, or 'ways of development', and show which way of empirical technology development should be chosen among other existing ones, so as to optimise the *creative* complexity development (as it is exemplified by the above results for micro-machine dynamics, Chapters 5, 7, 8), instead of the dominating conventional science practice of using the *already obtained* results of *purely empirical* technology development for justification of further financial investment into its useless plays with unrealistic unitary models of those 'practically important' phenomena (while the blind, complexity-destroying technology becomes indeed increasingly dangerous, despite the equally blind 'protective' efforts and 'ethical' commissions).

Although science content extension to its unreduced, dynamically multivalued version could, in principle, be performed, or at least started, within its existing organisational structure, it is clear that the truly new kind of science, including practically unlimited creativity and causally complete, universal understanding of unreduced reality (as it is outlined by the universal science of complexity [1]), can only be built within the *corresponding* new kind of *organisational structure* that can be described as open, *unreduced interaction* between many *independent, explicitly* creative and *permanently developing* enterprises based on definite (causally complete) *problem solution* by *individual* modes of intellectual work and *direct* links to 'lay', massive consumers and supporters of knowledge (see ref. [1] for more details). The self-developing, explicitly creative and *totally open* structure includes *both* properly 'scientific' units *and* 'organisational', 'management' elements of the *same, 'independent' kind*, which are *dynamically entangled* in their *intrinsically creative* activity (contrary to any unitary science organisation), so that 'organisational' units contribute con-structively not only to 'promotion'/realisation of fruitful ideas, but also to their essential content (and the reverse). Transition from unitary to complex-dynamical kind of organisation can be compared to a transition from the 'command' to 'free-market' economy, but where the essential change involves not only properly 'professional' ('economic' or 'scientific') aspects, but also the global system structure/dynamics itself, changing its unitary, rigidly fixed, industrial 'governance' for the unlimited, free, intrinsically creative complexity development of the driving interaction processes.

As a result, the universal science of complexity and its new, also *explicitly* complex-dynamical, creative structure provide a *qualitatively higher efficiency* with respect to the unitary science structure, which actually acts *against* any essential novelty and *suppresses* the opening possibilities for real, causally complete problem solutions (cf. [382]). Whereas the unitary system tries (in vain) to compensate this inbred deficiency by high *quantities* of invested resources, the unreduced complexity development solves the problem by the intrinsically high *quality* of its creation, where the massive, but artificially limited and therefore *fruitless* 'search' of unitary science does not appear in principle (apart from occasional, rare exceptions that replace exceptionally rare breakthroughs of unreduced creativity in the unitary science dynamics) and therefore a *much higher quality of results* is obtained now by using *much smaller quantities of resources*. The false, imitative 'miracles' of unitary science discussed above in relation to 'quantum information' concept are thus replaced by the real, causally understood miracle of unreduced, complex-dynamical creation in both studied system behaviour (like multivalued micro-machine dynamics) and internal science dynamics itself, extending and generalising previous similar transitions, such as that from illusive miracles of alchemy to the efficiency of modern chemical and nuclear processes or that from obscure magic of cabbala to the power of computer-assisted creation.

Although the difference of the new science structure from its conventional, unitary version can only be qualitatively big in the domains where the change has already occurred, the new kind of structure can proliferate *progressively* within the existing system, forming a fractal, finely structured pattern of the 'new phase' (i. e. it is the 'generalised phase transition' [1] of a higher 'order'), which greatly facilitates practical introduction of the new structure. Since the unreduced complexity is involved in any aspect of this change and the resulting superior level of knowledge (and civilisation in the whole), we call this emerging, mainly *intellectual and spiritual* (rather than 'social') transition *revolution of complexity*, emphasising thus its qualitative novelty with respect to the ending unitary system and its own imitations of change. Transition from the unrealistic unitary scheme of conventional quantum computation to the causally complete understanding and unlimited creation of real, dynamically multivalued and self-developing, 'living' micro-machines of every possible kind and application, described in more detail in this work, represents just one particular manifestation of that universal revolution of complexity [1].

# References

[1] A.P. Kirilyuk, *Universal Concept of Complexity by the Dynamic Redundance Paradigm: Causal Randomness, Complete Wave Mechanics, and the Ultimate Unification of Knowledge* (Naukova Dumka, Kiev, 1997). See also: *E-print physics/9806002 at http://arXiv.org*.

A.P. Kirilyuk, "The Last Scientific Revolution", in *Against the Tide. A Critical Review by Scientists of How Physics and Astronomy Get Done*, Eds. M. Lopez-Corredoira and C. Castro Perelman (Universal Publishers, Boca Raton, 2008), p. 179. E-print arXiv:0705.4562.

[2] A.P. Kirilyuk, "75 Years of Matter Wave: Louis de Broglie and Renaissance of the Causally Complete Knowledge", *E-print quant-ph/9911107 at http://arXiv.org*.

[3] A.P. Kirilyuk, "100 Years of Quanta: Complex-Dynamical Origin of Planck's Constant and Causally Complete Extension of Quantum Mechanics", *E-print quant-ph/0012069 at http://arXiv.org*.

[4] A.P. Kirilyuk, "75 Years of the Wavefunction: Complex-Dynamical Extension of the Original Wave Realism and the Universal Schrödinger Equation", *E-print quant-ph/0101129 at http://arXiv.org*.

[5] J. Horgan, *The End of Science. Facing the Limits of Knowledge in the Twilight of the Scientific Age* (Addison-Wesley, Helix, 1996).

J. Horgan, "The Final Frontier", *Discover Magazine*, October (2006), http://discovermagazine.com/2006/oct/cover.

[6] J. Horgan, "From Complexity to Perplexity", *Scientific American*, June (1995) 74.

[7] J. Horgan, *The Undiscovered Mind: How the Human Brain Defies Replication, Medication, and Explanation* (Touchstone/Simon & Schuster, New York, 1999).

[8] A.P. Kirilyuk, "Theory of charged particle scattering in crystals by the generalised optical potential method", *Nucl. Instr. Meth.* **B69** (1992) 200.

[9] A.P. Kirilyuk, "Quantum chaos and fundamental multivaluedness of dynamical functions", *Annales de la Fondation Louis de Broglie* **21** (1996) 455. E-print quant-ph/9511034, 35, 36 at http://arXiv.org.

[10] A.P. Kirilyuk, "Causal Wave Mechanics and the Advent of Complexity. IV. Dynamical origin of quantum indeterminacy and wave reduction", *E-print quant-ph/9511037 at http://arXiv.org*;

A.P. Kirilyuk, "Causal Wave Mechanics and the Advent of Complexity. V. Quantum field mechanics", *E-print quant-ph/9511038 at http://arXiv.org*.

A.P. Kirilyuk, "New concept of dynamic complexity in quantum mechanics and beyond", *E-print quant-ph/9805078 at http://arXiv.org*.

[11] A.P. Kirilyuk, "Double Solution with Chaos: Dynamic Redundance and Causal Wave-Particle Duality", *E-print quant-ph/9902015 at http://arXiv.org*.

[12] A.P. Kirilyuk, "Double Solution with Chaos: Completion of de Broglie's Nonlinear Wave Mechanics and its Intrinsic Unification with the Causally Extended Relativity", *E-print quant-ph/9902016 at http://arXiv.org*.

[13] A.P. Kirilyuk, "Universal gravitation as a complex-dynamical process, renormalised Planckian units, and the spectrum of elementary particles", *E-print gr-qc/9906077 at http://arXiv.org*.

A.P. Kirilyuk, "Quantum Field Mechanics: Complex-Dynamical Completion of Fundamental Physics and Its Experimental Implications", *E-print physics/0401164 at http://arXiv.org*.

A.P. Kirilyuk, "Electron as a Complex-Dynamical Interaction Process", *E-print physics/0410269 at http://arXiv.org*.

A.P. Kirilyuk, "Complex-Dynamical Approach to Cosmological Problem Solution", *E-print physics/0510240 at http://arXiv.org*.

A.P. Kirilyuk, "Consistent Cosmology, Dynamic Relativity and Causal Quantum Mechanics as Unified Manifestations of the Symmetry of Complexity", *E-print physics/0601140 at http://arXiv.org*.

A.P. Kirilyuk, "Complex-Dynamical Solution to Many-Body Interaction Problem and Its Applications in Fundamental Physics", *Nanosystems, Nanomaterials, Nanotechnologies* **10** (2012) 217. E-print arXiv:1204.3460 at http://arXiv.org.

A.P. Kirilyuk, "What Do They Actually Probe at LHC?", *E-print viXra:1210.0162 at http://viXra.org/*; hal-00740459 at http://hal.archives-ouvertes.fr/.

[14] B. Georgeot and D.L. Shepelyansky, "Quantum Chaos Border for Quantum Computing", *Phys. Rev. E* **62** (2000) 3504. E-print quant-ph/9909074 at http://arXiv.org.

B. Georgeot and D.L. Shepelyansky, "Emergence of Quantum Chaos in the Quantum Computer Core and How to Manage It", *Phys. Rev. E* **62** (2000) 6366. E-print quant-ph/0005015 at http://arXiv.org.

D.L. Shepelyansky, "Quantum Chaos and Quantum Computers", *Physica Scripta* **T90** (2001) 112. E-print quant-ph/0006073 at http://arXiv.org.

G. Benenti, G. Casati, S. Montangero and D.L. Shepelyansky, "Eigenstates of Operating Quantum Computer: Hypersensitivity to Static Imperfections", *Eur. Phys. J. D* **20** (2002) 293. E-print quant-ph/0112132 at http://arXiv.org.

B. Levi, B. Georgeot and D.L. Shepelyansky, "Quantum Computing of Quantum Chaos in the Kicked Rotator Model", *Phys. Rev. E* **67** (2003) 046220. E-print quant-ph/0210154 at http://arXiv.org.

M. Terraneo and D.L. Shepelyansky, "Dynamical Localization and Repeated Measurements in a Quantum Computation Process", *Phys. Rev. Lett.* **92** (2004) 037902. E-print quant-ph/0309192 at http://arXiv.org.

K.M. Frahm, R. Fleckinger and D.L. Shepelyansky, "Quantum chaos and random matrix theory for fidelity decay in quantum computations with static imperfections", *European Physical Journal D* **29** (2004) 139. E-print quant-ph/0312120 at http://arXiv.org.

A.A. Pomeransky, O.V. Zhirov and D.L. Shepelyansky, "Effects of decoherence and imperfections for quantum algorithms", *E-print quant-ph/0407264 at http://arXiv.org*.

J. Lages and D. L. Shepelyansky, "Suppression of quantum chaos in a quantum computer hardware", *Phys. Rev. E* **74** (2006) 026208. E-print cond-mat/0510392 at http://arXiv.org.

[15] R. Penrose, *Shadows of the Mind* (Oxford University Press, New York, 1994).

[16] B. d'Espagnat, *Le Réel Voilé* (Fayard, Paris, 1994).

B. d'Espagnat, "Quantum Theory: A Pointer to an Independent Reality", *E-print quant-ph/9802046 at http://arXiv.org*.

[17] P. Benioff, "The computer as a physical system: a microscopic quantum mechanical model of computers as represented by Turing machines", *J. Stat. Phys.* **22** (1980) 563.

P. Benioff, "Quantum mechanical Hamiltonian models of Turing machines that dissipate no energy", *Phys. Rev. Lett.* **48** (1982) 1581.

P. Benioff, *Ann. N.Y. Acad. Sci.* **480** (1986) 475.

[18] R.P. Feynman, "Simulating Physics with Computers", *Int. J. Theor. Phys.* **21** (1982) 467.

R.P. Feynman, "Quantum mechanical computers", *Found. Phys.* **16** (1986) 507.

[19] D.P. DiVincenzo, "Quantum Computation", *Science* **270** (1995) 255.

[20] C.H. Bennett, "Quantum Information and Computation", *Physics Today*, October (1995) 24.

[21] S. Lloyd, "Almost Any Quantum Logic Gate is Universal", *Phys. Rev. Lett.* **75** (1995) 346.

[22] A. Barenco, C.H. Bennett, R. Cleve, D.P. DiVincenzo, N. Margolus, P. Shor, T. Sleator, J.A. Smolin and H. Weinfurter, "Elementary gates for quantum computation", *Phys. Rev. A* **52** (1995) 3457.

[23] A. Ekert and R. Jozsa, "Shor's Quantum Algorithm for Factorising Numbers", *Rev. Mod. Phys.* **68** (1996) 733.

[24] S. Wiesner, "Simulations of Many-Body Quantum Systems by a Quantum Computer", *E-print quant-ph/9603028 at http://arXiv.org*.

[25] S. Lloyd, "Universal Quantum Simulators", *Science* **273** (1996) 1073.

D.S. Abrams and S. Lloyd, "Simulation of Many-Body Fermi Systems on a Universal Quantum Computer", *Phys. Rev. Lett.* **79** (1997) 2586. E-print quant-ph/9703054 at http://arXiv.org.

G. Blumfiel, "Quantum Leaps", *Nature* **491** (2012) 322.

[26] M.P. Ciamarra, "Quantum reversibility and a new model of quantum automaton", *E-print quant-ph/0102104 at http://arXiv.org*.

[27] P. Benioff, "Models of Quantum Turing Machines", *Fortsch. Phys.* **46** (1998) 423. E-print quant-ph/9708054 at http://arXiv.org.

P. Benioff, "Quantum Robots and Quantum Computers", in *Feynman and Computation*, Ed. A.J.G. Hey (Perseus books, 1999), p. 155. E-print quant-ph/9706012 at http://arXiv.org.

P. Benioff, "Quantum Robots and Environments", Phys. Rev. A **58** (1998) 893. *E-print quant-ph/9802067 at http://arXiv.org*.

P. Benioff, "Space Searches with a Quantum Robot", *AMS Contemporary Math Series* **305** (2002) 1. E-print quant-ph/0003006 at http://arXiv.org.

[28] A. Steane, "Quantum Computing", *Rep. Progr. Phys.* **61** (1998) 117. E-print quant-ph/9708022.

[29] B.M. Boghosian and W. Taylor IV, "Simulating Quantum Mechanics on a Quantum Computer", *Physica D* **120** (1998) 30. E-print quant-ph/9701019 at http://arXiv.org.

[30] *Physics World* **11**(3), March (1998), special issue on Quantum Information.

Ch. Monroe and M. Lukin, "Remapping the Quantum Frontier", *Physics World*, No. 8 (2008) 32.

[31] S.S. Somaroo, C.H. Tseng, T.F. Havel, R. Laflame and D.G. Cory, "Quantum Simulations on a Quantum Computer", *Phys. Rev. Lett.* **82** (1999) 5381. E-print quant-ph/9905045 at http://arXiv.org.

[32] D. Gottesman and I.L. Chuang, "Demonstrating the viability of universal quantum computation using teleportation and single-qubit operations", *Nature* **402** (1999) 390.

[33] N. Cerf and N. Gisin, "Les promesses de l'information quantique", *La Recherche*, No. 327 (Janvier 2000) 46.

[34] J. Gruska, *Quantum Computing* (McGraw-Hill, London, 1999).

[35] C.H. Bennett and D.P. DiVincenzo, "Quantum information and computation", *Nature* **404** (2000) 247.

[36] M.A. Nielsen and I.L. Chuang, *Quantum Computation and Quantum Information* (Cambridge University Press, 2000). See also E-print quant-ph/0011036 at http://arXiv.org.

[37] *The Physics of Quantum Information*, Eds. D. Bouwmeester, A. Ekert and A. Zeilinger (Springer-Verlag, Berlin, 2000).

[38] *Fort. Phys.* **48** (2000) 767-1138. Special issue: Experimental proposals for quantum computation.

D.P. DiVincenzo, "The Physical Implementation of Quantum Computation", *E-print quant-ph/0002077 at http://arXiv.org*.

*Practical Realisations of Quantum Information Processing*, Eds. P.L. Knight, E.A. Hinds and M.B. Plenio, *Phil. Trans. Roy. Soc. Lond. A* **361** (2003) No. 1808.

A. Olaya-Castro and N.F. Johnson, "Quantum Information Processing in Nanostructures", *E-print quant-ph/0406133 at http://arXiv.org*.

[39] E.H. Knill and M.A. Nielsen, "Theory of Quantum Computation", in *Encyclopedia of Mathematics, Supplement III*, Ed. M. Hazewinkel (Kluwer, 2002). E-print quant-ph/0010057 at http://arXiv.org.

E.H. Knill and M.A. Nielsen, "Quantum Information Processing", in *Encyclopedia of Mathematics, Supplement III*, Ed. M. Hazewinkel (Kluwer, 2002). E-print quant-ph/0010058 at http://arXiv.org.

[40] R.F. Werner, "Quantum Information Theory – an Invitation", *E-print quant-ph/0101061 at http://arXiv.org*.

[41] A. Cabello, "Bibliographic guide to the foundations of quantum mechanics and quantum information", *E-print quant-ph/0012089 at http://arXiv.org*.

[42] S. Lloyd, "Unconventional Quantum Computing Devices", *E-print quant-ph/0003151 at http://arXiv.org*. In: *Unconventional Models of Computation*, Eds. C.S. Calude, J. Casti and M.J. Dinneen (Springer, Singapor, 1998).

A.Yu. Kitaev, "Fault-tolerant quantum computation by anyons", *Annals of Physics* **303** (2003) 2. E-print quant-ph/9707021 at http://arXiv.org.

S. Bravyi and A. Kitaev, "Fermionic quantum computation", *Annals of Physics* **298** (2002) 210. E-print quant-ph/0003137.

S. Lloyd, "Quantum computation with abelian anyons", *E-print quant-ph/0004010 at http://arXiv.org*.

M.S. Shahriar, P.R. Hemmer, S. Lloyd, J.A. Bowers and A.E. Craig, "Solid State Quantum Computing Using Spectral Holes", *E-print quant-ph/0007074 at http://arXiv.org*.

S. Lloyd, "Hybrid quantum computing", *E-print quant-ph/0008057 at http://arXiv.org*.

L. Venema, "The Dreamweaver's Abacus", *Nature* **452** (2008) 803.

[43] E. Knill, R. Laflamme and G.J. Milburn, "A scheme for efficient quantum computation with linear optics", *Nature* **409** (2001) 46. E-print quant-ph/0006088 at http://arXiv.org.

[44] M.A. Man'ko, V.I. Man'ko and R. V. Mendes, "Quantum computation by quantum-like systems", *Phys. Lett. A* **288** (2001) 132. E-print quant-ph/0104023 at http://arXiv.org.

[45] A. Mizel, M.W. Mitchell and M.L. Cohen, "Ground State Quantum Computation", *E-print quant-ph/9908035 at http://arXiv.org*.

[46] D. Bacon, J. Kempe, D.A. Lidar, K.B. Whaley and D.P. DiVincenzo, "Encoded Universality in Physical Implimentations of a Quantum Computer", *E-print quant-ph/0102140 at http://arXiv.org*.

J. Kempe, D. Bacon, D.P. DiVincenzo and K.B. Whaley, "Encoded Universality from a Single Physical Interaction", *Quantum Information and Computation* **1** *(Special Issue) (2001) 33*. E-print quant-ph/0112013 at http://arXiv.org.

[47] C.A. Trugenberger, "Probabilistic Quantum Memories", *Phys. Rev. Lett.* **87** (2001) 067901. E-print quant-ph/0012100 at http://arXiv.org.

See also *E-print quant-ph/0210176 at http://arXiv.org*.

[48] S. Bandyopadhyay, "Prospects for a Quantum Dynamic Random Access Memory (Q-DRAM)", *E-print quant-ph/0101058 at http://arXiv.org*.

T.T. Wu and M.L. Yu, "Quantum memory: Write, read and reset", *E-print quant-ph/0208137 quant-ph/0012100*.

[49] A.D. Greentree, S.G. Schirmer and A.I. Solomon, "Robust quantum memory via quantum control", *E-print quant-ph/0103118 at http://arXiv.org*.

[50] S. Bettelli, L. Serafini and T. Calarco, "Toward an architecture for quantum programming", *Eur. Phys. J. D* **25** (2003) 181. E-print cs.PL/0103009 at http://arXiv.org.

[51] S. Lloyd, M.S. Shahriar and P.R. Hemmer, "Long Distance, Unconditional Teleportation of Atomic States Via Complete Bell State Measurements", *Phys. Rev. Lett.* **87** (2001) 167903; quant-ph/0003147.

[52] T.A. Brun, "Remotely prepared entanglement: a quantum web page", *Algorithmica* **34** (2002) 502. E-print quant-ph/0102046.

[53] G. Brassard, "Quantum Communication Complexity (A Survey)", *E-print quant-ph/0101005 at http://arXiv.org*.

[54] L. Goldenberg, L. Vaidman and S. Wiesner, "Quantum Gambling", *Phys. Rev. Lett.* **82** (1999) 3356. E-print quant-ph/9808001.

[55] D.A. Meyer, "Quantum Strategies", *Phys. Rev. Lett.* **82** (1999) 1052. E-print quant-ph/9804010 at http://arXiv.org.

J. Eisert, M. Wilkens and M. Lewenstein, "Quantum Games and Quantum Strategies", *Phys. Rev. Lett.* **83** (1999) 3077. E-print quant-ph/9806088 at http://arXiv.org.

J. Eisert and M. Wilkens, "Quantum Games", *J. Mod. Opt.* **47** (2000) 2543. E-print quant-ph/0004076 at http://arXiv.org.

S.C. Benjamin and P.M. Hayden, "Multi-Player Quantum Games", *Phys. Rev. A* **64** (2001) 030301. E-print quant-ph/0007038.

S.J. van Enk and R. Pike, "Classical Rules in Quantum Games", *Phys. Rev. A* **66** (2002) 024306. E-print quant-ph/0203133.

A.P. Flitney and D. Abbott, "An introduction to quantum game theory", *Fluct. Noise Lett.* **2** (2002) R175. E-print quant-ph/0208069.

A.P. Flitney and D. Abbott, "A semi-quantum version of the game of Life", in *Advances in Dynamic Games: Applications to Economics, Finance, Optimization, and Stochastic Control*, Eds. A. S. Nowack and K. Szajowski (Birkhauser, Boston, 2004) 649; quant-ph/0208149.

C.F. Lee and N. Johnson, "Let the quantum games begin", *Physics World*, October (2002) 25.

[56] A. Iqbal and A.H. Toor, "Evolutionary Stable Strategies in Quantum Games", *Phys. Lett. A* **280** (2001) 249. E-print quant-ph/0007100.

A. Iqbal and A.H. Toor, "Darwinism in quantum systems?", *Physics Letters A* **294** (2002) 261. E-print quant-ph/0103085 at http://arXiv.org.

R. Kay, N.F. Johnson and S.C. Benjamin, "Evolutionary quantum game", *J. Phys. A* **34** (2001) L547. E-print quant-ph/0102008.

C.F. Lee and N. Johnson, "Efficiency and formalism of quantum games", *Phys. Rev. A* **67** (2003) 022311. E-print quant-ph/0207012.

A. Iqbal, "Quantum games with a multi-slit electron diffraction setup", *Nuovo Cimento B* 118 (2003) 463. E-print quant-ph/0207078.

[57] E.W. Piotrowski and J. Sladkowski, "Quantum Market Games", *Physica A* **312** (2002) 208. E-print quant-ph/0104006 at http://arXiv.org.

E.W. Piotrowski and J. Sladkowski, "Quantum English Auctions", *E-print quant-ph/0108017 at http://arXiv.org*.

E.W. Piotrowski and J. Sladkowski, "Quantum-like approach to financial risk: quantum anthropic principle", *Acta Phys. Polon.* **B32** (2001) 3873. E-print quant-ph/0110046. See also *quant-ph/0201045*.

E.W. Piotrowski and J. Sladkowski, "Quantum bargaining games", *Physica A* **308** (2002) 391. E-print quant-ph/0106140.

E.W. Piotrowski and J. Sladkowski, "Quantum diffusion of prices and profits", *Physica A* **345** (2005) 185. E-print cond-mat/0207132.

E.W. Piotrowski and J. Sladkowski, "An invitation to Quantum Game Theory", *Int. J. Theor. Phys.* **42** (2003) 1089. E-print quant-ph/0211191 at http://arXiv.org.

M. Schaden, "Quantum Finance", *Physica A* **316** (2002) 511. E-print physics/0203006 at http://arXiv.org.

R.V. Mendes, "Quantum games and social norms.The quantum ultimatum game", *Quantum Information Processing* **4** (2005) 1. E-print quant-ph/0208167 at http://arXiv.org.

[58] S. Lloyd, "Quantum-Mechanical Computers and Uncomputability", *Phys. Rev. Lett.* **71** (1993) 943.

[59] L. Fortnow and J.D. Rogers, "Complexity limitations on quantum computation", *E-print cs.CC/9811023 at http://arXiv.org*.

[60] R. Cleve, "An Introduction to Quantum Complexity Theory", *E-print quant-ph/9906111 at http://arXiv.org*.

J. Watrous, "Quantum Computational Complexity", *E-print arXiv:0804.3401 at http://arXiv.org*.

[61] P. Vitányi, "Three Approaches to the Quantitative Definition of Information in an Individual Pure Quantum State", *E-print quant-ph/9907035 at http://arXiv.org*.

[62] S. Lloyd, "Universe as quantum computer", *Complexity* **3** (1997) 32. E-print quant-ph/9912088 at http://arXiv.org.

[63] P.A. Zizzi, "The Early Universe as a Quantum Growing Network", *E-print gr-qc/0103002 at http://arXiv.org*.

P.A. Zizzi, "Ultimate Internets", *E-print gr-qc/0110122 at http://arXiv.org*.

P. Zizzi, "Spacetime at the Planck Scale: The Quantum Computer View", *E-print gr-qc/0304032 at http://arXiv.org*.

[64] D.S. Abrams and S. Lloyd, "Nonlinear quantum mechanics implies polynomial-time solution for NP-complete and #P problems", *Phys. Rev. Lett.* **81** (1998) 3992. E-print quant-ph/9801041 at http://arXiv.org.

[65] R. Schack, "Using a quantum computer to investigate quantum chaos", *Phys. Rev. A* **57** (1998) 1634. E-print quant-ph/9705016.

T.A. Brun and R. Schack, "Realizing the quantum baker's map on an NMR quantum computer", *Phys. Rev. A* **59** (1999) 2649. E-print quant-ph/9807050 at http://arXiv.org.

[66] I. Kim and G. Mahler, "Quantum chaos in quantum Turing machines", *Phys. Lett. A* **263** (1999) 268. E-print quant-ph/9910068.

[67] M. Ohya and I.V. Volovich, "Quantum Computing, NP-complete Problems and Chaotic Dynamics", *E-print quant-ph/9912100 at http://arXiv.org*.

[68] S. Hameroff and R. Penrose, "Orchestrated reduction of quantum coherence in brain microtubules: A model for consciousness", in *Proceedings of the Workshop "Biophysical Aspects of Coherence" (1995), Neural Network World* **5** (1995) 793.

S. Hagan, S.R. Hameroff and J.A. Tuszynski, "Quantum Computation in Brain Microtubules: Decoherence and Biological Feasibility", *Phys. Rev. E* **65** (2002) 061901. E-print quant-ph/0005025 at http://arXiv.org.

S. Hameroff, "Quantum computation in brain microtubules? The Penrose-Hameroff "Orch OR" model of consciousness", *Phil. Trans. Roy. Soc. Lond. A* **356** (1998) 1869. See also http://www.quantumconsciousness.org/penrose-hameroff/quantumcomputation.html.

[69] A. Mershin, D.V. Nanopoulos and E.M.C. Skoulakis, "Quantum Brain?", *E-print quant-ph/0007088 at http://arXiv.org*.

M. Perus and H. Bischof, "A neural-network-like quantum information processing system", *E-print quant-ph/0305072 at http://arXiv.org*.

V.I. Yukalov and D. Sornette, "Scheme of thinking quantum systems", *Laser Phys. Lett.* **6** (2009) 833. E-print arXiv:0909.1186.

[70] E. Alfinito and G. Vitiello, "The dissipative quantum model of brain: how do memory localize in correlated neuronal domains", *E-print quant-ph/0006066 at http://arXiv.org*.

E. Pessa and G. Vitiello, "Quantum noise, entanglement and chaos in the quantum field theory of mind/brain states", *Mind and Matter* **1** (2003) 59. E-print q-bio.OT/0309009 at http://arXiv.org.

[71] H. Hu and M. Wu, "Spin-Mediated Consciousness: Theory, Experimental Studies, Further Development & Related Topics", *Medical Hypotheses* **63** (2004) 633. E-print quant-ph/0208068 at http://arXiv.org.

[72] A. Zeilinger, "A foundational principle for quantum mechanics", *Found. Phys.* **29** (1999) 631.

C. Brukner and A. Zeilinger, "Information and fundamental elements of the structure of quantum theory", *E-print quant-ph/0212084 at http://arXiv.org*.

[73] K. Svozil, "The information interpretation of quantum mechanics", *E-print quant-ph/0006033 at http://arXiv.org*.

[74] D. Deutsch, *The Fabric of Reality* (Penguin Press, London, 1997).

[75] D. Deutsch, "The Structure of the Multiverse", *E-print quant-ph/0104033 at http://arXiv.org*.

[76] A.M. Steane, "A quantum computer only needs one universe", *E-print quant-ph/0003084 at http://arXiv.org*.

[77] D. Deutsch, A. Ekert and R. Lupacchini, "Machines, Logic and Quantum Physics", *E-print math.HO/9911150 at http://arXiv.org*.

[78] P. Benioff, "A Simple Example of Definitions of Truth, Validity, Consistency, and Completeness in Quantum Mechanics", *Phys. Rev. A* **59** (1999) 4223. E-print quant-ph/9811055 at http://arXiv.org.

P. Benioff, "Use of Mathematical Logical Concepts in Quantum Mechanics: An Example", *J. Phys. A* **35** (2002) 5843. E-print quant-ph/0106153 at http://arXiv.org.

[79] S. Haroche and J.-M. Raimond, "Quantum Computing: Dream or Nightmare?", *Physics Today*, August (1996) 51.

[80] R. Landauer, "The physical nature of information", *Phys. Lett. A* **217** (1996) 188.

R. Landauer, "Information is physical", *Physics Today* **44**, May (1991) 23.

[81] M.A. Nielsen, "Computable Functions, Quantum Measurements, and Quantum Dynamics", *Phys. Rev. Lett.* **79** (1997) 2915.

M.A. Nielsen, "Quantum information science as an approach to complex quantum systems", *E-print quant-ph/0208078 at http://arXiv.org*. See also *quant-ph/0210005*.

M.A. Nielsen, C.M. Dawson, J.L. Dodd, A. Gilchrist, D. Mortimer, T.J. Osborne, M.J. Bremner, A.W. Harrow and A. Hines, "Quantum dynamics as a physical resource", *Phys. Rev. A* **67** (2003) 052301. E-print quant-ph/0208077 at http://arXiv.org.

[82] G. 't Hooft, "Quantum Gravity as a Dissipative Deterministic System", *Class. Quant. Grav.* **16** (1999) 3263. E-print gr-qc/9903084 at http://arXiv.org.

G. 't Hooft, "Determinism and Dissipation in Quantum Gravity", *E-print hep-th/0003005 at http://arXiv.org*.

[83] G. Vidal and J.I. Cirac, "Irreversibility in asymptotic manipulations of entanglement", *Phys. Rev. Lett.* **86** (2001) 5803. E-print quant-ph/0102036 at http://arXiv.org.

[84] H. De Raedt, K. Michielsen, A. Hams, S. Miyashita and K. Saito, "Quantum spin dynamics as a model for quantum computer operation", *Eur. Phys. J. B* **27** (2002) 15. E-print quant-ph/0104085.

H. De Raedt, K. Michielsen, S. Miyashita and K. Saito, "Physical Models for Quantum Computers", *Prog. Theor. Phys. Suppl.* **145** (2002) 233.

[85] G. Brassard, N. Lütkenhaus, T. Mor and B.C. Sanders, "Security Aspects of Practical Quantum Cryptography", *Phys. Rev. Lett.* **85** (2000) 1330. E-print quant-ph/9911054 at http://arXiv.org.

[86] G. Adenier, "Representation of Joint Measurement in Quantum Mechanics: A Refutation of Quantum Teleportation", *E-print quant-ph/0105031 at http://arXiv.org*.

D. Tommasini, "A single quantum cannot be teleported", *E-print quant-ph/0210060 at http://arXiv.org*.

[87] S. Kak, "Are Quantum Computing Models Realistic?", *E-print quant-ph/0110040 at http://arXiv.org*.

S. Kak, "Uncertainty In Quantum Computation", *E-print quant-ph/0206006 at http://arXiv.org*.

S. Kak, "General Qubit Errors Cannot Be Corrected", *Information Sciences* **152** (2003) 195. E-print quant-ph/0206144.

S. Kak, "Three Paradoxes of Quantum Information", *E-print quant-ph/0304060 at http://arXiv.org*.

S. Kak, "Statistical Constraints on State Preparation for a Quantum Computer", *Pramana* **57** (2001) 683. E-print quant-ph/0010109.

S. Kak, "On Initializing Quantum Registers and Quantum Gates", *Foundations of Physics* **29** (1999) 267. E-print quant-ph/9805002 at http://arXiv.org.

A. Rossi, "Energy-Information Coupling From Classical To Quantum", *E-print quant-ph/0211069 at http://arXiv.org*.

Q. Cai, "Accessible Information and Quantum Operations", *E-print quant-ph/0303117 at http://arXiv.org*.

T. Brun, H. Klauck, A. Nayak and Ch. Zalka, "Comment on "Probabilistic Quantum Memories"", *Phys. Rev. Lett.* **91** (2003) 209801. E-print quant-ph/0303091 at http://arXiv.org.

[88] M.I. Dyakonov, "Quantum computing: a view from the enemy camp", in *Future Trends in Microelectronics. The nano Millenium*, Eds. S. Luryi, J. Xu and A. Zaslavsky (Wiley, New York, 2002), p. 307. E-print cond-mat/0110326 at http://arXiv.org.

M.I. Dyakonov, "Is Fault-Tolerant Quantum Computation Really Possible?", in *Future Trends in Microelectronics. Up the Nano Creek*, Eds. S. Luryi, J. Xu and A. Zaslavsky (Wiley, New York, 2007), p. 4. E-print arXiv:1212.3562 at http://arXiv.org.

M.I. Dyakonov, "Revisiting the hopes for scalable quantum computation", *E-print arXiv:1210.1782 at http://arXiv.org*.

M.I. Dyakonov, "State of the art and prospects for quantum computing", in *Future Trends in Microelectronics. The nano Millenium*, Eds. S. Luryi, J. Xu and A. Zaslavsky (Wiley, New York, 2013), p. 266. E-print arXiv:1212.3562 at http://arXiv.org.

R. Alicki, "Remarks on the nature of quantum computation", *E-print quant-ph/0306103 at http://arXiv.org*.

G. Kalai, "How Quantum Computers Fail: Quantum Codes, Correlations in Physical Systems, and Noise Accumulation", *E-print arXiv:1106.0485 at http://arXiv.org*.

V.V. Aristov and A.V. Nikulov, "The fundamental obscurity in quantum mechanics. Why it is needed to shout "wake up"", *E-print arXiv:1108.2628 at http://arXiv.org*.

G.S. Paraoanu, "Quantum Computing: Theoretical versus Practical Possibility", *Phys. Perspect.* **13** (2011) 359. E-print arXiv:1110.3190 at http://arXiv.org.

[89] G. Castagnoli and D. Monti, "The Non-mechanistic Character of Quantum Computation", *E-print quant-ph/9811039 at http://arXiv.org*.

[90] W.G. Unruh, "Maintaining coherence in quantum computers", *Phys. Rev. A* **51** (1995) 992.

[91] I.L. Chuang, R. Laflamme, P.W. Shor and W.H. Zurek, "Quantum Computers, Factoring, and Decoherence", *Science* **270** (1995) 1633.

[92] J.P. Paz and W.H. Zurek, "Environment-Induced Decoherence and the Transition From Quantum to Classical", in *Coherent Matter Waves. Les Houches Session LXXII*, Eds. R. Kaiser, C. Westbrook and F. David (Springer, Berlin, 2001), p. 533. E-print quant-ph/0010011.

J.P. Paz, "Protecting the quantum world", *Nature* **412** (2001) 869.

[93] P.W. Shor, "Scheme for reducing decoherence in quantum computer memory", *Phys. Rev. A* **52** (1995) R2493.

A.R. Calderbank and P.W. Shor, "Good quantum error-correcting codes exist", *Phys. Rev. A* **54** (1996) 1098.

[94] A. Steane, "Multiple particle interference and quantum error correction", *Proc. Roy. Soc. Lond. A* **452** (1996) 2551.

A.M. Steane, "Overhead and noise threshold of fault-tolerant quantum error correction", *Phys. Rev. A* **68** (2003) 042322. E-print quant-ph/0207119 at http://arXiv.org.

A.M. Steane, "Quantum Computing and Error Correction", in *Decoherence and its implications in quantum computation and information transfer*, Eds. Gonis and Turchi (IOS Press, Amsterdam, 2001) p. 284. E-print quant-ph/0304016 at http://arXiv.org.

[95] E. Knill and R. Laflamme, "A theory of quantum error correcting codes", *Phys. Rev. A* **55** (1997) 900.

E. Knill, R. Laflamme and L. Viola, "Theory of Quantum Error Correction for General Noise", *E-print quant-ph/9908066*.

E. Knill, R. Laflamme, A. Ashikhmin, H. Barnum, L. Viola and W.H. Zurek, "Introduction to Quantum Error Correction", *E-print quant-ph/0207170 at http://arXiv.org*.

[96] C.H. Bennett, D.P. DiVincenzo, J.A. Smolin and W.K. Wootters, "Mixed state entanglement and quantum error-correcting codes", *Phys. Rev. A* **54** (1996) 3824.

[97] D.G. Cory, M.D. Price, W. Maas, E. Knill, R. Laflamme, W.H. Zurek, T.F. Havel and S.S. Somaroo, "Experimental Quantum Error Correction", *Phys. Rev. Lett.* **81** (1998) 2152.

[98] J. Preskill, "Battling Decoherence: The Fault-Tolerant Quantum Computer", *Physics Today*, June (1999) 24.

[99] P. Zanardi, "Stabilizing Quantum Information", *Phys. Rev. A* **63** (2001) 12301. E-print quant-ph/9910016 at http://arXiv.org.

P. Zanardi, "Stabilization of Quantum Information: A Unified Dynamical-Algebraic Approach", *E-print quant-ph/0203008 at http://arXiv.org*.

P. Zanardi and S. Lloyd, "Universal control of quantum subspaces and subsystems", *Phys. Rev. A* **69** (2004) 022313; quant-ph/0305013.

[100] S. Lloyd, "Quantum controllers for quantum systems", *E-print quant-ph/9703042 at http://arXiv.org*.

[101] L. Viola and S. Lloyd, "Dynamical suppression of decoherence in two-state quantum systems", *Phys. Rev.* A **58** (1998) 2733. E-print quant-ph/9803057 at http://arXiv.org.

[102] L. Viola, E. Knill and S. Lloyd, "Dynamical Decoupling of Open Quantum Systems", *Phys. Rev. Lett.* **82** (1999) 2417. E-print quant-ph/9809071 at http://arXiv.org.

L. Viola, S. Lloyd and E. Knill, "Universal Control of Decoupled Quantum Systems", *Phys. Rev. Lett.* **83** (1999) 4888. E-print quant-ph/9906094 at http://arXiv.org.

L. Viola, E. Knill and S. Lloyd, "Dynamical Generation of Noiseless Quantum Subsystems", *Phys. Rev. Lett.* **85** (2000) 3520. E-print quant-ph/0002072 at http://arXiv.org.

S. Lloyd and L. Viola, "Control of open quantum system dynamics", *E-print quant-ph/0008101 at http://arXiv.org*.

[103] D.A. Lidar, I.L. Chuang and K.B. Whaley, "Decoherence-free subspaces for quantum computation", *Phys. Rev. Lett.* **81** (1998) 2594. E-print quant-ph/9807004 at http://arXiv.org.

D. Bacon, D.A. Lidar and K.B. Whaley, "Robustness of decoherence-free subspaces for quantum computation", *Phys. Rev. A* **60** (1999) 1944. E-print quant-ph/9902041.

D. Bacon, J. Kempe, D.A. Lidar and K.B. Whaley, "Universal Fault-Tolerant Quantum Computation on Decoherence-Free Subspaces", *Phys. Rev. Lett.* **85** (2000) 1758. E-print quant-ph/9909058 at http://arXiv.org.

M.S. Byrd and D.A. Lidar, "Combined encoding, recoupling, and decoupling solution to problems of decoherence and design in solid-state quantum computing", *Phys. Rev. Lett.* **89** (2002) 047901. E-print quant-ph/0112054 at http://arXiv.org.

D.A. Lidar and L.-A. Wu, "Quantum Computers and Decoherence: Exorcising the Demon from the Machine", *E-print quant-ph/0302198 at http://arXiv.org*.

[104] A. Beige, D. Braun, B. Tregenna and P.L. Knight, "Quantum Computing Using Dissipation to Remain in a Decoherence-Free Subspace", *Phys. Rev. Lett.* **85** (2000) 1762.

[105] A.C. Doherty, S. Habib, K. Jacobs, H. Mabuchi and S.M. Tan, "Quantum feedback control and classical control theory", *Phys. Rev. A* **62** (2000) 012105. E-print quant-ph/9912107 at http://arXiv.org.

A. Doherty, J. Doyle, H. Mabuchi, K. Jacobs and S. Habib, "Robust control in the quantum domain", *E-print quant-ph/0105018 at http://arXiv.org*. Proceedings of the 39th IEEE Conference on Decision and Control (2000), p. 949.

[106] S.G. Schirmer, H. Fu and A.I. Solomon, "Complete controllability of quantum systems", *Phys. Rev. A* **63** (2001) 063410. E-print quant-ph/0010031 at http://arXiv.org.

H. Fu, S.G. Schirmer and A.I. Solomon, "Complete controllability of finite-level quantum systems", *Journ. Phys. A* **34** (2001) 1679. E-print quant-ph/0102017 at http://arXiv.org.

A.I. Solomon and S.G. Schirmer, "Limitations on Quantum Control", *Int. J. Mod. Phys. B* **16** (2002) 2107. E-print quant-ph/0110030.

S.G. Schirmer, A.I. Solomon and J.V. Leahy, "Degrees of controllability for quantum systems and application to atomic systems", *J. Phys. A* **35** (2002) 4125. E-print quant-ph/0108114 at http://arXiv.org.

[107] R.J. Nelson, Y. Weinstein, D. Cory and S. Lloyd, "Experimental Demonstration of Fully Coherent Quantum Feedback", *Phys. Rev. Lett.* **85** (2000) 3045.

Y.S. Weinstein, S. Lloyd, J. Emerson and D.G. Cory, "Experimental Implementation of the Quantum Baker's Map", *Phys. Rev. Lett.* **89** (2002) 157902. E-print quant-ph/0201064 at http://arXiv.org.

[108] I.L. Chuang, N. Gershenfeld and M. Kubinec, "Experimental Implementation of Fast Quantum Searching", *Phys. Rev. Lett.* **80** (1998) 3408.

L.M.K. Vandersypen, M. Steffen, G. Breyta, C.S. Yannoni, M.H. Sherwood and I.L. Chuang, "Experimental realization of Shor's quantum factoring algorithm using nuclear magnetic resonance", *Nature* **414** (2001) 883.

[109] S.M. Maurer, T. Hogg and B.A. Huberman, "Portfolios of Quantum Algorithms", *Phys. Rev. Lett.* **87** (2001) 257901. E-print quant-ph/0105071 at http://arXiv.org.

[110] N.J. Cerf and C. Adami, "Information theory of quantum entanglement and measurement", *Physica D* **120** (1998) 62.

[111] L. Viola, E. Knill and R. Laflamme, "Constructing Qubits in Physical Systems", *J. Phys. A* **34** (2001) 7067. E-print quant-ph/0101090.

[112] D.F.V. James, P.G. Kwiat, W.J. Munro and A.G. White, "Measurement of Qubits", *Physical Review A* **64** (2001) 052312. E-print quant-ph/0103121 at http://arXiv.org.

[113] M.B. Plenio and V. Vitelli, "The physics of forgetting: Landauer's erasure principle and information theory", *Contemporary Physics* **42** (2001) 25. E-print quant-ph/0103108 at http://arXiv.org.

[114] G. Chapline, "Is theoretical physics the same thing as mathematics?", *Phys. Rep.* **315** (1999) 95.

[115] H. Touchette and S. Lloyd, "Information-Theoretic Limits of Control", *Phys. Rev. Lett.* **84** (2000) 1156.

H. Touchette and S. Lloyd, "Information-Theoretic Approach to the Study of Control Systems", *Physica A* **331** (2004) 140. E-print physics/0104007 at http://arXiv.org.

[116] G. Brassard, S.L. Braunstein and R. Cleve, "Teleportation as a quantum computation", *Physica D* **120** (1998) 43.

[117] P. Benioff, "The Representation of Natural Numbers in Quantum Mechanics", *Phys. Rev. A* **63** (2001) 032305; quant-ph/0003063.

P. Benioff, "The Representation of Numbers by States in Quantum Mechanics", *E-print quant-ph/0009124 at http://arXiv.org*.

P. Benioff, "The Representation of Numbers in Quantum Mechanics", *Algorithmica* **34** (2002) 529. E-print quant-ph/0103078.

P. Benioff, "Efficient Implementation and the Product State Representation of Numbers", *Phys. Rev. A* **64** (2001) 052310.

[118] D. Aharonov, "A Quantum to Classical Phase Transition in Noisy Quantum Computers", *E-print quant-ph/9910081 at http://arXiv.org*.

[119] R.G. Newton, *Scattering Theory of Waves and Particles* (McGraw-Hill, New York, 1967).

[120] N.F. Mott and H.S.W. Massey, *The Theory of Atomic Collisions* (Clarendon Press, Oxford, 1965).

[121] P.H. Dederichs, "Dynamical Diffraction Theory by Optical Potential Methods", *Solid State Physics: Advances in Research and Applications*, Eds. H. Ehrenreich, F. Seitz and D. Turnbull (Academic Press, New York) **27** (1972) 136.

[122] B. Mandelbrot, *The Fractal Geometry of Nature* (Freeman, San Francisco, 1982).

[123] J. Feder, *Fractals* (Plenum Press, New York, 1988).

[124] H.-O. Peintgen, H. Jürgens and D. Saupe, *Chaos and Fractals. New Frontiers of Science* (Springer-Verlag, New York, 1992).

[125] T. Nakayama, K. Yakubo and R.L. Orbach, "Dynamical properties of fractal networks: scaling, numerical simulations, and physical realisations", *Rev. Mod. Phys.* **66** (1994) 381.

[126] R. Blumenfeld and B.B. Mandelbrot, "Lévy dusts, Mittag-Leffler statistics, mass fractal lacunarity, and perceived dimension", *Phys. Rev. E* **56** (1997) 112.

[127] M.V. Berry, "Quantum fractals in boxes", *J. Phys. A* **29** (1996) 6617.

[128] D. Wójcik, I. Bialynicki-Birula and K. Zyczkowski, "Time Evolution of Quantum Fractals", *Phys. Rev. Lett.* **85** (2000) 5022.

[129] H. Haken, *Rev. Mod. Phys.* **47** (1975) 67.

H. Haken, *Synergetics. An Introduction,* 3rd ed. (Springer-Verlag, Berlin, 1983).

H. Haken, *Advanced Synergetics* (Springer-Verlag, Berlin, 1983).

H. Haken, *Information and Self-Organisation: A Macroscopic Approach to Complex Systems* (Springer-Verlag, Berlin, 1988).

*Interdisciplinary Approaches to Nonlinear Complex Systems*, Eds. H. Haken and A. Mikhailov (Springer-Verlag, Berlin, 1993).

H. Haken, "Slaving principle revisited", *Physica D* **97** (1996) 95.

W. Ebeling and F. Schweitzer, "Self-Organization, Active Brownian Dynamics, and Biological Applications", *Nova Acta Leopoldina NF* **88** (2003) 169. E-print cond-mat/0211606 at http://arXiv.org.

[130] I. Prigogine and I. Stengers, *Order out of Chaos* (Heinemann, London, 1984).

G. Nicolis and I. Prigogine, *Exploring Complexity* (Freeman, San Francisco, 1989).

G. Nicolis, "Physics of far-from-equilibrium systems and self-organisation", in *The New Physics*, Ed. P. Davies (Cambridge University Press, 1989).

[131] P. Bak, *How Nature Works: The Science of Self-Organised Criticality* (Copernicus/Springer, 1996).

[132] A. Csilling, I.M. Jánosi, G. Pásztor and I. Scheuring, *Phys. Pev. E* **50** (1994) 1083.

[133] S.L. Perke and J.M. Carlson, *Phys. Rev. E* **50** (1994) 236.

[134] M. de Sousa Vieira and A.J. Lichtenberg, *Phys. Rev. E* **53** (1996) 1441.

[135] S. Boccaletti, C. Grebogi, Y.-C. Lai, H. Mancini and D. Maza, "The Control of Chaos: Theory and Applications", *Phys. Rep.* **329** (2000) 103.

[136] R.L. Viana and C. Grebogi, "Unstable dimension variability and synchronization of chaotic systems", *Phys. Rev. E* **62** (2000) 462. E-print chao-dyn/9911027 at http://arXiv.org.

[137] W. van der Water and J. de Weger, "Failure of chaos control", *Phys. Rev. E* **62** (2000) 6398.

[138] L.D. Landau and E.M. Lifshitz, *Mechanics* (Nauka, Moscow, 1988). Forth Russian edition.

[139] B.V. Chirikov, At. Energ. **6** (1959) 630. English translation: *J. Nucl. Energy Part C: Plasma Phys.* **1** (1960) 253.

B.V. Chirikov, *Phys. Rep.* **52** (1979) 263.

[140] D.F. Escande, *Phys. Rep.* **121** (1985) 165.

[141] A.J. Lichtenbegr and M.A. Lieberman, *Regular and Stochastic Motion* (Springer-Verlag, New York, 1983).

[142] G.M. Zaslavsky, *Chaos in Dynamical Systems* (Harwood Academic Publishers, London, 1985). Russian edition: Nauka, Moscow, 1984.

G.M. Zaslavsky, R.Z. Sagdeev, D.A. Usikov and A.A. Chernikov, *Weak Chaos and Quasi-Regular Patterns* (Cambridge University Press, 1991). Russian edition: Nauka, Moscow, 1991.

[143] H.G. Schuster, *Deterministic Chaos* (Physik-Verlag, Weinheim, 1984).

[144] M.C. Gutzwiller, *Chaos in Classical and Quantum Mechanics* (Springer-Verlag, New York, 1990).

[145] E. Ott, *Chaos in Dynamical Systems* (Cambridge University Press, 1993).

[146] W.H. Zurek, "Environment-induced super-selection rules", *Phys. Rev. D* **26** (1982) 1862.

[147] E. Joos and H.D. Zeh, "The emergence of classical properties through the interaction with the environment", *Z. Phys. B* **59** (1985) 229.

H.D. Zeh, "What is Achieved by Decoherence", *E-print quant-ph/9610014 at http://arXiv.org*.

[148] W.H. Zurek, "Decoherence and the transition from quantum to classical", *Phys. Today* **44** (1991) 36.

W.H. Zurek, "Decoherence, Einselection, and the Existential Interpretation (the Rough Guide)", *Phil. Trans. Roy. Soc. Lond. A* **356** (1998) 1793.

J.P. Paz and W.H. Zurek, "Quantum limit of decoherence: Environment induced superselection of energy eigenstates", *Phys. Rev. Lett.* **82** (1999) 5181. E-print quant-ph/9811026.

W. H. Zurek, "Decoherence, einselection, and the quantum origin of the classical", *Rev. Mod. Phys.* **75** (2003) 715. E-print quant-ph/0105127 at http://arXiv.org.

[149] D. Guilini, E. Joos, C. Kiefer, J. Kupsch, L.-O. Stamatescu and H. D. Zeh, *Decoherence and the Appearance of a Classical World in Quantum Theory* (Springer, Berlin, 1996). New edition: 2003.

[150] S. Lloyd, "Ultimate physical limits to computation", *Nature* **406** (2000) 1047. E-print quant-ph/9908043 at http://arXiv.org.

Also: "Seth Lloyd — How fast, how small, and how powerful?: Moore's law and the ultimate laptop", http://www.edge.org/3rd_culture/lloyd/lloyd_index.html;

"The Computational Universe: Seth Lloyd", http://www.edge.org/3rd_culture/lloyd2/lloyd2_index.html.

S. Lloyd, "Computational capacity of the Universe", *Phys. Rev. Lett.* **88** (2002) 237901. E-print quant-ph/0110141 at http://arXiv.org.

[151] Y.J. Ng, "From computation to black holes and space-time foam", *Phys. Rev. Lett.* **86** (2001) 2946. E-print gr-qc/0006105 at http://arXiv.org.

[152] S. Tomsovic, "Tunneling and Chaos", *E-print nlin.CD/0008031 at http://arXiv.org*.

S. Tomsovic and D. Ullmo, "Chaos-assisted tunneling", *Phys. Rev. E* **50** (1994) 145.

[153] T. Onishi, A. Shudo, K.S. Ikeda and K. Takahashi, "Tunneling mechanism due to Chaos in a Complex Phase Space", *E-print nlin.CD/0105067 at http://arXiv.org*.

A. Shudo and K.S. Ikeda, "Complex Classical Trajectories and Chaotic Tunneling", *Phys. Rev. Lett.* **74** (1995) 682.

[154] T. Giesel, G. Radons and J. Rubner, "Kolmogorov-Arnol'd-Moser Barriers in the Quantum Dynamics of Chaotic Systems", *Phys. Rev. Lett.* **57** (1986) 2883.

[155] *Chaos and Quantum Physics*, Ed. M.-J. Giannoni, A. Voros and J. Zinn-Justin (North-Holland, Amsterdam, 1991).

[156] Ph. Jacquod and D.L. Shepelyansky, "Emergence of Quantum Chaos in Finite Interacting Fermi Systems", *Phys. Rev. Lett.* **79** (1997) 1837.

[157] P. H. Song and D.L. Shepelyansky, "Quantum Computing of Quantum Chaos and Imperfection Effects", *Phys. Rev. Lett.* **86** (2001) 2162. E-print quant-ph/0009005 at http://arXiv.org.

B. Georgeot and D.L. Shepelyansky, "Exponential Gain in Quantum Computing of Quantum Chaos and Localization", *Phys. Rev. Lett.* **86** (2001) 2890. E-print quant-ph/0010005 at http://arXiv.org.

G. Benenti, G. Casati, S. Montangero and D. Shepelyansky, "Dynamical localization simulated on a few qubits quantum computer", *Phys. Rev. A* **67** (2003) 052312. E-print quant-ph/0210052.

[158] B. Georgeot and D.L. Shepelyansky, "Quantum Computing of Classical Chaos: Smile of the Arnold-Schrödinger Cat", *Phys. Rev. Lett.* **86** (2001) 5393. E-print quant-ph/0101004 at http://arXiv.org.

B. Georgeot and D.L. Shepelyansky, "Efficient quantum computing insensitive to phase errors", *E-print quant-ph/0102082 at http://arXiv.org.*

B. Georgeot and D.L. Shepelyansky, "Quantum computer inverting time arrow for macroscopic systems", *Eur. Phys. J. D* **19** (2002) 263. E-print quant-ph/0105149 at http://arXiv.org.

G. Benenti, G. Casati, S. Montangero and D.L. Shepelyansky, "Efficient Quantum Computing of Complex Dynamics", *Phys. Rev. Lett.* **87** (2001) 227901. E-print quant-ph/0107036 at http://arXiv.org.

C. Zalka, "Comment on "Stable Quantum Computation of Unstable Classical Chaos"", *E-print quant-ph/0110019 at http://arXiv.org.*

L. Diosi, "Comment on "Stable Quantum Computation of Unstable Classical Chaos"", *Phys. Rev. Lett.* **88** (2002) 219801. E-print quant-ph/0110026 at http://arXiv.org.

B. Georgeot and D.L. Shepelyansky, "Efficient quantum computation of high harmonics of the Liouville density distribution", *Phys. Rev. Lett.* **88** (2002) 219802. E-print quant-ph/0110142 at http://arXiv.org.

A. Maassen van den Brink, "Comments on 'Stable Quantum Computation of Unstable Classical Chaos', 'Efficient Quantum Computing Insensitive to Phase Errors', and 'Quantum Computer Inverting Time Arrow for Macroscopic Systems'", *E-print quant-ph/0112006 at http://arXiv.org*.

A.D. Chepelianskii and D.L. Shepelyansky, "Schrödinger cat animated on a quantum computer", *Phys. Rev. A* **66** (2002) 054301. E-print quant-ph/0202113 at http://arXiv.org.

M. Terraneo, B. Georgeot and D.L. Shepelyansky, "Strange attractor simulated on a quantum computer", *Eur. Phys. J. D* **22** (2003) 127. E-print quant-ph/0203062 at http://arXiv.org.

B. Georgeot and D.L. Shepelyansky, "Les ordinateurs quantiques affrontent le chaos", *E-print quant-ph/0307103 at http://arXiv.org*.

[159] D. Braun, "Quantum Chaos and Quantum Algorithms", *Phys. Rev. A* **65** (2002) 042317-1. E-print quant-ph/0110037 at http://arXiv.org.

[160] G.P. Berman, F. Borgonovi, F.M. Izrailev and V.I. Tsifrinovich, "Avoiding Quantum Chaos in Quantum Computation", *Phys. Rev. E* **65** (2001) 015204. E-print quant-ph/0012106 at http://arXiv.org.

G.P. Berman, F. Borgonovi, F.M. Izrailev and V.I. Tsifrinovich, "Onset of Chaos in a Model of Quantum Computation", *E-print quant-ph/0103009 at http://arXiv.org*.

G.P. Berman, F. Borgonovi, F.M. Izrailev and V.I. Tsifrinovich, "Delocalization border and onset of chaos in a model of quantum computation", *Phys. Rev. E* **64** (2001) 056226. E-print quant-ph/0104086.

G.P. Berman, F. Borgonovi, G. Celardo, F.M. Izrailev and D.I. Kamenev, "Dynamical fidelity of a solid-state quantum computation", *Phys. Rev. E* **66** (2002) 056206. E-print quant-ph/0206158.

[161] V.V. Flambaum and F.M. Izrailev, "Unconventional decay law for excited states in closed many-body systems", *Phys. Rev. E* **64** (2001) 026124. E-print quant-ph/0102088 at http://arXiv.org.

[162] J. Gong and P. Brumer, "Coherent Control of Quantum Chaotic Diffusion", *Phys. Rev. Lett.* **86** (2001) 1741. E-print quant-ph/0012150 at http://arXiv.org.

[163] A.R.R. Carvalho, P. Milman, R.L. de Matos Filho and L. Davidovich, "Decoherence, Pointer Engineering, and Quantum State Protection", *Phys. Rev. Lett.* **86** (2001) 4988. E-print quant-ph/0009024 at http://arXiv.org.

[164] T. Prosen, "On general relation between quantum ergodicity and fidelity of quantum dynamics: from integrable to ergodic and mixing motion in kicked Ising chain", *Phys. Rev. E* **65** (2002) 036208. E-print quant-ph/0106149 at http://arXiv.org.

T. Prosen and M. Žnidarič, "Can quantum chaos enhance stability of quantum computation?", *J. Phys. A* **34** (2001) L681. E-print quant-ph/0106150 at http://arXiv.org.

T. Prosen and M. Žnidarič, "Stability of quantum motion and correlation decay", *J. Phys. A* **35** (2002) 1455. E-print nlin.CD/0111014 at http://arXiv.org.

T. Prosen and T.H. Seligman, "Decoherence of spin echoes", *J. Phys. A* **35** (2002) 4707. E-print nlin.CD/0201038 at http://arXiv.org.

[165] S. Mancini, D. Vitali, R. Bonifacio and P. Tombesi, "Stochastic control of quantum coherence", *Europhys. Lett.* **60** (2002) 498. E-print quant-ph/0108011 at http://arXiv.org.

[166] V. Scarani, M. Ziman, P. Stelmachovic, N. Gisin and V. Buzek, "Thermalizing Quantum Machines: Dissipation and Entanglement", *Phys. Rev. Lett.* **88** (2002) 097905. E-print quant-ph/0110088 at http://arXiv.org.

[167] Y. Aharonov and J. Anandan, "Meaning of the Density Matrix", *E-print quant-ph/9803018 at http://arXiv.org*.

[168] J.F. Traub, "A Continuous Model of Computation", *Physics Today*, May (1999) 39. E-print physics/0106045 at http://arXiv.org.

[169] P. Gacs, "Quantum Algorithmic Entropy", *J. Phys. A* **34** (2001) 6859. E-print quant-ph/0011046 at http://arXiv.org.

[170] P. Gacs, J. Tromp and P. Vitanyi, "Algorithmic Statistics", *IEEE Trans. Inf. Theory* **47** (2001) 2443. E-print math.PR/0006233 at http://arXiv.org.

P. Vitányi, "Randomness", *E-print math.PR/0110086 at http://arXiv.org*.

P. Vitanyi, "Meaningful Information", *IEEE Trans. Inform. Th.* **52** (2006) 4617. E-print cs.CC/0111053 at http://arXiv.org.

P.M.B. Vitányi, "Algorithmic Chaos", *E-print nlin.CD/0303016 at http://arXiv.org*.

M. Li, J. Tromp and P. Vitanyi, "Sharpening Occam's Razor", *E-print cs.LG/0201005 at http://arXiv.org*.

G. Segre, "The information-theoretical viewpoint on the physical complexity of classical and quantum objects and their dynamical evolution", *Int. J. Theor. Phys.* **43** (2004) 1371; quant-ph/0210148.

[171] H. Buhrman, J. Tromp and P. Vitanyi, "Time and space bounds for reversible simulation", *J. Phys. A* **34** (2001) 6821.

[172] C.H. Bennett, P.W. Shor, J.A. Smolin and A.V. Thapliyal, "Entanglement-assisted capacity of a quantum channel and the reverse Shannon theorem", *E-print quant-ph/0106052 at http://arXiv.org*.

C.H. Bennett, G. Brassard, S. Popescu, B. Schumacher, J.A. Smolin and W.K. Wootters, "Purification of Noisy Entanglement and Faithful Teleportation via Noisy Channels", *Phys. Rev. Lett.* **76** (1996) 722. E-print quant-ph/9511027 at http://arXiv.org.

C.H. Bennett, A. Harrow, D.W. Leung and J.A. Smolin, "On the capacities of bipartite Hamiltonians and unitary gates", *IEEE Trans. Inf. Theory* **49** (2003) 1895. E-print quant-ph/0205057.

[173] A.S. Holevo, "On entanglement-assisted classical capacity", *E-print quant-ph/0106075 at http://arXiv.org*.

[174] C. Brukner, M. Zukowski and A. Zeilinger, "The essence of entanglement", *E-print quant-ph/0106119 at http://arXiv.org*.

T. Jennewein, G. Weihs, J.-W. Pan and A. Zeilinger, "Experimental Nonlocality Proof of Quantum Teleportation and Entanglement Swapping", *Phys. Rev. Lett.* **88** (2002) 017903. E-print quant-ph/0201134 at http://arXiv.org.

[175] E. Schrödinger, "Die gegenwärtige Situation in der Quantenmechanik", *Naturwissenschaften* **23** (1935) 807, 823, 844. English translation: *Proc. Am. Phil. Soc.* **124** (1980) 323; *Quantum Theory and Measurement*, Eds. J.A. Wheeler and W.H. Zurek (Princeton University Press, Princeton, 1983), p. 152.

[176] G. Ghirardi, L.Marinatto and T. Weber, "Entanglement and Properties of Composite Quantum Systems: a Conceptual and Mathematical Analysis", *J. Stat. Phys.* **108** (2002) 49. E-print quant-ph/0109017 at http://arXiv.org.

G. Ghirardi and L. Marinatto, "Entanglement and Properties", *Fortschr. Phys.* **51** (2003) 379. E-print quant-ph/0206021.

[177] M. Hillery, V. Buzek and M. Ziman, "Probabilistic implementation of universal quantum processors", *Phys. Rev. A* **65** (2002) 022301. E-print quant-ph/0106088 at http://arXiv.org.

[178] M.A. Nielsen, "Quantum computation by measurement and quantum memory", *Phys. Lett. A* **308** (2003) 96. E-print quant-ph/0108020.

[179] P. Wocjan, M. Rötteler, D. Janzing and T. Beth, "Universal Simulation of Hamiltonians Using a Finite Set of Control Operations", *E-print quant-ph/0109063 at http://arXiv.org*.

[180] M.A. Nielsen, M.J. Bremner, J.L. Dodd, A.M. Childs and C.M. Dawson, "Universal simulation of Hamiltonian dynamics for qudits", *Phys. Rev. A* **66** (2002) 022317. E-print quant-ph/0109064.

J.L. Dodd, M.A. Nielsen, M.J. Bremner and R.T. Thew, "Universal quantum computation and simulation using any entangling Hamiltonian and local unitaries", *Phys. Rev. A* **65** (2002) 040301(R). E-print quant-ph/0106064 at http://arXiv.org.

[181] Yu. Ozhigov and L. Fedichkin, "Quantum Computer with Fixed Interaction is Universal", *JETP Lett.* **77** (2003) 328. E-print quant-ph/0202030 at http://arXiv.org.

[182] W.H. Zurek, "Einselection and decoherence from an information theory perspective", *Annalen der Physik* **9** (2000) 855. E-print quant-ph/0011039 at http://arXiv.org.

[183] K. Svozil, "Irreducibility of n-ary quantum information", *E-print quant-ph/0111113 at http://arXiv.org*.

K. Svozil, "Quantum information in base n defined by state partitions", *Phys. Rev. A* **66** (2002) 044306. E-print quant-ph/0205031.

[184] K.H. Hoffmann, "Quantum thermodynamics", *Annalen der Physik* **10** (2000) 79.

[185] V. Capek, "Zeroth and Second Laws of Thermodynamics Simultaneously Questioned in the Quantum Microworld", *European Physical Journal B* **25** (2002) 101. E-print cond-mat/0012056 at http://arXiv.org.

V. Capek and J. Bok, "Stationary flows in quantum dissipative closed circuits as a challenge to thermodynamics", *E-print physics/0110018 at http://arXiv.org*.

V. Čápek and J. Bok, "Violation of the second law of thermodynamics in the quantum microworld", *Physica A* **290** (2001) 379.

V. Čápek and D.P. Sheehan, "Quantum mechanical model of a plasma system: a challenge to the second law of thermodynamics", *Physica A* **304** (2002) 461.

[186] A.E. Allahverdyan and Th.M. Nieuwenhuizen, "Extraction of work from a single thermal bath in the quantum regime", *Phys. Rev. Lett.* **85** (2000) 1799. E-print cond-mat/0006404 at http://arXiv.org.

Th.M. Nieuwenhuizen and A.E. Allahverdyan, "Statistical thermodynamics of quantum Brownian motion: Birth of perpetuum mobile of the second kind", *E-print cond-mat/0011389 at http://arXiv.org*.

A.E. Allahverdyan and Th.M. Nieuwenhuizen, "Invalidity of the Landauer inequality for information erasure in the quantum regime", *E-print cond-mat/0206052 at http://arXiv.org*.

A.E. Allahverdyan, R. Balian and Th.M. Nieuwenhuizen, "Quantum thermodynamics: thermodynamics at the nanoscale", *J. Mod. Optics* **51** (2004) 2703. E-print cond-mat/0402387 at http://arXiv.org.

[187] O. Choustova, "Pilot wave quantum model for the stock market", *E-print quant-ph/0109122 at http://arXiv.org*.

[188] A.F. Abouraddy, B.E.A. Saleh, A.V. Sergienko and M.C. Teich, "Quantum Holography", *Opt. Express* **9** (2001) 498. E-print quant-ph/0110075 at http://arXiv.org.

L.A. Lugiato, A. Gatti and E. Brambilla, "Quantum imaging", *J. Opt. B* **4** (2002) S176. E-print quant-ph/0203046.

[189] R. Somma, G. Ortiz, J.E. Gubernatis, E. Knill and R. Laflamme, "Simulating Physical Phenomena by Quantum Networks", *Phys. Rev. A* **65** (2002) 042323. E-print quant-ph/0108146 at http://arXiv.org. See also *quant-ph/0304063*.

[190] D.A. Meyer, "Quantum computing classical physics", *E-print quant-ph/0111069 at http://arXiv.org*.

D.A. Meyer, "Physical quantum algorithms", *Comput. Phys. Comm.* **146** (2002) 295.

[191] C. Brukner, J. Pan, C. Simon, G. Weihs and A. Zeilinger, "Probabilistic Instantaneous Quantum Computation", *Phys. Rev. A* **67** (2003) 034304. E-print quant-ph/0109022 at http://arXiv.org.

[192] M.V. Altaisky, "Quantum neural network", *E-print quant-ph/0107012 at http://arXiv.org*.

M.V. Altaisky, "On some algebraic problems arising in quantum mechanical description of biological systems", *E-print quant-ph/0110043 at http://arXiv.org*.

[193] A.A. Ezhov, "Role of interference and entanglement in quantum neural processing", *E-print quant-ph/0112082 at http://arXiv.org*.

[194] S.M. Hitchcock, "'Photosynthetic' Quantum Computers?", *E-print quant-ph/0108087 at http://arXiv.org*.

S.M. Hitchcock, "Time and Information, The Origins of 'Time' from Information Flow In Causal Networks and Complex Systems", *E-print quant-ph/0111025 at http://arXiv.org*.

[195] R.T. Cahill, "Smart Nanostructures and Synthetic Quantum Systems", *E-print quant-ph/0111026 at http://arXiv.org*.

R.T. Cahill, "Synthetic Quantum Systems", *Smart Materials and Structures* **11** (2002) 699. E-print physics/0209064 at http://arXiv.org.

R.T. Cahill and C.M. Klinger, "Bootstrap Universe from Self-Referential Noise", *Prog. Phys.* **2** (2005) 108. E-print gr-qc/9708013.

[196] P.R. Holland, *The Quantum Theory of Motion* (Cambridge University Press, 1995). First edition: (1993).

S. Teufel, K. Berndl, D. Dürr, S. Goldstein and N. Zanghì, "Locality and Causality in Hidden Variables Models of Quantum Theory", *Phys. Rev. A* **56** (1997) 1217. E-print quant-ph/9609005 at http://arXiv.org.

D. Dürr, S. Goldstein and N. Zanghì, "Bohmian Mechanics and the Meaning of the Wave Function", *E-print quant-ph/9512031 at http://arXiv.org*.

[197] E. Nelson, "Derivation of the Schrödinger Equation from Newtonian Mechanics", *Phys. Rev.* **150** (1966) 1079.

E. Nelson, *Dynamical Theories of Brownian Motion* (Princeton University Press, Princeton, 1967).

E. Nelson, *Quantum Fluctuations* (Princeton University Press, Princeton,1985).

[198] I. Stein, "The Nonstructure of Physical Reality: The Source of Special Relativity and Quantum Mechanics", *Physics Essays* **1** (1988) 155.

I. Stein, "Quantum Mechanics and the Special Theory of Relativity from a Random Walk", *Physics Essays* **3** (1990) 66.

[199] G.N. Ord and A.S. Deakin, "Random walks, continuum limits, and Schrödinger's equation", *Phys. Rev. A* **54** (1996) 3772.

G.N. Ord, "Fractal Space-Time and the Statistical Mechanics of Random Walks", *Chaos, Solitons and Fractals* **7** (1996) 821.

G.N. Ord and J.A. Gualtieri, "A realistic setting for Feynman paths", *Chaos, Solitons and Fractals* **14** (2002) 929.

[200] J.G. Gilson, "Stochastic Simulation of The Three Dimensional Quantum Vacuum", *E-print quant-ph/0112047 at http://arXiv.org*.

[201] G. Kaniadakis, "Statistical Origin of Quantum Mechanics", *Physica A* **307** (2002) 172. E-print quant-ph/0112049 at http://arXiv.org.

[202] M. Davidson, "The origin of the algebra of quantum operators in the stochastic formulation of quantum mechanics", *Letters in Mathematical Physics* **3** (1979) 367. E-print quant-ph/0112099 at http://arXiv.org.

M. Davidson, "A model for the stochastic origins of Schrödinger's equation", *J. Math. Phys.* **20** (1979) 1865. E-print quant-ph/0112157 at http://arXiv.org.

[203] L. Smolin, "Matrix models as hidden variables theories", *AIP Conf. Proc.* **607** (2002) 244. E-print hep-th/0201031 at http://arXiv.org.

[204] G. Groessing, "Quantum Cybernetics: A New Perspective for Nelson's Stochastic Theory, Nonlocality, and the Klein-Gordon Equation", *Phys. Lett. A* **296** (2002) 1. E-print quant-ph/0201035 at http://arXiv.org.

[205] C. Castro, J. Mahecha and B. Rodriguez, "Nonlinear QM as a fractal Brownian motion with complex diffusion constant", *E-print quant-ph/0202026 at http://arXiv.org*.

[206] R. Czopnik and P. Garbaczewski, "Quantum Potential and Random Phase-Space Dynamics", *Phys. Lett. A* **299** (2002) 447. E-print quant-ph/0203018 at http://arXiv.org.

H. Bergeron, "New derivation of quantum equations from classical stochastic arguments", *E-print quant-ph/0303153 at http://arXiv.org*.

[207] L. de Broglie, "Sur la thermodynamique du corpuscule isolé", *C.R. Acad. Sc.* **253** (1961) 1078.

L. de Broglie, "Quelques conséquences de la thermodynamique de la particule isolée", *C.R. Acad. Sc.* **255** (1962) 1052.

[208] L. de Broglie, *La thermodynamique de la particule isolée (thermodynamique cachée des particules)* (Gauthier-Villars, Paris, 1964).

[209] L. de Broglie, "Le mouvement brownien d'une particule dans son onde", *C.R. Acad. Sc. (série B)* **264** (1967) 1041.

[210] S. Lloyd, "The power of entangled quantum channels", *E-print quant-ph/0112034 at http://arXiv.org*.

[211] C. Simon, "Natural Entanglement in Bose-Einstein Condensates", *Phys. Rev. A* **66** (2002) 052323. E-print quant-ph/0110114 at http://arXiv.org.

A. Sørensen, L.-M. Duan, J.I. Cirac and P. Zoller, "Many-particle entanglement with Bose-Einstein condensates", *Nature* **409** (2001) 63.

[212] Ph. Ball, "Cool atoms make physics prize matter", *Nature* **413** (2001) 554.

"New State of Matter Revealed: Bose-Einstein Condensate", The 2001 Nobel Prize in Physics, http://www.nobelprize.org/nobel_prizes/physics/laureates/2001/press.html.

W. Ketterle, "Bose-Einstein condensation in dilute atomic gases: atomic physics meets condensed matter physics", *Physica B* **280** (2000) 11.

J.R. Anglin and W. Ketterle, "Bose-Einstein condensation of atomic gases", *Nature* **416** (2002) 211.

[213] R.W. Hill, C. Proust, L. Taillefer, P. Fournier and R.L. Greene, "Breakdown of Fermi-liquid theory in a copper-oxide superconductor", *Nature* **414** (2001) 711.

K.M. Lang, V. Madhavan, J.E. Hoffman, E.W. Hudson, H. Eisaki, S. Uchida and J.C. Davis, "Imaging the granular structure of high-$T_c$ superconductivity in underdoped $Bi_2Sr_2CaCu_2O_{8+\delta}$", *Nature* **415** (2002) 412.

[214] R.B. Laughlin and D. Pines, "The Theory of Everything", *Proc. Natl. Acad. Sci. USA* **97** (2000) 28.

[215] C. Monroe, "Quantum information processing with atoms and photons", *Nature* **416** (2002) 238.

[216] E.A. Donley, N.R. Claussen, S.L. Cornish, J.L. Roberts, E.A. Cornell and C.E. Wieman, "Dynamics of collapsing and exploding Bose-Einstein condensates", *Nature* **412** (2001) 295.

M. Greiner, O. Mandel, T.W. Hänsch and I. Bloch, "Collapse and revival of the matter wave field of a Bose-Einstein condensate", *Nature* **419** (2002) 51.

[217] B. Julsgaard, A. Kozhekin and E.S. Polzik, "Experimental long-lived entanglement of two macroscopic objects", *Nature* **413** (2001) 400.

[218] J.R. Friedman, V. Patel, W. Chen, S.K. Tolpygo and J.E. Lukens, "Quantum superposition of distinct macroscopic states", *Nature* **406** (2001) 43.

[219] S. Haroche, "Entanglement, Decoherence and the Quantum/Classical Boundary", *Physics Today*, July (1998) 36.

[220] S.L. Adler, "Why Decoherence has not Solved the Measurement Problem: A Response to P.W. Anderson", *Stud. Hist. Philos. Mod. Phys.* **34** (2003) 135. E-print quant-ph/0112095 at http://arXiv.org.

[221] G.J. Milburn, "Intrinsic decoherence in quantum mechanics", *Phys. Rev. A* **44** (1991) 5401.

H. Moya-Cessa, V. Buzek, M.S. Kim and P.L. Knight, "Intrinsic decoherence in the atom-field interaction", *Phys. Rev. A* **48** (1993) 3900.

[222] N.A. Zidan, M. Abdel-Aty and A.F. Obada, "Influence of intrinsic decoherence on entanglement degree in the atom-field coupling system", *Chaos, Solitons and Fractals* **13** (2002) 1421.

[223] D. Braun, F. Haake and W.T. Strunz, "Universality of Decoherence", *Phys. Rev. Lett.* **86** (2001) 2913.

W.T. Strunz, F. Haake and D. Braun, "Universality of decoherence in the macroworld", *E-print quant-ph/0204129 at http://arXiv.org*.

[224] Z.P. Karkuszewski, C. Jarzynski and W.H. Zurek, "Quantum Chaotic Environments, the Butterfly Effect, and Decoherence", *Phys. Rev. Lett.* **89** (2002) 170405. E-print quant-ph/0111002 at http://arXiv.org.

R.A. Jalabert and H.M. Pastawski, "Environment-Independent Decoherence Rate in Classically Chaotic Systems", *Phys. Rev. Lett.* **86** (2001) 2490. E-print cond-mat/0010094 at http://arXiv.org.

[225] W.H. Zurek, "Sub-Planck structure in phase space and its relevance for quantum decoherence", *Nature* **412** (2001) 712. E-print quant-ph/0201118 at http://arXiv.org.

A. Albrecht, "Quantum ripples in chaos", *Nature* **412** (2001) 687.

A. Jordan and M. Srednicki, "Sub-Planck Structure, Decoherence, and Many-Body Environments", *E-print quant-ph/0112139 at http://arXiv.org*.

[226] I. Prigogine, *La fin des certitudes. Temps, chaos et les lois de la nature* (Éditions Odile Jacob, Paris, 1996, 1998).

T.Y. Petrosky and I. Prigogine, "Laws and events: The dynamical basis of self-organisation", *Can. J. Phys.* **68** (1990) 670.

T. Petrosky and I. Prigogine, "Poincaré Resonances and the Extension of Classical Dynamics", *Chaos, Solitons and Fractals* **7** (1996) 441.

T. Petrosky and I. Prigogine, "Thermodynamic limit, Hilbert space and breaking of time symmetry", *Chaos, Solitons and Fractals* **11** (2000) 373.

I. Antoniou, I. Prigogine, V. Sadovnichii and S.A. Shkarin, "Time operator for diffusion", *Chaos, Solitons and Fractals* **11** (2000) 465.

[227] P. Cvitanović, "Chaotic field theory: a sketch", *Physica A* **288** (2000) 61. E-print nlin.CD/0001034 at http://arXiv.org.

[228] C. Beck, "Chaotic strings and standard model parameters", *Physica D* **171** (2002) 72. E-print hep-th/0105152 at http://arXiv.org.

[229] J.S. Nicolis, G. Nicolis and C. Nicolis, "Nonlinear dynamics and the two-slit delayed experiment", *Chaos, Solitons and Fractals* **12** (2001) 407.

[230] J. Ford, "Directions in Classical Chaos", in *Directions in Chaos*, Ed. Hao Bai-lin, Vol. 1 (World Scientific, Singapore, 1987), p. 1.

J. Ford, "Quantum Chaos, Is There Any?", in *Directions in Chaos*, Ed. Hao Bai-lin, Vol. 2 (World Scientific, Singapore, 1988), p. 128.

J. Ford, G. Mantica and G. H. Ristow, "The Arnol'd cat: Failure of the correspondence principle?", *Physica D* **50** (1991) 493.

J. Ford and M. Ilg, "Eigenfunctions, eigenvalues, and time evolution of finite, bounded, undriven, quantum systems are not chaotic", *Phys. Rev. A* **45** (1992) 6165.

J. Ford and G. Mantica, "Does quantum mechanics obey the correspondence principle? Is it complete?", *Am. J. Phys.* **60** (1992) 1086.

[231] B. Eckhardt, "Quantum Mechanics of Classically Non-integrable Systems", *Phys. Rep.* **163** (1988) 205.

[232] G. Casati, I. Guarneri and D. Shepelyansky, "Classical Chaos, Quantum Localization and Fluctuations: A Unified View", *Physica A* **163** (1990) 205.

[233] F. Haake, *Quantum Signatures of Chaos* (Springer-Verlag, Berlin, 1991).

[234] O. Bohigas, in *Chaos and Quantum Physics*, Ed. M.-J. Giannoni, A. Voros and J. Zinn-Justin (North-Holland, Amsterdam, 1991).

[235] M.V. Berry, "Some quantum-to-classical asymptotics", in *Chaos and Quantum Physics*, Ed. M.-J. Giannoni, A. Voros and J. Zinn-Justin (North-Holland, Amsterdam, 1991), p. 251.

M.V. Berry, "Quantum Chaology", *Proc. Roy. Soc. Lond. A* **413** (1987) 183.

M. Berry, "Quantum Chaology, Not Quantum Chaos", *Physica Scripta* **40** (1989) 335.

[236] B.V. Chirikov, "Time-dependent Quantum Systems", in *Chaos and Quantum Physics*, Ed. M.-J. Giannoni, A. Voros and J. Zinn-Justin (North-Holland, Amsterdam, 1991), p. 443.

B.V. Chirikov, F.M. Izrailev and D.L. Shepelyansky, "Quantum Chaos: Localization vs Ergodicity", *Physica D* **33** (1988) 77.

[237] M.V. Berry and J.P. Keating, *Proc. Roy. Soc. Lond. A* **437** (1992) 151.

[238] E.B. Bogomolny, "Semiclassical quantisation of multidimensional systems", *Nonlinearity* **5** (1992) 805.

E.B. Bogomolny, "On dynamical zeta function", *Chaos* **2** (1992) 5.

E. Bogomolny, "Quantum and Arithmetical Chaos", *E-print nlin.CD/0312061 at http://arXiv.org*.

[239] O. Bohigas, S. Tomsovic and D. Ullmo, *Phys. Rep.* **223** (1993) 45.

[240] R. Aurich, J. Bolte and F. Steiner, "Universal Signatures of Quantum Chaos", *Phys. Rev. Lett.* **73** (1994) 1356.

[241] I. Prigogine, "Dissipative processes in quantum theory", *Phys. Rep.* **219** (1992) 93.

[242] *Quantum Chaos - Quantum Measurement*, NATO AS1 Series C. Math. Phys. Sci. 357, Ed. P. Cvitanovich, I.C. Percival and A. Wirzba (Kluwer, Dordrecht, 1992).

*Quantum Chaos, Quantum Measurement*, Ed. H.A. Cerdeira, R. Ramaswamy, M.C. Gutzwiller and G. Casati (World Scientific, Singapore, 1991).

[243] K. Nakamura, *Quantum Chaos — A New Paradigm of Nonlinear Dynamics* (Cambridge University Press, 1993).

[244] *Quantum and Chaos: How Incompatible?* Proceedings of the 5th Yukawa International Seminar, Ed. K. Ikeda, *Progr. Theor. Phys. Suppl.* No. 116 (1994).

[245] *Quantum Chaos: Between Order and Disorder*, Eds. G. Casati and B. Chirikov (Cambridge University Press, 1995).

[246] G. Casati and B.V. Chirikov, "Comment on "Decoherence, Chaos, and the Second Law"", *Phys. Rev. Lett.* **75** (1995) 350.

[247] G. Casati and B.V. Chirikov, "Quantum chaos: unexpected complexity", *Physica D* **86** (1995) 220.

[248] G. Casati, "Quantum chaos", *Chaos* **6** (1996) 391.

[249] B.V. Chirikov, "Linear and Nonlinear Dynamical Chaos", *E-print chao-dyn/9705003 at http://arXiv.org*. Preprint Budker INP 95-100 (Novosibirsk, 1995).

B.V. Chirikov, "Pseudochaos in Statistical Physics", *E-print chao-dyn/9705004 at http://arXiv.org*. Preprint of the Budker Institute of Nuclear Physics, Budker INP 95-99 (Novosibirsk, 1995).

[250] B.V. Chirikov and F. Vivaldi, "An algorithmic view of pseudochaos", *Physica D* **129** (1999) 223.

[251] R.F. Fox and T.C. Elston, "Chaos and the quantum-classical correspondence in the kicked pendulum", *Phys. Rev. A* **49** (1994) 3683.

[252] M. Tegmark, "Does the Universe in Fact Contain Almost No Information?", *Found. Phys. Lett.* **9** (1996) 25.

[253] W.-M. Zhang and D. H. Feng, "Quantum nonintegrability in finite systems", *Phys. Rep.* **252** (1995) 1.

[254] T. Prozen, "Quantum surface of section method: eigenstates and unitary quantum Poincaré evolution", *Physica D* **91** (1996) 244.

[255] A.V. Andreev, O. Agam, B.D. Simons and B.L. Altshuler, "Quantum Chaos, Irreversible Classical Dynamics, and Random Matrix Theory", *Phys. Rev. Lett.* **76** (1996) 3947. E-print cond-mat/9601001 at http://arXiv.org.

O. Agam, A.V. Andreev and B.L. Altshuler, "Relations Between Quantum and Classical Spectral Determinants (Zeta-Functions)", *E-print cond-mat/9602131 at http://arXiv.org*.

A.V. Andreev, B.D. Simons, O. Agam and B.L. Altshuler, "Semiclassical Field Theory Approach to Quantum Chaos", *Nucl. Phys. B* **482** (1996) 536. E-print cond-mat/9605204 at http://arXiv.org.

[256] Y.V. Fyodorov, O.A. Chubykalo, F.M. Izrailev and G. Casati, "Wigner Random Banded Matrices with Sparse Structure: Local Spectral Density of States", *Phys. Rev. Lett.* **76** (1996) 1603.

[257] T. Kottos and D. Cohen, "Failure of random matrix theory to correctly describe quantum dynamics", *Phys. Rev. E* **64** (2001) 065202-R. E-print cond-mat/0105274 at http://arXiv.org.

[258] F. Borgonovi, "Localization in Discontinuous Quantum Systems", *Phys. Rev. Lett.* **80** (1998) 4653.

[259] G. Casati and T. Prosen, "The quantum mechanics of chaotic billiards", *Physica D* **131** (1999) 293.

G. Casati and T. Prosen, "Quantum localization and cantori in the stadium billiard", *Phys. Rev. E* **59** (1999) R2516.

G. Casati and T. Prosen, "Triangle Map: A Model of Quantum Chaos", *Phys. Rev. Lett.* **85** (2000) 4261. E-print nlin.CD/0009030 at http://arXiv.org.

T. Prosen and M. Žnidarič, "Anomalous diffusion and dynamical localization in a parabolic map", *Phys. Rev. Lett.* **87** (2001) 114101. E-print nlin.CD/0103001 at http://arXiv.org.

[260] G. Casati, G. Maspero and D.L. Shepelyansky, "Quantum fractal eigenstates", *Physica D* **131** (1999) 311.

G. Casati, G. Maspero and D.L. Shepelyansky, "Quantum Poincaré Recurrences", *Phys. Rev. Lett.* **82** (1999) 524; cond-mat/9807145.

B.V. Chirikov and D.L. Shepelyansky, "Asymptotic Statistics of Poincaré Recurrences in Hamiltonian Systems with Divided Phase Space", *Phys. Rev. Lett.* **82** (1999) 528.

G. Casati, "Quantum relaxation and Poincaré recurrences", *Physica A* **288** (2000) 49.

G. Benenti, G. Casati, I. Guarneri and M. Terraneo, "Quantum Fractal Fluctuations", *Phys. Rev. Lett.* **87** (2001) 014101. E-print cond-mat/0104450 at http://arXiv.org.

[261] D.A. Wisniacki, F. Borondo, E. Vergini and R.M. Benito, "Localization properties of groups of eigenstates in chaotic systems", *Phys. Rev. E* **63** (2001) 066220. E-print nlin.CD/0103031.

[262] V.Ya. Demikhovskii, F.M. Izrailev and A.I. Malyshev, "Manifestation of the Arnol'd Diffusion in Quantum Systems", *Phys. Rev. Lett.* **88** (2002) 154101. E-print quant-ph/0109147.

[263] R. Ketzmerick, L. Hufnagel, F. Steinbach and M. Weiss, "New Class of Eigenstates in Generic Hamiltonian Systems", *Phys. Rev. Lett.* **85** (2000) 1214. E-print nlin/0004005 at http://arXiv.org.

[264] G.A. Luna-Acosta, J.A. Méndez-Bermúdez and F.M. Izrailev, "Periodic Chaotic Billiards: Quantum-Classical Correspondence in Energy Space", *Phys. Rev. E* **64** (2001) 036206; cond-mat/0105108.

G.A. Luna-Acosta, J.A. Méndez-Bermúdez and F.M. Izrailev, "Quantum-classical correspondence for local density of states and eigenfunctions of a chaotic periodic billiard", *Phys. Lett. A* **274** (2000) 192. E-print nlin.CD/0002044 at http://arXiv.org.

[265] J. Wilkie and P. Brumer, "Quantum-Classical Correspondence via Liouville Dynamics: I. Integrable Systems and the Chaotic Spectral Decomposition", *Phys. Rev. A* **55** (1997) 27; chao-dyn/9608013.

J. Wilkie and P. Brumer, "Quantum-Classical Correspondence via Liouville Dynamics: II. Correspondence for Chaotic Hamiltonian Systems", *Phys. Rev. A* **55** (1997) 43; chao-dyn/9608014.

[266] G. Ordóñez, T. Petrosky, E.Karpov and I. Prigogine, "Explicit construction of a time superoperator for quantum unstable systems", *Chaos, Solitons and Fractals* **12** (2001) 2591.

T. Petrosky and V. Barsegov, "Quantum decoherence, Zeno process, and time symmetry breaking", *Phys. Rev. E* **65** (2002) 046102.

[267] J. Weber, F. Haake, P.A. Braun, C. Manderfeld and P. Šeba, "Resonances of the Frobenius-Perron Operator for a Hamiltonian Map with a Mixed Phase Space", *J. Phys. A* **34** (2001) 7195. E-print nlin.CD/0105047 at http://arXiv.org.

J. Weber, F. Haake and P. Šeba, "Frobenius-Perron Resonances for Maps with a Mixed Phase Space", *Phys. Rev. Lett.* **85** (2000) 3620. E-print nlin.CD/0001013 at http://arXiv.org.

C. Manderfeld, J. Weber and F. Haake, "Classical versus quantum time evolution of (quasi-) probability densities at limited phase-space resolution", *J. Phys. A* **34** (2001) 9893. E-print nlin.CD/0107020 at http://arXiv.org.

[268] A. Sugita and H. Aiba, "The second moment of Husimi distribution as a definition of complexity of quantum states", *Phys. Rev. E* **65** (2002) 036205. E-print nlin.CD/0106012 at http://arXiv.org.

A. Sugita, "Moments of generalised Husimi distribution and complexity of many-body quantum states", *J. Phys. A: Math. Gen.* **36** (2003) 9081. E-print nlin.CD/0112139 at http://arXiv.org.

[269] A. Jordan and M. Srednicki, "The Approach to Ergodicity in the Quantum Baker's Map", *E-print nlin.CD/0108024 at http://arXiv.org*.

[270] M.H. Partovi, "Entropic formulation of chaos for quantum dynamics", *Phys. Lett. A* **151** (1990) 389.

[271] A. Peres, "Instability of Quantum Motion of a Chaotic System," in *Chaos and Quantum Chaos: Proceedings of the Adriatico Research Conference on Quantum Chaos*, Eds. H. A. Cerdeira, R. Ramaswamy, M. C. Gutzwiller and G. Casati (World Scientific, Singapore, 1991), p. 73.

A. Peres, *Quantum Theory: Concepts and Methods* (Kluwer, Dordrecht, 1993).

[272] R. Schack, G.M. D'Ariano and C.M. Caves, "Hypersensitivity to perturbation in the quantum kicked top", *Phys. Rev. E* **50** (1994) 972.

R. Schack and C.M. Caves, "Information-theoretic characterization of quantum chaos", *Phys. Rev. E* **53** (1996) 3257.

[273] K. Shiokawa and B.L. Hu, "Decoherence, delocalization, and irreversibility in quantum chaotic systems", *Phys. Rev. E* **52** (1995) 2497.

[274] W.H. Zurek and J.P. Paz, "Decoherence, Chaos, and the Second Law", *Phys. Rev. Lett.* **72** (1994) 2508; *Phys. Rev. Lett.* **75** (1995) 351.

W.H. Zurek and J.P. Paz, "Quantum chaos: a decoherent definition", *Physica D* **83** (1995) 300.

W.H. Zurek, "Decoherence, chaos, quantum-classical correspondence, and the algorithmic arrow of time", *Phys. Scripta* **T76** (1998) 186; *Acta Phys. Polon. B* **29** (1998) 3689. E-print quant-ph/9802054 at http://arXiv.org.

S. Habib, K. Shizume and W.H. Zurek, "Decoherence, Chaos, and the Correspondence Principle", *Phys. Rev. Lett.* **80** (1998) 4361. E-print quant-ph/9803042 at http://arXiv.org.

Z.P.Karkuszewski, J. Zakrzewski and W.H. Zurek, "Breakdown of correspondence in chaotic systems: Ehrenfest versus localization times", *Phys. Rev. A* **65** (2002) 042113. E-print nlin.CD/0012048.

[275] R.A. Jalabert and H.M. Pastawski, "Environment-Independent Decoherence Rate in Classically Chaotic Systems", *Phys. Rev. Lett.* **86** (2001) 2490. E-print cond-mat/0010094 at http://arXiv.org.

F.M. Cucchietti, H.M. Pastawski and D.A. Wisniacki, "Decoherence as Decay of the Loschmidt Echo in a Lorentz Gas", *Phys. Rev. E* **65** (2002) 045206(R). E-print cond-mat/0102135 at http://arXiv.org.

D.A. Wisniacki, E.G. Vergini, H.M. Pastawski and F.M. Cucchietti, "Sensitivity to perturbations in a quantum chaotic billiard", *Phys. Rev. E* **65** (2002) 055206(R). E-print nlin.CD/0111051.

F.M. Cucchietti, C.H. Lewenkopf, E.R. Mucciolo, H.M. Pastawski and R.O. Vallejos, "Measuring the Lyapunov exponent using quantum mechanics", *Phys. Rev. E* **65** (2002) 046209; nlin.CD/0112015.

D.A. Wisniacki and D. Cohen, "Quantum irreversibility, perturbation independent decay, and the parametric theory of the local density of states", *Phys. Rev. E* **66** (2002) 046209. E-print quant-ph/0111125.

[276] V.I. Man'ko and R. Vilela Mendes, "Quantum sensitive dependence", *Phys. Lett. A* **300** (2002) 353. E-print quant-ph/0205148.

V.I. Man'ko and R. Vilela Mendes, "Lyapunov exponent in quantum mechanics. A phase-space approach", *Physica D* **145** (2000) 330. E-print quant-ph/0002049 at http://arXiv.org.

[277] Ph. Jacquod, P.G. Silvestrov and C.W.J. Beenakker, "Golden rule decay versus Lyapunov decay of the quantum Loschmidt echo", *Phys. Rev. E* **64** (2001) 055203(R). E-print nlin.CD/0107044.

P.G. Silvestrov, J. Tworzydlo and C.W.J. Beenakker, "Hypersensitivity to perturbations of quantum-chaotic wave-packet dynamics", *Phys. Rev. E* **67** (2003) 025204(R). E-print nlin.CD/0207002.

[278] T. Bhattacharya, S. Habib, K. Jacobs and K. Shizume, "The $\delta$-function-kicked rotor: Momentum diffusion and the quantum-classical boundary", *Phys. Rev. A* **65** (2002) 032115. E-print quant-ph/0105086 at http://arXiv.org.

[279] N.R. Cerruti and S. Tomsovic, "Sensitivity of Wave Field Evolution and Manifold Stability in Chaotic Systems", *Phys. Rev. Lett.* **88** (2002) 054103. E-print nlin.CD/0108016 at http://arXiv.org.

[280] G. Benenti and G. Casati, "Sensitivity of Quantum Motion for Classically Chaotic Systems", *Phys. Rev. E* **65** (2002) 066205. E-print quant-ph/0112060 at http://arXiv.org.

[281] E.B. Bogomolny, B. Georgeot, M.-J. Giannoni and C. Schmit, "Arithmetical Chaos", *Phys. Rep.* **291** (1997) 219. See also E-print nlin.CD/0312061, ref. [238].

[282] J.A. González, L.I. Reyes and L.E. Guerrero, "Exact solutions to chaotic and stochastic systems", *Chaos* **11** (2001) 1. E-print nlin.CD/0101049 at http://arXiv.org.

J.A. González, L.I. Reyes, J.J. Suárez, L.E. Guerrero and G. Gutiérrez, "Chaos-induced true randomness", *Physica A* **316** (2002) 259. Also: E-print nlin.CD/0202022 at http://arXiv.org.

[283] T. Kottos and U. Smilansky, "Quantum Chaos on Graphs", *Phys. Rev. Lett.* **79** (1997) 4794.

B. Gutkin and U. Smilansky, "Can One Hear the Shape of a Graph?", *J. Phys. A: Math. Gen.* **34** (2001) 6061. E-print nlin.CD/0105020.

[284] S.R. Jain, B. Gremaud and A. Khare, "Quantum Expression of Classical Chaos", *E-print nlin.CD/0107051 at http://arXiv.org*.

[285] R. Blümel, Yu. Dabaghian and R.V. Jensen, "One-dimensional quantum chaos: Explicitly solvable cases", *Phys. Rev. Lett.* **88** (2002) 044101. E-print quant-ph/0107092 at http://arXiv.org.

Yu. Dabaghian, R.V. Jensen and R. Blümel, "Integrability in 1D Quantum Chaos", *E-print quant-ph/0209042 at http://arXiv.org*.

[286] M. Azam, "Dynamics of the Subsets of Natural Numbers: A Nursery Rhyme of Chaos", *E-print nlin.CD/0108055 at http://arXiv.org*.

[287] O. Bohigas, P. Leboeuf and M.-J. Sanchez, "Spectral spacing correlations for chaotic and disordered systems", *Found. Phys.* **31** (2001) 489. E-print nlin.CD/0012049 at http://arXiv.org.

[288] S. Åberg, "Onset of Chaos in Rapidly Rotating Nuclei", *Phys. Rev. Lett.* **64** (1990) 3119.

[289] V.V. Flambaum and F.M. Izrailev, "Distribution of occupation numbers in finite Fermi-systems and role of interaction in chaos and thermalization", *Phys. Rev. E* **55** (1997) R13. E-print cond-mat/9610178 at http://arXiv.org.

V.V. Flambaum and F.M. Izrailev, "Statistical Theory of Finite Fermi-Systems Based on the Structure of Chaotic Eigenstates", *Phys. Rev. E* **56** (1997) 5144. E-print cond-mat/9707016.

V.V. Flambaum and F.M. Izrailev, "Entropy production and wave packet dynamics in the Fock space of closed chaotic many-body systems", *Phys. Rev. E* **64** (2001) 036220. E-print quant-ph/0103129.

F. Borgonovi, G.Celardo, F.M. Izrailev and G. Casati, "A semiquantal approach to finite systems of interacting particles", *Phys. Rev. Lett.* **88** (2002) 054101. E-print nlin.CD/0106036 at http://arXiv.org.

V.V. Flambaum and F.M. Izrailev, "Time dependence of occupation numbers and thermalization time in closed chaotic many-body systems", *E-print quant-ph/0108109 at http://arXiv.org*.

[290] I.V. Gornyi and A.D. Mirlin, "From quantum disorder to quantum chaos", *J. Low Temp. Phys.* **126** (2002) 1339; cond-mat/0107552.

[291] L.A. Caron, H. Jirari, H. Kröger, X.Q. Luo, G. Melkonyan and K.J.M. Moriarty, "Quantum chaos at finite temperature", *Phys. Lett. A* **288** (2001) 145. E-print quant-ph/0106130 at http://arXiv.org.

H. Kröger, "Quantum chaos via the quantum action", *E-print quant-ph/0212093 at http://arXiv.org*.

[292] V.E. Bunakov, I.B. Ivanov and R.B. Panin, "The critical perturbation parameter estimate for the transition from regularity to chaos in quantum systems", *E-print quant-ph/0105082 at http://arXiv.org*.

V.E. Bunakov and I.B. Ivanov, "The measure of chaoticity in stationary quantum systems", *J. Phys. A* **35** (2002) 1907.

I.B. Ivanov, "Unpredictability of wave function's evolution in nonintegrable quantum systems", *E-print quant-ph/0203019 at http://arXiv.org*.

[293] V.I. Kuvshinov and A.V. Kuzmin, "Heading towards chaos criterion for quantum field systems", *E-print nlin.CD/0111003 at http://arXiv.org*.

[294] K. Inoue, M. Ohya and I.V. Volovich, "Semiclassical Properties and Chaos Degree for the Quantum Baker's Map", *J. Math. Phys.* **43** (2002) 734. E-print quant-ph/0110135 at http://arXiv.org.

Y.S. Weinstein, S. Lloyd and C. Tsallis, "The Edge of Quantum Chaos", *Phys. Rev. Lett.* **89** (2002) 214101; cond-mat/0206039.

J. Emerson, Y.S. Weinstein, S. Lloyd and D. Cory, "Fidelity Decay as an Efficient Indicator of Quantum Chaos", *Phys. Rev. Lett.* **89** (2002) 284102. E-print quant-ph/0207099 at http://arXiv.org.

T.A. Brun, H.A. Carteret and A. Ambainis, "The quantum to classical transition for random walks", *Phys. Rev. Lett.* **91** (2003) 130602. E-print quant-ph/0208195 at http://arXiv.org.

S.M. Soskin, O.M. Yevtushenko and R. Mannella, "Drastic facilitation of the onset of global chaos in a periodically driven Hamiltonian system due to an extremum in the dependence of eigenfrequency on energy", *Phys. Rev. Lett.* **90** (2003) 174101; nlin.CD/0206002.

P. Bracken, P. Góra and A. Boyarsky, "A minimum principle for chaotic dynamical systems", *Physica D* **166** (2002) 63.

G. Blum, S.Snutzmann and U. Smilansky, "Nodal Domain Statistics: A Criterion for Quantum Chaos", *Phys. Rev. Lett.* **88** (2002) 114101.

[295] C.E. Shannon, "A mathematical theory of communication", *Bell Syst. Tech. J.* **27** (1948) 379, 623; reproduced at http://cm.bell-labs.com/cm/ms/what/shannonday/paper.html.

[296] N. Wiener, *Cybernetics or Control and Communication in the Animal and the Machine* (MIT Press and John Wiley & Sons, New York – London, 1961). First edition: 1948.

[297] L. Brillouin, *Science and Information Theory* (Academic Press, New York, 1956).

L. Brillouin, *Scientific Uncertainty and Information* (Academic Press, New York, 1964).

[298] J.S. Nicolis, *Dynamics of Hierarchical Systems: An Evolutionary Approach* (Springer-Verlag, Berlin, 1986).

[299] H. Haken, *Information and Self-Organisation: A Macroscopic Approach to Complex Systems* (Springer-Verlag, Berlin, 1988).

[300] W.H. Zurek, "Algorithmic randomness and physical entropy", *Phys. Rev. A* **40** (1989) 4731.

W.H. Zurek, "Thermodynamic cost of computation, algorithmic complexity and the information metric", *Nature* **341** (1989) 119.

W.H. Zurek, "Algorithmic randomness, physical entropy, measurement, and the Demon of Choice", *E-print quant-ph/9807007 at http://arXiv.org*.

[301] H. Ollivier and W.H. Zurek, "Quantum Discord: A Measure of the Quantumness of Correlations", *Phys. Rev. Lett.* **88** (2002) 017901.

W.H. Zurek, "Quantum Discord and Maxwell's Demons", *Phys. Rev. A* **67** (2003) 012320. E-print quant-ph/0301127 at http://arXiv.org.

[302] J. Bub, "Maxwell's demon and the thermodynamics of computation", *Studies in History and Philosophy of Modern Physics* **32** (2001) 569. E-print quant-ph/0203017 at http://arXiv.org.

C.H. Bennett, "Notes on Landauer's principle, Reversible Computation and Maxwell's Demon", *Studies in History and Philosophy of Modern Physics* **34** (2003) 501. E-print physics/0210005.

[303] R. Landauer, "Dissipation and noise immunity in computation and communication", *Nature* **335** (1988) 779.

[304] C.H. Bennett, "Universal computation and physical dynamics", *Physica D* **86** (1995) 268.

[305] *Complexity, Entropy and the Physics of Information*, Ed. W. Zurek (Addison-Wesley, Redwood City, 1990).

[306] B.B. Kadomtsev, "Dynamics and Information", *Uspekhi Fiz. Nauk* **164** (1994) 449.

[307] M.N. Izakov, "Self-organisation and information on planets and in ecosystems", *Uspekhi Fiz. Nauk* **167** (1997) 1087.

[308] C.D. Van Siclen, "Information entropy of complex structures", *Phys. Rev. E* **56** (1997) 5211.

[309] C.R. Shalizi and J.P. Crutchfield, "Computational Mechanics: Pattern and Prediction, Structure and Simplicity", *J. Stat. Phys.* **104** (2001) 816. E-print cond-mat/9907176 at http://arXiv.org.

[310] C. Adami and N.J. Cerf, "Physical complexity of symbolic sequences", *Physica D* **137** (2000) 62. E-print adap-org/9605002.

[311] J. Schmidhuber, "Algorithmic Theories of Everything", *E-print quant-ph/0011122 at http://arXiv.org*.

J. Schmidhuber, "A Computer Scientist's View of Life, the Universe, and Everything", in *Foundations of Computer Science: Potential - Theory - Cognition, Lecture Notes in Computer Science*, Ed. C. Freksa (Springer, Berlin, 1997) p. 201. E-print quant-ph/9904050.

J. Schmidhuber, "The New AI: General and Sound and Relevant for Physics", *E-print cs.AI/0302012 at http://arXiv.org*.

[312] G.J. Chaitin, "A Century of Controversy over the Foundations of Mathematics II", *E-print nlin.CD/0004007 at http://arXiv.org*.

G.J. Chaitin, "A Century of Controversy over the Foundations of Mathematics", *E-print chao-dyn/9909001 at http://arXiv.org*.

[313] A. Ostruszka, P. Pakonski, W. Slomczynski and K. Zyczkowski, "Dynamical entropy for systems with stochastic perturbation", *Phys. Rev. E* **62** (2000) 2018. E-print chao-dyn/9905041 at http://arXiv.org.

[314] G. Boffetta, M. Cencini, M. Falcioni and A. Vulpiani, "Predictability: a way to characterize complexity", *Phys. Rep.* **356** (2002) 367. E-print nlin.CD/0101029 at http://arXiv.org.

[315] C. Tsallis, "Entropic nonextensivity: a possible measure of complexity", *Chaos, Solitons and Fractals* **13** (2002) 371.

[316] J. Wheeler, in *Complexity, Entropy and the Physics of Information*, Ed. W. Zurek (Addison-Wesley, Redwood City, 1990), p. 3.

[317] C.A. Fuchs, "Quantum Foundations in the Light of Quantum Information", *E-print quant-ph/0106166 at http://arXiv.org*.

C.A. Fuchs, "Quantum Mechanics as Quantum Information (and only a little more)", *E-print quant-ph/0205039 at http://arXiv.org*.

[318] A. Galindo and M.A. Martin-Delgado, "Information and Computation: Classical and Quantum Aspects", *Rev. Mod. Phys.* **74** (2002) 347. E-print quant-ph/0112105 at http://arXiv.org.

[319] C. Brukner and A. Zeilinger, "Conceptual Inadequacy of the Shannon Information in Quantum Measurements", *Phys. Rev. A* **63** (2001) 022113. E-print quant-ph/0006087 at http://arXiv.org.

M.J.W. Hall, "Comment on "Conceptual Inadequacy of Shannon Information..." by C. Brukner and A. Zeilinger", *E-print quant-ph/0007116 at http://arXiv.org*.

C. Brukner and A. Zeilinger, "Quantum Measurement and Shannon Information, A Reply to M.J.W. Hall", *E-print quant-ph/0008091*.

C.G. Timpson, "On the Supposed Conceptual Inadequacy of the Shannon Information", *Stud. Hist. Phil. Mod. Phys.* **33** (2003) 441. E-print quant-ph/0112178 at http://arXiv.org.

C. Brukner and A. Zeilinger, "Information and fundamental elements of the structure of quantum theory", *E-print quant-ph/0212084 at http://arXiv.org*.

[320] S.E. Massen, Ch.C. Moustakidis and C.P. Panos, "Universal property of the information entropy in fermionic and bosonic systems", *E-print quant-ph/0201102 at http://arXiv.org*.

[321] I. Devetak and A.E. Staples, "The union of physics and information", *E-print quant-ph/0112166 at http://arXiv.org*.

[322] R. Alicki, "Information-theoretical meaning of quantum dynamical entropy", *Phys. Rev. A* **66** (2002) 052302. E-print quant-ph/0201012.

[323] M. Keyl, "Fundamentals of Quantum Information Theory", *Physics Reports* **369** (2002) 431. E-print quant-ph/0202122.

[324] I.V. Volovich, "Quantum Information in Space and Time", *E-print quant-ph/0108073 at http://arXiv.org*.

I.V. Volovich, "Towards Quantum Information Theory in Space and Time", *E-print quant-ph/0203030 at http://arXiv.org*.

I.V. Volovich, "Quantum Information and Spacetime Structure", *E-print quant-ph/0207050 at http://arXiv.org*.

[325] J.B. Hartle, "Spacetime Information", *Phys. Rev. D* **51** (1995) 1800. E-print gr-qc/9409005 at http://arXiv.org.

[326] T.D. Kieu, "Computing the Non-computable", *Contemporary Physics* **44** (2003) 51. E-print quant-ph/0203034 at http://arXiv.org.

T.D. Kieu, "Quantum Principles and Mathematical Computability", *E-print quant-ph/0205093 at http://arXiv.org*.

[327] R. Horodecki, M. Horodecki and P. Horodecki, "Balance of information in bipartite quantum-communication systems: Entanglement-energy analogy", *Phys. Rev. A* **63** (2001) 022310; quant-ph/0002021.

J. Oppenheim, K. Horodecki, M. Horodecki, P. Horodecki and R. Horodecki, "A new type of complementarity between quantum and classical information", *Phys. Rev. A* **68** (2003) 022307. E-print quant-ph/0207025 at http://arXiv.org.

M. Horodecki, K. Horodecki, P. Horodecki, R. Horodecki, J. Oppenheim, A. Sen and U. Sen, "Information as a resource in distributed quantum systems", *Phys. Rev. Lett.* **90** (2003) 100402. E-print quant-ph/0207168 at http://arXiv.org.

J. Oppenheim, M. Horodecki and R. Horodecki, "Are there phase transitions in information space?", *Phys. Rev. Lett.* **90** (2003) 010404. E-print quant-ph/0207169 at http://arXiv.org.

[328] M. Koashi and N. Imoto, "Quantum information is incompressible without errors", *Phys. Rev. Lett.* **89** (2002) 097904. E-print quant-ph/0203045 at http://arXiv.org.

[329] A. Valentini, "Subquantum Information and Computation", *Pramana - J. Phys.* **59** (2002) 269. E-print quant-ph/0203049 at http://arXiv.org.

[330] R.B. Griffiths, "The Nature and Location of Quantum Information", *Phys. Rev. A* **66** (2002) 012311. E-print quant-ph/0203058 at http://arXiv.org.

[331] R. Duvenhage, "The Nature of Information in Quantum Mechanics", *Found. Phys.* **32** (2002) 1399. E-print quant-ph/0203070 at http://arXiv.org.

[332] C.S. Calude, "Incompleteness, Complexity, Randomness and Beyond", *Minds and Machines: Journal for Artificial Intelligence, Philosophy and Cognitive Science* **12** (2002) 503. E-print quant-ph/0111118 at http://arXiv.org.

C. S. Calude and B. Pavlov, "Coins, Quantum Measurements, and Turing's Barrier", *Quantum Information Processing* **1** (2002) 107. E-print quant-ph/0112087 at http://arXiv.org.

[333] S.M. Hitchcock, "Is There a 'Conservation of Information Law' for the Universe?", *E-print gr-qc/0108010 at http://arXiv.org*.

[334] D.R. Brooks and E.O. Wiley, *Evolution as Entropy* (University of Chicago Press, Chicago, 1988). First edition: 1986.

[335] W. Dembski, *Intelligent Design. The bridge between science and technology* (InterVarsity Press, 1999).

W. Dembski, *The Design Inference: Eliminating Chance through Small Probabilities* (Cambridge University Press, 1998).

W. Dembski, *No Free Lunch: Why Specified Complexity Cannot be Purchased without Intelligence* (Rowman & Littlefield, 2001).

[336] S. Kauffman, *Investigations* (Oxford University Press, 2000).

[337] E. Schrödinger, *What is Life? The Physical Aspect of the Living Cell* (Cambridge University Press, 1944).

[338] G. Cattaneo, M.L. Dalla Chiara, R. Giuntini and R. Leporini, "An unsharp logic from quantum computation", *E-print quant-ph/0201013 at http://arXiv.org*.

[339] H. Bergson, *L'Évolution Créatrice* (Félix Alcan, Paris, 1907). English translation: *Creative Evolution* (Macmillan, London, 1911).

[340] M.C. Cross and P.C. Hohenberg, "Pattern formation outside of equilibrium", *Rev. Mod. Phys.* **65** (1993) 851.

[341] Y.L. Klimontovich, "S-Theorem", *Z. Phys. B* **66** (1987) 125.

Yu.L. Klimontovich, "Relative ordering criteria in open systems", *Uspekhi Fiz. Nauk* **166** (1996) 1231.

[342] P. Glansdorf and I. Prigogine, *Thermodynamic Theory of Structure, Stability and Fluctuations* (John Wiley & Sons, New York, 1971).

I. Prigogine, *Étude thermodynamique des phénomènes irreversibles* (Dunod, Paris et Desoer, Liège, 1947).

[343] A. Babloyantz, *Molecules, dynamics and life* (John Wiley & Sons, New York, 1986).

[344] J.R. Dorfman, P. Gaspard and T. Gilbert, "Entropy production of diffusion in spatially periodic deterministic systems", *Phys. Rev. E* **66** (2002) 026110. E-print nlin.CD/0203046 at http://arXiv.org.

[345] X. Calbet and R. Lopez-Ruiz, "Tendency to Maximum Complexity in a Non-Equilibrium Isolated System", *Phys. Rev. E* **63** (2001) 066116. E-print nlin.CD/0205022 at http://arXiv.org.

R. Lopez-Ruiz, "Complexity in Some Physical Systems", *Int. Journal of Bifurcation and Chaos* **11** (2001) 2669. E-print nlin.AO/0205021 at http://arXiv.org.

R. Lopez-Ruiz, H. Mancini and X. Calbet, "A Statistical Measure of Complexity", *Phys. Lett. A* **209** (1995) 321. E-print nlin.CD/0205033 at http://arXiv.org.

[346] G. Hu, Z. Zheng, L. Yang and W. Kang, "Thermodynamical second law in irreversible processes of chaotic few-body systems", *Phys. Rev. E* **64** (2001) 045102.

[347] S. Gupta and R.K.P. Zia, "Quantum Neural Networks", *E-print quant-ph/0201144 at http://arXiv.org*.

[348] F. Shafee, "Semiclassical Neural Network", *Stochastics and Dynamics* **7** (2007) 403. E-print quant-ph/0202015 at http://arXiv.org.

F. Shafee, "Entangled Quantum Networks", *E-print quant-ph/0203010 at http://arXiv.org*.

[349] E.C. Behrman, V. Chandrashekar, Z.Wang, C.K. Belur, J.E. Steck and S.R. Skinner, "A Quantum Neural Network Computes Entanglement", *E-print quant-ph/0202131 http://arXiv.org*.

[350] N.E. Mavromatos, A. Mershin and D.V. Nanopoulos, "QED-Cavity model of microtubules implies dissipationless energy transfer and biological quantum teleportation", *Int. J. Modern Physics B* **16** (2002) 3623. E-print quant-ph/0204021 at http://arXiv.org.

[351] T. Hogg and J.G. Chase, "Quantum Smart Matter", *E-print quant-ph/9611021 at http://arXiv.org*.

[352] Y. Benenson, T. Paz-Elizur, R. Adar, E. Keinan, Z. Livneh and E. Shapiro, "Programmable and autonomous computing machine made of biomolecules", *Nature* **414** (2001) 430.

[353] R.S. Braich, N. Chelyapov, C.Johnson, P.W.K. Rothemund and L. Adleman, "Solution of a 20-Variable 3-SAT Problem on a DNA Computer", *Science* **296** (2002) 499.

[354] L.H. Kauffman, "Biologic", *E-print quant-ph/0204007 at http://arXiv.org*.

[355] J. Wakeling and P. Bak, "Intelligent systems in the context of surrounding environment", *Phys. Rev. E* **64** (2001) 051920. E-print nlin.AO/0201046 at http://arXiv.org.

[356] K.E. Drexler, *Nanosystems: Molecular Machinery, Manufacturing, and Computation* (John Wiley & Sons, New York, 1992).

K.E. Drexler, *Engines of Creation: The Coming Era of Nanotechnology* (Anchor Press/Doubleday, New York, 1986).

[357] The detailed references to many nanotechnology sources can be found at the web sites of Foresight Institute, http://www.foresight.org/cms/resources/55 and Zyvex company, http://www.zyvex.com/nano/.

[358] Many popular reviews and professional discussions on nanotechnology can be found in *Scientific American*. See for example September 2001 issue:

G. Stix, "Little Big Science", *Scientific American* **285**, September (2001) 32.

K.E. Drexler, "Machine-Phase Nanotechnology", ibid., p. 74.

G.M. Whitesides, "The Once and Future Nanomachine", ibid., p. 78.

[359] A broad and ambitious research programme on nanotechnology, the National Nanotechnology Initiative, is established and generously supported by the US Government, see http://nano.gov.

[360] R. P. Feynman, "There's Plenty of Room at the Bottom", *Engineering and Science* (California Institute of Technology), February 1960, p. 22. Reproduced at http://www.zyvex.com/nanotech/feynman.html.

[361] L. Papp, S. Bumble, F. Friedler and L.T. Fan, "Characteristics of Molecular-biological Systems and Process-network Synthesis", *E-print physics/0203024 at http://arXiv.org*.

[362] C. Adami and J.P. Dowling, "Quantum Computation — The Ultimate Frontier", *E-print quant-ph/0202039 at http://arXiv.org*.

L. Fortnow, "Theory of Quantum Computing and Communication", *E-print quant-ph/0203074 at http://arXiv.org*.

J. Yepez, "Quantum computation for physical modeling", *Comput. Phys. Comm.* **146** (2002) 277.

M.A. Nielsen, "Rules for a Complex Quantum World", *Scientific American* **287**, November (2002) 66.

M. Mosca, R. Jozsa, A. Steane and A. Ekert, "Quantum-enhanced information processing", *Phil. Trans. Roy. Soc. Lond. A* **358** (2000) 261.

[363] E. Knill, R. Laflamme, H. Barnum, D. Dalvit, J. Dziarmaga, J. Gubernatis, L. Gurvits, G. Ortiz, L. Viola and W.H. Zurek, "Introduction to Quantum Information Processing", *E-print quant-ph/0207171 at http://arXiv.org*.

R. Laflamme, E. Knill, D.G. Cory, E.M. Fortunato, T. Havel, C. Miquel, R. Martinez, C. Negrevergne, G. Ortiz, M.A. Pravia, Y. Sharf, S. Sinha, R. Somma and L. Viola, "Introduction to NMR Quantum Information Processing", *E-print quant-ph/0207172 at http://arXiv.org*.

[364] D. Grundler, "Spintronics", *Physics World*, April (2002) 39.

D.D. Awschalom, M.E. Flatté and N. Samarth, "Spintronics", *Scientific American* **286**, June (2002) 66.

[365] G.J. Milburn, *Schrödinger's Machines: The Quantum Technology Reshaping Everyday Life* (W.H. Freeman & Co, 1997).

G.I. Milburn and P. Davies, *The Feynman Processor: Quantum Entanglement and the Computing Revolution* (Perseus Pr, 1999).

G.P. Dowling and G.J. Milburn, "Quantum Technology: The Second Quantum Revolution", *E-print quant-ph/0206091 at http://arXiv.org*.

[366] D. Zohar and I.N. Marshall, *Quantum Self: Human Nature and Consciousness Defined by the New Physics* (Quill, 1991).

D. Zohar and I. Marshall, *The Quantum Society: Mind, Physics, and a New Social Vision* (Quill, 1995).

[367] J. Satinover, *The Quantum Brain: The Search for Freedom and the Next Generation of Man* (John Wiley, 2001).

[368] R. Kurzweil, *The Age of Spiritual Machines: When Computers Exceed Human Intelligence* (Penguin USA, 2000). See also: "The Singularity: A Talk with Ray Kurzweil", http://www.edge.org/3rd_culture/kurzweil_singularity/kurzweil_singularity_index.html.

[369] P.W. Anderson, "Brainwashed by Feynman?", *Physics Today* **53**, February (2000) 11.

[370] S. Wolfram, *A New Kind of Science* (Wolfram Media, Champaign, 2002).

[371] "Physics and Computation of Complex Systems", Santa Fe Institute Research Theme, http://www.santafe.edu/research/themes/physics-and-computation-complex-systems/. See also ref. [309] for a conceptual example.

Dirk Helbing, Steven Bishop and Paul Lukowicz, "FuturICT", *E-print arXiv:1211.2313 at http://arXiv.org*.

[372] A. Connes, "La réalité mathématique archaïque", *La Recherche*, No. 332 (Juin 2000) 109.

[373] J.B. Barbour, *The End of Time: The Next Revolution in Physics* (Oxford University Press, 2001).

[374] B. Greene, *The Elegant Universe: Superstrings, Hidden Dimensions, and the Quest for the Ultimate Theory* (Vintage Books, 2000). First edition: 1999.

[375] C. Rovelli and L. Smolin, "Spin Networks and Quantum Gravity", *Phys. Rev. D* **52** (1995) 5743. E-print gr-qc/9505006 at http://arXiv.org.

F. Markopoulou and L. Smolin, "Causal evolution of spin networks", *Nucl. Phys. B* **508** (1997) 409. E-print gr-qc/9702025 at http://arXiv.org.

F. Markopoulou and L. Smolin, "Quantum geometry with intrinsic local causality", *Phys. Rev. D* **58** (1998) 084032. E-print gr-qc/9712067 at http://arXiv.org.

L. Smolin, "Strings as perturbations of evolving spin-networks", *Nucl. Phys. Proc. Suppl.* **88** (2000) 103. E-print hep-th/9801022 at http://arXiv.org.

L. Smolin, "A holographic formulation of quantum general relativity", *Phys. Rev. D* **61** (2000) 084007. E-print hep-th/9808191 at http://arXiv.org.

Y. Ling and L. Smolin, "Supersymmetric Spin Networks and Quantum Supergravity", *Phys. Rev. D* **61** (2000) 044008. E-print hep-th/9904016 at http://arXiv.org.

F. Markopoulou and L. Smolin, "Holography in a quantum spacetime", *E-print hep-th/9910146 at http://arXiv.org*.

L. Smolin, "The cubic matrix model and a duality between strings and loops", *E-print hep-th/0006137 at http://arXiv.org*.

Y. Ling and L. Smolin, "Holographic Formulation of Quantum Supergravity", *Phys. Rev. D* **63** (2001) 064010. E-print hep-th/0009018 at http://arXiv.org.

L. Smolin, "The exceptional Jordan algebra and the matrix string", *E-print hep-th/0104050 at http://arXiv.org*.

[376] C. Rovelli, "Covariant hamiltonian formalism for field theory: symplectic structure and Hamilton-Jacobi equation on the space G", *Lect. Notes Phys.* **633** (2003) 36. E-print gr-qc/0207043.

E. Hawkins, F. Markopoulou and H. Sahlmann, "Evolution in Quantum Causal Histories", *Class. Quant. Grav.* **20** (2003) 3839. E-print hep-th/0302111 at http://arXiv.org.

C.J. Isham,"A New Approach to Quantising Space-Time: I. Quantising on a General Category", *Adv. Theor. Math. Phys.* **7** (2003) 331. E-print gr-qc/0303060 at http://arXiv.org.

[377] R. Aldrovandi and A.L. Barbosa, "Spacetime algebraic skeleton", *E-print gr-qc/0207044 at http://arXiv.org*.

[378] *The New Physics*, Ed. P. Davies (Cambridge University Press, 1989).

[379] C.P. Snow, *The Two Cultures* (Cambridge University Press, 1993). First edition: 1959.

[380] J. Brockman, *The Third Culture: Beyond the Scientific Revolution* (Touchstone Books, 1996).

J. Brockman, "The Third Culture", *Edge* (2001), http://www.edge.org/3rd_culture/index.html.

J. Brockman, "The New Humanists", *Edge*, 31 December (2001), http://www.edge.org/conversation/the-new-humanists.

[381] L. de Broglie, "Nécessité de la liberté dans la recherche scientifique", In *Certitudes et Incertitudes de la Science* (Albin Michel, Paris, 1966). Original edition: 1962.

L. de Broglie, "Les idées qui me guident dans mes recherches", In *Certitudes et Incertitudes de la Science* (Albin Michel, Paris, 1966). Original edition: 1965.

[382] T. Kuhn, *The Structure of Scientific Revolutions* (Chicago University Press, 1970). First edition: 1962.

[383] G.Lochak, *Louis de Broglie. Un prince de la science* (Flammarion, Paris, 1992).

[384] J. Maddox, "Restoring good manners in research", *Nature* **376** (1995) 113.

J. Maddox, "The prevalent distrust of science", *Nature* **378** (1995) 435.

[385] J. Ziman, "Is science losing its objectivity?", *Nature* **382** (1996) 751.

[386] A. Sangalli, "They burn heretics, don't they?", *New Scientist*, 6 April (1996) 47.

[387] D. Braben, "The repressive regime of peer-review bureaucracy?", *Physics World*, November (1996) 13.

[388] Y. Farge, "Pour davantage d'éthique dans le monde de la recherche", *Science Tribune*, October 1996, http://www.tribunes.com/tribune/art96/farg.htm.

[389] A.A. Berezin, "Hampering the progress of science by peer review and by the 'selective' funding system", *Science Tribune*, December 1996, http://www.tribunes.com/tribune/art96/bere.htm.

[390] *The Flight from Science and Reason*, Eds. P.R. Gross, N. Levitt and M.W. Lewis (New York Academy of Sciences, 1996). Also: *Annals of the New York Academy of Sciences*, vol. **775**.

[391] J. Bricmont, "Le relativisme alimente le courant irrationnel", *La Recherche*, No. 298, mai (1997) 82.

[392] A. Sokal and J. Bricmont, *Impostures intellectuelles* (Odile Jacob, Paris, 1997).

[393] P.A. Lawrence and M. Locke, "A man for our season", *Nature* **386** (1997) 757.

[394] J. de Rosnay, "Du pasteur au passeur", *Le Monde de l'Education, de la Culture et de la Formation*, No. 245, fevrier (1997) 20.

[395] O. Postel-Vinay, "La recherche menacée d'asphyxie", *Le Monde de l'Education, de la Culture et de la Formation*, No. 245, fevrier (1997) 47.

O. Postel-Vinay, "La défaite de la science française", *La Recherche*, No. 352, avril (2002) 60.

O. Postel-Vinay, "L'avenir de la science française", *La Recherche*, No. 353, mai (2002) 66.

*Sciences à l'école: De désamour en désaffection*, Le Monde de l'Education, Octobre (2002), 25-42.

[396] C. Wennerås and A. Wold, "Nepotism and sexism in peer-review", *Nature* **387** (1997) 341.

[397] S. Fuller, *The Governance of Science: Ideology and the Future of the Open Society* (Open Society Press, Buckingham, 1999).

S. Fuller, *Thomas Kuhn: A Philosophical History for Our Times* (University of Chicago Press, 2000).

[398] M. Gibbons, "Science's new social contract with society", *Nature* **402** Supp (1999) C81.

[399] H. Nowotny, P. Scott and M. Gibbons, *Rethinking Science: Knowledge and the Public in an Age of Uncertainty* (Polity Press, 2001).

[400] D.S. Greenberg, *Science, Money, and Politics: Political Triumph and Ethical Erosion* (University of Chicago Press, 2001).

[401] C. Chiesa and L. Pacifico, "Patronage lies at the heart of Italy's academic problems", *Nature* **414** (2001) 581.

[402] M. Lopez-Corredoira, "What is research?", *E-print physics/0201012 at http://arXiv.org*. Ciencia Digital 8 (2000).

*Against the Tide. A Critical Review by Scientists of How Physics and Astronomy Get Done*, Eds. M. Lopez-Corredoira and C. Castro Perelman (Universal Publishers, Boca Raton, 2008).

M. Lopez-Corredoira, *The Twilight of the Scientific Age* (Brown Walker Press, Boca Raton, 2013). See also: *E-print arXiv:1305.4144 at http://arXiv.org*.

[403] P.A. Lawrence, "Rank injustice", *Nature* **415** (2002) 835.

P.A. Lawrence, "The politics of publication", *Nature* **422** (2003) 259.

[404] G. Brumfiel, "Misconduct in physics: Time to wise up?", *Nature* **418** (2002) 120. "Reputations at risk", Editorial, Physics World, August (2002).

E. Check, "Sitting in judgement", *Nature* **419** (2002) 332.

D. Adam and J. Knight, "Publish, and be damned ...", *Nature* **419** (2002) 772.

[405] K. Svozil, "Censorship and the peer review system", *E-print physics/0208046 at http://arXiv.org*.

[406] B.C. Chauhan, "Science in Trauma", *E-print physics/0210088 at http://arXiv.org*.

[407] M. Shermer, "The Physicist and the Abalone Diver", *Scientific American* **287**, October (2002) 42.

[408] M. Shermer, "The Shamans of Scientism", *Scientific American* **286**, June (2002) 35.

[409] S. Weinberg, *Facing Up: Science and Its Cultural Adversaries* (Harvard University Press, 2001).

[410] S. Nagel, "Physics in Crisis", *FermiNews* **25**, No. 14, August 30 (2002), http://www.fnal.gov/pub/ferminews/ferminews02-08-30/p1.html. Also: *Physics Today* **55**, September (2000) 55.

P. Rodgers, "Hanging together", *Physics World*, October (2002), Editorial.

P. Woit, *Not Even Wrong: The Failure of String Theory and the Search for Unity in Physical Law* (Basic Books, New York, 2006).

H. Ghassib, "Is Physics Sick? [In Praise of Classical Physics]", *E-print arXiv:1305.4144 at http://arXiv.org*.

P. Wells, "Perimeter Institute and the crisis in modern physics", *Maclean's*, 5 September (2013), http://www2.macleans.ca/2013/09/05/perimeter-institute-and-the-crisis-in-modern-physics/. See also http://www.math.columbia.edu/~woit/wordpress/?p=6238.

A. Unzicker, *Bankrupting Physics: How Today's Top Scientists are Gambling Away Their Credibility* (Palgrave Macmillan, New York, 2013).

[411] L. Smolin, "Response to 'The New Humanists' by J. Brockman", *Edge*, 15 April (2002), http://www.edge.org/discourse/humanists/smolin.humanists.html.

L. Smolin, *The Trouble With Physics: The Rise of String Theory, The Fall of a Science, and What Comes Next* (Mariner Books, New York, 2007).

[412] See e. g. the official European Commission's research website, http://ec.europa.eu/research/participants/portal/page/fp7_documentation.

Cf. also: "Frameworks can be too rigid", *Nature* **420** (2002) 107.

[413] R.B. Griffiths, *J. Stat. Phys.* **36** (1984) 219.

R.B. Griffiths, "Review of R. Omnes, The Interpretation of Quantum Mechanics", *E-print quant-ph/9505008 at http://arXiv.org*.

R.B. Griffiths, "Consistent Quantum Reasoning", *E-print quant-ph/9505009 at http://arXiv.org*.

R.B. Griffiths, "Consistent Histories and Quantum Reasoning", *E-print quant-ph/9606004 at http://arXiv.org*.

R.B. Griffiths, "Choice of Consistent Family, and Quantum Incompatibility", *Phys. Rev. A* **57** (1998) 1604. E-print quant-ph/9708028 at http://arXiv.org.

R.B. Griffiths, "Probabilities and Quantum Reality: Are There Correlata?", *E-print quant-ph/0209116 at http://arXiv.org*.

[414] M. Gell-Mann and J.B. Hartle, "Classical Equations for Quantum Systems", *Phys. Rev. D* **47** (1993) 3345. E-print gr-qc/9210010 at http://arXiv.org.

M. Gell-Mann and J.B. Hartle, "Equivalent Sets of Histories and Multiple Quasiclassical Realms", *E-print gr-qc/9404013 at http://arXiv.org*.

M. Gell-Mann and J.B. Hartle, "Strong Decoherence", *E-print gr-qc/9509054 at http://arXiv.org*.

J.B. Hartle, "Bohmian Histories and Decoherent Histories", *E-print quant-ph/0209104 at http://arXiv.org*.

[415] R. Omnès, *The Interpretation of Quantum Mechanics* (Princeton University Press, 1994).

R. Omnes, "Decoherence: An Irreversible Process", *E-print quant-ph/0106006 at http://arXiv.org*.

R. Omnes, "Decoherence, irreversibility and the selection by decoherence of quantum states with definite probabilities", *Phys. Rev. A* **65** (2002) 052119. E-print quant-ph/0304100 at http://arXiv.org.

[416] A. Shimony, "Conceptual foundations of quantum mechanics", in *The New Physics*, Ed. P. Davies (Cambridge University Press, 1989).

[417] M. Tegmark, "The Interpretation of Quantum Mechanics: Many Worlds or Many Words?", *Fortschr. Phys.* **46** (1998) 855. E-print quant-ph/9709032 at http://arXiv.org.

[418] M. Tegmark and J.A. Wheeler, "100 Years of the Quantum", *Scientific American*, February (2001) 68. E-print quant-ph/0101077 at http://arXiv.org.

P. Ball, "Physics: Quantum quest", *Nature* **501** (2013) 154.

[419] H.D. Zeh, "The Problem of Conscious Observation in Quantum Mechanical Description", *Found. Phys. Lett.* **13** (2000) 221. E-print quant-ph/9908084 at http://arXiv.org.

[420] H.D. Zeh, "The Wave Function: It or Bit?", *E-print quant-ph/0204088 at http://arXiv.org*. Contribution to the Ultimate Reality conference and book, see ref. [441].

[421] D. Aerts, "Foundations of quantum physics: a general realistic and operational approach", *Int. J. Theor. Phys.* **38** (1999) 289. E-print quant-ph/0105109 at http://arXiv.org.

D. Aerts, "The hidden measurement formalism: what can be explained and where quantum paradoxes remain", *Int. J. Theor. Phys.* **37** (1998) 291. E-print quant-ph/0105126 at http://arXiv.org.

D. Aerts, "Quantum Structures: An Attempt to Explain the Origin of their Appearance in Nature", *Int. J. Theor. Phys.* **34** (1995) 1165. E-print quant-ph/0111071 at http://arXiv.org.

B. Coecke, D.J. Moore and S. Smets, "Logic of Dynamics and Dynamics of Logic; Some Paradigm Examples", *E-print math.LO/0106059 at http://arXiv.org*.

L. Gabora and D. Aerts, "Contextualizing Concepts using a Mathematical Generalization of the Quantum Formalism", *Journal of Experimental and Theoretical Artificial Intelligence* **14** (2002) 327. E-print quant-ph/0205161 at http://arXiv.org.

D. Aerts, "Being and Change: Foundations of a Realistic Operational Formalism", *E-print quant-ph/0205164 at http://arXiv.org*.

D. Aerts, "Reality and Probability: Introducing a New Type of Probability Calculus", *E-print quant-ph/0205165 at http://arXiv.org*.

S. Abramsky and B. Coecke, "Physical Traces: Quantum vs. Classical Information Processing", *E-print cs/0207057 at http://arXiv.org*.

[422] A. Khrennikov, "Classical and quantum mechanics on information spaces with applications to cognitive, psychological, social and anomalous phenomena", *Found. Phys.* **29** (1999) 1065. E-print quant-ph/0003016 at http://arXiv.org.

A. Khrennikov, "'Quantum probabilities' as context depending probabilities", *E-print quant-ph/0106073 at http://arXiv.org*.

A. Khrennikov, "Interference of probabilities and number field structure of quantum models", *E-print quant-ph/0107135 at http://arXiv.org*.

A. Khrennikov, "Quantum-like formalism for cognitive measurements", *E-print quant-ph/0111006 at http://arXiv.org*.

A. Khrennikov and J. Summhammer, "Dialogue on Classical and Quantum between mathematician and experimenter", *E-print quant-ph/0111130 at http://arXiv.org*.

A. Khrennikov, "Växjö Interpretation of Quantum Mechanics", *E-print quant-ph/0202107 at http://arXiv.org*.

A. Khrennikov and Ya. Volovich, "Discrete Time Leads to Quantum-Like Interference of Deterministic Particles", *E-print quant-ph/0203009 at http://arXiv.org*.

A. Khrennikov and E. Loubenets, "On relations between probabilities under quantum and classical measurements", *Found. Phys.* **34** (2004) 689. E-print quant-ph/0204001 at http://arXiv.org.

A. Khrennikov, "Fundamental Principle for Quantum Theory", *E-print quant-ph/0204008 at http://arXiv.org*.

C.A. Fuchs, "The Anti-Växjö Interpretation of Quantum Mechanics", *E-print quant-ph/0204146 at http://arXiv.org*.

A. Khrennikov, "On the cognitive experiments to test quantum-like behaviour of mind", *BioSystems* **84** (2006) 225; quant-ph/0205092.

A. Bulinski and A. Khrennikov, "Nonclassical Total Probability Formula and Quantum Interference of Probabilities", *E-print quant-ph/0206030 at http://arXiv.org*.

A. Khrennikov, "Reconstruction of quantum theory on the basis of the formula of total probability", *AIP Conf. Proc.* **750** (2004) 187. E-print quant-ph/0302194 at http://arXiv.org.

[423] *Foundations of Probability and Physics*, Proceedings of the Conference Växjö-2000 (Sweden), Ed. A. Khrennikov (WSP, Singapore, 2001). See also http://lnu.se/forskargrupper/icmm/conferences.

A. Khrennikov, "Foundations of Probability and Physics, Round Table", *E-print quant-ph/0101085 at http://arXiv.org*.

A. Khrennikov, "Quantum theory: Reconsideration of foundations", *E-print quant-ph/0302065 at http://arXiv.org*.

[424] K.A. Kirkpatrick, "'Quantal' behaviour in classical probability", *Found. Phys. Lett.* **16** (2003) 199. E-print quant-ph/0106072.

[425] C. Anastopoulos, "Quantum processes on phase space", *Annals Phys.* **303** (2003) 275. E-print quant-ph/0205132 at http://arXiv.org.

C. Anastopoulos, "Quantum theory without Hilbert spaces", *Found. Phys.* **31** (2001) 1545. E-print quant-ph/0008126 at http://arXiv.org.

C. Anastopoulos, "Quantum vs stochastic processes and the role of complex numbers", Int. J. Theor. Phys. 42 (2003) 1229. E-print gr-qc/0208031 at http://arXiv.org.

[426] V.E. Shemi-zadeh, "To Physical Foundation of Quantum Mechanics", *E-print quant-ph/0206195 at http://arXiv.org*.

[427] T. Dass, "Towards an Autonomous Formalism for Quantum Mechanics", *E-print quant-ph/0207104 at http://arXiv.org*.

[428] V.P. Belavkin, "Quantum Causality, Stochastics, Trajectories and Information", *Rep. Prog. Phys.* **65** (2002) 353; quant-ph/0208087.

M.B. Mensky, "Quantum continuous measurements, dynamical role of information and restricted path integrals", *E-print quant-ph/0212112 at http://arXiv.org*.

[429] J.B. Hartle, "The State of the Universe", *E-print gr-qc/0209046*.

J.B. Hartle, "Theories of Everything and Hawking's Wave Function of the Universe", *E-print gr-qc/0209047 at http://arXiv.org*.

[430] G. 't Hooft, "Determinism Beneath Quantum Mechanics", *E-print quant-ph/0212095 at http://arXiv.org*.

G. 't Hooft, "How Does God Play Dice? (Pre-)Determinism at the Planck Scale", *E-print hep-th/0104219 at http://arXiv.org*.

[431] A. Bohm, M. Loewe and B. Van de Ven, "Time Asymmetric Quantum Theory - I Modifying an Axiom of Quantum Physics", *Fortsch. Phys.* **51** (2003) 551. E-print quant-ph/0212130 at http://arXiv.org.

[432] I.V. Volovich, "Seven Principles of Quantum Mechanics", *E-print quant-ph/0212126 at http://arXiv.org*.

[433] P. Grangier, "Contextual objectivity: a realistic interpretation of quantum mechanics", *Eur. J. Phys.* **23** (2002) 331. E-print quant-ph/0012122 at http://arXiv.org.

P. Grangier, "Reconstructing the formalism of quantum mechanics in the contextual objectivity point of view", *E-print quant-ph/0111154 at http://arXiv.org*.

P. Grangier, "Contextual objectivity and quantum holism", *E-print quant-ph/0301001 at http://arXiv.org*.

[434] A.M. Steinberg, "Speakable and Unspeakable, Past and Future", *E-print quant-ph/0302003 at http://arXiv.org*.

[435] M. Paty, "Are quantum systems physical objects with physical properties?", *Eur. J. Phys.* **20** (1999) 373.

[436] F. Laloë, "Do we really understand quantum mechanics? Strange correlations, paradoxes, and theorems", *Am. J. Phys.* **69** (2001) 655. E-print quant-ph/0209123 at http://arXiv.org.

[437] K. Hess and W. Philipp, "A possible loophole in the theorem of Bell", *Proc. Natl. Acad. Sci. USA* **98** (2001) 14224.

K. Hess and W. Philipp, "Bell's theorem and the problem of decidability between the views of Einstein and Bohr", *Proc. Natl. Acad. Sci. USA* **98** (2001) 14228.

W.C. Myrvold, "A Loophole in Bell's Theorem? Parameter Dependence in the Hess-Philipp Model", *E-print quant-ph/0205032 at http://arXiv.org*.

R.D. Gill, G. Weihs, A. Zeilinger and M. Zukowski, "No time loophole in Bell's theorem; the Hess-Philipp model is non-local", *Proc. Natl. Acad. Sci. USA* **99** (2002) 14632. E-print quant-ph/0208187.

[438] D. Collins, N. Gisin, N. Linden, S. Massar and S. Popescu, "Bell Inequalities for Arbitrary High-Dimensional Systems", *Phys. Rev. Lett.* **88** (2002) 040404.

A.G. Valdenebro, "Proof of Bell's Inequalities for Generalised Hidden Variables Models", *E-print quant-ph/0211188 at http://arXiv.org*.

G. Adenier, "A Refutation of Bell's Theorem", *E-print quant-ph/0006014 at http://arXiv.org*.

A. Khrennikov and I. Volovich, "Local Realism, Contextualism and Loopholes in Bell's Experiments", *E-print quant-ph/0212127*.

M. Clover, "Quantum Mechanics and Reality are Really Local", *E-print quant-ph/0312198 at http://arXiv.org*.

[439] A. Kent, "Locality and causality revisited", in *Non-locality and Modality*, NATO Science Series II **64**, Eds. T. Placek and J. Butterfield (Kluwer Academic Publishers, Dordrecht, 2002), p. 163. E-print quant-ph/0202064 at http://arXiv.org.

A. Kent, "Causal Quantum Theory and the Collapse Locality Loophole", *Phys. Rev. A* **72** (2005) 012107. E-print quant-ph/0204104.

A. Kent, "Nonlinearity without Superluminality", *Phys. Rev. A* **72** (2005) 012108. E-print quant-ph/0204106 at http://arXiv.org.

[440] A. Khrennikov, "Event-independence, collective-independence, EPR-Bohm experiment and and incompleteness of quantum mechanics", *E-print quant-ph/0205081 at http://arXiv.org*.

[441] *The Ultimate Reality*, research initiative of the John Templeton Foundation, http://www.metanexus.net/ultimate_reality/info.htm, including the conference in honor of J.A. Wheeler, http://www.metanexus.net/archive/ultimate_reality/agenda.htm and the book of conference talks, *Science and Ultimate Reality: Quantum Theory , Cosmology and Complexity*, Eds. J.D. Barrow, P.C.W. Davies and C.L. Harper, Jr. (Cambridge University Press, 2003).

[442] H.P. Stapp, "Values and the Quantum Conception of Man", *E-print quant-ph/9506035 at http://arXiv.org*.

H.P. Stapp, "Chance, Choice, and Consciousness: The Role of Mind in the Quantum Brain", *E-print quant-ph/9511029 at http://arXiv.org*.

H.P. Stapp, "Decoherence, Quantum Zeno Effect, and the Efficacy of Mental Effort", *E-print quant-ph/0003065 at http://arXiv.org*.

H.P. Stapp, "Quantum theory and the role of mind in nature", *E-print quant-ph/0103043 at http://arXiv.org*.

[443] M. Dugic, M.M. Cirkovic and D.R. Rakovic, "On a Possible Physical Metatheory of Consciousness", *Open Systems & Information Dynamics* **9** (2002) 153. E-print quant-ph/0212128 at http://arXiv.org.

[444] R. Nakhmanson, "Physical interpretation of quantum mechanics", *Phys.-Usp.* **44** (2001) 421. E-print physics/0111109.

R. Nakhmanson, "Quantum mechanics as a sociology of matter", *E-print quant-ph/0303162 at http://arXiv.org*.

[445] D.V. Juriev, "Quantum String Field Psychophysics of Nastroenie", *E-print physics/0105066 at http://arXiv.org*.

[446] R.E. Zimmermann, "Beyond the Physics of Logic: Aspects of Transcendental Materialism or URAM in a Modern View", *E-print physics/0105094 at http://arXiv.org at http://arXiv.org*.

[447] K.K. Yee, "Introduction to Spin and Lattice Models in the Social Sciences", *E-print nlin.AO/0106028 at http://arXiv.org*.

[448] M.K. Samal, "Speculations on a Unified Theory of Matter and Mind", *E-print physics/0111035 at http://arXiv.org*.

[449] I. Vakarchuk, "Schrödinger's Cat and the Problem of Two Cultures", *E-print quant-ph/0205147 at http://arXiv.org*.

[450] S. Prvanovic, "Mona Lisa — ineffable smile of quantum mechanics", *E-print physics/0302089 at http://arXiv.org*.

[451] R. Descartes, *Discours de la Méthode. Plus: La Dioptrique, les Météores et la Géometrie* (Fayard, Paris, 1987). First edition: (1637).